\documentclass[a4paper,11pt]{article}
\usepackage[left=2cm,right=2cm]{geometry}
\usepackage[graphicx]{realboxes}
\usepackage{epstopdf,epsfig}
\pdfoutput=1
\usepackage{rotating}
\usepackage{lineno}
\usepackage{soul}
\usepackage{verbatim}
\usepackage{authblk}
\usepackage{natbib}
\usepackage{amsbsy,amsmath,amssymb}
\usepackage{pbox}
\usepackage{pstool}
\usepackage{subfigure}
\usepackage{makecell}
\usepackage{mathrsfs}
\usepackage[hyperindex,breaklinks,bookmarks=false]{hyperref}
\usepackage[hyphenbreaks]{breakurl}
\usepackage{setspace}
\usepackage{floatrow}
\newfloatcommand{capbtabbox}{table}[][\FBwidth]
\newcommand{\R}{\mathcal{R}}

\title{\textbf{A review of turbulent skin-friction drag reduction by near-wall transverse forcing}}

\author[$\dagger$]{Pierre Ricco}
\author[$\star$]{Martin Skote}
\author[$\ddag$]{Michael A. Leschziner}

\affil[$\dagger$]{Department of Mechanical Engineering, University of Sheffield, United Kingdom}
\affil[$\star$]{School of Aerospace, Transport and Manufacturing, Cranfield University, United Kingdom}
\affil[$\ddag$]{Department of Aeronautics, Imperial College London, United Kingdom}

\date{Published in {\it Progress in Aerospace Sciences} [Vol. 123, (100713), 2021]\\ DOI: \href{https://doi.org/10.1016/j.paerosci.2021.100713}{10.1016/j.paerosci.2021.100713}}

\begin{document}

\maketitle

\begin{abstract}
\noindent
\textit{
The quest for reductions in fuel consumption and CO$_2$ emissions in transport has been a powerful driving force for scientific research into methods that might underpin drag-reducing technologies for a variety of vehicular transport on roads, by rail, in the air, and on or in the water. In civil aviation, skin-friction drag accounts for around 50\% of the total drag in cruise conditions, thus being a preferential target for research.  With laminar conditions excluded, skin friction is intimately linked to the turbulence physics in the fluid layer closest to the skin. Hence, research into drag reduction has focused on methods to depress the turbulence activity near the surface.  The most effective method of doing so is to exercise active control on the near-wall layer by subjecting the drag-producing flow in this layer to an unsteady and/or spatially varying cross-flow component, either by the action of transverse wall oscillations, by embedding rotating discs into the surface or by plasma-producing electrodes that accelerate the near-wall fluid in the transverse direction. In ideal conditions, drag-reduction margins of order of 50\% can thereby be achieved. The present article provides a near-exhaustive review of research into the response of turbulent near-wall layers to the imposition of unsteady and wavy transverse motion. The review encompasses experiments, simulation, analysis and modelling, mainly in channel flows and boundary layers. It covers issues such as the drag-reduction margin in a variety of actuation scenarios and for a wide range of actuation parameters, the underlying physical phenomena that contribute to the interpretation of the origin of the drag reduction, the dependence of the drag reduction on the Reynolds number, passive control methods that are inspired by active control, and a forward look towards possible future research and practical realizations. The authors hope that this review, by far the most extensive of its kind for this subject, will be judged as a useful foundation for future research targeting friction-drag reduction. 
}
\end{abstract}

\newpage

\tableofcontents

\section{Introduction}

Fuel consumption is amongst the most important issues exercising civil-aviation operators and is often critical to aircraft-procurement decisions. From a societal and environmental perspective, the principal concern about fuel consumption in aviation revolves around CO$_2$ emissions and its impact on global warming, as amply demonstrated by the many reports issued by the  Intergovernmental Panel on Climate Change (IPCC, e.g., Aviation and the Global Atmosphere, \cite{penner-etal-1999}). In cruise conditions, fuel consumption is almost linearly dependent on the aerodynamic drag, and this correlation explains the quest for innovative, technologically realizable approaches to flow control that achieve a reduction in drag. Indeed, drag reduction is regarded by many as the holy grail of aerodynamics in civil aviation, especially in the context of long-haul flights utilizing large aircraft.

There are numerous statistics on the future sectorial distribution of CO$_2$ emissions, and figures vary from one set of statistics to another, depending on the different scenarios of how the various sectors will evolve over the next decade or two. Nevertheless, there is a fair consensus that the present contribution of civil aviation is around 3\% of the total emission, relative to around 20\% of the total emitted by road transport as a whole. We remark that, due to the current pandemic, the discussed extrapolated statistics are based on data available until the end of 2019. Studies on future trends are, unavoidably, highly speculative, but suggest that the contribution of aviation to total emissions rises rapidly and could even reach 25\% and more by 2050 \citep{graver-etal-2019}. In fact, the proportion of emission due to aviation may well be substantially higher because the road-transport emission is bound to decrease due to the accelerated replacement of internal-combustion engines by electric propulsion. According to a report by QinetiQ \citep{horton-2006}, commissioned by the UK Department of Trade and Industry and leaning on potential technology-driven efficiency-gain figures adopted by the IPCC, it is estimated that global aviation-produced CO$_2$ emissions will rise from 500 Mtonnes in 2002 to $970-1600$ Mtonnes by 2030, depending on economic circumstances and technological changes. This figure now appears to be seriously adrift, as the most recent report by the International Council on Clean Transportation \citep{graver-etal-2019} states that the emissions by civil aviation in 2018 already amounts to approximately 918 Mtonnes. In the same vein, estimates by the International Civil Aviation Organisation \cite{ICAO-2009} indicate increases in fuel-burn by a factor of $4-8$ by 2050, the multiplier value depending upon the level of technological innovation. These estimates pose a major challenge to civil aeronautics, and they put aviation at the centre of the debate on environmental damage by technological activity.

The pressure to reduce fuel consumption and emissions has led the Advisory Council for Aeronautics Research in Europe (ACARE) to adopt extremely ambitious savings targets by 2050 (\url{http://www.acare4europe.org/documents/latest-acare-documents/acare-flightpath-2050}). Specifically in relation to CO$_2$, the ACARE's Flightpath 2050 initiative envisions a 75\% reduction of CO$_2$ per passenger-mile, relative to the 2000 level, to be procured from both operational and technological advances. From a technological perspective, it is very difficult to imagine how even a modest proportion of this margin can be achieved without recourse to highly innovative, unconventional approaches. Statistics covering the years between 1960 and 2000 \citep{albritton-1997} demonstrate that the fuel-burn per passenger-mile reduced by approximately 60\% in that time interval (reliable and accurate statistics are difficult to obtain), the main driver being jet-engine efficiency improvements. It is clear that the rate of reduction in fuel-burn has however substantially slowed to roughly 1\% in 10 years \citep{kharina-rutherford-2015}, with further positive increments requiring very high bypass-ratio (and thus large-diameter) engines.  

It is a popular assumption that weight savings due to the extensive use of composite materials are set to result in the next major downward trend in fuel-burn per passenger \citep{harris-2001}, and both Boeing and Airbus are engaged in major efforts to replace metal components by composites \citep{marsh-2015}, with the aim of reducing the empty-aircraft weight. It is difficult to find explicit and trustworthy statements on weight reductions achieved with composites in the civil-aviation sector. A rare quantitative estimate of the weight reduction achievable with composites in large civilian aircraft is 20\%, a figure given in the book ``Manufacturing Processes for Advanced Composites'' \citep{campbell-2003}. However, at the time of writing, composite components are limited by low tolerance to impact damage and sudden post-elastic failure at close to maximum tolerable load, both limitations requiring high safety factors and thicker, and thus heavier, components. Other problems with composite wings and bodies include high scrappage rates, and thus high production costs, anisotropic material properties, extreme difficulties with local-damage repair, and intolerance to physically intrusive control sensors and actuators being embedded into the skin. Anecdotal evidence suggests that the only promising route to significant weight savings would involve the introduction of active load-alleviation control, which would reduce the need to over-dimension wings and wing roots, the latter being especially problematic for composites, in response to rare extreme-load events at the edge of the flight envelope.

\begin{figure}[t]
\centering
\includegraphics[width=0.8\textwidth]{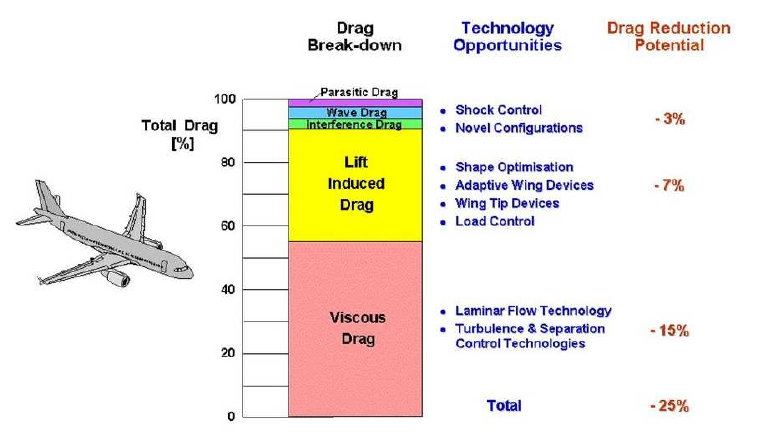}
\caption{Breakdown of the contributions to the total drag for a commercial aircraft flying in cruise conditions and the potential drag-reduction levels that could be achieved by various flow-control strategies. Taken from \cite{abbas-etal-2017} with permission from Springer Nature.}
\label{fig:aircraft-drag}
\end{figure}

From an aerodynamic perspective, reductions in drag translate directly to reductions in fuel-burn.  Specifically, at cruise conditions in long-haul flight, a 1\% reduction in drag results, conservatively, in a 0.75\% reduction in fuel-burn.  A single trans-continental B747 flight produces roughly 400 tonnes of CO$_2$.  Hence, a 1\% reduction in drag will save, roughly, $2-3$ tonnes of CO$_2$ per flight. If applied across all civil-aviation flights, this reduction would translate to approximately 10 Mtonnes per year (at year 2000 traffic level). Also, in cruise conditions, $50-60\%$ of the total drag is due to viscous friction, mostly induced by turbulent boundary layers on the skin of the aircraft, the remainder combining shape, induced, trim, and wave drag. The breakdown of the contributions to the total drag for a commercial aircraft flying in cruise conditions is shown in Fig. \ref{fig:aircraft-drag}. Hence, friction drag offers, by a substantial margin, the greatest potential for drag reduction.  

A relevant question underlying the present review is, therefore, whether the drag-reduction methods discussed in this paper can be more than academically interesting and form the foundation for, or assist the creation of, realizable technologies that could contribute to a reduced level of fuel-burn.

There are two main routes to achieving friction-drag reduction: a delay in transition -- i.e., extending the low-drag laminar portion of the boundary layer -- and a reduction in the turbulent drag. Delaying laminar-to-turbulent transition -- for example through the suppression of Tollmien-Schlichting waves -- is a well-trodden path, but one that is still vigorously pursued.  
Passive approaches, for which no external energy is supplied for the flow control, often involve shape optimisation to suppress adverse pressure gradients and the strategic placement of micro-grooves or micro-roughness elements, e.g., \cite{saric-etal-2003}.
Alternatively, active control may be applied via suction, which reduces the thickness of the boundary layer and makes it more stable, or plasma actuators, which locally accelerate the boundary layer \citep{arnal-etal-1997, joslin-1998, moreau-2007}, thus rendering it less prone to separation.
  
\begin{figure}
\centering
\includegraphics[width=0.7\textwidth]{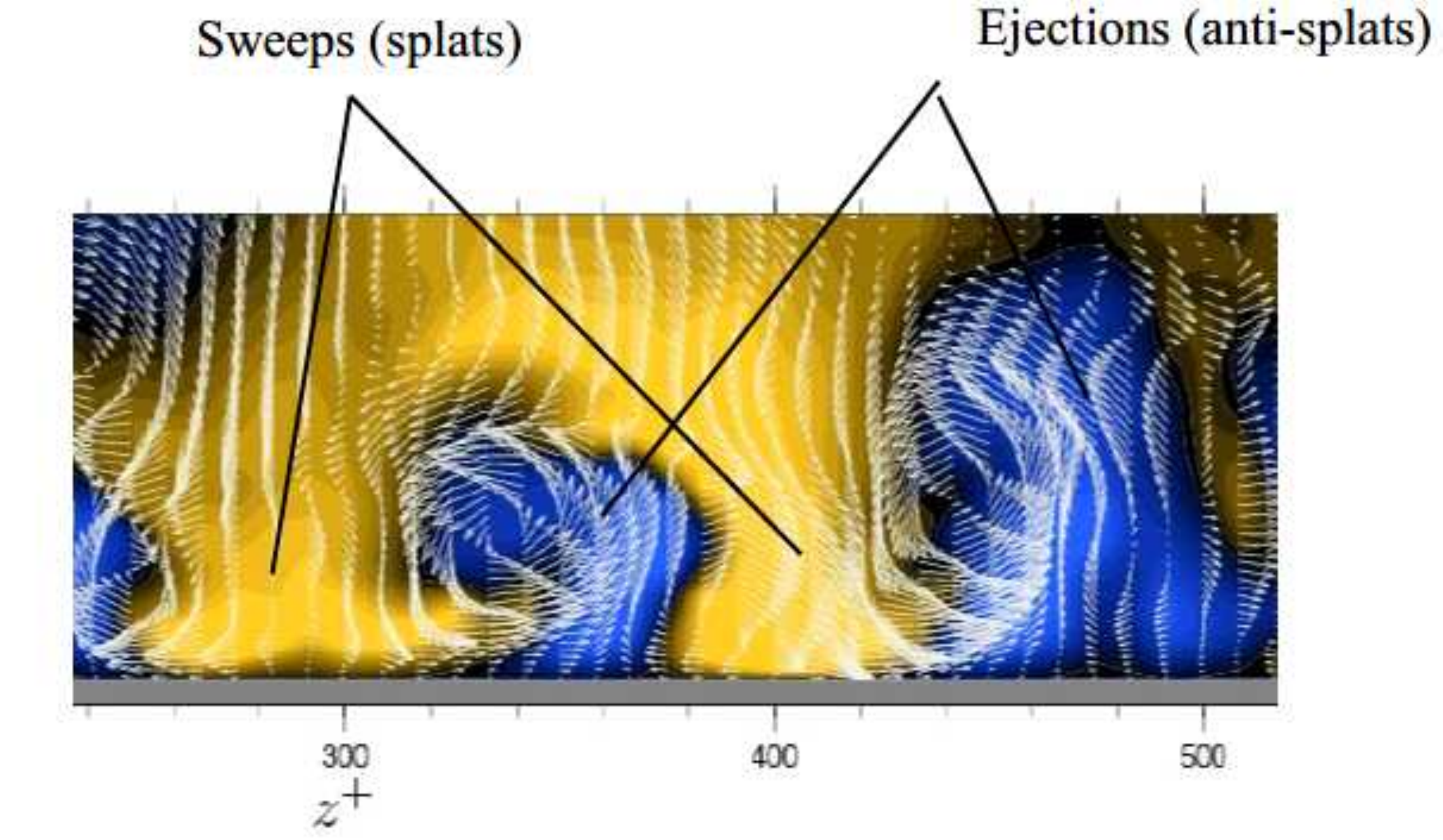}
\caption{Turbulent ejections and sweeps close to the wall, within a spanwise layer of thickness $y^+=y u_\tau /\nu=100$, derived from direct numerical simulations of a channel flow at $Re_\tau =u_\tau h/\nu=500$. The image shows a transverse cut normal to the mean-flow direction.
Colours indicate streamwise-velocity excess/defect relative to the mean level: blue indicates regions of lower velocity, yellow indicates regions of higher velocity.  Note that the spanwise length scale separating regions of high/low streamwise velocity is of order $z^+=100$. Adapted from \citep{touber-leschziner-2012}.}
\label{fig:intro2}
\end{figure}

Once a boundary layer has become fully turbulent, flow-control strategies often aim at reducing the mixing of streamwise momentum across the near-wall layer, illustrated in Fig. \ref{fig:intro2}.
This objective is essentially achieved by reducing the wall-normal turbulent fluctuations, thereby enhancing the contribution of viscous transport over a thicker near-wall layer than that in the uncontrolled state. The reduction of wall-normal fluctuations results, in a time-averaged sense, in a lower turbulent shear stress, and hence in a lower skin-friction drag due to a reduced velocity gradient at the wall. 

One theoretically interesting approach to turbulent-drag reduction is ``opposition control'' \citep{choi-moin-kim-1994, fukagata-kasagi-2003}. This strategy involves the detection of wall-normal motions at a given height, or their wall-pressure signature, followed by a local and dynamically varying application of an opposing injection, via suction or body-force-induced opposing momentum.  While this method demonstrates that the damping of wall-normal fluctuations, and thus the turbulent shear stress, is an effective means of reducing the drag, it is not practically realizable. A passive technique that is demonstrably effective is to use flow-aligned riblets of various shapes having a height and inter-crest distance of the order of the thickness of the viscous sublayer. Riblet-shaped surfaces yield, in ideal laboratory conditions, a maximum friction-drag reduction of about 8\%. The fundamental physical mechanism by which these riblets affect drag is not well understood, although several explanations -- based on trapped streamwise vortices, the inapplicability of the conventional Stokes law and the formation of quasi two-dimensional spanwise vortices -- have been advanced in the literature \citep{choi-moin-kim-1993}. A significant problem with the practical implementation of riblets is the need to maintain a high level of geometric fidelity and freedom from contamination, especially with respect to the sharpness on the riblet crests, a major challenge when the riblet height and inter-riblet distance  are fractions of a millimetre. Other passive approaches, investigated by computational methods (direct numerical simulations, DNS) and/or experimental means, include skewed wavy surfaces \citep{ghebali-etal-2017} and circular or tear-shaped dimples \citep{lienhart-etal-2008, tay-etal-2017, tay-etal-2018}. These surfaces have yielded either no drag reduction at all or very modest levels of order of a few percents, the highest values pertaining to carefully configured tear-drop shaped dimples.

Given that a selective control of the flow within specific layers above the wall -- for example, by Lorentz forcing -- is not a realistic proposition, what is left is to attempt to influence the near-wall turbulence by wall motion.  While some attempts have been made to use active dimples \citep{dearing-lambert-morrison-2007} to induce vortical motions into the near-wall region that would disrupt the streaky structure, most efforts in this area have been directed towards the imposition of wall-parallel oscillatory motion, either by spanwise oscillations of the wall \citep{jung-mangiavacchi-akhavan-1992}, streamwise-traveling waves of spanwise wall velocity \citep{quadrio-ricco-viotti-2009}, or plasma-induced near-wall forcing \citep{jukes-choi-2009}. 

The implementation of any of these methods in aviation poses formidable engineering challenges, which may or may not be overcome, whatever the pressure on fuel-burn savings may become in the future.  Moreover, it is well understood \citep{frohnapfel-hasegawa-quadrio-2012} that the balance between the cost associated with actuation and the drag reduction is very subtle and depends strongly on many factors, such as the actuation parameters, the Reynolds number of the actuated boundary layer and the turbulence state in non-equilibrium conditions provoked by strong pressure gradients, separation and reattachment. Nevertheless, it is important that all scientific means are engaged to gain an understanding of the fundamental physics involved, so as to develop a clear view of what the most promising engineering options are. This reasoning is the rationale that underpins the present review.    

The present article provides a wide-ranging overview of the numerous experimental, computational, and modelling studies that deal with the effects of temporally and/or spatially varying in-plane spanwise (transverse) wall motion on turbulent wall-bounded flows. The review includes a variety of forcing scenarios, oscillatory as well as unidirectional in transient conditions, in which the baseline wall-bounded turbulent flows are either statistically streamwise-homogeneous, as in a fully-developed channel or pipe flows, or spatially developing, as in free-stream boundary layers. The primary emphasis is on the drag-reducing effect of the forcing, but much attention is also devoted to the interpretation of the physical processes that are associated with the drag reduction as the forcing provokes major alterations in the strain and turbulence properties of the near-wall layer -- principally, the viscous sublayer and the buffer layer.

Excluded from the review are internal flows in which the drag reduction is caused by curvature strain (for example, drag reduction in flows over spanwise-convex surfaces \citep{rao-keshavan-1972}) or system rotation, either in cylindrical geometries (spinning pipes \citep{white-1964} or annular flows \citep{jung-sung-2006}) or in rectilinear geometries (rotating channel flows \citep{johnston-etal-1972,oberlack-etal-2006} or Couette flows \citep{komminaho-1996}). In all these cases, the drag reduction or drag increase is caused by mechanisms that are entirely different from those in which the near-wall layer is perturbed by near-wall spanwise straining. Specifically, rotation gives rise to extra strains or body forces that act in combination with other terms (e.g., Reynolds-stress components) to yield negative or positive source or sink terms, depending on the sign of the curvature strain relative to other strains in the equations governing the Reynolds stresses. This mechanism causes the Reynolds stresses to be either damped or amplified. A particular example is a channel flow subject to orthogonal-mode rotation (the angular-speed vector being normal to the flow direction and parallel to the walls), an idealised representation of a turbomachine-rotor passage. In this case, the friction on the suction side decreases, while that on the pressure side increases, which is solely due to the angular rotation of the entire flow configuration. The exclusion of scenarios in which the bulk flow is forced reflects the explicitly stated focus of the review, and is rooted in the premise that any reduction in frictional drag in an aeronautical context will have to target the near-wall region. Also excluded from the review are fully-developed flows subjected to unidirectional spanwise forcing, in which case the friction drag either increases monotonically in flows in cylindrical geometries, or remains unchanged in flows in planar geometries, after the flow has readjusted to the uniform forcing following the end of the initial temporal transient.

The review aims to provide the reader with a comprehensive appreciation of the state of the art, focusing on  
\begin{itemize}
	    \item the early observations inspiring the investigation of oscillatory wall motion as a means of controlling drag;
	    \item the physical interactions that are deemed responsible for drag reduction; 
	    \item the energy balance involving the power saved thanks to the altered turbulent flow and the power utilized for the actuation;
	    \item the influence of the Reynolds number and the associated structural properties of the controlled boundary layer;
	    \item alternative control methodologies that have been inspired by oscillatory wall actuation;
	    \item future research directions and realizability prospects.
\end{itemize}

As might be expected, any survey of substantial research extending over a period of more than thirty years follows a number of reviews that have discussed turbulent drag reduction by spanwise wall forcing. Notable earlier reviews are those of \cite{choi-2000}, \cite{karniadakis-choi-2003}, \cite{quadrio-2011} and \cite{abdulbari-etal-2013}. The nature, emphasis and extent of coverage of these surveys vary substantially.  
The review of \cite{choi-2000} gives a broad account of drag-reduction research in Europe up to 1998, and includes a short section on experimental and computational studies on wall oscillation in boundary layers, channels and pipes.  The review by \cite{karniadakis-choi-2003} covers a selection of research studies on riblets, wall oscillation and travelling waves induced by near-wall body-forcing, with particular emphasis on the possible physical mechanisms behind the drag-reduction effect. The review of \cite{quadrio-2011} is devoted entirely to drag reduction by wall motion and mostly covers computational studies on spanwise oscillation and spanwise- and streamwise-travelling waves. It addresses selective topics that pertain to the effect of the Reynolds number, the important issue of scaling for a meaningful physical interpretation of the results, and the energy balance. 
Finally the review by \cite{abdulbari-etal-2013} provides a broad, multi-topic coverage on drag reduction by riblets, dimples, oscillating walls, compliant surfaces and micro-bubbles. The discussion of spanwise wall motion is brief, covering some major observations derived from DNS studies on the optimum choice of actuation parameters, and power-saving issues. More recent reviews on flow control that discuss spanwise wall oscillations are by \cite{asidin-etal-2019} and \cite{zhang-etal-2020}. In addition, a few books on turbulence and flow control contain chapters devoted to the oscillating-wall technique, e.g., \cite{fan-dong-2016,soldati-monti-2014,tardu-2017}.
  
Our review starts with an introduction to the basic ideas behind spanwise forcing in Section \ref{sec:basic}. Overviews of experimental and numerical investigations are given in Section \ref{sec:exp} and Section \ref{sec:num}, respectively. Section \ref{sec:phys} includes discussions on the physical mechanisms, while Section \ref{sec:redep} presents the dependence of the drag reduction and the flow physics on the Reynolds number. The results obtained via modelling the drag-reduction flows are discussed in Section \ref{sec:models} and Section \ref{sec:extensions} is devoted to flows where the spanwise forcing is induced by other means, such as body forces, rotating discs or riblets, and to transitional and compressible flows. In Section \ref{sec:future} some ideas regarding future research directions and technological applications are presented.

The authors hope that the present review, with its emphasis on a comprehensive coverage of wall-turbulence altered by oscillatory near-wall transverse actuation, will be a valuable contribution to the literature within the general area of flow control.

\section{Basic ideas}
\label{sec:basic}

The preceding introduction set out the general context within which the friction drag reduction by spanwise wall motion is reviewed herein. The purpose of the present section is to provide some basic facts, definitions and qualitative explanations intended to give readers some insight into the principal physical and phenomenological aspects that pertain to the control methods being reviewed.

In the simplest actuated configuration, the entire wall oscillates purely in time in the spanwise direction according to:

\begin{equation}
\label{eq:oscillation}
    W_w(t) = W_m \cos \left( \frac{2 \pi t}{T}\right).
\end{equation}
The key consequence of the wall oscillation \eqref{eq:oscillation} is the formation of an unsteady spanwise shear layer -- a ``Stokes layer'' -- that perturbs the wall turbulence sustained by the mean streamwise strain, so as to produce a reduction in the turbulence intensity. This attenuation of turbulent activity is the cause of the decrease in wall-friction drag. At a given Reynolds number, the magnitude of the reduction depends on the two forcing parameters $W_m$ and $T$, i.e., the amplitude of the wall velocity and the period of oscillation, respectively. To be more precise, the governing parameters are the respective non-dimensional, wall-scaled amplitude and period, 

\begin{equation}
\label{eq:oscillation-parameters}
    W_m^+ = \frac{W_m}{u_\tau} \quad \mbox{and} \quad  T^+ = \frac{T u_\tau^2}{\nu},
\end{equation}
where $u_\tau=\sqrt{\tau_w/\rho}$ is the wall-friction velocity, $\tau_w$ is the space- and time-averaged wall-shear stress, $\rho$ is the density of the fluid, and $\nu$ is the kinematic viscosity of the fluid. A related parameter is the peak-to-peak wall displacement $D_m^+=W_m^+ T^+/\pi$.

Any reference made herein to ``drag reduction'' requires this quantity to be defined. Unless otherwise stated, it is the gross aerodynamic friction-drag reduction, i.e., the reduction of the streamwise wall-shear stress caused by the wall actuation. If the skin-friction coefficient is defined as $C_f = 2 \tau_w/\rho U_b^2$ in channel and pipe flows, where $U_b$ is the streamwise mean (bulk) velocity, the percentage skin-friction drag reduction is defined as:

\begin{equation}
\label{eq:drag-reduction}
    \mathcal{R}(\%) = 100 \cdot \frac{C_{f,0}-C_f}{C_{f,0}},
\end{equation}
where $C_{f,0}$ is the skin-friction coefficient in the canonical case. In the case of unconfined, statistically two-dimensional boundary layers, the free-stream velocity $U_\infty$ is used in lieu of $U_b$ in the definition of $C_f$, and the drag reduction $\mathcal{R}$ depends on the streamwise direction. It is appropriate to point out here that a skin-friction drag reduction may not necessarily translate to an overall drag reduction as the form drag may be adversely altered by the control strategy.

The power saved as a consequence of the reduction in drag is counteracted by the need to expend power on moving the wall. This actuation power $\mathcal{P}_z$ can be expressed as a percentage of $\mathcal{P}_x$, the power spent on driving the fluid along the streamwise direction, 

\begin{equation}
\label{eq:power-spent}
    \mathcal{P}_z(\%) = \frac{100 \mu}{\mathcal{P}_x (t_f-t_i)} \cdot \int_{t_i}^{t_f} \int_S W_w \frac{\partial W}{\partial y}\Big|_{y=0}  \mathrm{d}S \mathrm{d}t,
\end{equation}
where $y$ is the wall-normal coordinate measured from the wall, $W_w$ is defined in \eqref{eq:oscillation}, $W$ is the instantaneous and local spanwise velocity, $\mu$ is the dynamic viscosity of the fluid, $t_i$ and $t_f$ are the initial and final time values for the temporal averaging, and $S$ is the surface area. For a channel flow, the power to push the fluid along the streamwise direction is $\mathcal{P}_x=2 \tau_w U_b$.
The power \eqref{eq:power-spent} ignores any power loss below the surface due to mechanical and electrical systems used for the actuation. For constant-mass-flow-rate conditions, the drag reduction in \eqref{eq:drag-reduction} coincides with the percentage power saved for propelling the fluid along the streamwise direction, and therefore the net power saved is:

\begin{equation}
\label{eq:power-net}
    \mathcal{P}_{net}(\%) = \mathcal{R} - \mathcal{P}_z.
\end{equation}

It is observed, mostly as an outcome of DNS of turbulent channel flows, that the drag reduces monotonically as $W_m$ increases, while the drag-reduction margin reaches a maximum for $T^+$ close to 100.  For this period, the thickness of the unsteady transverse Stokes layer is $y^+ = y u_\tau/\nu \approx 15-20$, i.e., the strain in this layer is confined to the viscous sublayer and, at most, to the lower portion of the buffer layer. The maximum drag-reduction levels are about $\mathcal{R}=40\%$ at $T^+=100$. If $T^+$ is increased beyond 100, the Stokes layer penetrates into the buffer layer, eventually propagating into the fully turbulent layer. In this case, the Stokes layer enhances the production of turbulence above the viscous sublayer, thus counteracting the beneficial effects in the viscous sublayer. Indeed, no drag reduction occurs at all for $T^+>400$, and for sufficiently high values ($T^+ \approx 1000$), i.e., when approaching a steady, unidirectional actuation -- the drag increases by $5-10\%$. The physical interactions that are (or presumed to be) responsible for the observed reductions in the turbulent activity and drag are discussed in Section \ref{sec:phys}.

Several studies have revealed that, with the exception of low values of $W_m$, for which $\mathcal{P}_{net}\approx10\%$, the actuation power required to oscillate the wall according to equation \eqref{eq:oscillation} exceeds the power saved by the drag-reduction margin, leading to a negative net energy balance. Hence, although this simple mode of actuation provides a useful foundation for analysing the fundamentals of the interaction, as it transpires in the physical interpretation discussed in Section \ref{sec:phys}, it is not a viable basis for a technology targeting practical drag reduction. 

\begin{figure}
\centering
\includegraphics[width=0.7\columnwidth]{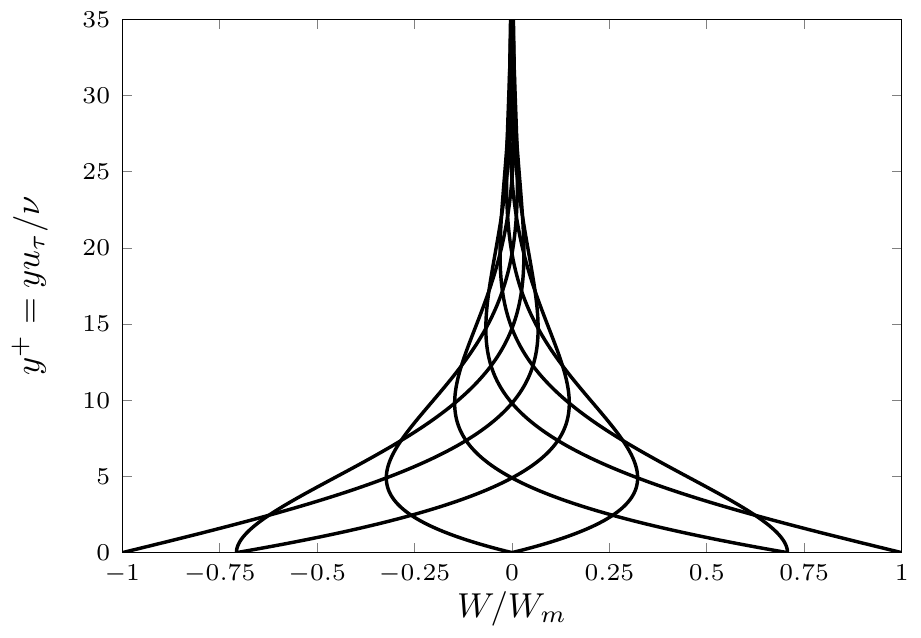}
\caption{Profiles of the Stokes layer generated by the wall motion \eqref{eq:oscillation} and given by equation \eqref{eq:stokes} scaled in inner units of a turbulent channel flow with Reynolds number $Re_\tau=u_\tau h/\nu=350$ and period $T^+=125$.}
\label{fig:stokes}
\end{figure}

As the Stokes layer is (or should be) confined to the viscous sublayer where the turbulent fluctuations are low, it is instructive to study the corresponding simplified case that arises when the flow is purely laminar. In this case, the spanwise motion flow is decoupled from the streamwise flow and can be considered in isolation. When the wall oscillates according to \eqref{eq:oscillation} below an unbounded still fluid, the laminar-flow solution, due to \cite{stokes-1851}, is:

\begin{equation}
\label{eq:stokes}
    W(y,t) = W_m e^{-y \sqrt{\pi/\nu T}} \cos \left( \frac{2 \pi t}{T} - y \sqrt{\frac{\pi}{\nu T}}\right).
\end{equation}
The maximum amplitude decays exponentially from the wall and the phase of the oscillation changes with height. Fig. \ref{fig:stokes} shows profiles of the Stokes layer, where the wall-normal coordinate is scaled in inner viscous units. For relevant drag-reducing cases, these profiles agree very well with the turbulent profiles averaged along the homogeneous spatial directions and over the oscillation phases. This assumption has been shown to hold remarkably well for periods of oscillations smaller or comparable with the optimal value \citep{choi-xu-sung-2002,quadrio-sibilla-2000}, for a wide range of streamwise-flow Reynolds numbers and especially in the very proximity of the wall  \citep{yao-etal-2019}.
This match occurs because the spanwise Reynolds-stress term involving $\overline{vw}$, where $v$ and $w$ are the turbulent velocity fluctuations in the wall-normal and spanwise directions, respectively \citep{ricco-quadrio-2008}, is much smaller than the corresponding viscous term (it can be shown that $\overline{vw}$ declines as $\mathcal{O}(y^3)$ close to the wall). As a result of this agreement, the laminar solutions can be, and have been, used to compute the power spent to generate the spanwise motion and hence the net energy balance with excellent accuracy. 

Solution \eqref{eq:stokes} also gives an accurate estimate of the thickness $\delta_z$ of the spanwise shear layer at which the maximum velocity decays to a specified maximum velocity $W_\delta$, often expressed as a percentage of the bulk velocity $U_b$ or set equal to a representative value for the near-wall spanwise turbulent fluctuations,

\begin{equation}
\label{eq:delta}
    \delta_z = \ln \left( \frac{W_m}{W_\delta} \right) \sqrt{\frac{\nu T}{\pi}}.
\end{equation}
Formula \eqref{eq:delta} gives the explicit dependence of the Stokes-layer thickness on $T$ and thus on the drag reduction, as $\delta_z$ has been shown to be linearly related to the drag-reduction margin as long as the Stokes layer is confined within the viscous sublayer \citep{quadrio-ricco-2011}. 
Formulae \eqref{eq:stokes} and \eqref{eq:delta} are also valid for the unsteady layers produced by the wall motion \eqref{eq:oscillation} in the laminar Poiseuille channel and pipe flows, as long as $T$ is sufficiently small for the Stokes layer to be much thinner than the channel height or the pipe radius. These cases are of interest because they feature large drag-reduction levels.

A more complicated form of actuation than that given by the wall-motion \eqref{eq:oscillation} involves streamwise-travelling waves of spanwise wall velocity, first studied by \cite{quadrio-ricco-viotti-2009}. This actuation is described by:

\begin{equation}
\label{eq:waves}
    W_w(x,t) = W_m \cos(\kappa_x x - \omega t).
\end{equation}
At a fixed Reynolds number, the drag reduction thus depends on the streamwise wavenumber $\kappa_x$ and the frequency $\omega$. The flow reduces to the spanwise-uniform case when $\kappa_x=0$ and to the steady-wave case, first investigated by \cite{viotti-quadrio-luchini-2009} and \cite{yakeno-etal-2009}, when $\omega=0$. The solutions of the laminar flow produced by the wall motion \eqref{eq:waves} can be found in \cite{ricco-hicks-2018}.
At low Reynolds numbers, the streamwise-wavy wall motion \eqref{eq:waves} performs better than the spanwise-uniform wall oscillation \eqref{eq:oscillation}, yielding higher gross drag-reduction margins, i.e., up to $\mathcal{R}\approx45\%$, as well as net energy savings as large as $\mathcal{P}_{net}\approx20\%$ for waves travelling forward with a small phase speed. Backward-travelling waves always lead to drag reduction, while forward-travelling waves with a phase speed that is comparable with the so-called convection velocity of the near-wall turbulent structures \citep{quadrio-luchini-2003} produce an increase of drag of the order of 20\%. 
A key result that highlights the time-space analogy of the turbulence response to the wall motion \eqref{eq:waves} is that at $\omega=0$ the optimal wavelength is $\lambda_x^+=2 \pi/\kappa_x^+=\mathcal{O}(1000)$. In a Lagrangian frame of reference fixed with the near-wall turbulence, this wavelength indeed corresponds to the optimal period $T^+=\mathcal{O}(100)$ for the purely temporal motion \eqref{eq:oscillation} because the convection velocity in the viscous sublayer and the buffer layer, where the low-streaks occur, is $\mathcal{U}_c^+\approx10$.

We conclude this section with a few comments intended to clarify a distinction between two constraints adopted when simulating drag-reduction scenarios in statistically streamwise-homogeneous flows: the ``constant mass-flow-rate'' (CFR) and ``constant pressure-gradient'' (CPG) conditions.  
In simulations of channel and pipe flow, which must inevitably be restricted to a finite segment, streamwise-homogeneity allows having identical outlet and inlet conditions, i.e., data at inlet and outlet planes coincide. This condition however poses the problem that the flow needs to be driven along the segment by an appropriate constraint -- the two options being to fix the flow rate (CFR) or the pressure gradient (CPR).  
When a baseline flow at a given Reynolds number is subjected to some drag-reduction actuation, the CFR condition yields, as an output of the simulation, a reduction in wall-shear stress and streamwise pressure gradient. The drag reduction is thus simply characterized by the reduction in wall-shear stress at the given Reynolds number. This approach is the most natural to adopt, and it is also the only one that is applicable to free flows, in which case the flow rate in the boundary layer is fixed by the outer stream. A disadvantage of the CFR approach is that the physical (turbulent) state of the baseline and actuated near-wall layer is dictated by the friction Reynolds number $Re_\tau$ (based on the wall-friction velocity and the channel half height or the pipe radius) and not by the bulk Reynolds number $Re_b$ (based on the bulk velocity and the channel height or pipe diameter). In a strict physical sense, therefore, the two flows are not entirely equivalent, as any universal representation of near-wall turbulence (e.g., the log-law) is based on scaling with the wall-shear stress.  
When, in contrast, the CPR condition is imposed, the consequence is that the flow in the passage adjusts itself to a higher flow rate and thus higher bulk Reynolds number. However, a clear advantage is that the canonical state of the near-wall layer in the two flows is retained and the inner-viscous-unit scaling is invariant because the wall-shear stress does not change. This scaling allows meaningful comparisons between the near-wall canonical and actuated flows.
The drag reduction is implicitly given by the increase of the bulk Reynolds number and can be quantified by comparing the difference in power between the actuated and unactuated flows for the fixed head loss (i.e., pressure gradient). The CPG condition is also pertinent to experimental studies of water channel or pipe flows, where a head loss between two large reservoirs is maintained fixed. The reduced drag then leads to a higher discharge given by the higher mass-flow rate.
A novel condition has been advanced by \cite{frohnapfel-hasegawa-quadrio-2012}, for which the combined power required for driving the fluid along the streamwise direction and for activating the wall motion is maintained constant, i.e., ``constant power input'' (CPI). When the flow is subjected to a successful drag-reduction technique, the systems benefits from both an increased mass-flow rate and a reduced wall-shear stress.
\section{Experiments}
\label{sec:exp}

The need to impose a transverse wall motion, or some equivalent forcing condition very close to the surface, added to the requirement of accurately quantifying relatively small drag-reduction margins, poses major challenges to any experimental exploration of drag reduction by wall motion. Experiments are, generally, far more challenging than numerical simulations, except for Reynolds-number constraints that the latter face. Practical constraints render the experimental investigation of fully-developed turbulent channel flows almost impossible, while this geometry, in contrast, is instead frequently used in DNS studies, targeting the quantification of drag reduction for a wide range of wall-actuation parameters. The large majority of experimental studies have, therefore, been undertaken on spatially evolving turbulent boundary layers and pipe flows, the former entailing a wall section that is laterally oscillated by mechanical or electro-magnetic means, or, in one case, by an electro-active foil forming the actuated wall. The attraction of pipe flows arises from the possibility of imposing the spanwise-homogeneous wall motion by means of streamwise-sequential pipe segments that rotate in a steady or oscillatory fashion. In turbulent boundary layers an additional complication arises from the necessity to measure the wall-shear stress through the mean-flow gradient at the wall, while in pressure-driven channel and pipe flows the friction drag can be found from the mean streamwise pressure gradient under CFR conditions. Confined flows are also attractive because, under CPG conditions, the increased bulk velocity can be used to quantify the beneficial effects of the control.

This section reviews the large majority of published experimental studies on turbulent flows altered by spanwise wall motion. These studies are summarized in Table \ref{table:experimental-studies}. Other experimental studies on wall-bounded flows modified by techniques inspired by the wall motion are discussed in Section \ref{sec:extensions}.

{\scriptsize
\begin{sidewaystable}
\begin{tabular}{|c|c|c|c|c|c|}
\hline
\centering
\textbf{Work} & \textbf{Geometry} & \textbf{Forcing} & \textbf{Fluid} & \textbf{Reynolds number} & {\bf max} $\boldsymbol{\R(\%)}$\\
\hline
\cite{lohmann-1976} & B.L. & Spinning cylinder  & Air & $Re_\tau>1000$ &  \begin{tabular}{@{}c@{}}Evidence of \vspace{-0.1cm} \\  drag reduction\end{tabular}  \\
\hline
\citet{bradshaw-pontikos-1985} & B.L. & 35$^\circ$-angled suction slots   & Air & \pbox{20cm}{$Re_\theta=3300$ ($Re_\tau=1340$)} & Not reported\\
\hline
\pbox{20cm}{\cite{laadhari-skandaji-morel-1994} \\ \cite{skandaji-1997}}   & B.L. & Spanwise wall oscillations    & Air & $Re_\theta=770-1600$ & 36\%\\
\hline
\cite{raskob-sanderson-1997}                                & B.L. & Spanwise wall oscillations        & Air & $Re_x=3.5\cdot10^5$-$2.2\cdot10^6$ & 22\%\\
\hline
\pbox{20cm}{\cite{trujillo-bogard-ball-1997} \\ \cite{trujillo-1999}}     & B.L. & Spanwise wall oscillations        & Water & \pbox{20cm}{$Re_\theta=1400$ \\ ($Re_\tau=633-962$)} & 35\%\\
\hline
\cite{choi-graham-1998}                                     & Pipe flow      & Circumferential wall oscillations & Water & \pbox{20cm}{$Re_D=23300-36300$ \\ ($Re_\tau=651-962$)}  & 24\%\\
\hline
\pbox{20cm}{\cite{choi-debisschop-clayton-1998} \\ \cite{choi-clayton-2001} \\ \cite{choi-2002}} & B.L. & Spanwise wall oscillations & Air & $Re_\theta=1190$ ($Re_\tau=549$) & 45\%\\
\hline
\cite{wu-2000} & B.L. & Spanwise wall oscillations & Water & $Re_\theta=1400$ ($Re_\tau=633$) & 32\%\\
\hline
\pbox{20cm}{\cite{dicicca-etal-2002} \\ \cite{iuso-etal-2003}} & B.L. & Spanwise wall oscillations & Water & $Re_\theta=1160$ ($Re_\tau=537$) & Not reported\\
\hline
\pbox{20cm}{\cite{kiesow-plesniak-1998,kiesow-plesniak-2001} \\ \cite{kiesow-plesniak-2003}} & B.L. & Spanwise-running belt  & Water & $Re_\theta=1450$ ($Re_\tau=653$) & Not reported\\
\hline
\pbox{20cm}{\cite{ricco-2000,ricco-2004} \\ \cite{ricco-wu-2004-a}} & B.L. & Spanwise wall oscillations & Water & \pbox{20cm}{$Re_\theta=500-1400$ \\ ($Re_\tau=257-633$)} & 32\%\\
\hline
\cite{auteri-etal-2010}  & Pipe flow & Circumferential wall waves  & Water & $Re_D=4900$ ($Re_\tau=175$) & 33\%\\
\hline
\cite{gouder-potter-morrison-2013}  & B.L. & Spanwise wall oscillations & Air & $Re_\theta=2430$ ($Re_\tau=1025$) & 16\%\\
\hline
\cite{gatti-etal-2015}  & Duct flow & Spanwise wall oscillations & Air & 
\pbox{20cm}{$Re_b=4520-15140$ \\ ($Re_\tau=150-440$)} & 2.4\%\\
\hline
\pbox{20cm}{\cite{bird-santer-morrison-2018} \\ \cite{bird-santer-morrison-2018-ijss}}  & B.L. & \pbox{20cm}{Streamwise-travelling spanwise \\ wall velocity waves} & Air & $Re_\tau=1125$ & 21.5\%\\
\hline
\pbox{20cm}{\cite{kempaiah-etal-2020}}  & B.L. & \pbox{20cm}{Spanwise wall oscillations} & Air & $Re_\tau=570$ & 15\%\\
\hline
\end{tabular}
\caption{Experimental studies of wall-bounded turbulent flows altered by spanwise wall motion. The Reynolds numbers are defined as follows: $Re_\tau=u_\tau h/\nu$ (where $h$ is the half-channel height, $u_\tau=\sqrt{\tau_w/\rho}$ is the friction velocity, $\tau_w$ is the time and space averaged wall-shear stress, $\rho$ and $\nu$ are the density and the kinematic viscosity of the fluid), $Re_\theta=U_\infty \theta/\nu$ (where $U_\infty$ is the free-stream velocity of the boundary layer and $\theta$ is the momentum thickness), and $Re_x=U_\infty x/\nu$ for boundary layers (where $x$ is the streamwise coordinate measured from the leading-edge of the plate), $Re_b=U_b h/\nu$ for the duct flow (where $U_b$ is the bulk velocity), and $Re_D=U_b D/\nu$ for pipe flows (where $D$ is the diameter of the pipe). B.L. stands for boundary layers exposed to a free stream. The friction Reynolds numbers in parenthesis are estimated according to the formulas $Re_\tau=1.118Re_\theta^{0.875}$ for boundary layers \citep{ricco-quadrio-2008} and $Re_D = 4\sqrt{2} Re_\tau \left[2 \log_{10} \left( 4\sqrt{2} Re_\tau \right) - 0.8 \right]$\citep{pope-2000}. In this table and in the following ones, when multiple Reynolds numbers are considered in a study, the maximum drag-reduction margin occurs at the lowest value}. 
\label{table:experimental-studies}
\end{sidewaystable}
}

\subsection{Early experimental studies of flows with spanwise actuation}
\label{sec:earlyexp}

Interest in the effects of spanwise wall motion on streamwise wall-bounded turbulent flows can be traced back to the experimental efforts by \cite{furuya-etal-1966}, \cite{lohmann-1976}, \cite{bissonnette-mellor-1974}, \cite{driver-hebbar-1987}, and \cite{hebbar-driver-1987} (\cite{higuchi-rubesin-1979} reported tests of turbulence models on some of these experimental results), in which turbulent flows along the axial direction of a cylinder were subjected to a spinning section of the cylinder. 
Although the accuracy of the wall-shear-stress measurements might have been low, due to the indirect estimation of the stress, a consistent outcome of these studies was that the spinning surface enhanced the turbulence intensity and thus the Reynolds stresses. \cite{lohmann-1976}, whilst reporting a drag increase when the forcing was applied, also recorded, confusingly, that the mean axial velocity decreased near the wall. This effect intensified at larger rotational velocity of the cylinder (refer to their figure 6 on page 358), thus suggesting drag reduction. 

It is important to clarify at this juncture that, in the present context of spanwise forcing where the bulk-flow streamlines remain aligned along the streamwise direction, a reduction in drag can only be achieved if the near-wall layer and the turbulent structures within this layer -- the low-speed streaks, in particular -- are subjected to a spanwise strain that varies in time and/or along the streamwise direction. In contrast, the application of a homogeneous and steady spanwise straining inevitably results in a rise in the turbulence intensity and drag, following a transient decay immediately after the start of the forcing. It is for this reason that a periodically varying and/or streamwise wavy spanwise wall motion is imposed for achieving large and sustained drag-reduction margins in most investigations.

The wind-tunnel study by \citet{bradshaw-pontikos-1985} was the first to report the effective drag-reducing action of a near-wall spanwise shear on a nominally two-dimensional turbulent boundary layer. The aim of that study was to reproduce and extend the earlier experimental campaign by \cite{vandenberg-etal-1975}, in which the turbulent wall-shear stress over a swept wing was reduced with respect to the two-dimensional value by the spanwise flow induced by the sweep. \cite{bradshaw-pontikos-1985} used suction slots inclined at 35 degrees relative to the spanwise direction in an effort to emulate the spanwise shear produced on a swept wing.  
Remarkably, \cite{bradshaw-pontikos-1985} recognized that it is the rate of change of the spanwise mean-velocity gradient, $\mathrm{d} W/\mathrm{d} y$ (in space in their case), not simply the gradient itself, that is the cause for the sensitive response of the intensity of the tilted near-wall eddy structures (and the low-speed streaks) to the spanwise strain and thus ultimately for the drag reduction. This result is, therefore, the first confirmation of the statement made above about the crucial importance of the rate of change of the spanwise strain. \citet{bradshaw-pontikos-1985} realized that the spanwise shear had to vary in space to obtain a sustained drag-reduction effect and avoid an asymptotic realignment of the flow to a two-dimensional state. These insightful observations by \citet{bradshaw-pontikos-1985} were arguably the principal source of inspiration for subsequent work on forcing by spanwise-wall motion, where the rate of change of the spanwise-velocity gradient is provoked by temporal and/or streamwise-sinusoidal wall oscillations.

\subsection{Experimental studies of planar flows with streamwise-uniform forcing}
\label{sec:exp-planar}

The first experimental study to investigate the effect of spanwise-sinusoidal wall oscillations on a turbulent boundary layer was conducted by \cite{laadhari-skandaji-morel-1994}, with the aim of verifying the DNS results by \cite{jung-mangiavacchi-akhavan-1992} for a forced turbulent channel flow. \cite{laadhari-skandaji-morel-1994}'s {\em Physics of Fluids} publication only provides a condensed extract of the subsequent Ph.D. thesis by \citet{skandaji-1997}, which contains a wealth of results unpublished in the archival literature. In their wind tunnel, they studied turbulent boundary layers in the range of momentum-thickness Reynolds numbers, $Re_\theta=770-1600$ (only the data for $Re_\theta=950$ were published in \cite{laadhari-skandaji-morel-1994}), flowing over a rigid plate that was oscillated transversally by a crank-shaft mechanism. They measured the flow with the aid of three-dimensional hot-wire probes and found that the wall-shear stress reduced by a maximum of $\R=36\%$. Reductions as large as 50\% in the Reynolds stress $\overline{uv}$ and in the root-mean-square values of the three velocity components were also recorded ($u$ and $v$ denote the streamwise and wall-normal velocity fluctuations). 

The wind-tunnel study by \cite{raskob-sanderson-1997} (private communication, courtesy of Dr Raskob) has not been published, but includes the first reference to the interesting result that a minimal distance $\Delta z^+\approx80$ of the maximum spanwise displacement is required for drag reduction, a result later confirmed through a scaling-factor analysis by \cite{quadrio-ricco-2004} and in a DNS study of a channel flow modified by streamwise-travelling waves by \cite{quadrio-ricco-2011}. For the optimal period of oscillation, $T^+=100$, this displacement is comparable to the spacing of the near-wall low-speed streaks, $\lambda_z^+\approx100$, and induces spanwise-velocity oscillations in the buffer layer that are comparable with the root-mean-square of the spanwise turbulent fluctuations, suggesting that the wall turbulence, and thus the wall-shear stress, are unaffected when the imposed spanwise velocity is lower than that representing the spanwise-velocity fluctuations.

The results by \cite{laadhari-skandaji-morel-1994} were confirmed by \cite{trujillo-bogard-ball-1997}, who carried out an experimental study of a turbulent boundary layer flowing in a water flume and modified by spanwise wall oscillations. A crank-slider mechanism produced the wall oscillation beneath the boundary layer at $Re_\theta=1400$, and a maximum wall velocity, $W_m^+=17$, was reached. By employing laser Doppler velocimetry (LDV) and how-wire probes, \cite{trujillo-bogard-ball-1997} measured a maximum drag-reduction margin of $\R=27\%$, later corrected to $\R=35\%$ in the Ph.D. thesis of \cite{trujillo-1999} to account for the initially neglected cooling effect of the Stokes layer on the hot-wire film. For the first time, \cite{trujillo-bogard-ball-1997} also measured the optimal oscillation frequency at a fixed maximum wall displacement, although this result was not reported as significant because the change in the drag-reduction margin at the maximum value was judged to be within the experimental uncertainty. The drag reduction was thus thought to scale with the maximum wall velocity, irrespectively of the frequency.

The wind-tunnel study of a turbulent boundary layer at $Re_\theta=1190$ by \cite{choi-debisschop-clayton-1998} and \cite{choi-clayton-2001} reported the first images of the tilting of the streaks due to the oscillatory wall motion and the highest measured drag-reduction level by spanwise wall oscillations, at $\R=45\%$. The wall motion was generated by a crank-shaft mechanism using a flywheel and a set of counter balances. \cite{choi-2002} also reported a reduction in the turbulent fluctuations, an increased frequency of sweeping motions at $y^+=2$, and phase-averaged spanwise velocity profiles that resembled the laminar Stokes-layer profiles. \cite{dicicca-etal-2002} and \cite{iuso-etal-2003} carried out experiments in the Hydra water flume at Politecnico di Torino on flows at Reynolds numbers that were very similar to \cite{choi-debisschop-clayton-1998}'s and focussed on the flow modifications produced by the oscillations, quantifying the altered strength, spacing, and width of the low-speed streaks.

\cite{trujillo-bogard-ball-1997}'s water flume was also used by \cite{wu-2000}, \citet{ricco-2000}, \cite{ricco-wu-2004-a}, and \cite{ricco-2004}. The M.Sc. thesis by \cite{wu-2000} presents drag-reduction results at the highest maximum spanwise wall velocity ever imposed in a laboratory, namely, $W_m^+=56$. This velocity was achieved by increasing the actuation frequency at a fixed maximum displacement. \cite{wu-2000} reported an initial increase in the drag-reduction margin with increasing $W_m^+$, followed by constant drag-reduction values at higher wall velocities. This result is fully consistent with the drag-reduction map of figure 1 in \citep{quadrio-ricco-2004} (also shown in Fig. \ref{fig:QRmapD} of this review): along a hyperbola of constant maximum wall displacement, the drag-reduction margin increases up to the optimum period at fixed displacement as the period decreases, but at lower periods the drag-reduction value remains almost constant, because the effects of increasing maximum spanwise velocity and decreasing period balance each other. \cite{wu-2000}'s maximum velocity was not large enough to detect a drop in the drag-reduction margin, as predicted by \cite{quadrio-ricco-2004} at vanishingly low periods of oscillation for fixed wall displacement. \cite{ricco-wu-2004-a} computed the drag reduction by directly measuring the mean velocity in the viscous sublayer by LDV and hot-film anemometry. They reported the streamwise evolution of the drag reduction along the oscillating wall and downstream of the end of the actuated surface, showing that the flow readjusted to the unforced conditions at a short distance downstream of the plate, while a longer distance was needed by the flow to fully establish to the drag-reduction condition downstream of the start of the plate.

As regards to the effects of the spanwise motion on the flow structures, \cite{ricco-2004} recorded flow visualizations of a turbulent boundary layer at $Re_\theta=1400$ from the top and the end of the water flume, using hydrogen bubbles generated by multiple platinum wires. These flow visualizations brought out the spanwise tilting of the low-speed streaks, their reduced length and increased spanwise spacing, as well as their relative spanwise motion with respect of the vortical structures present in the buffer layer. A simple model, based on the Stokes-layer displacement and the turbulent convection velocity, was shown to accurately predict the inclination angle of the streaks.
As the visualized streaks were dragged laterally by the wall oscillation, the number and intensity of the ejections were significantly reduced, and the turbulence-induced momentum exchange close to wall thus dropped accordingly. 
\cite{ricco-2004} also observed that, downstream of the oscillating wall, the scenario was reversed, as the oscillatory motion of the streaks penetrated upward to the quasi-streamwise vortices by viscous diffusion. There is reason to believe that this upward transfer of oscillatory motion from the wall was the cause of the drag increase, as computed by \cite{skote-mishra-wu-2019} downstream of the drag-reduction decay region measured by \cite{ricco-wu-2004-a}. Further flow visualizations were carried out by \cite{ricco-2000} by means of fluorescent dye injected into the turbulent boundary layer through a thin spanwise slit located at the wall. The low-speed streaks were observed distinctly in the fixed-wall conditions, although the flow altered by the wall motion could only be visualized when the wall oscillated in the spanwise direction at low speeds because the dye was ejected laterally at high speeds, hence disrupting the flow.

\begin{figure}
\centering
\includegraphics[width=0.49\columnwidth]{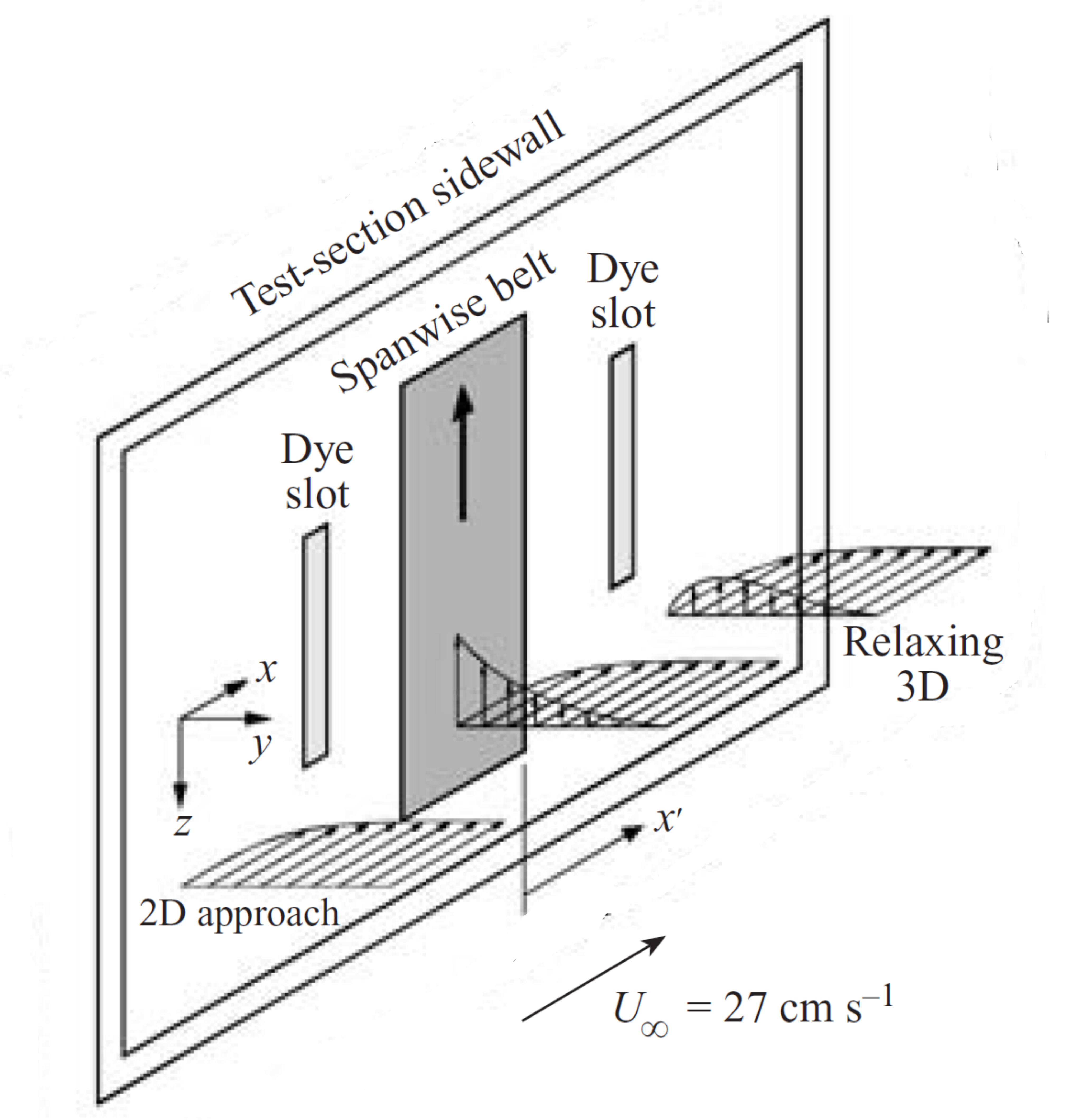}
\includegraphics[width=0.49\columnwidth]{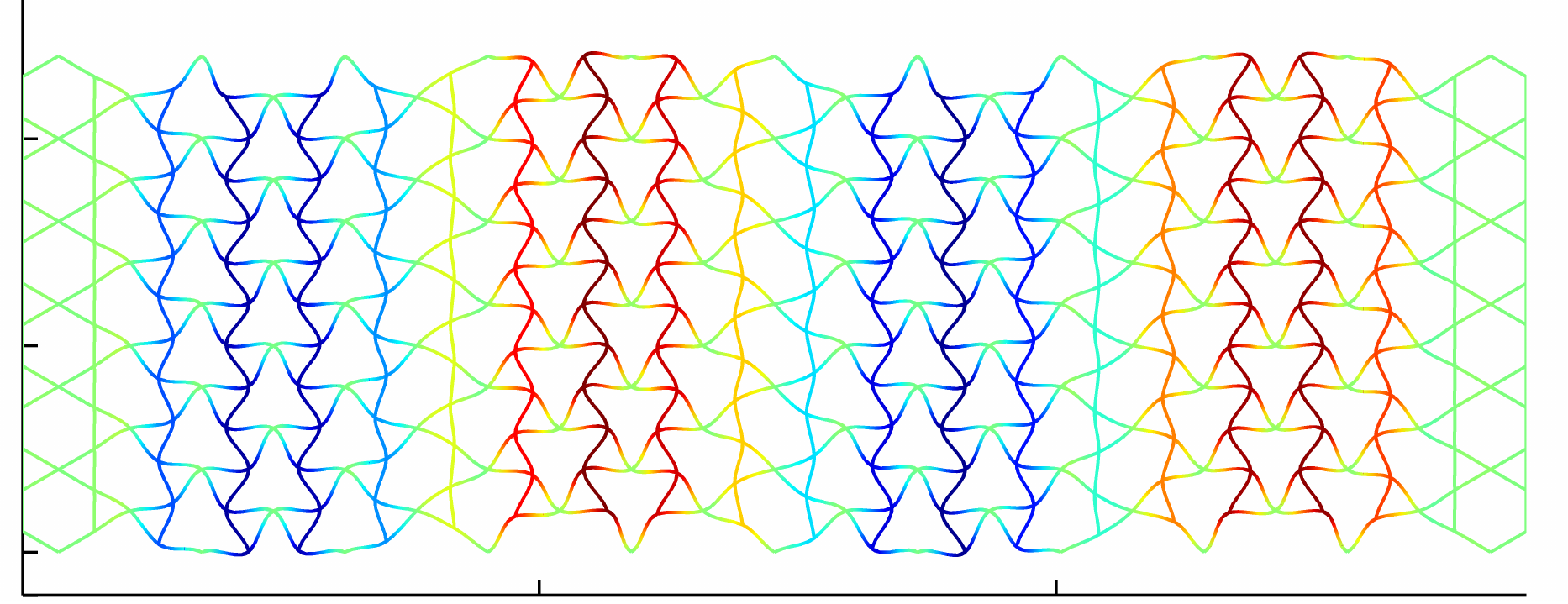}
\put(-50,90){$\rightarrow$ flow}
\caption{Alternative method for the imposition of wall motion. Left: schematic of the wind-tunnel apparatus used by \cite{kiesow-plesniak-1998,kiesow-plesniak-2001,kiesow-plesniak-2003} to study the effect of continuous spanwise wall motion on a turbulent boundary layer; right: schematic of the wall motion of a Kagome lattice employed by \cite{bird-santer-morrison-2018} for generating the wall waves studied numerically by \cite{quadrio-ricco-viotti-2009}. The blue and red regions represent the velocity of the wall in opposite directions. Taken from \cite{kiesow-plesniak-2003}, with permission from Cambridge University Press (CUP) (left) and by courtesy of J. Morrison (right).}
\label{fig:plesniak-morrison}
\end{figure}

The reduced length and inclination of the streaks observed by \cite{ricco-2000,ricco-2004} confirmed the findings of the water-flume study of \cite{kiesow-plesniak-1998,kiesow-plesniak-2001,kiesow-plesniak-2003}, in which a turbulent boundary layer was sheared along the spanwise direction by a belt running at constant speed, depicted in Fig. \ref{fig:plesniak-morrison} (left). \cite{kiesow-plesniak-2003} did not measure the friction drag, but notable differences relative to the oscillating-wall flow were observed, such as the enhanced turbulence intensity and the insensibility of the streak spacing to the actuation. These latter results observed by \cite{kiesow-plesniak-2003} match those of turbulent flows along the axial direction of unidirectionally spinning cylinders, confirming that a spanwise strain that alternates in time and/or along the streamwise direction is a key factor for drag reduction.

The study of \cite{gouder-potter-morrison-2013} of an actuated wind-tunnel boundary layer demonstrated that it is possible to impose spanwise oscillatory wall motion by means of an electroactive polymer sheet and an electromagnetic actuator. Both methods allowed measurements of the turbulent fluctuations at wall-normal locations as low as $y^+=4.5$. A maximum local wall-shear-stress reduction of $\R=16\%$ was estimated from the near-wall mean-velocity gradient, and a global drag reduction over the entire oscillating plate was measured for the first time by a drag balance. 

\cite{gatti-etal-2015} reported the only experimental study in which spanwise wall oscillations were imposed on a turbulent air flow in a duct having a 12:1 width-to-height aspect ratio. This work is also the first one to demonstrate that drag reduction can be achieved despite challenging practical constraints, such as the unevenness and limited extent of the activated surface and the presence of small gaps. The wall motion was achieved by an electroactive polymeric surface and a maximum drag-reduction margin of $\R=2.4\%$ was measured. \cite{gatti-etal-2015} also performed DNS of a channel flow at a similar Reynolds number and identical oscillation parameters as in the experiments. By reducing the dimensions of the oscillating segment in the DNS, they showed that these geometrical restrictions could partially explain the low value of the drag reduction in the experiments.
Additional factors such as the influence of the side walls, that were excluded in the DNS, were also deemed responsible for the low drag-reduction levels in the experiments. Particle Image Velocimetry was successfully used for the first time by \cite{kempaiah-etal-2020} to measure the reduction in wall-shear stress in a wind-tunnel turbulent flow forced by spanwise wall oscillations. A maximum reduction of $\R=15\%$ was computed for $T^+=94$ and $W_m^+=3.34$, in very good agreement with the channel-flow DNS data of \cite{quadrio-ricco-2004}.

\subsection{Experimental studies of pipe flows with circumferential wall oscillations}
\label{sec:-exp-pipe}

Only two experimental studies have tested spanwise wall oscillations in turbulent pipe flows. \cite{choi-graham-1998} provided the first verification of turbulent drag reduction by imposing spatially uniform circumferential oscillations in pipe flows at a maximum bulk Reynolds numbers $Re_D=36300$, based on the pipe diameter and the mean velocity. They observed a plateau in the pressure drop (equivalent to the maximum drag-reduction margin) as the oscillation period was varied. They were not able to conclusively confirm the existence of an optimal period because the differences of the drag-reduction values around the maximum were within the experimental uncertainty.

\cite{auteri-etal-2010} proved experimentally for the first time that the streamwise-travelling waves of spanwise wall velocity given by equation \eqref{eq:waves} and studied earlier numerically by \cite{quadrio-ricco-viotti-2009}, leads to levels of drag reduction that are even higher than those achieved with uniform spanwise oscillations. A sophisticated mechanical system of pipe sections spinning and oscillating in synchronized fashion was built to generate the waves. The measured drag reduction was lower than that numerically computed. This discrepancy was attributed to the discrete form of the actuation, as well as to the differences in the geometries and the influence of the spatial transient in the experiment. The existence of an optimum period of oscillation for maximum drag reduction was proved experimentally for the first time, in satisfactory agreement with the DNS results. 

\subsection{Experimental studies of flows altered by travelling waves of spanwise wall velocity}

As discussed in Section \ref{sec:-exp-pipe}, \cite{auteri-etal-2010} were the first to experimentally demonstrate the drag reduction via the travelling waves given by equation \eqref{eq:waves}. The only other experimental realization of this travelling-wave actuation was carried out by \cite{bird-santer-morrison-2018}. A turbulent boundary layer on a flat plate at $Re_\tau=1125$ was forced by forward- and backward-travelling waves of spanwise wall displacement using a Kagome lattice covered with a thin smooth layer of silicon rubber, as shown in Fig. \ref{fig:plesniak-morrison} (right) and described in \cite{bird-santer-morrison-2015,bird-santer-morrison-2018-ijss}. A maximum drag-reduction level of $\R=21.5\%$ was measured for backward-travelling waves, matching the DNS results of \cite{quadrio-ricco-viotti-2009} and confirming that these waves are more effective in reducing the wall-shear stress than streamwise-uniform oscillations. A low drag-reduction value of $\R=2.6\%$ was reported for waves travelling forward at a phase speed close to one of the two phase speeds that give loci of zero drag reduction on either side of the drag-increase ``blue valley'' region in figure 2 of \cite{quadrio-ricco-viotti-2009} (also shown in Fig. \ref{fig:QRVmap} of this review), again in very good agreement with \cite{quadrio-ricco-viotti-2009}'s numerical results.  

\section{Numerical simulations}
\label{sec:num}

In this section the discussion focuses on the numerical procedures together with an overview on the computational studies in channel, pipe and boundary-layer flows.

\subsection{Computational aspects}

\subsubsection{Navier-Stokes solvers}

Since the effect of wall oscillations on the skin friction involves intricate changes to the turbulent regeneration process, only DNS can predict the drag reduction with a high degree of accuracy, although wall-resolved Large Eddy Simulations (LES) can provide qualitatively credible results. The most efficient numerical methodology adopted for DNS studies of channel flow is based on spectral schemes, and a substantial proportion of studies reviewed herein are of this type. Quadrio and Ricco and their many co-workers have used a pseudo-spectral DNS code based on mixed discretisation, combining Fourier expansions for the homogeneous (streamwise and spanwise) directions with a fourth-order accurate, compact, finite-difference scheme over a non-uniform wall-normal mesh. Other researchers have used fully spectral methods, with Chebyshev polynomials in the wall-normal direction, in order to preferentially cluster collocation points near the walls. Second-order and fourth-order finite-difference and finite-volume methods have also been used in a number of studies. For pipe flow, the spectral-element method has been preferred because it avoids numerical instabilities caused by excessively fine grid clustering near the singularity at the centre of the pipe, arising from the use of regular cylindrical-polar grids.

For boundary-layer flows, various studies have used finite-volume methods \citep{lardeau-leschziner-2013}, spectral schemes (Skote and co-workers in Table \ref{tab:DNS-bl}) and finite-difference approximations \citep{kannepalli-piomelli-2000}). Care must be taken to ensure that the reference boundary layer is correctly represented upstream of the onset of the wall forcing, whether the simulation includes the transitional upstream flow or fully turbulent inflow conditions are used. The requirements placed on simulations and possible pitfalls arising therefrom have been discussed by, among others, \cite{schlatter-orlu-2010}.

As regards to time-advancement, semi-implicit schemes, combining the explicit third-order Runge-Kutta method for the convection terms with the Crank-Nicolson implicit method for the viscous terms, have been popular. A second widely used approach is the fractional-step method, in which either the implicit Crank-Nicolson or an explicit discretisation has been implemented for the diffusion and convective terms. The mesh-resolution requirements vary significantly across the range of numerical schemes used. Conventional finite-difference and finite-volume schemes, especially low-order ones, require significantly finer grids than spectral methods to achieve a similar level of accuracy. In addition, the distribution of solution points in the wall-normal direction varies substantially from one scheme to another. Hence, the number of numerical and computational parameters is too large for a fair comparison in respect of accuracy and efficiency. When adopting a spectral-element method, the accuracy is also dependent on the order of the approximating polynomials. Hence, in view of the many mesh-related parameters at play, it is very difficult to attach any significance to the stated node numbers used in simulations, and they are therefore excluded from the tables.

Ideally, exhaustive mesh- and domain-independence studies should be conducted for the whole set of parameters. This goal is however often impossible because of computing-resource and cost limitations. In the specific case of channel flow, the width and length of the simulation domain rarely exceed $2\pi h$ and $4\pi h$, respectively, where $h$ is the channel half-height. The domain-size challenge is illustrated by a DNS at $Re_\tau=1000$ (corresponding to a Reynolds number based on the channel height and the bulk velocity of about 40000), using a computational box of dimensions $6h \times 12h$, requiring in excess of half a billion points when a fourth-order finite-difference or finite-volume scheme is employed.

\subsubsection{Computational requirements}
\label{sec:domain}

The dimension of the computational box that has been varied most extensively in simulations of channel flows with wall oscillations is its length $L_x$, the smallest one being $L_x=1.193h$ used by \cite{gatti-quadrio-2013} and the largest one being $L_x=36.7h$ used by \cite{quadrio-etal-2009a}.  
The choice of the small box in the former case was motivated by the desire to examine, at tenable resource requirements, flows at the large Reynolds number $Re_\tau=2108$ over a range of actuation parameters, while the latter size was required to accommodate travelling-wave actuation with a large wavelength.

\begin{figure}
\begin{floatrow}

\capbtabbox{
  \begin{tabular}{|c|c|c|}
\hline
\textbf{Study} & $L_x$ & $\R(\%)$\\
\hline
\cite{jung-mangiavacchi-akhavan-1992} & 5.03 & 40\\
\hline
\cite{quadrio-ricco-2004} & 21.0 & 32.8 \\
\hline
\cite{quadrio-ricco-viotti-2009} & 18.85 & 34 (at $T^+_{opt}=106$) \\
\hline
\cite{ricco-etal-2012} (CPG) & 18.85 & 33 (at $T^+_{opt}=70$) \\
\hline
\cite{touber-leschziner-2012} & 12.57 & 38.5 \\
\hline
\cite{gatti-quadrio-2013} & 6.28 & 38.4 \\
\hline
\cite{hurst-etal-2014} & 12.57 & 36.4 \\
\hline
\cite{gatti-quadrio-2016} & 5.0 & 37.4 \\
\hline
\cite{yao-etal-2019} & 12.57 & 35.3\\
\hline
\cite{yuan-etal-2019} & 21.0 & 35.0\\
\hline
\end{tabular}
}{
\caption{Drag reduction in channel flow at ${Re_\tau}=200$ with $W_m^+=12$ and $T^+=100$ (purely temporal oscillations). $L_x$ is the length of the computational box in the streamwise direction.}
\label{tab:DNS-channel-DR}
}
\ffigbox[\Xhsize]
{
\includegraphics[width=0.37\textwidth]{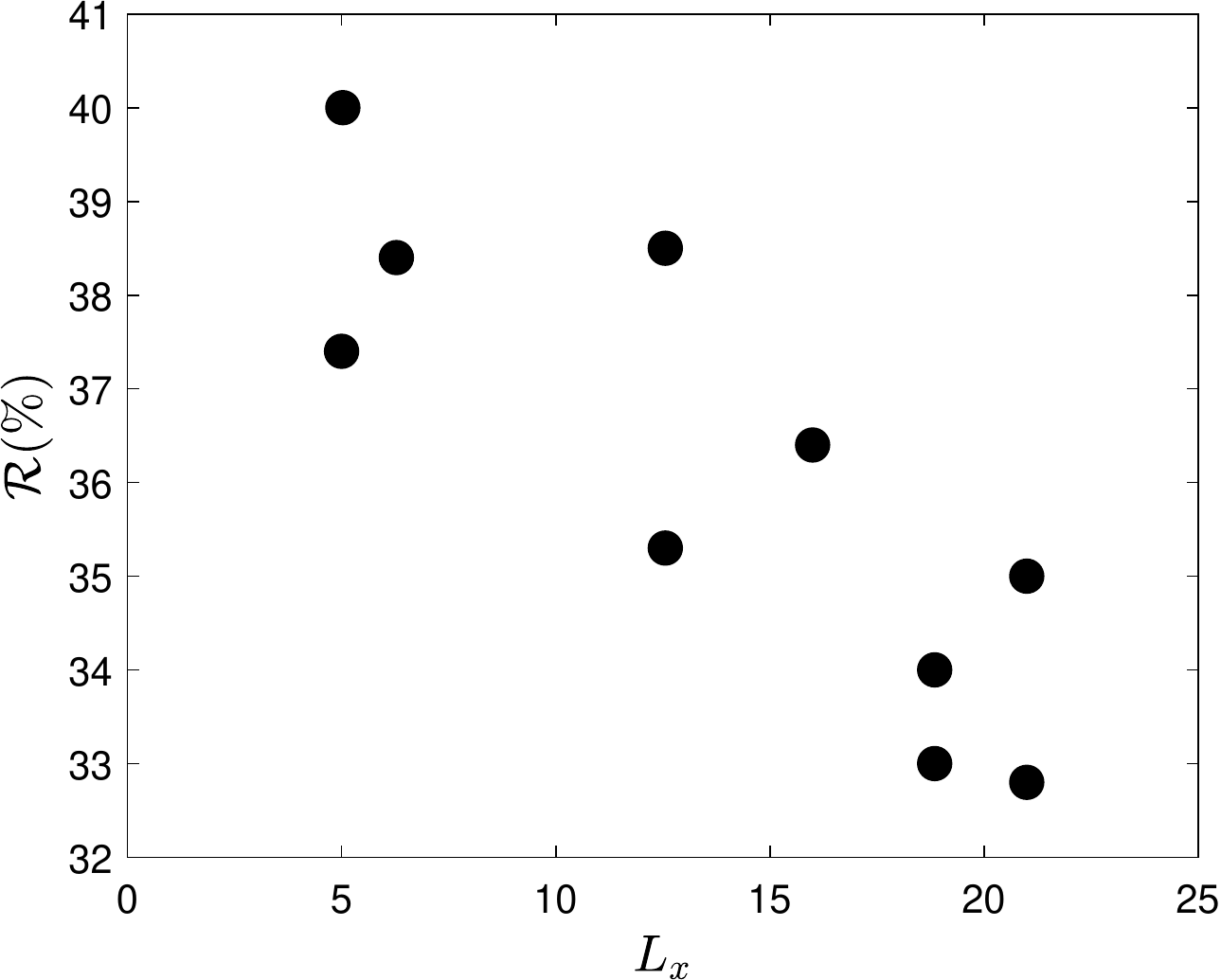}
}
{\caption{drag-reduction margin versus $L_x$ for the ten cases in Table \ref{tab:DNS-channel-DR}.}
\label{fig:chaDR-Lx}
}
\end{floatrow}
\end{figure}

In Table \ref{tab:DNS-channel-DR}, nine simulations of channel flow under the CFR constraint with identical Reynolds number, ${Re_\tau}=200$, and (almost identical) oscillation parameters, $W_m^+=12$ and $T^+=100$, are presented. 
A tenth simulation by \cite{ricco-etal-2012}, subject to the CPG constraint, is also included. The numerical method, order of accuracy and mesh resolution vary between the simulations. The only major variable among the flows that can be correlated to the drag-reduction margin is the size of the computational box, and in particular its length $L_x$.
In Fig. \ref{fig:chaDR-Lx}, the drag reduction values are plotted as a function of $L_x$ for the ten cases (although the width also varies). The influence of the box length was investigated by \cite{gatti-quadrio-2013}, who found that the skin-friction coefficient did not vary greatly with $L_x L_z / h^2$ at ${Re_\tau}=200$, unless the dimensions were very low. The simulation results presented in Fig. \ref{fig:chaDR-Lx} however indicate a qualitative, but distinct dependence of the drag-reduction margin on $L_x$. \cite{gatti-quadrio-2013} also observed that the sensitivity of the drag-reduction margin on $L_x$ is highest when the forcing parameters are near the optimal values for drag reduction, which applies to the conditions of Table \ref{tab:DNS-channel-DR}.

The extension of the purely temporal wall forcing to streamwise-travelling waves given by equation \eqref{eq:waves} with three parameters to optimize (frequency, wavelength, amplitude) has constrained the range of simulations that could be performed at any but modest Reynolds-number values. In recent years, however, simulations have slowly progressed towards larger values (refer to further details in Section \ref{sec:redep}). Apart from the steeply rising resolution requirements as the Reynolds number increases, another challenge posed by such simulations is the large number of time steps and/or large streamwise box size needed to ensure that a sufficient number of actuation periods or wavelengths is included in the simulation, so as to secure the reliability of the statistical analysis. \cite{gatti-quadrio-2013, gatti-quadrio-2016} met this challenge by using very small computational domains when performing simulations at $Re_\tau$ up to 2108. They argued that the box size had a limited effect on the drag-reduction margin mainly because the influence of the computational domain was confined to optimal drag-reduction cases.
Nevertheless, Fig. \ref{fig:chaDR-Lx} suggests that a truly domain-size-independent result for the drag-reduction margin at close-to-optimum parameter values requires large domains and that any reported maximum drag-reduction value needs to be viewed with caution. 

\subsubsection{Numerical issues related to the geometry}

Differently from experimental studies, the geometry of choice for numerical investigations is often the infinitely long channel and, to a lesser extent, the pipe. The reason for preferring these geometrically streamwise-constant internal flows is that the boundary conditions are easier to handle numerically, both in the wall-normal direction (no-slip walls) and in the wall-parallel directions (periodicity). Converged turbulence statistics in channels and pipes are much easier to obtain because of the statistical homogeneity in two coordinate directions (in most cases), compared with external boundary layers for which only the spanwise direction is statistically homogeneous. It follows that for boundary-layer flows the duration of the simulation is typically much longer than for channel and pipe flows in order to obtain fully converged statistics.

The more rapid convergence of the statistics in channel flow has allowed researchers to produce maps that show the dependence of the drag-reduction margin on the oscillation parameters. Furthermore, the analysis of the drag-reduction levels and their dependence on the Reynolds number is facilitated by the channel half-height being an unambiguous outer length scale and the flow being statistically homogeneous in the streamwise direction. The same studies are more difficult to undertake in boundary-layer flows because of the spatially varying thickness of the shear layer. Furthermore, spatial transients raise additional questions about the drag-reduction development, which can only be answered subject to extremely large computational domains. 

Fortunately, the basic physical mechanisms and their dependence on the wall-forcing parameters depend only weakly on whether or not the flow is internal or external. Hence, the extensive channel-flow investigations presented in the literature are pertinent to, and valuable for, efforts towards practical implementations in external-flow applications. With this commonality acknowledged, it is nevertheless important to bring out and carefully discuss the unique characteristics of internal and external flows and their control.

\subsection{Numerical studies of internal flows}
\label{sec:internal}

In this section, the numerical simulations of internal flows with spatially-uniform spanwise wall forcing are discussed, with a focus on the drag-reduction margin, the influence of the oscillation parameters, the flow geometry and the flow conditions (CFR, CPG, CPI). Channel flows with spanwise oscillations are presented in Section \ref{sec:channel-temp} and pipe flows with circumferential oscillations are discussed in Section \ref{sec:pipe-temp}. Channel flows altered by streamwise-travelling or standing waves of spanwise wall velocity are considered in Section \ref{sec:channel-form}. 

\subsubsection{Numerical studies of channel flows with streamwise-homogeneous wall oscillations}
\label{sec:channel-temp}

This section reviews the studies on the effects of purely temporal, streamwise-homogeneous oscillations in channel flows. Table \ref{tab:DNS-channel-temp} summarized these studies. Not all papers listed in the table are reviewed in detail; some articles that are specifically devoted to elucidating the physical interactions are further discussed in Section \ref{sec:phys}.

{\scriptsize
\begin{sidewaystable}
\centering
\begin{tabular}{|c|c|c|c|c|c|}
\hline
\centering
\textbf{Study}                                & Constraint & ${Re_\tau}$ & $W_m^+$ & $T^+$ & max ${\R(\%)}$\\
\hline
\hline
\cite{jung-mangiavacchi-akhavan-1992}         & CFR & 200 & 12 & 100 & 40 \\
\hline
\cite{baron-quadrio-1996}                     & CFR & 200 &  4, 9, 13, 17 & 100 & 40\\
\hline
\cite{miyake-tsujimoto-takahashi-1997}        & CPG (CPI?) & 150 &  12 & 100 & 35\\
\hline
\cite{choi-xu-sung-2002}                      & CFR & 100 & 1, 5, 10, 20 & $1-200$ & 45.4 ($T_{opt}^+=150$)\\
                                              &     & 200 & 5, 10, 20 & $50-200$ & 39.2 \\
                                              &     & 400 & 5, 10, 20 & $50-200$ & 34.1\\
\hline
\cite{quadrio-ricco-2003}                     & CFR & 200 & $3-27$ & $50-200$ & 50+ (transient)\\
\hline
\cite{quadrio-ricco-2004}                     & CFR & 200 & $1.5-27$ & $5-750$ & 44.7\\
\hline
\cite{xu-huang-2005}                          & CFR & 173 &  15 & 90 & 36\\
\hline
\cite{zhou-ball-2006,zhou-ball-2008}          & CFR/CPG & 180 & $2-70$ & $10-200$ & 88 \\
\hline
\cite{ricco-quadrio-2008}                     & CFR & 200 & $D_m^+=100, 200, 300$   & 28 cases & 36\\
                                              &     & 400 & 12 & 30, 125, 200 & 28.1 \\
\hline
\cite{ricco-etal-2012}                        & CPG & 200 & 12 & $0-500$ & 33 ($T_{opt}^+=70$)\\
\hline
\cite{touber-leschziner-2012}                 & CFR & 200 & 12  & $50-1000$ & 38.5 \\
                                              &     & 500 & 12 & 100, 200 & 32.4 \\
                                              &     & 1000 &12  & 100 & 29 \\
\hline
\cite{yakeno-etal-2014}                       & CPG & 150 & 3, 7, 12 & $16-500$ & 42.8 ($T_{opt}^+=75$)\\
\hline
\cite{agostini-touber-leschziner-2014}        & CFR & 1000 & 12 & 100, 200 & 29 \\
\hline
\cite{hasegawa-quadrio-fronhapfel-2014}       & CPI & 200 & $\gamma$ varied & 9 cases & NR\\
\hline
\cite{gatti-etal-2015}                        & CFR & 200 & 3.6  & 53    & 10 \\
\hline
\cite{ge-jin-2017}                            & CPG & 180 & 12  & 100    & 31 \\
\hline
\cite{gatti-etal-2018}                        & CPI & 200 & 4.47 & 125.5 & 17.2 \\
\hline
\cite{yao-etal-2019}                          & CFR & 200 & 12 & $16-628$ & 35.3\\ 
                                              &     & 497 & 12 & $16-628$ & 28\\
                                              &     & 1000 & 12 & $16-628$ & 25.9\\
                                              &     & 1998 & 12 & $63-157$ & 23.2\\
\hline
\cite{yuan-etal-2019}                         & CFR & 200 & 3, 6, 12, 18 & 50, 100, 200 & 41.6 \\
\hline
\cite{yang-hwang-tsfp-2019}                   & CFR & 800 & 12 & 100 & 25 \\
\hline
\end{tabular}
\caption{Numerical studies of channel flows with purely temporal wall oscillation (NR=not reported).}
\label{tab:DNS-channel-temp}
\end{sidewaystable}
}

The first DNS studies of turbulent channel flows over an oscillating wall were performed by \cite{jung-mangiavacchi-akhavan-1992}, \cite{akhavan-jung-mangiavacchi-1993}, and \cite{akhavan-jung-mangiavacchi-1993b}. Inspiration for this study came from earlier experiments \citep{bradshaw-pontikos-1985,driver-hebbar-1987} and simulations \citep{moin-etal-1990} of three-dimensional wall-bounded flows in which the onset of spanwise flow resulted in a temporary drag reduction. \cite{jung-mangiavacchi-akhavan-1992} examined the consequences of imposing an oscillatory spanwise cross-flow for a range of $T^+$ values, and compared the effects of this forcing to those given by oscillating the wall. Only one case was reported for the wall motion, at $T^+=100$, from which \cite{jung-mangiavacchi-akhavan-1992} concluded that the results of the two forcing methods were essentially identical. The results reported by \cite{jung-mangiavacchi-akhavan-1992} included a rough estimate of the optimal oscillation period, $T^+= 100$, for which the maximum drag-reduction margin was $\R=40\%$, and the observation that the percentage drop in the magnitude of the spanwise and wall-normal turbulent fluctuations was larger than that of the streamwise fluctuation. At the lowest frequency, $T^+= 500$, drag increase was observed. 

\cite{baron-quadrio-1996} reported results from a DNS of a channel flow at $Re_\tau=200$ subjected to spanwise wall oscillations with a period $T^+ = 100$ and wall velocities between $W_m^+=4$ and $17$. The main observation was that the actuation caused a shift of all turbulent quantities away from the wall towards the interior of the channel, signifying a thickening of the viscous sublayer. Other effects were the reductions of all the turbulence-budget terms in the near-wall region and of the level of dissipation at the wall. The maximum drag reduction was $\R=40\%$ and the optimal oscillation amplitude in terms of the overall energy balance, including the power spent on actuating the wall, was found to be at their lowest wall velocity, $W_m^+=4$. 

\cite{miyake-tsujimoto-takahashi-1997} performed DNS of channel flow at $Re_\tau=150$ with one oscillating wall at $T^+=100$ and $W_m^+=12$. The transient flow development from the start of the oscillation was tracked and, after five oscillation periods, a drag-reduction margin of $\R=35\%$ was reached over the moving wall.
They claimed that the simulations were carried out under CPG conditions, but both a decrease of the wall-shear stress and an increase of the bulk velocity were reported as effects of the wall motion. This result suggests that, for the first time, a constant streamwise power input, $\mathcal{P}_x=2 \tau_w U_b$, might have instead been the constraint because the fully-developed friction reduction, averaged over the two walls, was about 17\%, i.e., of very similar magnitude of the increase of the wall-normal integral of the mean velocity profile. This simulation thus appears to be more akin to the CPI simulation discussed at the end of Section \ref{sec:basic}, although the constant power fixed by \cite{frohnapfel-hasegawa-quadrio-2012} under the CPI constraint also included the power used to move the wall. In Table \ref{tab:DNS-channel-temp}, we have labelled the simulation of \cite{miyake-tsujimoto-takahashi-1997} with a tentative ``CPI?''.

\cite{choi-xu-sung-2002} also performed DNS of channel flow, covering the ranges $T^+=1-200$ and $W_m^+=1-20$ for $Re_\tau=100, 200$ and $400$. At $T^+=100$ and maximum amplitude $W_m^+=20$, the maximum drag reduction recorded were 44.5\%, 39.2\% and 34.1\% for the three Reynolds numbers, respectively. For $Re_\tau=100$ the optimal period shifted to $T^+=150$, for which the drag-reduction margin reached 45.4\%. 
A major outcome of this study was an empirical correlation that links the drag-reduction margin to a parameter that combines the friction Reynolds number, the wall-scaled amplitude, depth of penetration of the Stokes layer, and acceleration of the Stokes layer at $y^+=5$.

\cite{quadrio-ricco-2003} conducted DNS of channel flow at $Re_\tau=200$ and $W_m^+=3-27$ and analysed the transient behaviour in the  first few wall-oscillation cycles, recording a maximum drag-reduction level in excess of 50\%. They noted that the spanwise velocity profile closely followed the analytical laminar solution even during the first oscillation period following the start of the actuation. The streamwise shear stress instead adjusted much more slowly, the duration of its transient depending on the wall-velocity magnitude. The streamwise stress was found to exhibit a non-monotonic development in time, with a pronounced minimum prior to approaching the fully developed condition.  

\begin{figure}
\centering
\includegraphics[width=0.6\textwidth]{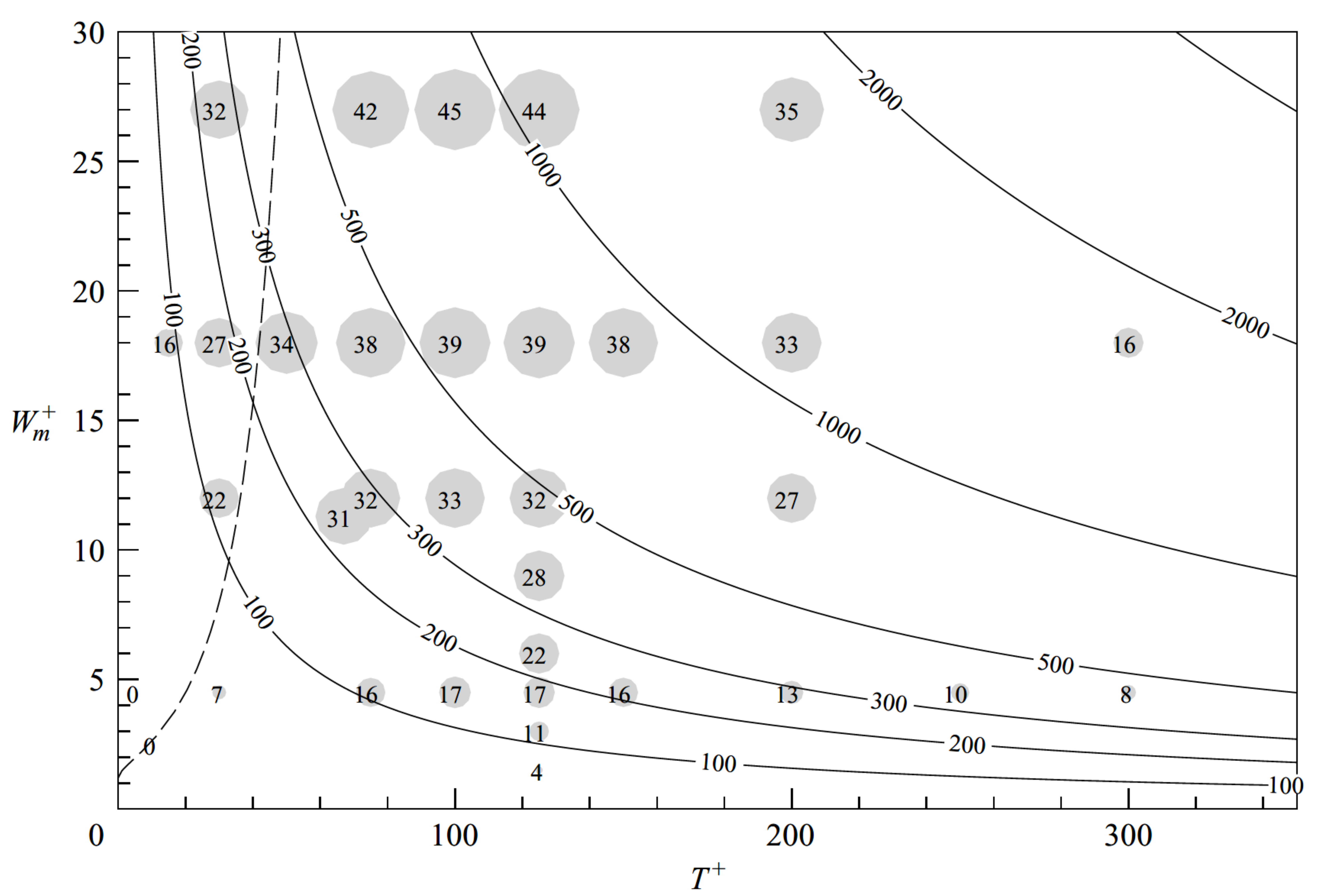}
\caption{Drag-reduction margin as a function of $W_m^+$ and $T^+$ at $Re_\tau=200$ . The size of the circles is proportional to the drag-reduction margin. The hyperbolas represent constant maximum displacement $D_m^+$. The dashed curve represents the optimum $T^+$ at a fixed $D_m^+$. Taken from \cite{quadrio-ricco-2004}, with permission from Cambridge University Press (CUP).}
\label{fig:QRmapD}
\end{figure}

In a later paper \cite{quadrio-ricco-2004} reviewed much of the earlier DNS and experimental work published at that time, and reported results for 37 new simulations at $Re_\tau=200$ with wall-velocity amplitudes in the range $W_m^+=0-27$, in an effort to derive a definitive statement on the dependence of the drag reduction on the oscillation parameters. The misconception of the independence of the drag-reduction margin on $T^+$ at fixed $W_m^+$ was attributed to differences in the Reynolds number, the influence of transients and the non-optimality of the actuation parameters. A monotonic increase in the drag-reduction margin with $W_m^+$ was found, but at a progressively declining rate, suggesting a saturation level, as shown in Fig. \ref{fig:QRmapD}. The maximum computed drag reduction was $\R=45\%$ and the scaling parameter first proposed by \cite{choi-xu-sung-2002} was shown to predict the drag-reduction margin very well for periods smaller than the optimum.

The drag-reduction map shown in Fig. \ref{fig:QRmapD} clarifies that the optimum period at fixed $W_m^+$ is within the range $T^+=100-125$ for the entire range of wall velocity at the prescribed low Reynolds number. 
\cite{quadrio-ricco-2004} noted that the wall oscillation becomes decoupled from the near-wall turbulence dynamics at high $T^+$ values. When $T^+$ is larger than the optimum, which coincides with the double of a pseudo-Lagrangian time scale representing a typical survival time of the longest-lived and statistically significant turbulent structures \citep{quadrio-luchini-2003}, the near-wall streaks have enough time to develop their inner dynamics between successive sweeps of the Stokes layer. During this time interval, the near-wall turbulence tends to readapt to its natural equilibrium state and to restore the unperturbed value of friction drag. When instead $T^+$ is close to the optimum, the forcing time matches the characteristic time of the streak dynamics and this mechanism is effective in disrupting the streaks, which do not have sufficient time to readjust to their natural life cycle. The near-wall flow is therefore more homogeneous and the near-wall turbulence intensity is attenuated. This paradigm offered an explanation for the existence of the optimum period.
The study also examined the net-energy savings when the actuation power is taken into account. A net saving was only attained at low velocity amplitudes and the maximum net saving was about $\mathcal{P}_{net}=7\%$ at $W_m^+=4.5$, at which the gross drag reduction was $\R=17\%$ at best. 
\cite{quadrio-ricco-2004} further explained that the optimal period at fixed $W_m^+$ typically emerges from numerical investigations where the actuation parameters can be varied independently, while another optimal period was reported when $D_m^+$ was fixed, the usual constraint in experiments. The latter optimal period was found to be lower than the nearly constant optimum period at fixed $W_m^+$ and to depend on the maximum displacement itself. 
The dashed line in Fig. \ref{fig:QRmapD} shows the optimum $T^+$ at fixed $D_m^+$ and was obtained from an implicit expression derived from the scaling parameter used for the drag-reduction estimation. The drag reduction was found to be almost independent of $T^+$ near the optimum at fixed $D_m^+$. For this reason, experimental campaigns, conducted at constant $D_m^+$, only reported the monotonic increase of the drag-reduction margin with $T^+$ and failed to capture this optimum. The almost constant drag reduction occurs because of the competing effect between lowering $T^+$, which gives low drag reduction, and increasing $W_m^+$, which is instead beneficial. 

\cite{ricco-quadrio-2008} extended the study of \cite{quadrio-ricco-2004} to further elucidate the differences between the two optimal oscillation periods, at constant $W_m^+$ and at constant $D_m^+$. Simulations with $D_m^+$ fixed at $Re_\tau=200$ and $400$ were performed to better predict the optimum $T^+$ at fixed $D_m^+$. They further exploited the scaling parameter used in \cite{quadrio-ricco-2004} to predict the power spent and the net energy saved with good accuracy.

\cite{xu-huang-2005} performed a channel-flow DNS at $Re_\tau=173$, $W_m^+=15$ and $T^+=90$, and reported a drag-reduction margin of 36\%. The Reynolds-stress budgets were analysed during the first two oscillation cycles to elucidate how the transfer of turbulent kinetic energy was inhibited by the oscillating wall, leading to drag reduction.

\cite{zhou-ball-2006} and \cite{zhou-ball-2008} reported simulation results of a channel flow subjected to wall oscillations under CPG and CFR conditions at $Re_\tau=180$. With oscillation parameters set at $W_m^+=12.48$ and $T^+=100$, the drag-reduction margin was $\R=29.9\%$ and $\R=27.2\%$ for CFR and CPG, respectively, with the difference attributed to the change in Reynolds number. Both drag-reduction values were lower than the value of $\R=33\%$ obtained by \cite{quadrio-ricco-2004}. 
They explored a large range of values of wall velocity and oscillation periods. The optimal oscillation periods were estimated to be $T^+=100$ and $T^+=50$ for the CFR and CPG configuration, respectively, with the difference explained by the altered flow structures in the two different flows, in particular a smaller thickness of the Stokes layer required in the CPG case. 
They revisited the scaling of the drag-reduction margin with the maximum wall velocity and computed cases with values as high as $W_m^+=70$, finding that an asymptotic drag-reduction level was attained beyond $W_m^+\approx 20$, within the range of actuation periods $T^+=10-143$, under the CFR condition. When the CPG constraint was used, the drag reduction continued to increase until almost $\R=90\%$, but this finding was undoubtedly due to relaminarization at such low levels of friction Reynolds number. \cite{zhou-ball-2008} also moved the wall obliquely with respect to the streamwise direction, concluding that purely spanwise forcing gives the best performance in terms of drag reduction. Saturation occurred at the inclination angle of $60^\circ$, beyond which the drag-reduction margin matched the value given by purely spanwise oscillations.

A further set of simulations were performed by \cite{ricco-etal-2012} for channel flow at $Re_\tau=200$ under the CPG constraint. The optimum period at $W_m^+=12$ was found to be $T^+=70$, i.e., lower than $T^+=100-125$, accepted as the optimum under the CFR constraint. A second objective pursued by \cite{ricco-etal-2012} was to examine the turbulence phenomena provoked by the actuation and identify the fundamental origin of the drag reduction. Using the CPG data for the case with $W_m^+=12$ and $T^+=100$, resulting in $\R=31\%$, \cite{ricco-etal-2012} examined the energy budgets to elucidate the role of, and changes in, the turbulence dissipation rate. The physical mechanism is described in more detail in Section \ref{sec:prod_and_diss}.

\cite{touber-leschziner-2012} performed DNS in a channel flow at $Re_\tau=500$ and $W_m^+=12$ for $T^+=50, 100, 200, 400$ and $1000$, and also reported results from wall-resolved large-eddy simulations (LES) at $Re_\tau=1000$. The accuracy of the LES, at least in terms of the drag-reduction level, was confirmed by comparing the LES result to the DNS result at $T^+=100$, the difference being 2\%. The primary purpose of this study was not, however, to quantify the drag-reduction margins ($\R=32\%$ at $Re_\tau=500$ and $\R=27\%$ at $Re_\tau=1000$), but to undertake a wide-ranging analysis of the turbulence statistics at $Re_\tau=500$ to gain a better view of the root causes of the drag reduction. To this end, \cite{touber-leschziner-2012} decomposed all velocity components into the mean, periodic and purely stochastic parts to examine the turbulence statistics and the visualized fields. By conducting phase-wise averaging, the contributions of the phase fluctuations to the total fluctuations were shown to be small for the streamwise component and non-existent for the wall-normal component. They noted that the streamwise Reynolds stress near the wall was lower for $T^+=200$ than for the optimal period $T^+=100$, while the spanwise Reynolds stress increased and rose above the nominal level for $T^+=200$. 
One interesting observation derived from the visualization is that the streaks were not simply realigned periodically in response to the actuation, but they weakened substantially when the Stokes strain changed rapidly, and strengthened again when the Stokes strain changed at a low rate. This behaviour was due to the strain variations in the buffer layer, which contributed to the production-rate of turbulence, implying that the production of the stresses in the buffer region was the main driver of the drag-reduction mechanism. The Reynolds number was sufficiently high to bring to light distinctive large-scale footprints of the energetic outer log-layer structures -- or ``super-streaks'' -- imprinted onto the small-scale near-wall structures. A decomposition of the large- and small-scale motions was included to identify the effects of the former on the skin friction.

\cite{agostini-touber-leschziner-2014}, following on from the previous study by \cite{touber-leschziner-2012}, focused on the mechanisms underpinning drag oscillations at the non-optimal actuation period $T^+=200$ and $Re_\tau=1020$. For this condition, the drag experienced significant periodic fluctuations, synchronised with the actuation. Such fluctuations were instead insignificant at the near-optimum value $T^+=100$. The behaviour at $T^+=200$ motivated \cite{agostini-touber-leschziner-2014} to examine a wide range of phase-locked statistics, including Reynolds-stress budgets during the drag-increase and drag-decrease portions of the actuation cycle in an effort to pin down cause-and-effect processes that drive both drag decrease and drag increase.

\cite{yakeno-etal-2014} varied the actuation period in the range $T^+=16-500$ in a DNS study of channel flow under the CPG constraint at $Re_\tau=150$ and found that a period $T^+=75$ produced a larger drag-reduction level than $T^+=100$, in accordance with the results of \cite{ricco-etal-2012}. The main purpose of the study was, however, to further investigate the turbulence mechanisms responsible for the drag reduction by conditionally averaging events that contribute to the wall-shear stress. One outcome of their analysis was an empirical model that correlated the drag reduction linearly to the spanwise shear at two different wall-normal positions, one where streamwise vortices interacted with the wall ($y^+=10$) and the other where streamwise vortices were spanwise-tilted by the spanwise shear ($y^+=15$). The model reproduced the DNS-recorded response of the increased bulk velocity in the range $W_m^+=3-12$ at $Re_\tau=150$ and 200. 

\cite{hasegawa-quadrio-fronhapfel-2014} performed channel-flow DNS with wall oscillation at $Re_\tau=200$, using the CPI constraint. They introduced a parameter $\gamma$ that quantified how much of the total available power was consumed by the control. \cite{hasegawa-quadrio-fronhapfel-2014} concluded that the optimum allocation of power from pumping to the control mechanism was $\R=10\%$ ($\gamma=0.1$), which yielded a 3\% increased flow rate.

Channel-flow simulations at $Re_\tau=200$ with wall oscillation at $W_m^+=3.6$ and $T^+=53$ were compared with duct-flow measurements by \cite{gatti-etal-2015}. The DNS with the complete wall in oscillation produced a drag-reduction margin of 10\%, while the drag-reduction levels declined to $\R=8.3\%$ and $\R=6.3\%$ for an oscillating segment with reduced length and width, respectively. For the former restriction of the oscillating segment,
the spatial evolution of the drag reduction downstream of the moving wall was compared with results in the literature, and good agreement with previous numerical and experimental data was achieved, in particular with those measured by \cite{ricco-wu-2004-a} and those computed via DNS by \cite{skote-2012}. Although the flow conditions among the datasets were different, the meaningful comparison was possible by normalizing the available data by the respective fully-developed drag-reduction margins.

In a study by \cite{gatti-etal-2018}, a channel-flow DNS with wall oscillation at $Re_\tau=200$ with $W_m^+=4.47$ and $T^+=125.5$ was performed under CPI conditions to investigate the energy-flux balance and to compare the oscillatory and opposition control methods. The underlying assumption was that two flows with identical CPI constraints should also have the same total energy dissipation rate. The streamwise pressure gradient (i.e., the pumping power) was therefore adjusted according to the value of the control power, such that the sum remained equal. The conclusion was that the two dissipation components, associated with the mean flow and the turbulence field, respectively, could not, on their own, characterize the success of the control. By introducing a decomposition of the mean velocity into the laminar profile and a deviation from it, together with an extended energy box, \cite{gatti-etal-2018} showed that the factors determining the success of the control strategy were the pumping power used for producing the turbulence and the total power dissipated by turbulence through viscous effects.

\cite{yao-etal-2019} reported the results of wall oscillations in channel flow up to $Re_\tau=2000$ and showed that the optimal period decreased from $T^+=100$ at $Re_\tau=200$ to $T^+=90$ at $Re_\tau=1000$ and to $T^+=79$ at $Re_\tau=2000$. The maximum drag reduction obtained at $Re_\tau=2000$ was $\R=23.2\%$, i.e., marginally lower than the value of $\R=24\%$ obtained by \cite{gatti-quadrio-2013} at $Re_\tau=2100$, as discussed in Section \ref{sec:channel-form}. \cite{yao-etal-2019} however pointed out that the simulations of \cite{gatti-quadrio-2013} were performed with a small computational box, which, according to them, tends to result in excessive drag-reduction levels. 

Channel-flow simulations at $Re_\tau=200$ with 12 combinations of $W_m^+$ and $T^+$ were performed by \cite{yuan-etal-2019}. The results compared successfully with those of \cite{quadrio-ricco-2004}. To analyse the link between drag reduction and the turbulence mechanisms during the oscillation of the wall, \cite{yuan-etal-2019} used the relationship between the transport of vorticity and the Reynolds shear stress, concluding that the drag reduction was linked to the increase in turbulence-energy dissipation. 

\cite{yang-hwang-tsfp-2019} performed one DNS of channel flow at $Re_{\tau}=800$ with $W_m^+=12$ and $T^+=100$, resulting in a drag-reduction margin of 25\%. The oscillation parameters were varied for a number of simulations of ``exact coherent states'', as a model of turbulence at larger Reynolds numbers in order to understand the impact of the wall oscillation on the turbulence structures.

The most recent study of a turbulent channel flow by \cite{dong-etal-2019}, based on the LES WALE model, confirmed previous known drag-reduction results and showed for the first time that the pressure fluctuations are also attenuated alongside the velocity fluctuations when the wall oscillates. 

\subsubsection{Numerical studies of pipe flows with circumferential wall oscillations}
\label{sec:pipe-temp}

In this section, numerical investigations pertaining to pipe flows forced by purely temporal oscillations are reviewed and summarized in Table \ref{tab:DNS-pipe-temp}. The number of pipe-flow studies is far lower than those for the channel-flow geometry. In pipe flows, the drag-reduction margins and their dependence on the oscillation parameters and the Reynolds number are similar to those obtained in channel flow. As in Section \ref{sec:channel-temp}, articles predominantly devoted to elucidating the physical interactions are further discussed in Section \ref{sec:phys}.

{\scriptsize
\begin{sidewaystable}
\begin{tabular}{|c|c|c|c|c|c|}
\hline
\centering
\textbf{Study}                       & Constraint & ${Re_\tau}$ & $W_m^+$ & $T^+$ & max ${\R(\%)}$\\
\hline
\hline
\cite{quadrio-sibilla-2000}          & CFR        & 172  & 14 (3, 6, 8, 12) & 100 (50, 150)              & 40\\
(Finite difference)                  &            &      &                  &                           &   \\
\hline
\cite{nikitin-2000}                  & CFR        & 133  & 3, 6, 9          & 105 ($T^+$ varied -- NR) & 8, 20, RLM \\
\hline
\cite{choi-xu-sung-2002}             & CFR        & 150  & 5, 10, 20        & $5-200$                     & RLM \\
(Spectral element)                   &            &      &                  &                           &    \\
\hline
\cite{duggleby-ball-paul-2007}       & CPG        & 150  & 20               & 50                        & 38 \\
(Spectral element)                   &            &      &                  &                           &  \\
\hline
\cite{coxe-peet-adrian-2019}         & CPG        & 170, 360 &  10          & 100                       & 37, 36 \\
(Spectral element)                   &            &      &                  &                           &  \\
\hline
\end{tabular}
\caption{DNS studies of pipe flows with purely temporal wall oscillation (NR=not reported, RLM=relaminarization).}
\label{tab:DNS-pipe-temp}
\end{sidewaystable}
}

The DNS performed by \cite{quadrio-sibilla-2000} at $Re_{\tau}=172$ with $W_m^+=14$ and $T^+=100$ demonstrated a wall-normal displacement of the low-speed streaks relative to the quasi-streamwise vortices, in line with the earlier observations by \cite{baron-quadrio-1996} for channel flows. The effects of the wall oscillations on the flow structures were especially evident in the viscous sublayer and in the near-wall buffer layer. Based on their observations, \cite{quadrio-sibilla-2000} proposed a phenomenological set of interactions between the quasi-streamwise vortices and the streamwise-velocity fluctuations, leading to a qualitative explanation for the drag reduction. Their results were compared with the experimental data of \cite{choi-graham-1998}, and the conclusion was that the lower drag-reduction margin obtained in the experiments was due to the non-optimal oscillation-period used. \cite{quadrio-sibilla-2000}  also noted that the spanwise velocity profile closely followed the Stokes-layer solution \eqref{eq:stokes}.

Another DNS investigation of an oscillating pipe flow was conducted by \cite{nikitin-2000} at $Re_\tau=133$. A drag-reduction margin of $\R=20\%$ for $T^+=105$ and $W_m^+=9$ was obtained, but the computations with larger oscillation periods did not yield an increase in drag, in contrast to the channel-flow results by \cite{jung-mangiavacchi-akhavan-1992}. He attributed this difference to the centrifugal forces present in the pipe, hence explaining a key difference between the two geometries. 

\cite{choi-xu-sung-2002} performed DNS of pipe flow at $Re_\tau=150$, with $T^+=5-200$ and $W_m^+=5-20$. At $T^+=100$, the maximum drag reduction was $\R=35\%$ for $W_m^+=10$, while relaminarization occurred at large periods and wall velocities. Based on a conditional statistical analysis of the quasi-streamwise vortices, \cite{choi-xu-sung-2002} offered a phenomenological interpretation for the mechanism responsible for drag reduction that was very similar to the conceptual model of \cite{quadrio-sibilla-2000}. 

\cite{duggleby-ball-paul-2007} performed simulations of a pipe flow subjected to circumferential oscillations under CPG condition to maintain a constant friction Reynolds number, $Re_\tau=150$. The actuation parameters, $W_m^+ = 20$ and $T^+=50$, were chosen such that a maximum drag reduction (without relaminarization) was achieved. 
They remarked that this $T^+$ value was much lower than that identified by \cite{quadrio-ricco-2004} to yield the maximum drag reduction for channel flow at $Re_\tau=200$ under CFR condition ($T^+=100-125$), possibly due to the CPG constraint. This result is also consistent with the findings by \cite{zhou-ball-2008} who estimated that the optimal oscillation periods were $T^+=100$ and $T^+=50$ for the CFR and CPG configuration, respectively. \cite{duggleby-ball-paul-2007} reported a 27\% increase in mean velocity, which corresponds to a drag reduction of $\R=38\%$. 
\cite{coxe-peet-adrian-2019} performed DNS of a pipe flow under CPG conditions at $Re_\tau=170$ and 360 with $W_m^+=10$ and $T^+=100$, and explained that the wall-shear stress was reduced because the oscillating wall affected the near-wall streamwise vortex pairs asymmetrically. The flow changes reported by \cite{duggleby-ball-paul-2007} and \cite{coxe-peet-adrian-2019} are discussed further in Section \ref{sec:phys}.

\subsubsection{Numerical studies of flows altered by travelling waves of spanwise wall velocity}
\label{sec:channel-form}

Investigations on flows forced by streamwise-travelling waves of spanwise wall velocity, given by equation \eqref{eq:waves}, are reviewed in this section and summarized in Table \ref{tab:DNS-channel-form}. Over recent years, the number of studies of this type of forcing has surpassed those of flows altered by purely temporal oscillations, arguably because of the larger drag-reduction and, more important, the net power-saving margins that can be obtained by the wavy wall motion given by equation \eqref{eq:waves}.

{\scriptsize
\begin{sidewaystable}
\begin{tabular}{|c|c|c|c|c|c|}
\hline
\centering
\textbf{Study} & \textbf{Forcing} & \textbf{Constraint} & $Re_\tau$ & $W_m^+$ &  max ${\R(\%)}$\\
\hline
\hline
\cite{viotti-quadrio-luchini-2009}  & SW & CFR & 200 & $1-27$ & 52 \\
\hline
\cite{quadrio-ricco-viotti-2009}    & TW (250 cases) & CFR & 200 & 12 ($2-30$) &  48 (60)\\
                                    & TW (1 case) & CFR &100 & 12        &  RLM \\
                                    & TW (1 case) & CFR &400 & 12        &  42 \\
\hline
\cite{yakeno-etal-2009}             & SW/OW       & CFR/CPG & 150 & $1-10$      & 55 \\
\hline
\cite{quadrio-ricco-2011}           & TW          & CPG   & 200 & 12        &  45\\
\hline
\cite{gatti-quadrio-2013}           & TW (Along 5 & CFR   & 199  & 12 & 49\\
                                    & lines in $\omega^+ - \kappa_x^+$ plane) & CFR & 951  & 12 & 42\\
                                    & TW (12 cases)  & CFR & 2108 & 12 & 37\\
\hline
\cite{hurst-etal-2014}              & TW (86 cases)  & CFR  & 200  & 12 & 50 \\
                                    & TW (90 cases)  & CFR  & 400  & 12 & 44 \\
                                    & TW (106 cases) & CFR  & 800  & 12 & 40  \\
                                    & OW $T^+=100$   & CFR  & 1600 & 12 & 22 \\
                                    & SW $\lambda_x^+=800$ & CFR  & 1600 & 12 & 33 \\
\hline
\cite{gatti-quadrio-2016}           & TW (2010 cases)  & CFR     & 199  & 12 ($2-20$) & 50\\
                                    & TW (2010 cases)  & CFR     & 948  & 12 ($2-20$) & 39\\
                                    & TW (10 cases)    & CFR/CPG & 200  & 7         & 25\\
                                    & TW (10 cases)    & CFR/CPG & 1000 & 7         & 17\\
\hline
\cite{gatti-etal-2018a}             & TW (1 case)  & CPG & 200  & 7         & 36.4 \\
                                    & TW (1 case)  & CPG & 1000 & 7         & 26.6 \\
                                    & TW (1 case)  & CFR & 1000 & 7         & 27.7 \\
\hline
\end{tabular}
\caption{Numerical studies of channel flows with streamwise-traveling waves of spanwise wall velocity, given by \eqref{eq:waves}. OW= oscillating wall ($\kappa_x=0$), SW= standing waves ($\omega=0$), TW= streamwise-traveling waves ($\omega\neq0$, $\kappa_x\neq0$).}
\label{tab:DNS-channel-form}
\end{sidewaystable}
}

\begin{figure}
\centering
\includegraphics[width=0.6\textwidth]{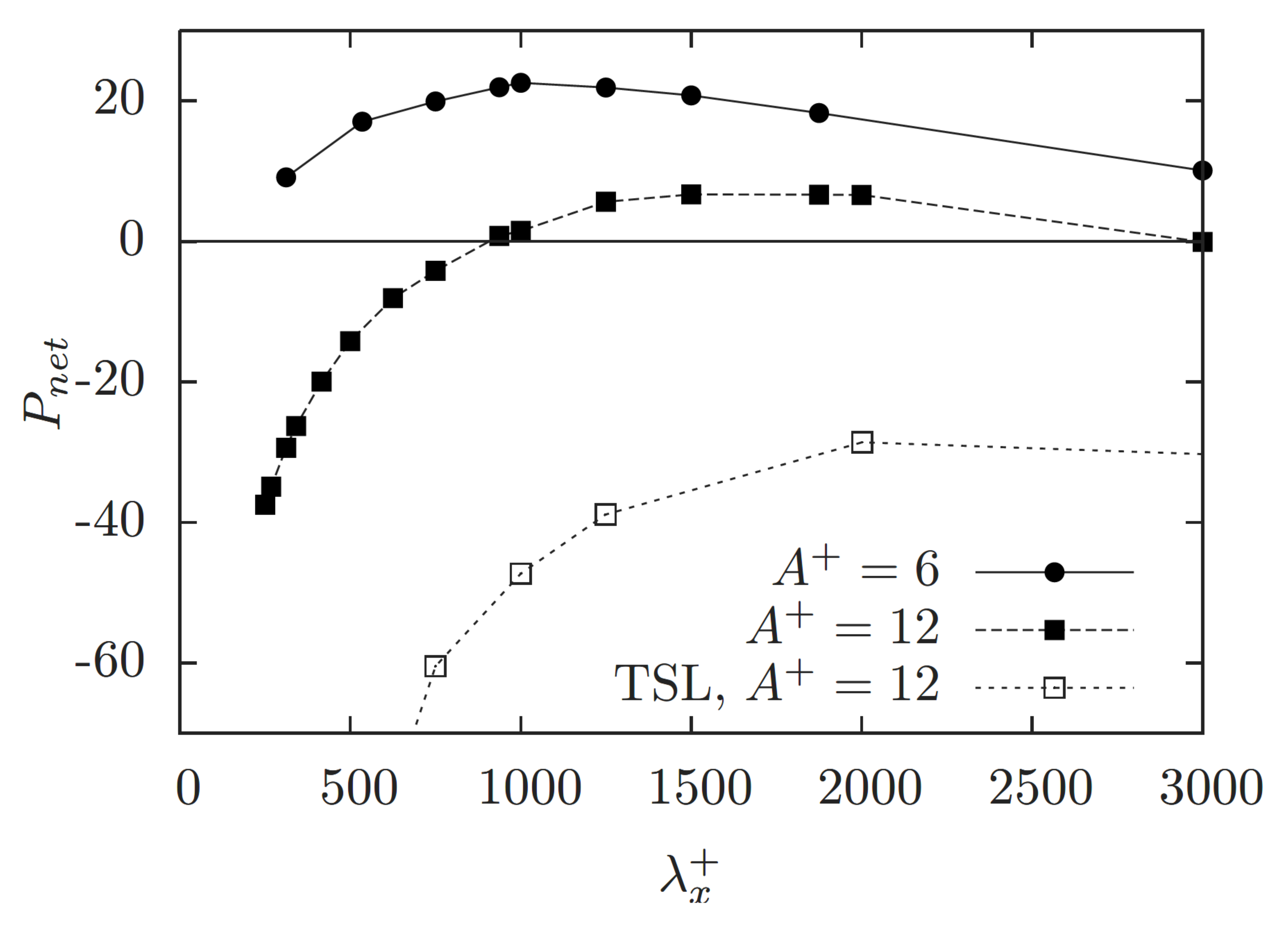}
\caption{Net power saved as a function of the streamwise wavelength for different forcing amplitudes. The temporal forcing case is denoted by TSL, for which the period of oscillation is converted to the wavelength via the convection velocity $U^+_c=10$. $A^+=W_m^+$. Taken from \cite{viotti-quadrio-luchini-2009}, with permission from AIP Publishing.}
\label{fig:VQLmap}
\end{figure}

The particular case of spanwise standing-wave forcing given by equation \eqref{eq:waves}, with $\omega=0$, was investigated by \cite{viotti-quadrio-luchini-2009} in a channel flow and by \cite{skote-2011} for a boundary-layer flow, both at the same Reynolds number, $Re_\tau=200$. The simulations by \cite{viotti-quadrio-luchini-2009} covered the range $\lambda^+_x=2\pi/\kappa_x^+=200-3000$ and the optimal value was found to be $\lambda^+_x=1200$. A drag-reduction margin $\R=45\%$ was reached for $W_m^+=12$, i.e., significantly larger than the maximum value achieved with purely temporal actuation at the same $W_m^+$ value. The maximum drag-reduction margin was $\R=52\%$ at $W_m^+=20$, which is about 30\% larger than that attained via temporal forcing.
For $W_m^+=6$, for which $\R=33\%$, a much improved net energy saving was obtained, namely $\mathcal{P}_{net}=22\%$, again in excess of $\mathcal{P}_{net}=7\%$ obtained by temporal oscillations. The actuation power computed from the DNS showed excellent agreement with that computed by the analytical solution of the laminar spatial Stokes layer. \cite{viotti-quadrio-luchini-2009} also demonstrated that the optimal $\lambda_x^+ = 1200$ was related to the optimal period for the temporal forcing ($T^+=100-125$) through the convective velocity $U^+_c=10$ of the turbulent structures in the buffer layer. The improved energy budget is illustrated in Fig. \ref{fig:VQLmap}, where the spatial forcing is compared with the temporal one.  

\begin{figure}
\centering
\includegraphics[width=0.8\textwidth]{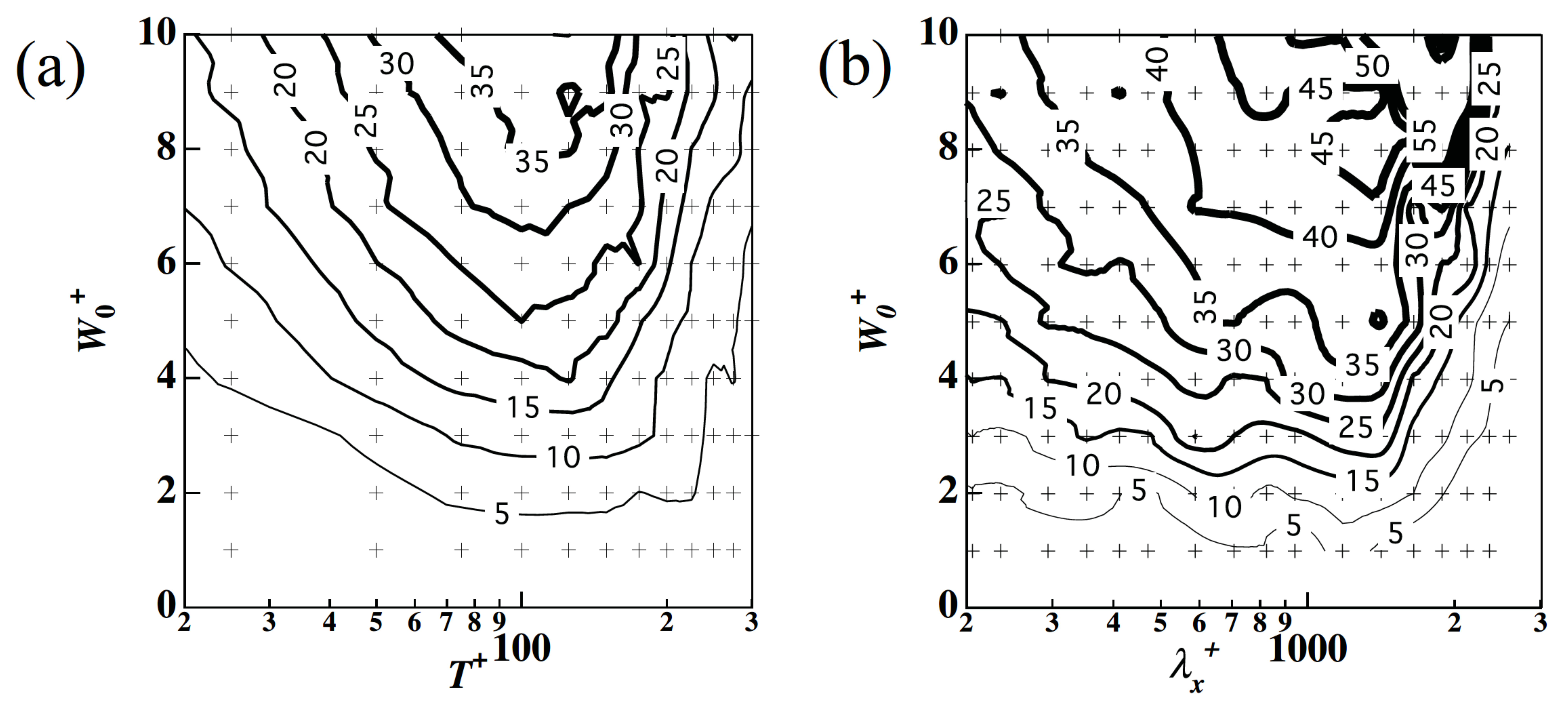}
\caption{Maps of drag-reduction margin at $Re_\tau=150$ versus amplitude ($W_0^+=W_m^+$) obtained by forcing the flow via \eqref{eq:waves} and (a) forcing period for purely temporal forcing ($\kappa_x=0$); (b) wavelength for purely spatial forcing (standing wave, $\omega=0$). Taken from \cite{yakeno-etal-2009}, with permission from the authors.}
\label{fig:YHKmap}
\end{figure}

\cite{yakeno-etal-2009} used DNS results for channel flow under CFR condition at $Re_\tau=150$ to construct the drag-reduction maps shown in Fig. \ref{fig:YHKmap} for purely spatial and purely temporal forcing. The juxtaposition of the two graphs of Fig. \ref{fig:YHKmap} brings to light the striking correspondence between the temporal period and the streamwise wavelength, supporting the proposition that the equivalence between the two modes of actuation are linked by the convective velocity $U^+_c=10$ in the buffer layer. As expected, the maximum drag-reduction levels, $\R=35\%$ and $\R=45\%$ for the temporal and spatial forcing, respectively, were obtained for $T^+=100$ in the former case and for $\lambda_x^+ \approx 1000$ in the latter. To exclude the Reynolds-number effects, \cite{yakeno-etal-2009} imposed a constant pressure gradient (CPG) for two simulations with purely temporal and purely spatial oscillations with $W_m^+=7$, and  $T^+=125$, $\lambda_x^+=1178$, respectively. The comparison between the two types of forcing revealed that the phase change of the Reynolds shear stress is larger in the spatial forcing.

The combination of temporal and spatial forcing, given by \eqref{eq:waves}, has been extensively investigated since it was first studied by \cite{quadrio-ricco-viotti-2009} and is now understood to offer, for particular combinations of period and wavelength, the maximum drag-reduction margins for a given wall-velocity amplitude. \cite{quadrio-ricco-viotti-2009} used DNS to investigate a channel flow at $Re_\tau=200$ with $W_m^+=12$ and a wide range of frequency and wavelength values. 
\begin{figure}[ht]
\centering
\includegraphics[width=0.7\textwidth]{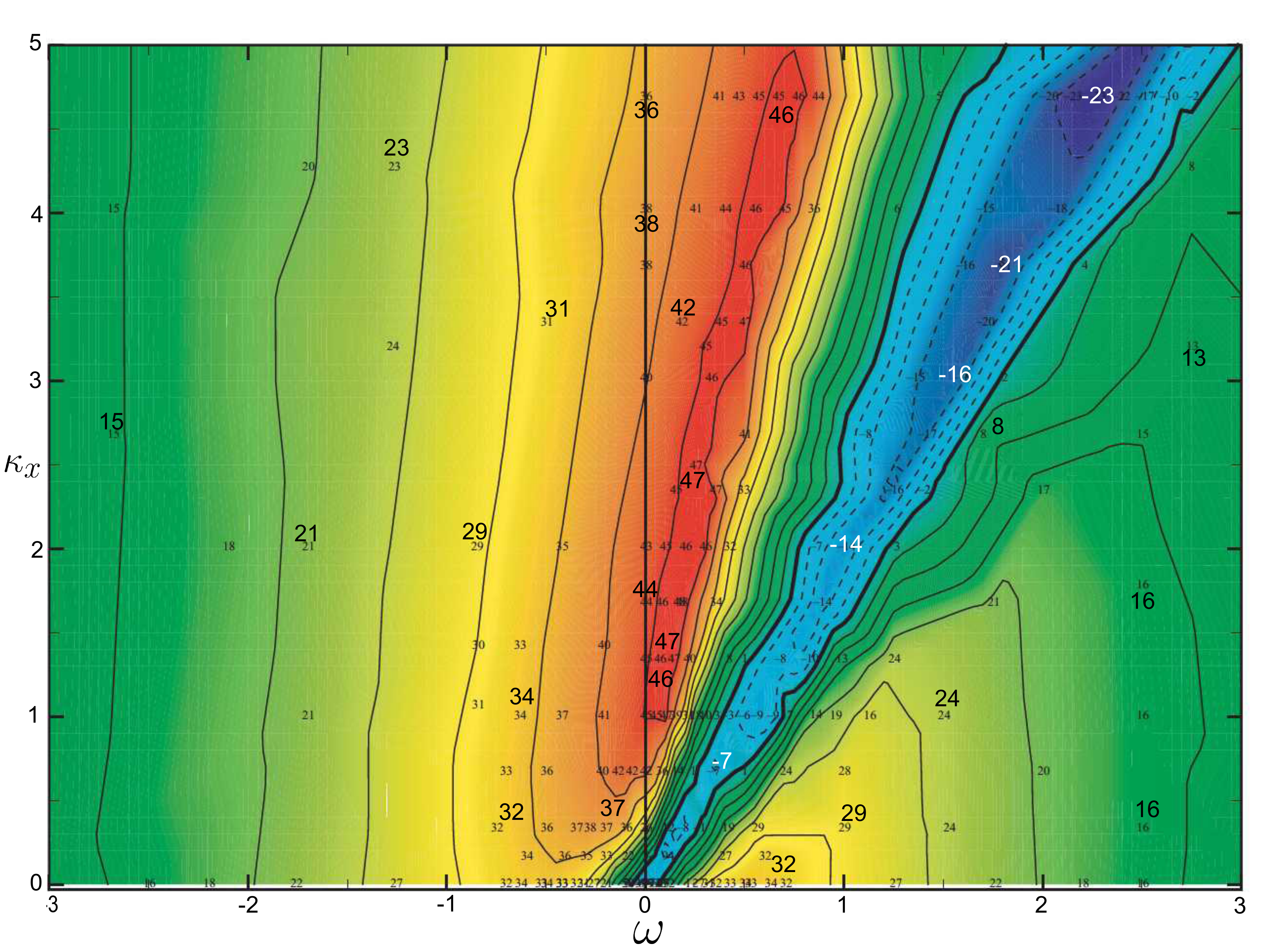}
\caption{Drag-reduction margin as a function of period and wavelength of the streamwise-travelling wave given by \eqref{eq:waves}. The frequency and wavenumber are scaled
with the maximum Poiseuille velocity ($U_p$) and channel half-height ($h$). Taken from \cite{quadrio-ricco-viotti-2009} with permission from CUP.}
\label{fig:QRVmap}
\end{figure}
The often-dubbed ``Quadrio map'', shown in Fig. \ref{fig:QRVmap}, emerged from this investigation, which entailed 250 simulations. The maximum drag-reduction achieved was $\R=48\%$ at $T^+\approx350$ and $\lambda_x^+ \approx 750$, which is larger than the maxima $\R=34\%$ for purely temporal oscillations with $T^+\approx106$ and $\R=45\%$ for standing streamwise waves with $\lambda_x^+ \approx 1257$. 
Although the increase in gross drag reduction relative to the steady-forcing case was modest at $W_m^+=12$, the net power saved was substantially larger, namely $\mathcal{P}_{net}=18\%$, obtained for $T^+\approx352$ and $\lambda_x^+ \approx 1257$, while $\mathcal{P}_{net}=5\%$ was computed in the steady-forcing case. 
Although this point in the period-wavelength plane was not identical to the conditions at which the gross drag reduction reached its maximum, it lay within the fairly flat region around the maximum. Hence, the travelling-wave actuation was distinguished by the close proximity of the maximum gross and maximum net power saved at $W^+_m=12$, whereas these two powers were distinguished in the case of purely spatial actuation. \cite{quadrio-ricco-viotti-2009} also showed that the drag-reduction margin increased monotonically with $W_m^+$, while the net energy gain increased to $\mathcal{P}_{net}=26\%$ near the optimum conditions for the lower wall-velocity $W_m^+=6$, for which $\R=37\%$.
\cite{quadrio-ricco-viotti-2009} defined a time scale $\mathscr{T}^+$ during which near-wall turbulent structures cover a distance of one forcing-wavelength ($\lambda_x^+$) at the relative velocity $\lvert U_t^+ - U_c^+\rvert$, where $U_t^+=\omega^+/\kappa^+$ is the phase speed of the travelling wave. Hence, $\mathscr{T}^+ \equiv \lvert \lambda_x^+ / (U_t^+ - U_c^+) \rvert$ is an equivalent oscillation period experienced by an observer travelling at the velocity of the turbulent fluctuations. 
\cite{quadrio-ricco-viotti-2009} also discovered a region in the parameter space that led to a significant drag increase for forward-travelling waves, shown in blue colour in the map of Fig. \ref{fig:QRVmap}. This region is confined by two lines in the $\omega-\kappa_x$ space for which the phase speed is constant and the maximum drag-increase margin of about $\R=-20\%$ occurs when the wall waves travel at about $U_t^+=10$, i.e., at the convection velocity of the near-wall turbulent structures. On the contrary, backward-travelling waves led to drag reduction at any wave speed.

Upon deriving a solution for the generalized laminar Stokes-layer generated by the wall waves \eqref{eq:waves}, \cite{quadrio-ricco-2011} correlated the drag-reduction margin and the thickness of the Stokes layer, $\delta_l^+$, based on the noted agreement between the laminar solution and the averaged spanwise velocity profile. 
\begin{figure}[ht]
\centering
\includegraphics[width=0.7\textwidth]{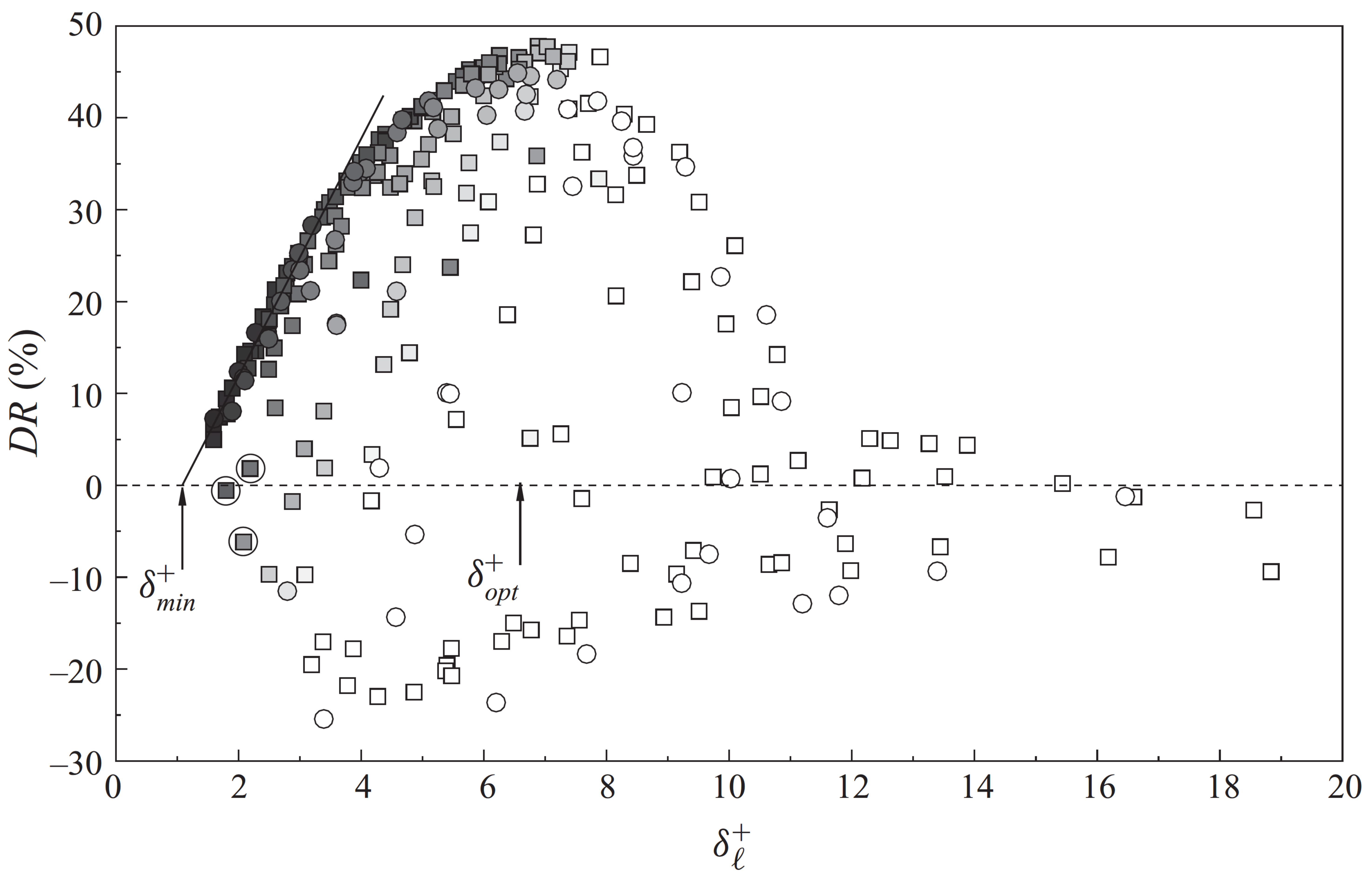}
\caption{Drag-reduction data as function of $\delta_l^+$ for CFR (squares) and CPG (circles) simulations. The oblique straight line shows the linear correlation. The arrows indicate the minimal condition for drag reduction, $\delta_l^+ \approx 1= \delta_{min}^+$, and the optimal Stokes-layer thickness, $\delta_l^+ \approx 6.5= \delta_{opt}^+$. Taken from \cite{quadrio-ricco-2011}, with permission from CUP.}
\label{fig:dr-gsl}
\end{figure}
As shown in Fig. \ref{fig:dr-gsl}, by plotting $\R$ vs. $\delta_l^+$, a linear relationship emerged up to $\R=35\%$ and $\delta_l^+=4$, subject to two conditions: (i) the wave phase speed was sufficiently different from the convection velocity of the near-wall turbulent fluctuations, i.e., $\lvert U_t^+ - U_c^+\rvert \gg 0$; and (ii) the time interval between two instants at which a near-wall structure passed over the wave at the same phase was shorter than the life-time of the structure, i.e., $\mathscr{T}^+ \ll 120$. A minimal $\delta_l^+=1$ is required to obtain drag reduction and the maximum drag-reduction margin was obtained for $\delta_l^+=6.5$. By using the equivalent forcing period $\mathscr{T}^+$ found by \cite{quadrio-ricco-viotti-2009}, \cite{quadrio-ricco-2011} created a schematic diagram to distinguish four distinct drag-reduction behaviours. When $U_t^+$ was comparable to $U_c^+$, a lock-in effect occurs so that the spanwise laminar solution failed to represent the spanwise turbulent flow and the friction drag increased. 

\cite{gatti-quadrio-2013} increased the Reynolds number for DNS of channel flows with travelling waves to $Re_\tau=2100$. Very small computational boxes, namely $1.19h \times 0.59h$, were employed, although the authors argued that the errors were small because both the actuated and the baseline simulations were affected by similar accuracy limitations.
The drag-reduction margin declined with the Reynolds number, with the most severe reductions occurring when the forcing parameters were optimal. As the Reynolds number increased, the optimal parameters shifted towards larger frequencies and wavenumbers. For purely temporal oscillations the drag-reduction margin decreased from $\R=39\%$ at $Re_\tau=200$ to $\R=29\%$ at $Re_\tau=1000$, and to $\R=24\%$ at $Re_\tau=2100$, as the optimal $T^+$ decreased to 90. For a fixed wavelength $\lambda_x^+=1250$, zero frequency (standing-wave actuation) was optimal at $Re_\tau=1000$ ($\R=37\%$) and $Re_\tau=2100$ ($\R=29\%$), whereas at $Re_\tau=200$ the optimum forcing was a travelling wave with $T^+=105$ for which the maximum was $\R=49\%$.

\cite{hurst-etal-2014} simulated a channel flow with travelling waves at $Re_\tau=200,400$ and $800$ with $W_m^+=12$, $T^+=35-628$ and $\lambda_x^+=393-3142$. The maximum drag-reduction margin attained for $Re_\tau=200$ was $\R=50\%$ at $T^+=300$ and $\lambda_x^+=800$, a level that decreased to $\R=40\%$ for $Re_\tau=800$, with the optimum parameters shifting modestly.
In the case of purely temporal forcing, the optimal period decreased with the Reynolds number, and \cite{hurst-etal-2014} concluded that the Reynolds number effect was stronger for the purely temporal forcing compared to the purely spatial actuation, in marked contrast to the findings by \cite{gatti-quadrio-2013}.
They explored the possibility of whether the drag-reduction margin could be correlated as $\R\propto Re_\tau^{-\alpha}$, and observed this scaling to be non-trivial, with $\alpha$ varying between 0.1 and 0.44, depending upon the combination of actuation parameters. The issue of Reynolds-number dependence is the subject of Section \ref{sec:redep}. 

In \cite{gatti-quadrio-2016}, the travelling-wave forcing in channel flows was compared at $Re_\tau=200$ and $Re_\tau=1000$ for a wide range of frequencies ($\omega^+= -0.5 - 1.0$, i.e., reaching periods as low as $T^+=2\pi$ and including backward travelling waves with $T^+=4\pi$), wavenumbers ($\kappa_x^+= 0-0.05$, i.e., including the standing wave and reaching the smallest wavelength of $\lambda_x^+=126$) and amplitudes ($W_m^+= 2-20$), thus creating an extensive sweep of the actuation-parameter space. A more detailed study of the effects of varying the period and wavelength at amplitude $W_m^+=12$ revealed that the maximum drag reduction for $Re_\tau=200$ was $\R=50\%$ at $T^+=320$ and $\lambda_x^+=997$, while at $Re_\tau=1000$ it decreased to $\R=39\%$ for $T^+=125$ and $\lambda_x^+=322$, in agreement with the maximum $\R=40\%$ reported by \cite{hurst-etal-2014} at $Re_\tau=800$ and confirming the result by \cite{gatti-quadrio-2013}. 
For purely temporal oscillations the drag-reduction margin decreased from $\R=37\%$ at $Re_\tau=200$ to $\R=28\%$ at $Re_\tau=1000$, as the optimal $T^+$ reduced to 75. By interpolating the data between the discrete amplitudes, \cite{gatti-quadrio-2016} created a drag-reduction map at constant $W_m^+=12$, scaled with the actual friction velocity. The maximum drag-reduction margin turned out to be lower, $\R=45\%$, compared to $\R=50\%$ at the corresponding condition scaled with the reference friction velocity. The effect of Reynolds number on the drag reduction and the shift of the location of maximum to larger frequencies and wavenumbers remained the same with the two scaling options, with the shift being less pronounced than in the actual wall-friction scaling. \cite{gatti-quadrio-2016} thus concluded that the scaling of the actuation-parameters based on the actual friction velocity should be adopted when comparing results at different Reynolds numbers. The relatively small computational domains used for the 4020 simulations were compared with 20 simulations performed with larger domains, as well as under CPG condition, to validate the conclusions. Their results are further discussed in Section \ref{sec:redep}. 

\cite{gatti-etal-2018a} compared the travelling-wave forcing ($W_m^+= 7, T^+=263$, $\lambda_x^+=628$) in a channel flow at $Re_\tau=1000$ with another strategy in the form of ``opposition control'', i.e., a body-force control designed to strongly attenuate the wall-normal fluctuations up to a specified height above the wall. The vertical upward shift ($\Delta B^+$) of the logarithmic part of the velocity profile, signifying the drag-reduction level, was utilized to create a model that predicted the turbulent spectra in channels where the flow control is confined to the near-wall region. The model was based on the observation that the turbulence in the outer layer was only influenced by the near-wall control through $\Delta U^+$. The changes in the turbulent spectra were similar above the buffer layer in the two control strategies. Hence, the effect of the near-wall control on the outer flow could be predicted by combining a turbulent field at a lower Reynolds number (due to the shift $\Delta U^+$ acting as a virtual wall) with a laminar flow added below the turbulent bulk flow. 

The only numerical studies of pipe flows altered by travelling waves were by \cite{biggi-2012} under CFR conditions and \cite{xie-2014} under CPI conditions, both at $Re_\tau=200$. It was found that the travelling-wave forcing performed better in pipe flow than in channel flow, with relaminarization more likely to occur at $Re_\tau=200$.  

Turbulent drag reduction by spanwise wall velocities that are modulated in time and along the spanwise direction itself,
\begin{equation}
W_w(z,t) = W_m \cos(\kappa_z z - \omega t),
\label{eq:twz}
\end{equation}
was first studied by \citet{zhao-wu-luo-2004}, and subsequently by \cite{xie-2014}, who showed that the maximum drag-reduction margin given by the forcing \eqref{eq:twz} occurs for $\kappa_z=0$.

\subsection{Numerical studies of external flows}
 
This section discusses the numerical studies of free-stream boundary layers over oscillating walls, summarized in Table \ref{tab:DNS-bl}.
A common feature of all simulations is that they were all performed over a streamwise-restricted segment within which the wall oscillates, analogous to experimental campaigns. The drag-reduction margin rises from zero at the onset of the oscillations and reaches its maximum value within $3-5$ boundary-layer thicknesses, followed by a slightly faster readjustment to the canonical state downstream of the actuated surface. In comparison with channel flow at a corresponding Reynolds number, the maximum drag-reduction margin for the boundary layer is, in general, slightly lower. Due to the higher computational cost required for simulating spatially developing boundary layers, wide-ranging actuation-parameter studies, as carried out for channel flows, have not been conducted, and the optimum actuation conditions are therefore not well established. As in channel flows, a monotonic increase of the drag-reduction level with forcing amplitude $W_m^+$ occurs and an optimal period $T^+$ at fixed wall velocity exists. Furthermore, the difference in performance between purely temporal and purely spatial forcing is similar to that in channel flow. The travelling-wave forcing has been shown (albeit for one case only) to yield similar results to those in channel flow. The most important difference between channel and boundary-layer flows is the spatial development of the latter, which greatly influences the net energy savings, as discussed in Section \ref{sec:spat}.

Unlike channel flows, the reference boundary layer develops downstream with a decreasing friction velocity. When specifying the oscillation parameters (wall velocity, oscillation period, wavenumber) in viscous units, the friction velocity at the streamwise position where the oscillating segment starts in the simulations is normally used for scaling. 
In most simulations, as the flow evolves downstream the oscillation parameters are kept constant in physical units. Therefore, in reference viscous units, $W_m^+$ increases downstream, while $T^+$ decreases. The drag-reduction margin may grow downstream compared to a forcing with a constant $W_m^+$, since the drag-reduction margin increases with $W_m^+$, while the decreasing $T^+$ has a limited effect if the oscillation period is chosen around its optimum value. At the same time, the Reynolds number based on the friction velocity and boundary-layer thickness slightly increases downstream, which has a very mild adverse impact on the drag-reduction effect. Also, as observed earlier for channel flow, the optimal $T^+$ decreases as the Reynolds number increases. This effect may occur in open turbulent boundary layers as well, the consequence being that the drag-reduction margin is affected favourably if the period $T$ is unvaried downstream. For most of the boundary layers investigated, by both DNS and experiments, the variation of the Reynolds number and friction velocity over the limited streamwise length of the oscillating wall has however not resulted in large variations in the oscillation parameters. Therefore, the effects of the spatially changing conditions on the drag reduction have not been quantified. When plotted as a function of the streamwise coordinate, the drag-reduction margin initially rapidly increases from zero to the maximum value, while further downstream it shows a very mild decline. In view of the above considerations, either the average friction velocity over the region of approximately constant drag reduction or the maximum value has been used to quantify the drag-reduction margin.

{\scriptsize
\begin{sidewaystable}
\centering
\begin{tabular}{|c|c|c|c|c|c|c|c|c|c|c|c|}
\hline
\centering
\textbf{Study} & \textbf{Forcing} & ${Re_{\theta}}$ & ${Re_{\tau}}$ & $W_m^+$ & $\lambda_x^+$ & $T^+$ & max 
$\R(\%)$\\
\hline
\hline
\cite{yudhistira-skote-2011} & Temporal & 532 & 271 & 17, 26 & - & 118 & 37, 40\\
\hline
\cite{skote-2011} & Spatial & 532 & 271 & 17 & 1300 & - & 46\\
\hline
\cite{skote-2012} & Temporal & 505 & 259 & 6, 12 & - & 132 & 19, 30  \\
 & & 505 & 259 & 11.3 & - & 67 & 29  \\
\hline
\cite{skote-2013} & Spatial & 505 & 259 & 6, 12 & 1320 & - & 25, 41 \\
\hline
\cite{lardeau-leschziner-2013} & Temporal & 1260 & 577 & 12 & - & 80, 100, 120, 200 & 25\\
\hline
\cite{skote-mishra-wu-2015} &  Temporal & 963 & 456 & 10.7 & - & 76.4 & 26 \\
\hline
\cite{mishra-skote-2015} & Spatial & 450 & 234 & $2-20$ & 85, 171, 341 & - & 36  \\
& (Positive-half square waves) & & & & & &   \\
\hline
\cite{skote-tsfp9} & Traveling Wave & 505 & 259 & 12 & 384 & 176 & 42  \\
\hline
\cite{skote-mishra-wu-2019} & Temporal & 1400 & 633 & 11.3 & - & 67 & 26\\
& & 505 & 259 & 12 & - & 30, 176 & 20, 26 \\
\hline
\end{tabular}
\caption{Boundary-layer simulations with different forms of oscillations. The friction Reynolds numbers are estimated according to the formula $Re_\tau=1.118Re_\theta^{0.875}$ \citep{ricco-quadrio-2008}.}
\label{tab:DNS-bl}
\end{sidewaystable}
}

\subsubsection{Numerical studies of free-stream boundary layers}

\cite{yudhistira-skote-2011} performed the first DNS of a turbulent boundary layer over a temporally oscillating wall, showing that the spatial development of the drag-reduction margin was similar to that observed in the experiments by \cite{ricco-wu-2004-a}, although the Reynolds number in the DNS ($Re_\theta=532$) was almost three times lower than in the experiments ($Re_\theta=1400$). The sensitivity of the Reynolds shear-stress profile to the sampling period over which the turbulence statistics were derived was investigated. In earlier experimental studies \citep{trujillo-bogard-ball-1997, ricco-wu-2004-a, laadhari-skandaji-morel-1994}, the Reynolds-stress profiles were observed to exhibit a pronounced second peak between the maximum value and the free stream, while this peculiar peak was absent in the DNS results. By reducing the time-period over which the statistics were computed, a peak also occurred in the simulation, which suggested that the peak in the experimental data was arguably due to an insufficiently long sampling time period. The empirical parameter that was used by \cite{ricco-quadrio-2008} to successfully link the drag reduction to properties of the Stokes layer in channel flows was also shown by \cite{yudhistira-skote-2011} to yield very good prediction in the boundary-layer case.

\begin{figure}[t]
\centering
\includegraphics[width=0.5\textwidth]{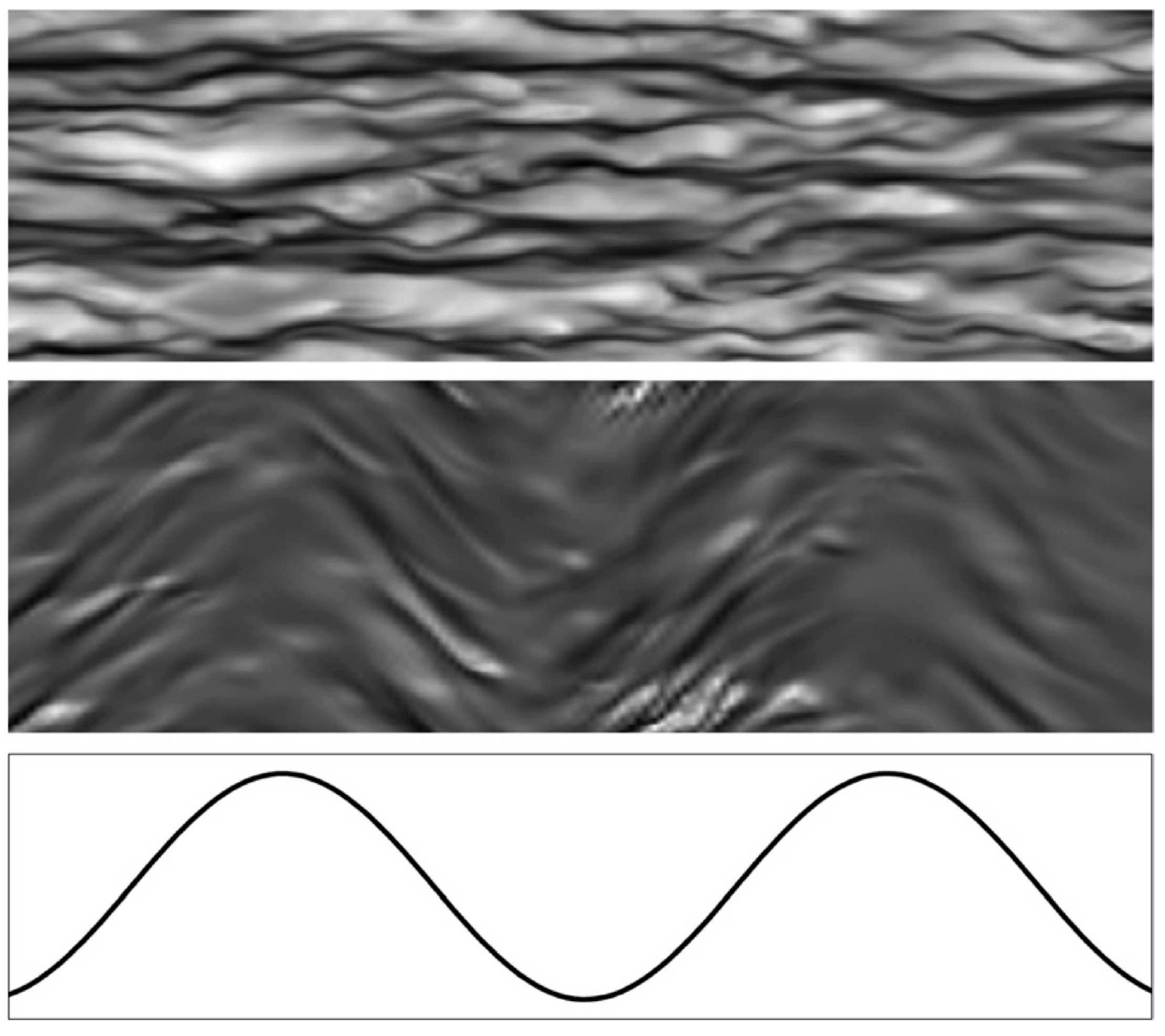}
\caption{Instantaneous velocity fields in a plane parallel to the wall at $y^+=11$ for the reference boundary layer (top) and the oscillating case (middle) forced with the wall-velocity distribution (bottom). The dark grey areas indicates low-speed regions and lighter shades indicate high-speed patches. In viscous units, the part shown is 770 wide and 2500 long. The flow is from left to right.
Taken from \cite{skote-2011}, with permission from AIP Publishing.}
\label{fig:skote-spatial}
\end{figure}

The steady streamwise-varying forcing, described by equation \eqref{eq:waves} with $\omega=0$, was found by \cite{skote-2011} to yield a maximum drag-reduction margin of $\R=46\%$, while the net power savings 
was negative at $\mathcal{P}_{net}=-19\%$ with $W_m^+=17$ and $\lambda_x^+=1320$. The spanwise velocity profile was shown to follow the laminar solution given by \cite{viotti-quadrio-luchini-2009}. The DNS was performed at the same Reynolds number, $Re_\theta=532$, as the temporal-oscillation cases studied by \cite{yudhistira-skote-2011}, who obtained a maximum of $\R=37\%$ for the same wall velocity $W_m^+=17$ and a oscillation period of $T^+=118$. \cite{skote-2011} found that the drag-reduction level exhibited an oscillatory variation with a wavelength that was half that of the wall-actuation wave. The maximum drag reduction occurred where the wall velocity was at its maximum and the minimum drag reduction correlated with the positions where the wall velocity was zero. These results can be linked to the maximum weakening of the streaks, which occurred at the maximum wall velocity, as shown by the flow visualizations of Fig. \ref{fig:skote-spatial}.

In the simulations of \cite{skote-2012} for boundary layers with purely temporal forcing, two wall-velocity amplitudes were used, $W_m^+=6$ and $12$, both with oscillation period $T^+=132$. The lower amplitude was chosen to investigate whether a positive net-energy balance could be obtained, in contrast to the negative value reported by \cite{yudhistira-skote-2011} and \cite{skote-2011}. Similar to the channel-flow case investigated by \cite{quadrio-ricco-2004}, a positive net energy was indeed obtained for the smaller amplitude. 
\cite{skote-2012} also performed a simulation at identical oscillation parameters, $T^+=67$ and $W_m^+=11.3$, as those used in the experiment of \cite{ricco-wu-2004-a}. The length of the oscillation segment in the DNS was similar to that in experimental setup. The drag reduction obtained, at 28.6\%, was larger than the value of 23\% recorded in the experiment, possibly because of the substantially lower Reynolds number in the DNS ($Re_\theta = 505$ relative to $Re_\theta = 1400$ in the experiments). Indeed, in a later study, \cite{skote-mishra-wu-2019} conducted the DNS at the same Reynolds number as that used in \cite{ricco-wu-2004-a} and obtained drag-reduction margins that matched the experimental values.

By comparing the boundary-layer results with those of \cite{quadrio-ricco-2004} for channel flow, and by taking the different Reynolds numbers into account, \cite{skote-2012} concluded that the drag-reduction margin for the boundary layer was slightly lower than that for the channel flow. Scaling of the Reynolds stresses with the actual friction velocity produced a wall-normal stress profile that was virtually unaffected compared to the reference profile, a shear-stress profile that was modestly reduced, and a longitudinal-normal stress that was reduced significantly. The results were in qualitative agreement with those of the CPG channel-flow investigation by \cite{ricco-etal-2012}. \cite{skote-2012} further noted that the longitudinal Reynolds-stress peak was located inside the Stokes layer, while the wall-normal and shear-stress peaks were further away from the wall. Furthermore, an analysis of the initial response, following the start of actuation, analogous to the one presented by \cite{quadrio-ricco-2003} for a channel flow, revealed a very similar development of both flows, with the exception of the streamwise shear-stress, which did not follow the non-monotonic development observed in channel flow. In particular, the complicated variations of the spanwise Reynolds-stresses during the first period of the wall oscillation were shown to be almost identical to those in the channel-flow case. 

Direct numerical simulations of boundary layers over an oscillating wall with $W_m^+=12$ and periods $T^+=80, 100, 120, 200$ were conducted by \cite{lardeau-leschziner-2013}. The focus of their study was on the transient flow following the onset of the actuation as the flow evolved in the downstream direction. The phase-averaged statistics revealed a rapid decline in drag towards the low-drag state in the form of a wave-like propagation with an amplitude that decayed downstream.
Only for the case with $T^+=200$ did the undershoot in skin friction surpass the long-term low-drag value. \cite{lardeau-leschziner-2013} speculated that the drag-reduction margin for the boundary layer may be $5\%-7\%$ lower than in the equivalent channel flow, while the optimal period was probably considerably lower at $T^+=60-70$, because their drag reduction at $T^+=80$ was larger than that at $T^+=100$. 
\cite{skote-2012} instead observed that the drag reduction is similar for $T^+=67$ and $T^+=132$, in agreement with the observation of \cite{quadrio-ricco-2004} that the drag reduction in channel flow was fairly uniform over this range of periods, as shown in Fig. \ref{fig:QRmapD}. There is, however, no contradiction in these seemingly different observations since the simulation by \cite{lardeau-leschziner-2013} was performed at a higher Reynolds number. This difference has the proven effect of shifting the optimum towards shorter periods (the optimum period is $T^+\approx 75$ at $Re_\tau = 1000$, \cite{gatti-quadrio-2016}).

\begin{figure}[t]
\centering
\includegraphics[width=0.5\textwidth]{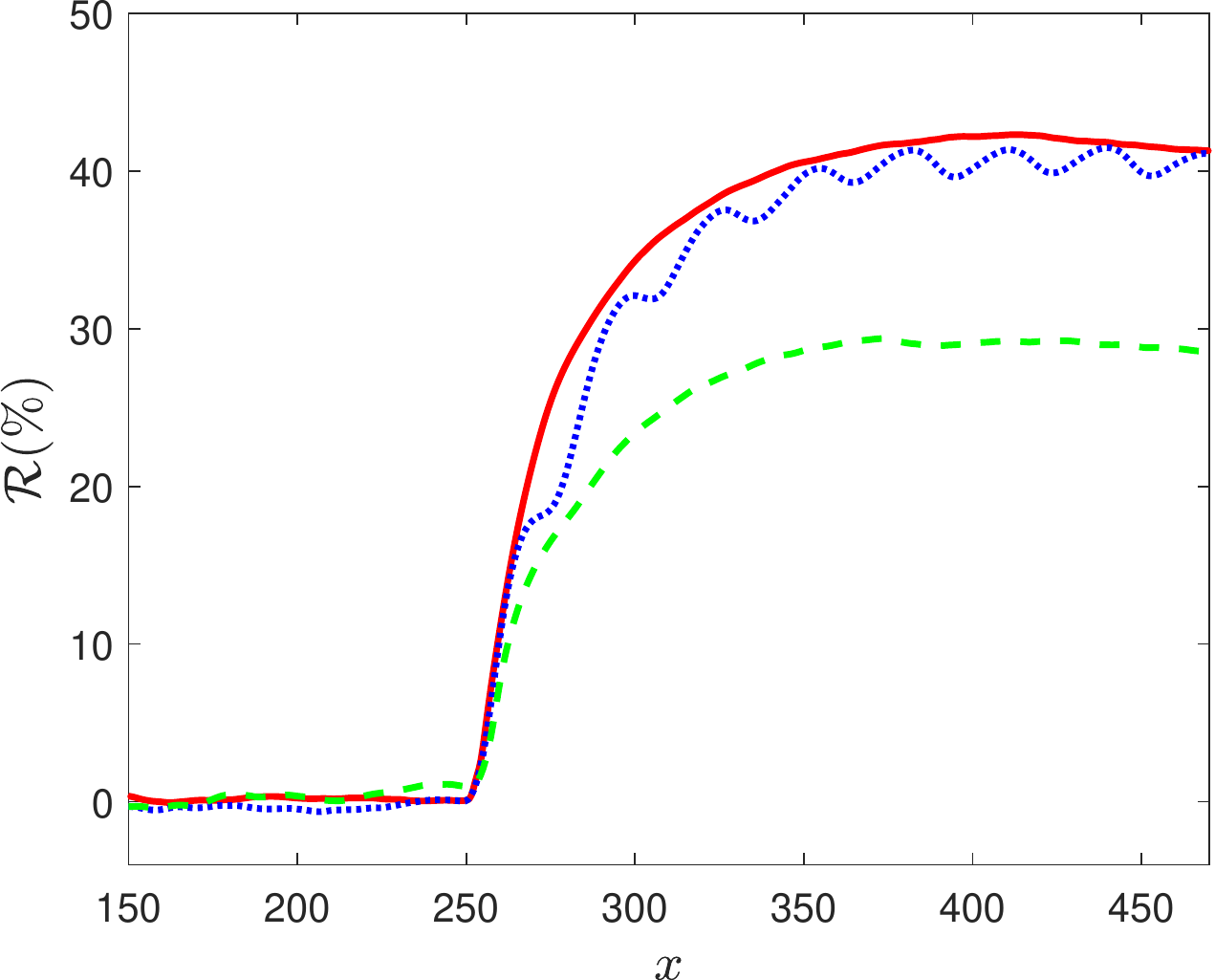}
\caption{Drag reduction as a function of streamwise distance for boundary layers with wall velocity $W_m^+=12$.
The streamwise coordinate $x$ is scaled with the displacement thickness of the laminar boundary layer at the inflow location ($x=0$). The boundary layer is fully turbulent at $x=150$, and the wall motion starts at $x=250$ (which corresponds to $Re_\theta=505$). Green dashed line: temporal forcing with $T^+=132$ \citep{skote-2012}; blue dotted line: spatial forcing with $\lambda_x^+=1320$ \citep{skote-2013}; red solid line: forcing in the form of a travelling-wave with $T^+=176$ and $\lambda_x^+=384$ \citep{skote-tsfp9}. Adapted from \cite{skote-tsfp9}.}
\label{fig:bl-trav}
\end{figure}

\cite{skote-2013} compared results for two boundary layers at $Re_\theta=505$ subjected to purely temporal and purely spatial wall oscillations, both with $W_m^+=12$. In the spatial case $\lambda_x^+=1320$, while in the temporal case $T^+=132$, the two quantities related by the convective velocity $U_c^+=10$, following the time-space analogy of \cite{quadrio-ricco-viotti-2009}. The drag-reduction margins for the two forcing types are shown in Fig. \ref{fig:bl-trav} (blue-dotted and green-dashed for spatial and temporal forcing, respectively), excluding the readjustment after the oscillating wall segment. The comparison revealed that only the spanwise component of the Reynolds stress exhibited major differences between the two forcing scenarios. This result allowed \cite{skote-2013} to explain the improved energy budget for the steady actuation relative to the temporal oscillation via the turbulence production of the spanwise Reynolds stress,

\begin{equation}
{P}_{33}={P}_{vw}+{P}_{uw}=-2\overline{vw}\frac{\partial \overline{W}}{\partial y}-2\overline{uw}\frac{\partial \overline{W}}{\partial x},
\label{eq:prod33}
\end{equation}
where $u$, $v$ and $w$ are the streamwise, wall-normal and spanwise velocity fluctuations, respectively. Both ${P}_{vw}$ and ${P}_{uw}$ are zero for the unforced boundary layer, the latter is zero in the temporal-forcing case, and both terms are non-zero for the spatial-forcing case. 
Contrarily to the channel-flow case of \cite{quadrio-ricco-2003}, \cite{touber-leschziner-2012} showed that, by reference to phase-averaged statistics, $P_{33}$ was non-zero for temporally forced channel flow. In a similar phase-locked statistical analysis for a spatially-actuated boundary layer flow, \cite{skote-2013} showed that ${P}_{vw}$ was positive, while ${P}_{uw}$ was negative. Hence, the latter term caused ${P}_{33}$ to be smaller in the spatial forcing case, contributing significantly to the drag reduction.
Similar to the behaviour observed in the channel-flow simulations by \cite{viotti-quadrio-luchini-2009}, the net energy saving was larger in the low-amplitude case, $W_m^+=6$, than in the high-amplitude case, $W_m^+=12$, although the actual drag-reduction margin was lower in the former case.

\begin{figure}[t]
\centering
\includegraphics[width=0.5\textwidth]{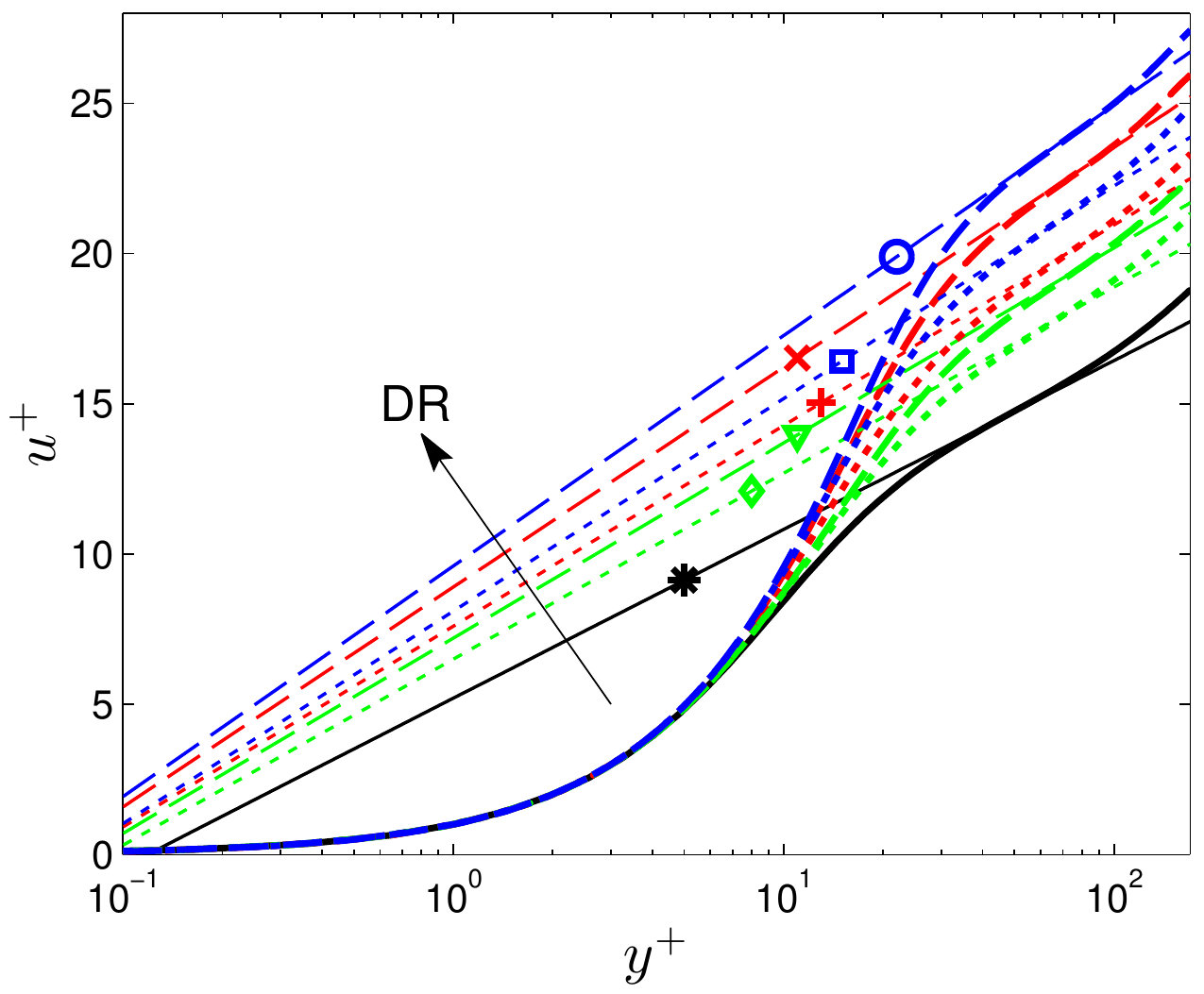}
\caption{Velocity profiles scaled with actual friction velocity. Thick lines from DNS data taken from a number of simulations with varying degree of drag-reduction margin (18\%-46\%) from purely temporal $(\cdots)$ or purely spatial ($-~-$) wall oscillations. The slope is predicted by ${\left(\kappa \sqrt{1-\R/100}\right)}^{-1}$ (the thin lines). The black solid line is the reference case. DR = drag reduction.
Taken from \cite{skote-2014}, with permission from Elsevier.}
\label{fig:skote-scaling}
\end{figure}

\cite{skote-2014} developed a theory that explained the progressive increase in the log-law slope observed in Fig. \ref{fig:skote-scaling}, as the drag-reduction margin increased in boundary-layer flows subjected to increasingly intense forcing. The velocity profiles in the figure were taken from earlier simulations of boundary layers with both temporal and spatial forcing, and these profiles were obtained by averaging over the streamwise stretch of approximately constant drag reduction, after discarding the initial spatial development. The theory is based on the assumption of a non-uniform adjustment of the inner and outer portions of the log-law layer to the actuation. While the near-wall part of the velocity profile was assumed to adjust rapidly to the new low-drag state (governed by the actual friction velocity), the outer part adjusted more slowly, being influenced by the outer part of the baseline boundary layer (governed the reference friction velocity). As a result, the velocity profile is governed by both the drag-adjusted actual velocity scale and the velocity scale associated with the canonical state for which the log-law slope was $1/\kappa$. In this non-equilibrium state of the boundary layer, the inverse slope of the logarithmic part of the velocity profile was theoretically derived to be $\kappa \sqrt{1-\R/100}$. 
Fig. \ref{fig:skote-scaling} illustrates how the mixed velocity scaling manifests itself through changes of the slope of the velocity profile for different drag-reduction levels. The theory predicted that, further downstream (or equivalently, after a sufficient amount of time has elapsed in the channel flow), where the outer part of the boundary layer has also  adjusted to the new wall friction, the slope of the logarithmic part would revert to its canonical value $1/\kappa$. In this latter fully developed flow, the sole velocity scale can be assumed to be the actual friction velocity.
The slope of the log layer can therefore serve as a measure of how adjusted is the flow to the actuation. None of the boundary-layer simulations were over a long enough actuated streamwise region to provide evidence for a full readjustment. As evident from Fig. \ref{fig:skote-scaling}, \cite{skote-2014} demonstrated that the theory was equally applicable to both steady streamwise-varying and temporal streamwise-constant forcing, i.e., the only factor determining the slope was the drag-reduction margin. 

The simulation by \cite{skote-mishra-wu-2015} (also presented in \cite{skote-iTi}) involved a relatively long oscillating wall-segment and was primarily conducted to investigate the Reynolds-number dependence of the drag reduction, further discussed in Section \ref{sec:redep}. The very slowly adjustment towards an equilibrium, as conjectured above, was confirmed by this long simulation.  

\cite{skote-tsfp9} presented the only case, at the time of writing, of a boundary layer forced by the travelling-wave wall motion given by equation \eqref{eq:waves}. The oscillation parameters, $T^+=176$, $\lambda_x^+=384$ and $W_m^+=12$, were chosen from a practical implementation point of view, since the DNS results were intended for comparisons with future experimental data. The maximum drag-reduction margin was $\R=42\%$, marginally lower than $\R=44\%$ obtained for channel flow by \cite{quadrio-ricco-viotti-2009} with $\omega=0.3$ and $\kappa_x=3.3$ at approximately the same Reynolds number. The drag reduction slightly decreased downstream, similar to the behaviour in the earlier purely temporal- and purely spatial-forcing simulations, shown in Fig. \ref{fig:bl-trav}. As in earlier simulations, the oscillation parameters were kept constant in the downstream direction in physical coordinates, and therefore they varied in viscous units when scaled with the local skin friction. Since the simulation box was small, the Reynolds-number variation over the streamwise length was however limited, and the oscillation parameters remained within their near-optimal range. 

\begin{figure}[t]
\centering
\includegraphics[width=0.5\textwidth]{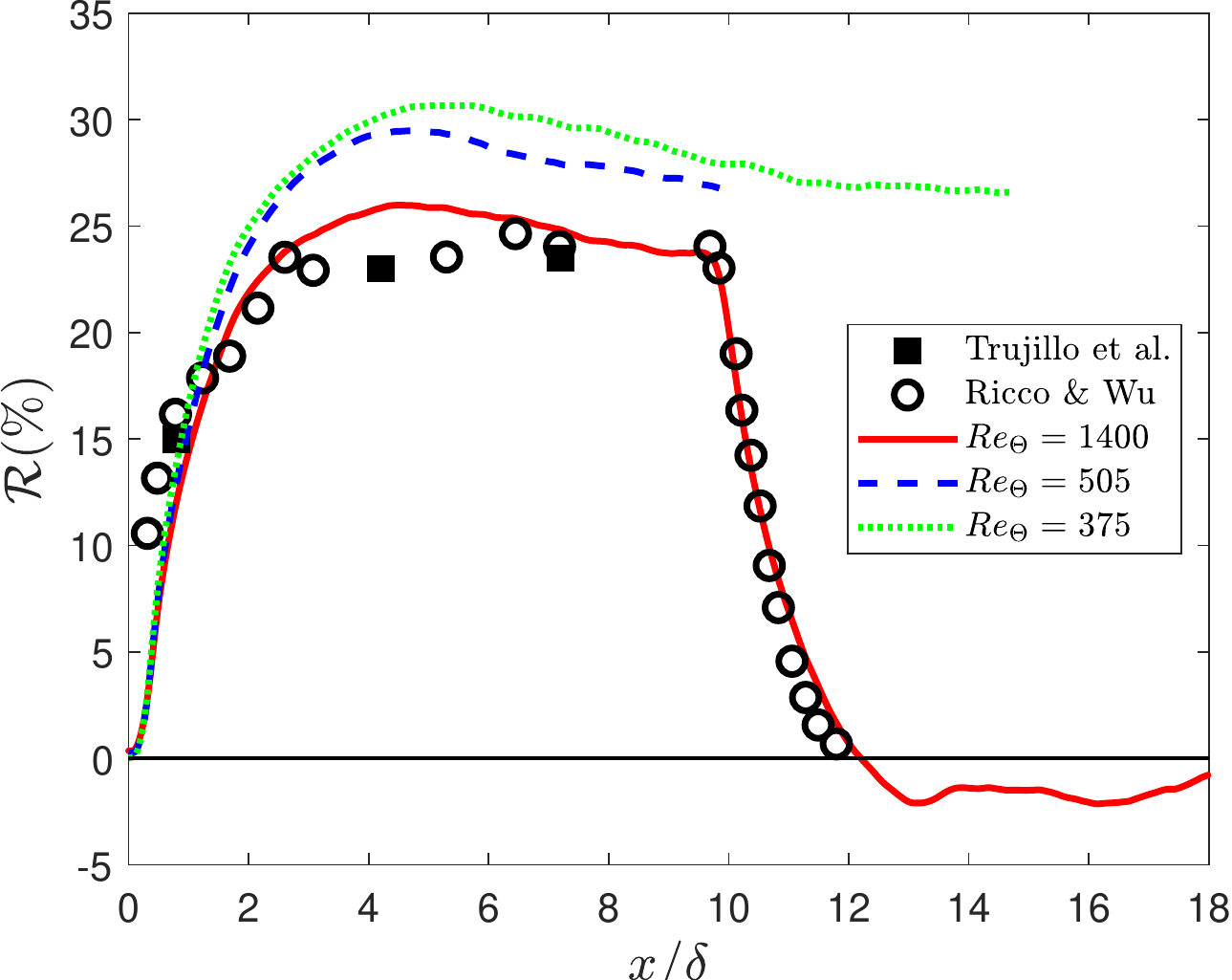}
\caption{The drag reduction as a function of streamwise distance for boundary layers at three Reynolds numbers with identical forcing parameters, $T^+=67$ and $A^{+}=11.3$. The streamwise coordinate $x$ is scaled by the boundary thickness $\delta$, given by the height at which the streamwise velocity is 99\% of the free-stream velocity. ($-$) DNS results at $Re_\theta=1400$ as compared to experimental results (o) from \cite{ricco-wu-2004-a} and  ($\blacksquare$) from \cite{trujillo-bogard-ball-1997}. The two broken lines represent DNS results at $Re_\theta=375$ ($\cdots$) and $Re_\theta=505$ ($-~-$). Adapted from figure 2 in \cite{skote-mishra-wu-2019}.}
\label{fig:skote-dr}
\end{figure}

\cite{skote-mishra-wu-2019} performed a simulation with purely temporal wall oscillations at $Re_\theta = 1400$ with $T^+=67$ and $W_m^+=11.3$, which were identical to those used in the experimental study by \cite{ricco-wu-2004-a}. Again, the oscillation parameters were kept constant in the downstream direction in dimensional terms (as in experiments), and therefore they varied in viscous units when scaled locally. The variation of the friction velocity was however minor over the oscillating region, and the amplitude calculated at the end of the oscillating segment was $W_m^+=11.6$ and $T^+=64$ when scaled with the local friction velocity. Hence, the influence of the downstream variation of the locally scaled parameters, as well as the Reynolds number, on the drag-reduction margin was insignificant. As shown in Fig. \ref{fig:skote-dr}, the DNS and experimental results for the drag-reduction level show a very good agreement. Also included in the figure are DNS results at two lower Reynolds numbers, further discussed in Section \ref{sec:spat}. The scaling analysis performed in \cite{skote-2014} was verified by using the indicator-function $y^+ \mathrm{d} u^+/\mathrm{d} y^+$, representing the inverse of the von K\'{a}rm\'{a}n constant in the log-law. \cite{skote-mishra-wu-2019} noted that the distance from the onset of the oscillation to the point where the maximum drag reduction is reached (the spatial transient) is detrimental from an energy-saving point of view, since energy is spent on moving the wall while the benefit of the drag reduction is limited to the spatial transient. Additional simulations at $Re_\theta=505$ with $T^+=30$ and $T^+=176$ showed that a linear relationship exists between $T^+$ and the distance from the onset of the oscillations to the point where the maximum drag reduction occurred. 

\subsubsection{The maximum drag-reduction margin in boundary layers}
\label{sec:maxdr}

\begin{figure}[t]
\centering
\includegraphics[width=0.5\textwidth]{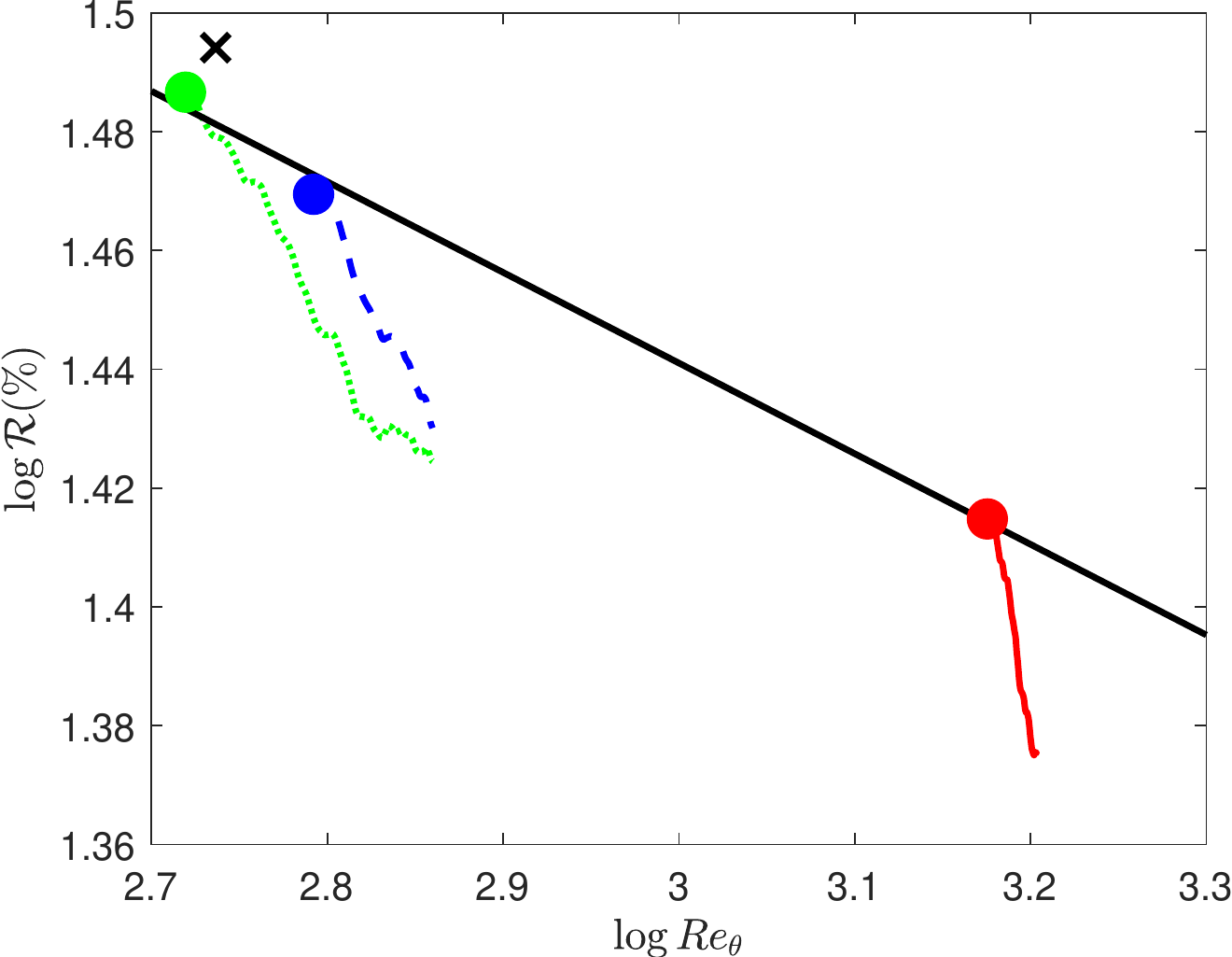}
\caption{The local drag reduction for boundary layers at three Reynolds numbers with identical forcing parameters $T^+=67$ and $A^{+}=11.3$. The circles represent the maximum drag-reduction margin. Coloured lines are as in Fig. \ref{fig:skote-dr}. The straight black line represents $Re_\theta^{-0.153}$ and is illustrating the decline of the maximum drag-reduction margin with increasing Reynolds number. The cross ($\times$) is channel-flow value from \cite{quadrio-ricco-2004} for identical forcing parameters. Adapted from figure 10 in \cite{skote-mishra-wu-2019}.}
\label{fig:skote-psave}
\end{figure}

In this section the only existing simulations with the same oscillation parameters at different Reynolds numbers are used to show that the maximum drag reduction obtained in a boundary-layer flow decreases with increasing Reynolds number in a manner similar to that in channel flow, although the drag-reduction margin is slightly larger in the latter case, as discussed earlier. The maximum drag-reduction values are shown as circles in Fig. \ref{fig:skote-psave} for the three simulations at $Re_\theta=375$ (green), $Re_\theta=505$ (blue) and $Re_\theta=1400$ (red), taken from \cite{skote-mishra-wu-2019}. All three cases were forced by identical oscillations ($T^+=67$ and $A^{+}=11.3$). The trend of the maximum drag reduction with increasing Reynolds number, identified by the black line in Fig. \ref{fig:skote-psave}, followed $\R \propto Re_\theta^{-0.153}$, the exponent having been obtained from the three boundary-layer simulations considered.  To compare with channel-flow results, the relation $\R \propto Re_\tau^{-\alpha}$, which is discussed in Section \ref{sec:redep}, was used by \cite{skote-mishra-wu-2019}. \cite{hurst-etal-2014} and \cite{yao-etal-2019} found that $\alpha$ varies between 0.1 and 0.46 in channel flow, depending on the Reynolds number and oscillation parameters. At $T^+=67$, $\alpha$ was however confined between 0.15 and 0.17, as observed from figure 4b in \cite{yao-etal-2019}. The exponent $-0.153$ obtained from the boundary layer simulations (as described above) can be compared with the corresponding channel flow estimations, which yielded an exponent between $-0.14$ and $-0.16$ when using the conversion $Re_\tau \propto Re_\theta^{0.923}$ \citep{eitel-amor-etal-2014}. Hence, the decline of the maximum drag-reduction in boundary-layer flow (indicated by the black line in Fig. \ref{fig:skote-psave}) was very similar to that derived from channel-flow simulations at a similar oscillation period. This finding highlights, once again, the equivalence between the two geometries when comparing the maximum drag-reduction in the boundary layer with the single drag-reduction margin obtained from channel flow, at least for the range of Reynolds numbers hitherto considered.

\subsubsection{Spatial development of drag reduction in boundary layers}
\label{sec:spat}

This section discusses four aspects of the spatial development of the drag-reduction level as the boundary layer evolves in the streamwise direction. The first is the propagation of disturbances upstream of the start of the actuation, observed in the experiments of \cite{choi-debisschop-clayton-1998}. Neither the numerical simulations nor later experiments by \cite{ricco-wu-2004-a} however support this observation, leading to the conclusion that the upstream influence is insignificant. 

A second aspect is the spatial variation of the boundary layer downstream of the end-point of the actuation, within a stretch in which the drag reduction declines quickly, as exemplified in Fig. \ref{fig:skote-dr}. The figure however also suggests the occurrence of a change of sign of the averaged wall-shear stress, i.e., a drag increase relative to the canonical value, once the influence of the oscillation has ceased. This drag increase has also been observed in other simulations of turbulent boundary layers over oscillating walls \citep[][figure 2(a)]{lardeau-leschziner-2013}, as well as with forcing by opposition control \citep[][figure 7]{xia-etal-2015}. The overshoot observed downstream of the oscillating segment was explained, at least partially, by both \cite{lardeau-leschziner-2013} and \cite{skote-mishra-wu-2019}, by the difference in the boundary-layer thickness, and hence Reynolds number, between the reference and the controlled flows at the same streamwise position. Even if this difference in the boundary-layer thickness is taken into account, there is still a small region of actual drag increase a short distance from the end-point of the oscillating segment. This may be related to an upward diffusion process observed in the experiments by \cite{ricco-2004} and discussed in Section \ref{sec:exp-planar}.

The third aspect pertaining to the spatial development concerns the drag reduction between the beginning of the oscillating wall to the point where the maximum drag-reduction margin occurs. This region, over which the drag reduction rises rapidly to the maximum level, constitutes a short spatial transient and has been shown by \cite{skote-2012} to be similar to the initial temporal transients discussed by \cite{quadrio-ricco-2003}. A more complicated phase-wise development was however presented by \cite{lardeau-leschziner-2013}, in which a distinctive oscillatory response was reported based on phase-locked statistics. Nevertheless, the maximum drag reduction is invariably reached within $4-5$ boundary-layer thicknesses, as also revealed by the experimental data of \cite{ricco-wu-2004-a}, shown in Fig. \ref{fig:skote-dr}. 
 
The drag reduction from the position of its maximum to the end of the oscillating wall-segment constitutes the fourth aspect peculiar to boundary layers. Fig. \ref{fig:skote-psave} shows the dependence of the drag reduction on the Reynolds number as the boundary layer evolves downstream. The results shown in the figure are for the three boundary layers also considered in Fig. \ref{fig:skote-dr}, plotted in logarithmic scale, but without the sections in which the drag reduction rises to its maximum level and declines after the termination of the actuation.
As shown in Fig. \ref{fig:skote-psave}, the local drag reduction downstream of the position of the maximum drag-reduction margin, within the actuated segment, decreases more rapidly with increasing Reynolds number than the variation pertaining to the maximum value (indicated by the straight black line). A definitive explanation for this decline has not been offered to date and its quantification is, at present, hindered by the need for very long computational boxes and for the very high-resource requirements that such boxes demand. Whether the decreasing trend reflects a persistent transient or is an effect that continues further downstream can only be determined with much longer simulation boxes than those used so far. While a simulation by \cite{skote-mishra-wu-2015} for a fairly long domain did not provide a definite answer, it suggested that the rate of decline diminishes downstream.
In contrast to channel flow, where the Reynolds shear stress dictates the drag reduction as given by equation \eqref{eq:fik}, the convection and spatial development have an impact on the skin-friction reduction in boundary layers, which may clarify the origin of the decline of the drag reduction shown in Fig. \ref{fig:skote-psave}. The influence of the oscillation period (and wavelength) on the downstream development also needs to be elucidated.

The persistent decline in drag reduction with downstream distance, shown in Fig. \ref{fig:skote-psave}, was further investigated by \cite{skote-mishra-wu-2019}, and the conclusion was that they were the cause for the lower power savings than those derived from channel-flow simulations. The effects were aggravated by the local actuation power increasing downstream because the forcing amplitude $W_m$ was kept constant, thus resulting in a rise of $W_m^+$. When a simulation with constant $W_m^+$ was performed, by adjusting $W_m$ as the boundary layer developed, the local power required was found to be constant. In channel flow, where both $W_m^+$ and $T^+$ are constant, the actuation power instead decreases with the Reynolds number. The two effects of the local drag-reduction margin decreasing downstream and the local actuation power increasing combine to render the net power saving less favourable than in an ideal case with no spatial transients and homogeneous streamwise flow. 

Although the net power saving under ideal conditions increased with the Reynolds number, as shown by \cite{skote-mishra-wu-2019}, it was argued that a positive balance would not be found unless the oscillation
parameters -- the most important one being the wall-velocity amplitude -- were judiciously chosen and dynamically adjusted in the downstream direction. By re-analysing a simulation performed by \cite{skote-2012} with the relatively low amplitude, $W_m^+=6$, for which a positive net energy saving was obtained, \cite{skote-mishra-wu-2019} showed that increasing the length of the oscillating segment had a negative impact on the net energy balance. Hence, shorter actuated segments, spaced at optimal intervals and actuated optimally in accordance with local boundary-layer conditions, would be preferable in practical scenarios.  

\subsection{Flows subjected to non-sinusoidal periodic forcing and in non-canonical geometries}
\label{sec:nonosc}

All previously reviewed studies imposed sinusoidally varying temporal and/or spatial forcing on channel, pipe and boundary-layer flows. This section discusses investigations where non-sinusoidal forcing was used in these standard geometries or sinusoidal oscillations were applied to non-canonical flows.

\subsubsection{Flows with non-sinusoidal periodic forcing}

\begin{figure}[t]
\centering
\includegraphics[width=0.6\textwidth]{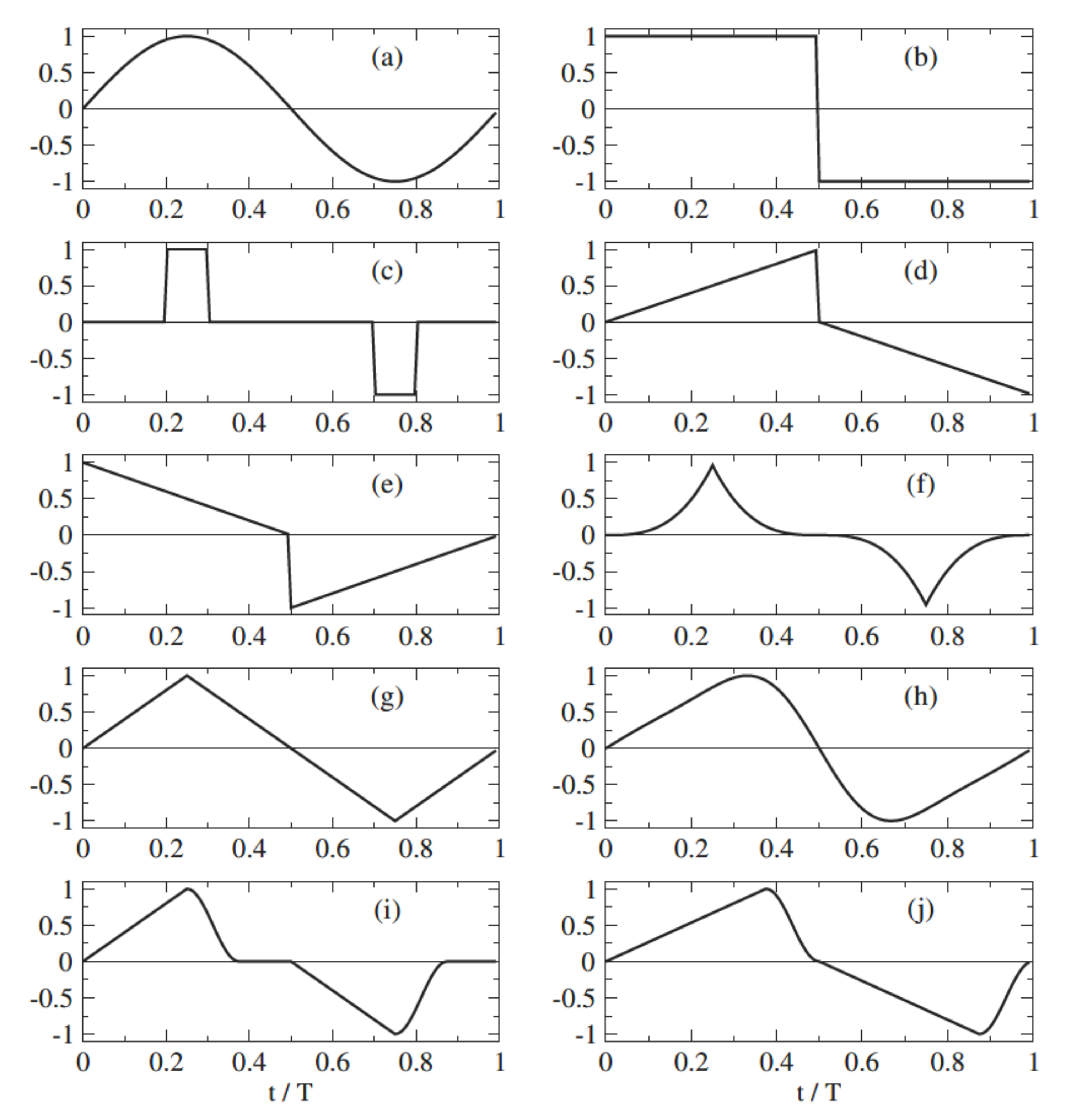}
\caption{Alternative temporal waveforms used in actuated flows. Taken from \cite{cimarelli-etal-2013}, with permission from AIP Publishing.}
\label{fig:waveforms}
\end{figure}

The study by \cite{cimarelli-etal-2013} explored different temporal waveforms, as shown in Fig. \ref{fig:waveforms}, using three periods, $T^+=62.5, 125, 250$, and three amplitudes, $W_m^+=2.25, 4.5, 9$. The profile (b) was found to outperform the sinusoidal actuation in terms of drag-reduction margin, but not in respect of net energy savings. All waveforms gave maxima of the net energy saving at $W_m^+=4.5$ and $T^+=125$, except (c) and (f), for which the optimum conditions were not in the tested parameter range. At the optimum parameters, $W_m^+=4.5$ and $T^+=125$, the sinusoidal waveform led to the best net energy savings, at $\mathcal{P}_{net}=7.8\%$, in agreement with \cite{quadrio-ricco-2004}, who reported $\mathcal{P}_{net}=7.3\%$. 
\cite{ cimarelli-etal-2013} also developed a predictive model that allowed the net energy saving to be estimated for any waveform. The model links the drag reduction with an effective penetration depth of the laminar Stokes layer, based on the distance from the wall where the induced velocity variance is lower than some threshold value. For a generic wave form decomposed as a Fourier series, the threshold value was calculated numerically through a sum over the harmonics. 
Under the constraint that the first harmonic was the most energetic, an analytical expression could be derived. With this model, a new waveform was designed for the low wall-velocity amplitude, $W_m^+=2.25$, which proved to yield a larger net energy saving, $\mathcal{P}_{net}=6\%$, than the simple sinusoidal forcing with $\mathcal{P}_{net}=5\%$.

The previously discussed difficulties in obtaining a positive net energy budget in boundary layers led \cite{mishra-skote-2015} to investigate spatial wall forcing in the form of a square wave, implemented by short moving segments. Only the positive part of the square-wave was enforced, thus arranging the wall as a series of alternating forced and unforced segments. The use of a smooth step function, so as to remove discontinuities in the forcing, was found to avoid oscillatory features at the step changes of the actuation velocity. An optimal set of wall oscillation parameters yielded $\mathcal{P}_{net}=18\%$.

\subsubsection{Non-canonical flows subjected to oscillatory forcing}

\cite{straub-etal-2017} performed simulations of a forced duct flow at $Re_\tau=180$, with a 3:1 width-to-height aspect ratio, in which the control was applied to the lower and upper walls. Oscillatory actuation was applied either over the entire surface of the walls or only to restricted segments. Due to the secondary motions in the duct flow, the oscillating-wall technique was much less successful than in channel flow. For example, by actuating the full extent of the walls at $W_m^+=12$ and $T^+=125$, \cite{straub-etal-2017} obtained a drag increase of almost 10\%, with all four walls contributing to the increase. The use of a lower forcing amplitude ($W_m^+=4.5$) and a restriction of the oscillating section to the middle part of the walls (with half of the surface oscillating) yielded $\R= 4.5\%$. 

In the study of \cite{zhao-huang-xu-2019}, the turbulent flow along the outer surface of a cylinder was altered by streamwise-travelling waves of wall circumferential velocity. 
This flow geometry was motivated by the cylindrical geometry of many flying objects (e.g., aircraft fuselages and missiles) and by their direction of motion being broadly aligned with the axis of symmetry. In this case the flow was however constrained by an outer cylinder, treated as an impermeable shear-free boundary, and driven by a pressure gradient along the gap. Hence, the flow was streamwise homogeneous, in contrast to the spinning cylinder flow discussed in Section \ref{sec:earlyexp}. The friction Reynolds number, based on the radial width between the inner and outer cylinders, was $Re_\tau=272$, while the gap-to-inner-radius ratio was three. 
The drag reduction was investigated as a function of the period and the wavelength for $W_m^+=16$. The maximum drag-reduction level was $\R=48.4\%$, comparable to the channel-flow value. For steady streamwise oscillations, the optimal wavelength, $\lambda_x^+ \approx 425$, was shorter than the corresponding optimum value in channel flow, $\lambda_x^+ \approx 1250$. This result was traced to the influence of centrifugal instabilities due to the curvature of the moving wall, as was also verified by a simulation with smaller curvature, yielding a larger optimum for $\lambda_x^+ \approx 842$. The maximum drag increase obtained was around $\R=-48\%$, i.e., much larger than for channel flow ($\R=-23\%$). This disagreement was explained by the rotation-induced vortices contributing to the increase in drag.

\cite{li-etal-2019} performed DNS of channel flow at $Re_\tau=180$ with a porous layer over the oscillating wall, using a Lattice Boltzmann method. Although the drag-reduction margin was lower than in the impermeable-wall case, the net energy savings, $\mathcal{P}_{net}=13\%$ for $W_m^+=4.5$ and $T^+=125$, increased above the corresponding value for a smooth wall, $\mathcal{P}_{net}=9\%$, for certain values of the Darcy number. This result was explained by the tendency of the fluid inside the porous medium to promote the oscillating motion during the decelerating phase, hence lowering the energy required for actuating the wall.

The DNS study by \citet{banchetti-etal-2020} of a channel flow at $Re_\tau=200$ with a bump on one of the walls showed that streamwise-traveling waves of spanwise wall velocity are also effective in reducing the drag of turbulent flows over smooth bumps and exhibiting mild separation.

\subsection{Flows subjected to non-periodic forcing}

This section considers three-dimensional flows, which are suddenly perturbed and in a transient non-equilibrium state. The discussion is confined to studies that specifically explore drag reduction in such flows.

The investigation by \cite{moin-etal-1990} of a channel flow subjected to a suddenly imposed spanwise pressure gradient revealed a reduction in the Reynolds shear stress, as well as a misalignment between the Reynolds-stress vector and the mean-velocity-gradient vector. They noted that their flow resembled, in terms of its response to the pressure gradient, the swept-wing experiments by \cite{bradshaw-pontikos-1985}, with the distinction that the simulation by \cite{moin-etal-1990} was streamwise-homogeneous, hence not involving the skewing that occurred in the experimental boundary layer of \cite{bradshaw-pontikos-1985}. \cite{moin-etal-1990} concluded that the decrease in turbulent kinetic energy was due to a lower production, which itself resulted from a drop in the pressure-strain correlation (\cite{durbin-1993} developed a turbulence model to reproduce these results).

\cite{coleman-kim-le-1996} investigated the impact of a suddenly imposed movement of the wall in a turbulent channel flow by DNS. They noted that the forcing in this case was equivalent to the imposed pressure gradient studied by \cite{moin-etal-1990} since the strain imposed close to the wall and flow response were the same, whether the fluid moved by a spanwise pressure gradient or by the motion of the wall. To avoid a singularity in the wall-normal direction created by a sudden start of the actuation, \cite{coleman-kim-le-1996} imposed a thin Stokes layer at an early stage of its evolution as the initial spanwise velocity profile. By investigating the enstrophy profiles, they argued that the drop in turbulent kinetic energy was due to the increased dissipation. Moreover, they noted that the largest drag-reduction margin occurred when the spanwise shear was applied in the region $y^+=5-15$, qualitatively consistent with the conditions that arise in the oscillating-wall case at the optimal period. \cite{coleman-kim-le-1996} also reported that the near-wall streaks were modified by the moving wall and that the interaction with the quasi-streamwise vortices was altered, although the vortices were not directly affected by the imposed spanwise shear.

In follow-up studies, \cite{coleman-kim-spalart-1996} and \cite{coleman-kim-spalart-2000} emulated swept-wing boundary layers with channel-flow DNS simulations. They applied irrotational temporal deformations to a channel flow by rotating the mean spanwise vorticity, hence creating a spanwise shear throughout the channel. The straining deformation generated a mean cross flow in the channel. They distinguished between the shear-driven and pressure-driven three-dimensional flow, with the former always experiencing reduced Reynolds shear stresses and drag, while the response of the latter depended on the streamwise pressure gradient. Although \cite{moin-etal-1990} imposed a spanwise pressure gradient, their flow qualified as a shear-driven case because the sole function of the pressure gradient was to create a skewed shear layer close to the wall.

\cite{le-coleman-kim-2000} studied the near-wall turbulence structures in a channel flow subjected to a spanwise-moving wall. A conditional quadrant analysis of the DNS data indicated that positive and negative streamwise vortices were affected differently, and hence the spanwise symmetry of the structures was significantly altered. The asymmetry developed due to changes in the vortical structures, while the streak length was reduced due to the different alignment of the vortices. 

\begin{figure}[t]
\begin{center}
\subfigure[]{\label{fig:kannepalli}\includegraphics[width=0.6\textwidth]{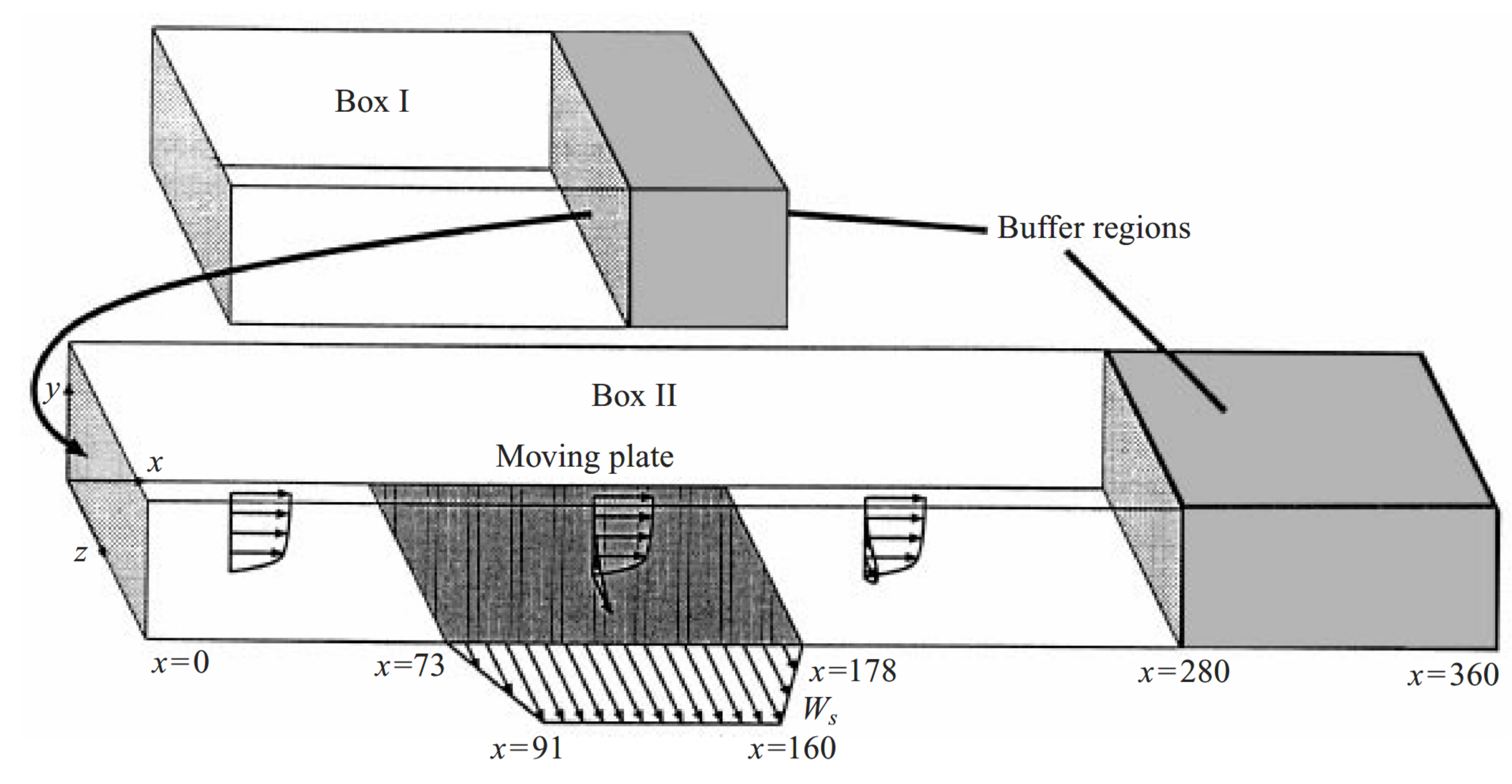}}
\subfigure[]{\label{fig:kannepalli-streaks}\includegraphics[width=0.6\textwidth]{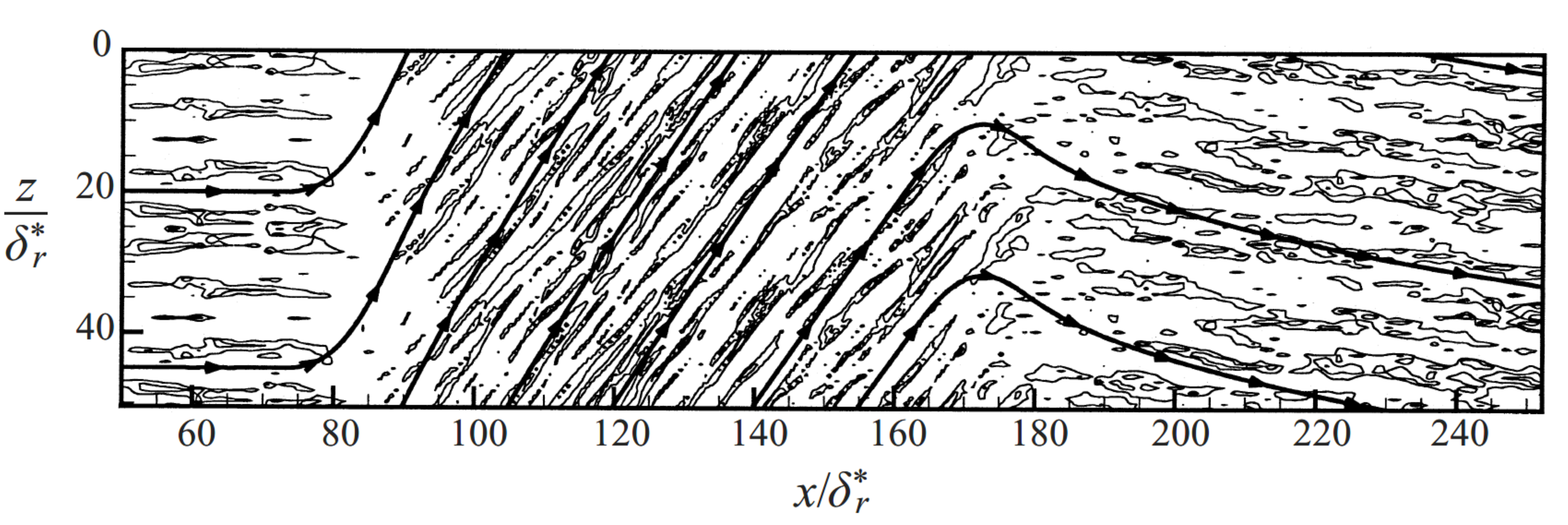}}
\caption{LES of Kannepalli and Piomelli; (a) Computational setup (compare with the experimental setup in Fig. \ref{fig:plesniak-morrison}): in Box I the initial laminar flow undergoes transition; a time sequence of turbulent data is used as inflow for the main computation in Box II;  in the buffer region at the end of the computational box, the governing equations are parabolized; (b) Low-speed streaks in a flow altered by a constant wall velocity in the segment $73 \leq x/\delta_r^* \leq 178$, where $\delta_r^*$ is the displacement thickness at the inflow of Box I (compare with the spatially oscillating wall velocity case in Fig. \ref{fig:skote-spatial}). Taken from \cite{kannepalli-piomelli-2000}, with permission of CUP.}
\end{center}
\end{figure}

\cite{kannepalli-piomelli-2000} performed a LES of a flow similar to that in the experiments by \cite{kiesow-plesniak-2003}, i.e., a turbulent boundary layer subjected to a moving section of the wall, as shown in Fig. \ref{fig:kannepalli}. The analysis of the budget for the turbulent kinetic energy revealed that a reduced level of production of the kinetic energy was the cause for the decline. The low-speed streaks were observed to shorten and realign to the direction of the mean shear, as shown in Fig. \ref{fig:kannepalli-streaks}. The region of reduced streamwise skin friction was located immediately downstream of the upstream end of the moving plate, where the low-speed streaks were weakened the most. The streamwise drag subsequently increased above its nominal level further downstream, before being marginally reduced again at the end of the moving section.

\cite{howard-sandham-2000} performed a simulation of a channel where both walls suddenly moved in the same spanwise direction at a constant velocity, using an initial spanwise-velocity profile that was identical to that used by \cite{coleman-kim-le-1996}. The near-wall quasi-streamwise vortices were reduced in strength and moved closer to the wall. The turbulence level and the drag decreased during the transient phase in which the flow adjusted to the forced state. The inability of turbulence models to capture the reduction in mean turbulence kinetic energy was demonstrated, and several modifications of the models were suggested.

A three-dimensional boundary layer at $Re_\theta=1200$ subjected to streamwise-varying spanwise pressure gradients was investigated by \cite{bentaleb-leschziner-2013} using DNS. The motivation was to create a database for comparison with turbulence-model predictions. The forcing consisted of three concurrent streamwise segments of positive, zero, and negative spanwise pressure gradients. Only a very small decline in the streamwise wall-shear stress occurred at the beginning of the first segment, while it increased above the nominal boundary value in the second segment before decreasing again in the third segment. The turbulent kinetic energy and most of the Reynolds stress components, however, dropped below the levels of the reference boundary layer. The exception was the spanwise normal and spanwise shear stresses that arose due to the increased spanwise shear strain. The decline in the wall-normal Reynolds stress caused a drop in the streamwise Reynolds shear stress and indicated that the near-wall streaks and the regeneration mechanism of the near-wall turbulence were disrupted by the spanwise pressure gradient. 

In the study of \cite{he-etal-2018}, the initial response of a channel flow to a sudden imposition of a spanwise pressure gradient was investigated. The channel flow, at a Reynolds number up to $Re_\tau=934$, was subjected to a constant spanwise pressure gradient, thus producing a developing Stokes flow. The spanwise flow was in agreement with the solution of the extended Stokes first problem. The Stokes layer first yielded a lower streamwise skin friction, followed a buffeted laminar boundary layer that underwent bypass transition. As the flow was turbulent from the beginning, the term ``turbulent-turbulent transition'' was used to characterize the flow. 

\cite{lozano-duran-2020} adopted the same conditions investigated by \cite{moin-etal-1990}, but at the higher Reynolds number of $Re_\tau=1000$. The reduction in the Reynolds shear stress progressed together with the growth of the spanwise shear layer. By performing a numerical experiment in which a free-slip boundary condition was applied in the spanwise direction, they showed that the spanwise acceleration by the spanwise pressure gradient alone did not reduce the Reynolds shear stress. Hence, the spanwise shear layer was responsible for the reduction, not the pressure gradient per se. By their multiscale analysis, \cite{lozano-duran-2020} concluded that larger eddies were more sensitive to the spanwise pressure gradient than the smaller ones. Based on these observations, they developed a model for the structural properties, after scrutinizing previous models \citep[refer to the review by][]{johnston-flack-1996}.

\section{Flow physical mechanisms and interactions}
\label{sec:phys}

The foregoing sections gave ample evidence that the focus of most research on drag reduction by spanwise motion has been on the quantification of the drag-reduction margin for a broad range of actuation scenarios and parameters. Studies that instead primarily aim to interpret the basic physical mechanisms responsible for the flow alteration and the drag reduction are rarer. 

It is fair to state that there is no unambiguous, or indeed generally accepted, paradigm that explains the physical mechanisms behind drag reduction in a fully satisfactory and convincing manner. When faced with disparate interpretations, the reader will hopefully accept that there are some reasonably credible propositions, mostly advanced on the basis of the analysis of flows altered by the streamwise-uniform actuation \eqref{eq:oscillation}, rather than by the wavy wall motion \eqref{eq:waves}. The existing conjectures -- some mutually supportive, others competing, or even conflicting -- have attributed the drag reduction to one or the combination of the following phenomena: (i) a reduced turbulence production; (ii) a rising dissipation; (iii) interactions of the Stokes layer with, and directional distortions in, the turbulent enstrophy components; (iv) hysteresis effects during the actuation cycle; (v) a reduction of sweep and ejection events; and (vi) structural features associated with vortical near-wall motions. A main route to explaining the physical processes has entailed the interpretation of statistical quantities, either time-averaged, phase-averaged, or conditionally averaged. Physical insight has also been gained by studying the stability of the near-wall streaks by use of the linearized Navier-Stokes equations, as further discussed in Section \ref{sec:models}.

\subsection{Mean-flow behaviour}
\label{sec:mean-flow}

\begin{figure}
    \centering
    \includegraphics[trim= 0 0 0 0, clip=true,width=0.55\textwidth]{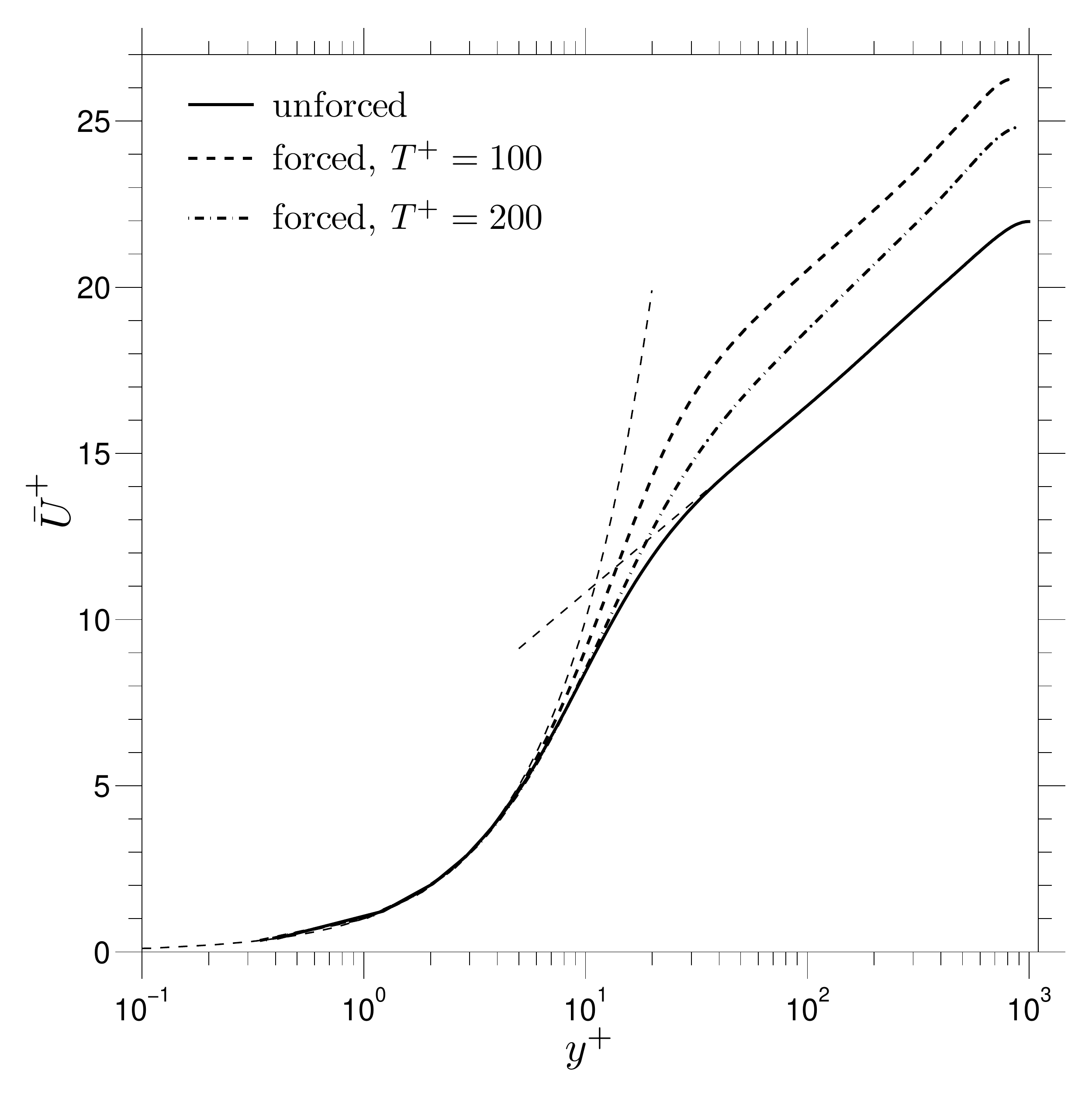}
    \caption{Log-law mean velocity profiles for unactuated and actuated channel flow at $Re_\tau=1000$.  Actuation is by spanwise-homogeneous oscillatory transverse wall motion without streamwise waves. Taken from \cite{agostini-touber-leschziner-2014}, with permission from CUP.}
    \label{Fig:5}
\end{figure}

An observation put forward in many studies in which drag is reduced by some wall-based control, passive or active, is that the mean-flow velocity $\overline{U}$, when scaled with the friction velocity of the actuated flow, shows the logarithmic behaviour typical of unactuated flows, i.e.,
\begin{equation}
\overline{U}^+=\frac{1}{\kappa}\ln y^+ + B,
\label{eq:log-law}
\end{equation}
where $\kappa$ is the von K{\'a}rm{\'a}n constant and $B$ is the additive constant. In drag-reducing conditions, the additive constant $B$ increases as the profile shifts upward relatively to the canonical case, while the buffer layer thickens, extending to larger $y^+$ values. 
These results are illustrated in Fig. \ref{Fig:5} by the mean channel-flow profiles at $Re_\tau=1000$ in the fixed-wall configuration and with spanwise wall motion at two frequencies \citep{agostini-leschziner-2014b}. The plot provides an indication that the slope of the logarithmic portion may not be constant. This point was also made in the light of a recent DNS study by \cite{yao-etal-2019} on a channel flow and a theoretical analysis by \cite{skote-2014} on a boundary layer, the latter albeit including the observation that the slope of the log-portion is restored to its canonical value if the flow has completely adjusted to the forcing.
Nevertheless, there is little doubt that the main feature is the upward shift in the logarithmic law. \cite{hurst-etal-2014,chung2016dns} reported two sets of DNS-derived log-law distributions for two actuation conditions at two near-optimum oscillating period and suggested that the upward shift in the log-law tended to become constant as the Reynolds number increased, while the other actuation parameters remained invariant. \cite{gatti-quadrio-2016} proposed that the lower limit beyond which this upward shift is essentially constant was $Re_\tau \approx 1000$. 

Even if the relevant issue of the variability in the log-law slope is set aside, the observation of constancy of the upward log-law shift needs to be viewed with caution, because it is only based on two Reynolds-number values, $Re_\tau$=800 and 1600. Further related evidence was provided by \cite{gatti-quadrio-2016}, who reported a large number of results for $Re_\tau$ in the range 200-1000. They argued that the von K\'{a}rm\'{a}n constant was close to 0.39 for all Reynolds numbers, but it must be pointed out that the $y^+$-range over which the slope is approximately constant is very limited at such low Reynolds numbers.  

Simulations for spatially developing boundary layers by \cite{skote-2014} and \cite{skote-mishra-wu-2015} similarly showed a distinctive upward shift in the log-law in response to the wall actuation. While this shift has been used by the authors to investigate the Reynolds-number dependence of the drag reduction, this investigation needs to be viewed with caution because the maximum friction Reynolds number reached at the end of the simulation domain was below 700. The question is, thus, whether a constant, Reynolds-number-independent, shift in the log-law is a sound assumption. This point is further discussed in Section \ref{sec:re-deltaB}.

\subsection{Time history of the wall-friction drag}
\label{sec:genobs}

\begin{figure}
\centerline{\includegraphics[trim= 0 0 0 0, clip=true,width=0.5\textwidth]{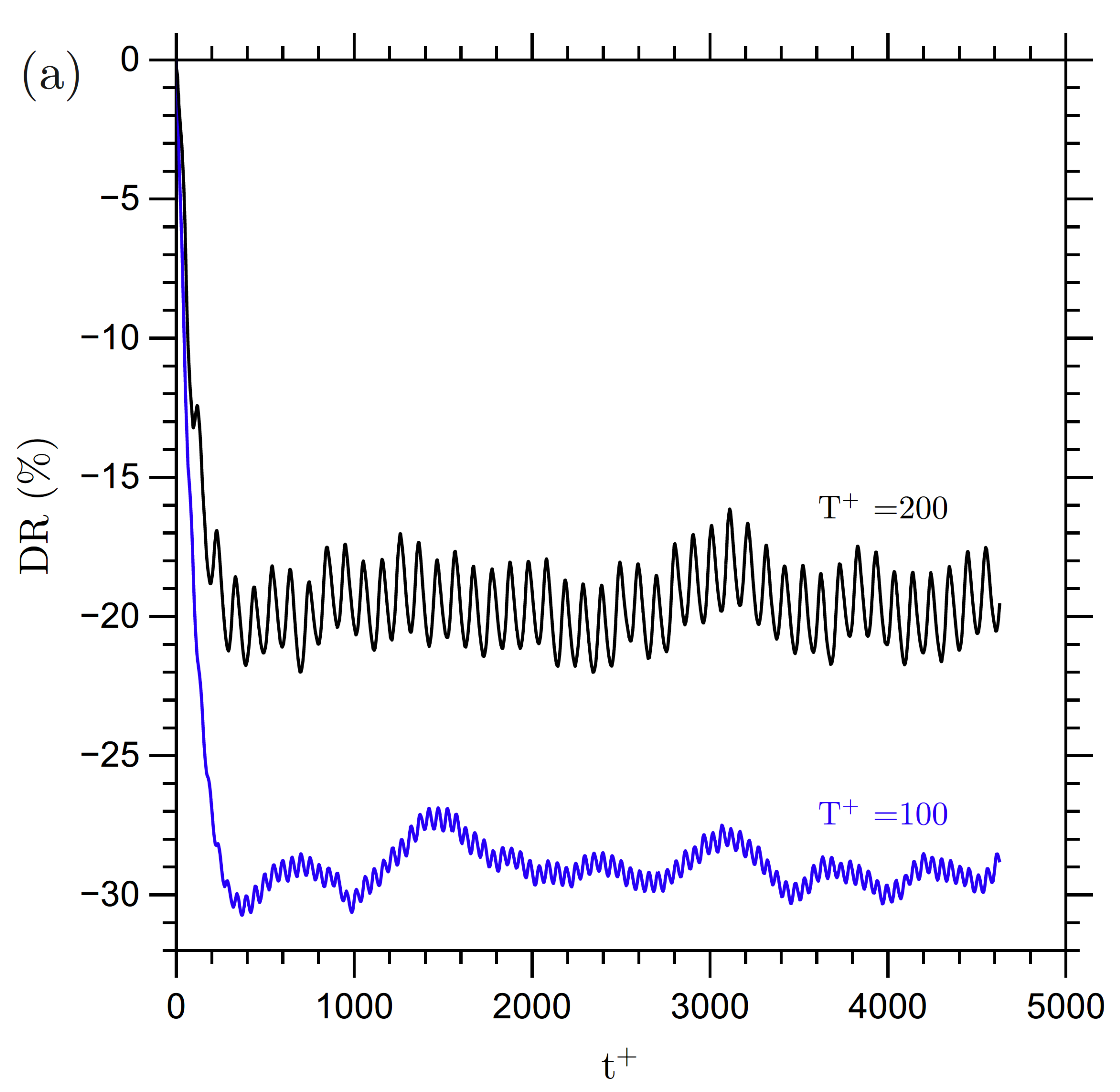}}
\caption{Temporal variation of wall-averaged drag reduction (DR) following a sudden start of actuation in a channel flow with purely spanwise actuation, $Re_\tau=1000$. Reproduced from \cite{agostini-touber-leschziner-2014}, with permission from CUP.}
\label{fig:DR}
\end{figure}

The time histories of the drag reduction, illustrated in Fig. \ref{fig:DR} for a channel flow at $Re_\tau=1000$, show the presence of two types of fluctuations following the initial transient phase after the actuation commences. The fluctuations showing a well-defined frequency of oscillations are directly related to the wall motion, while the larger, more irregular, fluctuations have time scales that are much larger than the oscillating period. At the near-optimum condition, $T^+=100$, the drag reaches the lowest value ($\mathcal{R} \approx 30\%$) and the oscillations that are linked directly to the wall actuation are insignificant compared with the average drag-reduction margin. As the oscillating period increases from the optimal value, the drag reduction declines and the direct influence of the forcing becomes more evident through the larger drag fluctuations. The period of the fluctuations is induced directly by the wall motion and its value is half of the wall-oscillating period. In the fixed-wall case, no time modulation of the spatially-averaged wall-shear stress should occur for a sufficiently large computational box. However, in the moving-wall case, it remains to be verified whether the slower-scale drag fluctuations, also clearly illustrated in Fig. \ref{fig:DR}, are caused by a nonlinear interaction between the oscillating Stokes strain and the long structures appearing the log-law layer, especially at larger Reynolds number, as discussed in Section \ref{sec:re-physics}, or whether they are a consequence of the finite computational box. Careful tests on the influence of the domain size on these time modulations will clarify this point.

The differences between the two time histories are due to the Stokes-layer thickness and to the variation of the Stokes strain, i.e., the wall-normal derivative of the spanwise velocity. For periods smaller than and close to the optimum period, the unsteady Stokes strain is confined in the viscous sublayer and in the lower part of the buffer layer, i.e., $y^+<15$, and the phase- and space-averaged spanwise profiles show excellent agreement with the Stokes-layer laminar solution \eqref{eq:stokes}. This scenario was confirmed by the water-channel experiments of \cite{ricco-2004}, i.e., the optimal condition for drag reduction occurs when the low-speed streaks, populating the viscous sublayer, are cyclically dragged and smeared laterally being fully embedded in the Stokes layer, while the quasi-streamwise vortices, mostly present in the buffer layer, are not directly sheared along the span.

As the period increases from the optimal value, the enhanced amplitude of the oscillation is indicative of the penetration of the Stokes layer in the upper portion of the buffer layer and of the increased strain-induced turbulence production. The space-averaged spanwise profiles differ from the laminar solution because the additional Reynolds stresses, given by the cross-flow correlation $\overline{vw}$, modify the mean spanwise profile \citep{ricco-quadrio-2008}. The total strain rate increases and decreases periodically at locations from the wall where most of the turbulent kinetic energy is generated. The shear-induced turbulence generation is thus cyclically affected by the spanwise Stokes layer, therefore causing a distinct temporal modulation of the wall drag. For $T^+\approx 1000$, the penetration of the Stokes strain into the upper layer is sufficiently large to cause a 6\% mean-drag increase relative to the fixed-wall value. These observations do not, however, explain the processes responsible for the drag reduction. The quest for an exhaustive explanation of the drag-reduction mechanism has branched into many different paths, discussed in the following sections.

\subsection{Statistics of the velocity and vorticity fluctuations}
\label{sec:rms}

The study of near-wall turbulence statistics is key to understanding the physical changes brought about by the wall motion. Two different velocity scales can be used for the inner scaling of the turbulence statistics in a channel flow under CFR conditions. Scaling with the friction velocity of the reference flow brings out the absolute changes of the quantities, while scaling with the actual friction velocity leads to the correct non-dimensionalisation of the mean flow near the wall, and thus allows for a comparison between the near-wall drag-reduced statistics and the statistics of the unactuated turbulent flow at the same friction Reynolds number. Under CPG conditions, the inner velocity scaling is unique because the mean wall-shear stress is constant. 

\begin{figure}[t]
\centering
\includegraphics[width=0.47\textwidth]{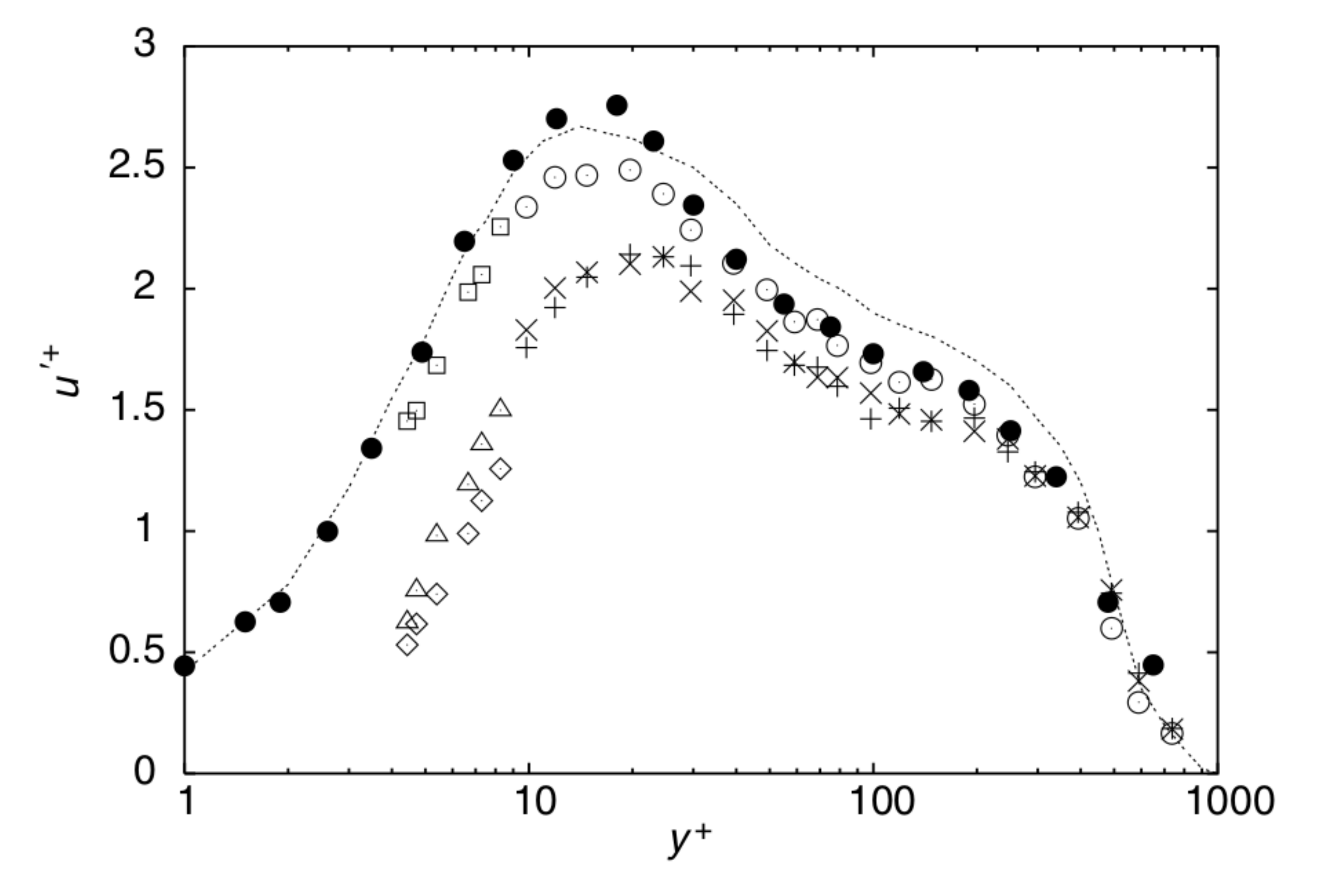}
\includegraphics[width=0.49\textwidth]{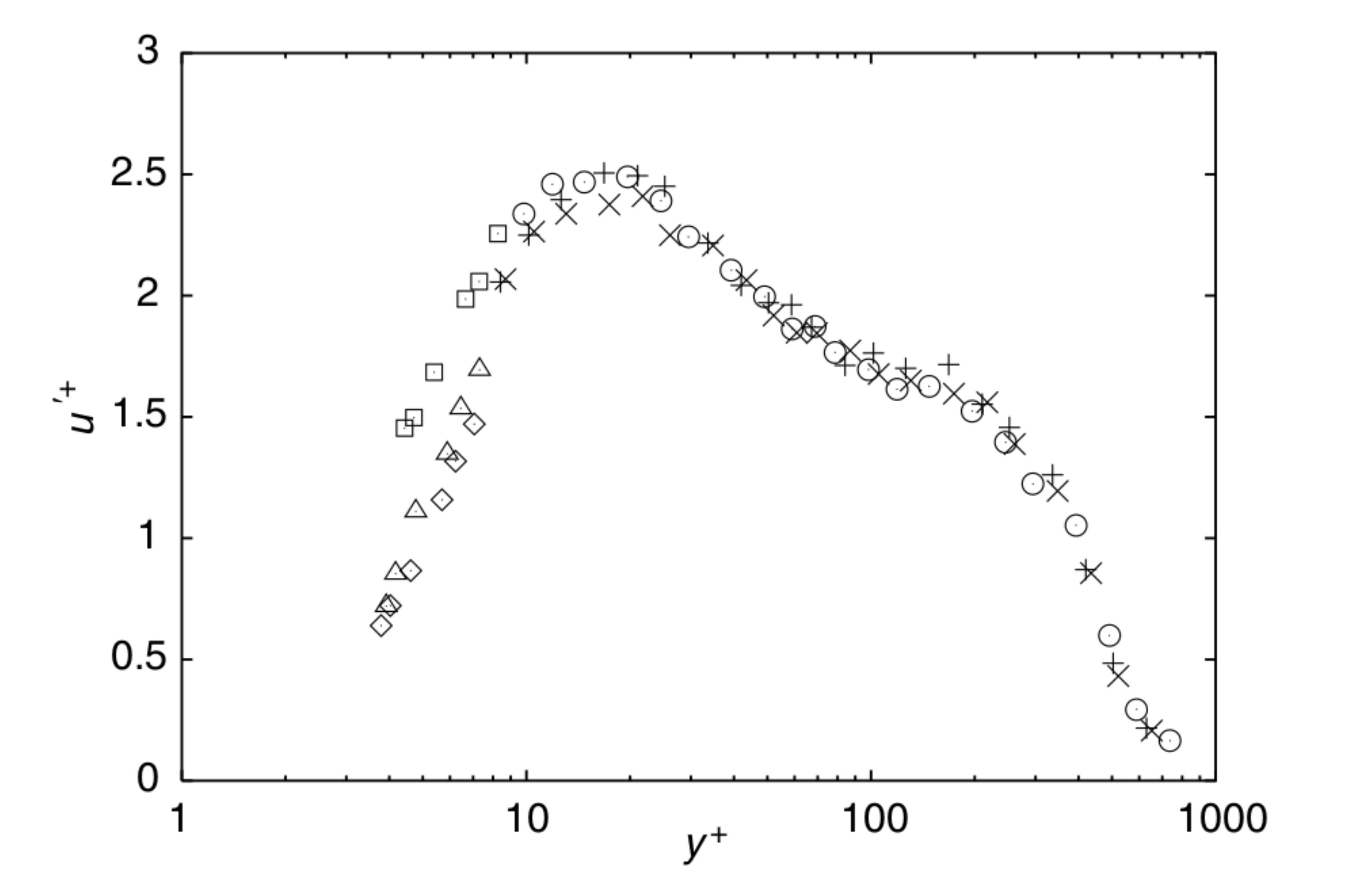}
\caption{Root-mean-square of the streamwise velocity fluctuations measured in a water-channel turbulent flow altered by spanwise wall oscillations \citep{ricco-wu-2004-a}. The white circles and the white squares indicate
 the reference values (black symbols and line are for validation with earlier experiments and DNS), while the other white symbols denote the values in the oscillating-wall case. Left: scaling by the reference wall-friction velocity; right: scaling by the drag-reduced wall-friction velocity. Taken from \cite{ricco-wu-2004-a}, with permission from Elsevier.}
\label{fig:ricco-wu-2004-u}
\end{figure}

As shown in Fig. \ref{fig:ricco-wu-2004-u} for a turbulent boundary layer, the root-mean-square of the streamwise velocity fluctuations is strongly affected by the actuation. When the scaling with the reference friction velocity is adopted, as in Fig. \ref{fig:ricco-wu-2004-u} (left), the peak is attenuated and shifted away from the wall by the wall oscillation. The wall-normal velocity fluctuations are also strongly damped. When the drag-reduced wall-unit scaling is used, as in Fig. \ref{fig:ricco-wu-2004-u} (right), the wall-normal velocity profiles agree more closely to the reference profile than when scaled with the reference friction velocity, showing an excellent match for $y^+>20$. Closer to the wall, the reduction in streamwise-velocity fluctuations is proportionally larger than the reduction of the mean-flow streamwise velocity. This non-dimensionalization reveals that the wall oscillation, besides reducing the wall-shear stress and thus the Reynolds number, significantly alters the character of the turbulent flow up to the buffer region.

\begin{figure}[t]
\centering
\includegraphics[width=\textwidth]{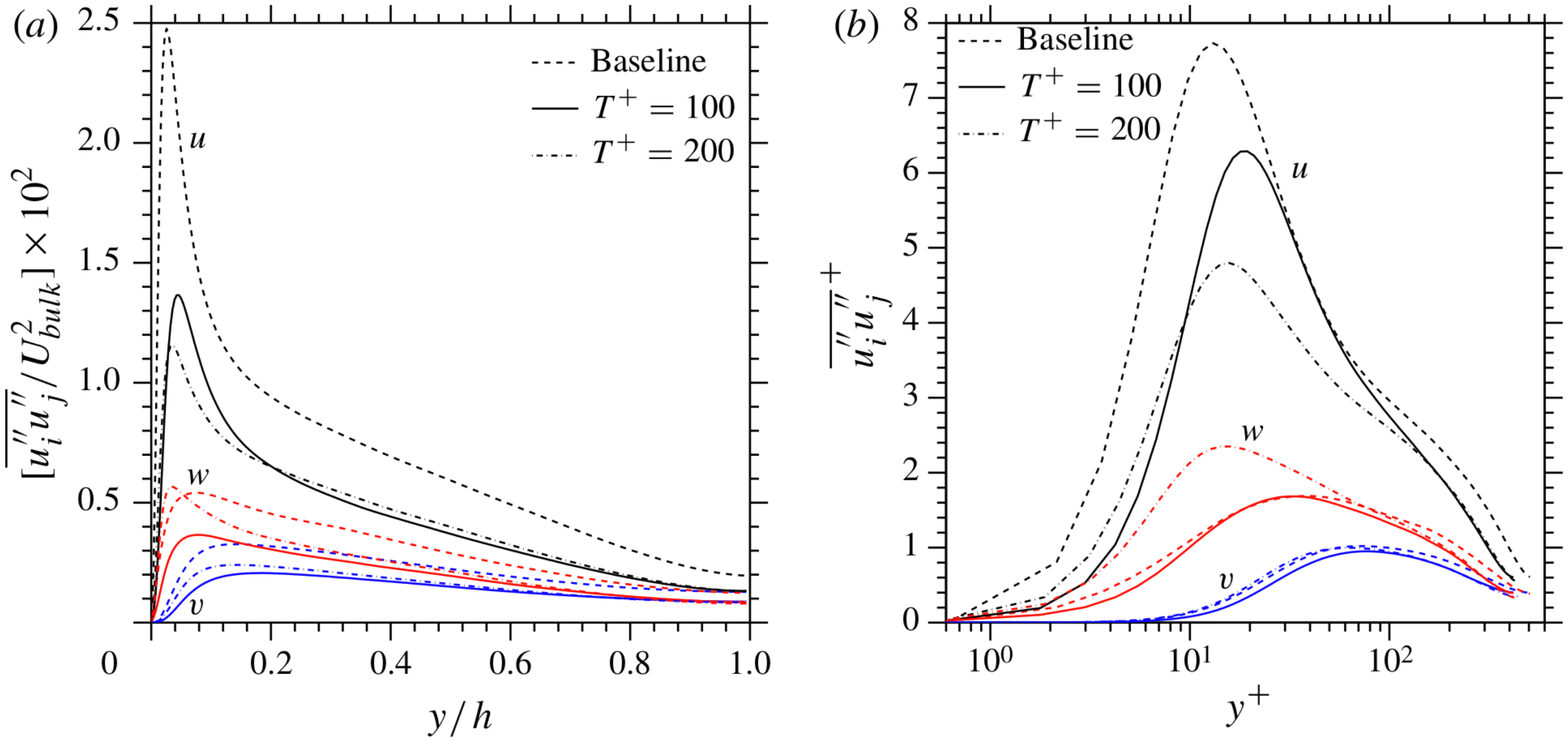}
\caption{Wall-normal profiles of Reynolds normal stresses in a channel flow at $Re_\tau = 500$ and two periods of oscillation, $T^+=100$ and 200: (a) outer-scale units; (b) inner-scale units. Taken from \cite{touber-leschziner-2012}, with permission from CUP.}
\label{fig:touber-leschziner-rms}
\end{figure}

The experimental results of the streamwise velocity by \cite{ricco-wu-2004-a} in Fig. \ref{fig:ricco-wu-2004-u} were confirmed by the numerical results of \cite{touber-leschziner-2012}, shown in Fig. \ref{fig:touber-leschziner-rms}. The stresses are purely stochastic in nature, i.e., they exclude any contributions that arise from actuation-induced periodic (deterministic) perturbations. The periodic oscillations associated with the wall forcing must be removed to obtain the stochastic velocity fluctuations, which is especially important for the spanwise velocity component that includes the large Stokes-layer oscillations \citep{quadrio-ricco-2011,touber-leschziner-2012}. In Fig. \ref{fig:touber-leschziner-rms}, the wall-normal velocity component is attenuated drastically, indicating that the ejections and sweeps are significantly damped near the wall. The behaviour of the spanwise component is more complex and depends on $T^+$. At near-optimum conditions, $T^+=100$, this component also reduces, but non-optimum actuation at $T^+=200$ enhances the generation of the spanwise component, which thus increases, as shown in Fig. \ref{fig:touber-leschziner-rms}. Increased peak values of the transverse velocity fluctuations have also been observed in a channel flow with oscillating walls under CPG conditions \citep{ricco-etal-2012} and in a boundary layer forced by a standing wave \citep{skote-2013}.

\begin{figure}[t]
\centering
\includegraphics[width=0.49\textwidth]{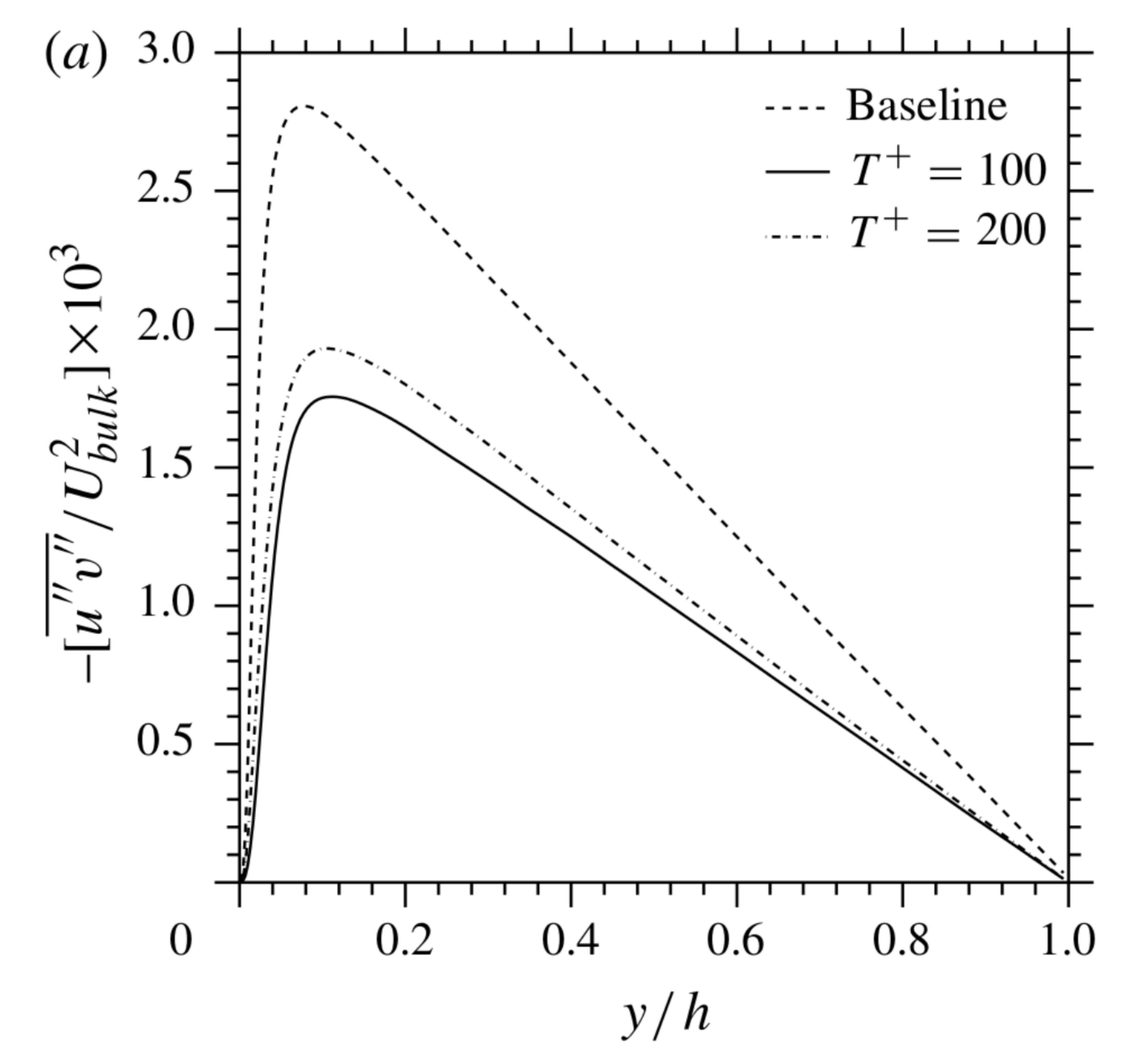}
\includegraphics[width=0.49\textwidth]{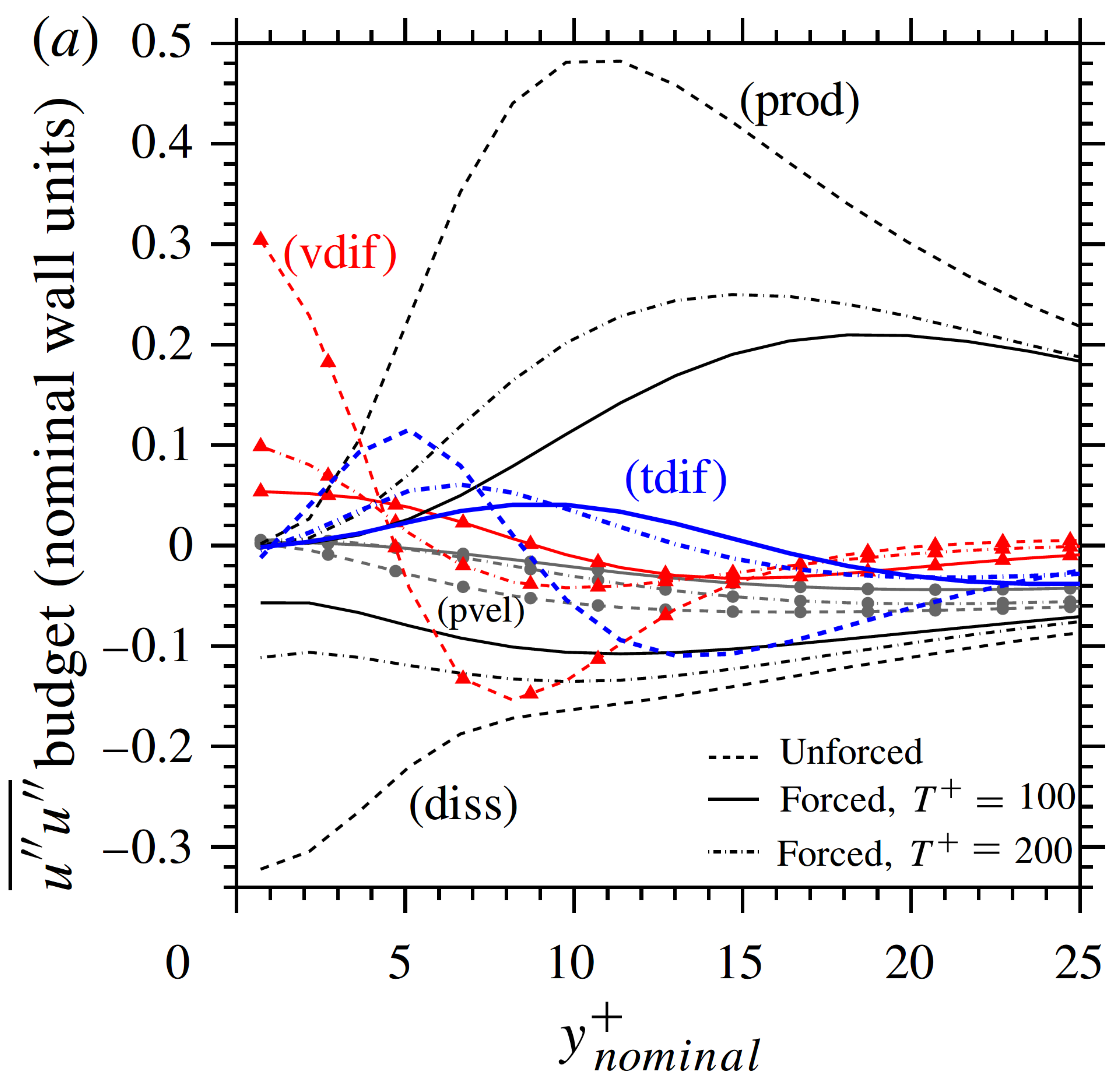}
\caption{Wall-normal profiles in a channel flow at $Re_\tau = 500$ and two periods of oscillation, $T^+=100$ and 200. Left: Reynolds shear stresses; right: Budget of the terms governing the streamwise stress (prod=production, diss=dissipation, tdif=turbulent diffusion, pvel=pressure-velocity interaction). Reproduced from \cite{touber-leschziner-2012}, with permission from CUP.}
\label{fig:touber-leschziner-uv}
\end{figure}

There is unambiguous evidence that the spanwise actuation causes the near-wall Reynolds shear stresses $\overline{uv}$ and turbulent kinetic energy to decline along with the reduction in drag. The attenuation of the Reynolds stresses is shown in Fig. \ref{fig:touber-leschziner-uv} (left). The joint probability density function contracts substantially \citep{zhou-ball-2008, touber-leschziner-2012, yuan-etal-2019}, revealing a drastic reduction in the streamwise and wall-normal turbulent fluctuations and the shear stress. 

\cite{touber-leschziner-2012} analysed the effect of the wall motion on the transport equation for the streamwise stress $\overline{uu}$. Fig. \ref{fig:touber-leschziner-uv} (right) shows the profiles of the production, dissipation, turbulent diffusion, viscous diffusion, and pressure-velocity interaction of $\overline{uu}$. The production in the buffer layer and the dissipation in the viscous sublayer are dramatically reduced.

The correlations of the vorticity fluctuations are relevant because they provide information about the vortex dynamics, which is relevant to turbulent mixing. The streamwise component is linked to the quasi-streamwise vortices, while the wall-normal and spanwise components near the wall are related to the intensity and regeneration of the low-speed streaks, respectively. 

\begin{figure}[t]
    \centering
    \includegraphics[width=0.9\textwidth]{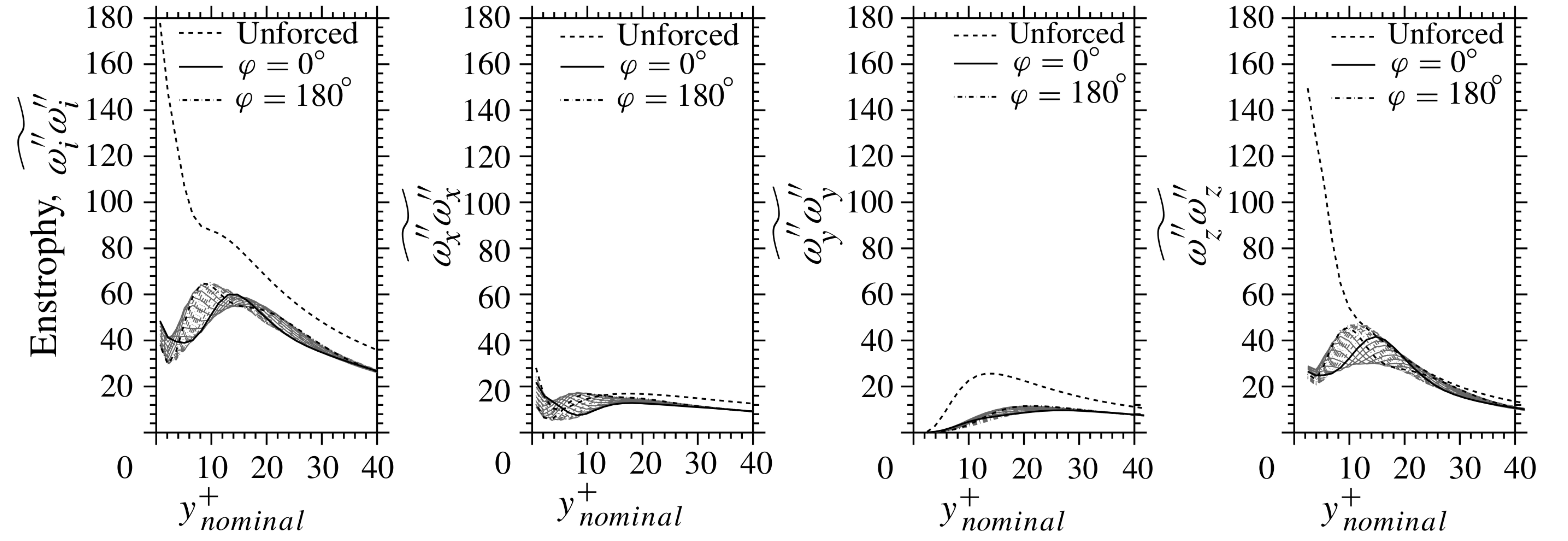}
    \caption{Profiles of enstrophy and of its contributions for $Re_\tau=500$ and $T^+=100$, phase-averaged over fifty periods of oscillation. Taken from \cite{touber-leschziner-2012}, with permission from CUP.}
    \label{fig:tou-les-2012-enstrophy}
\end{figure}

Under CFR conditions, \cite{touber-leschziner-2012} reported the phase-averaged correlations of the vorticity fluctuations for a drag-reduction case, shown in Fig. \ref{fig:tou-les-2012-enstrophy}. As the actuation is close to the optimum condition, the variations about the time-averaged profiles are modest. The spanwise vorticity correlation, $\overline{\omega_z \omega_z}$, are attenuated the most. As the intensity of the low-speed streaks is reduced, so is the wall-normal vorticity correlation, $\overline{\omega_y \omega_y}$, because this quantity is representative of the energy of the near-wall streaks. The streamwise component, $\overline{\omega_x \omega_x}$, indicative of the quasi-streamwise vortices and of the intensity of the uplift ejections/sweeps that give rise to the low-speed streaks, also diminishes, but its reduction is somewhat weaker than the other components. This disparity is surprising, as the wall oscillation is expected to disrupt and attenuate the low-speed streaks, the wall-normal fluctuations and the quasi-streamwise vortices in roughly equal measure. This suggests that the drag-reducing mechanisms are more involved than simply the direct damping action on all the near-wall structures. \cite{quadrio-ricco-2011} also reported the turbulent vorticity components, altered by streamwise-travelling waves in a channel flow under CPG conditions. The streamwise vorticity component increases near the wall in the forced case. The wall-normal and spanwise vorticity components are respectively enhanced and attenuated up to the viscous sublayer, while the opposite occurs at larger distances. This counter-intuitive increase of vorticity fluctuations is explained in Section \ref{sec:prod_and_diss} in terms of the increased turbulent dissipation that occurs under CPG conditions. The study of vorticity is also relevant because the square of the amplitude of the vorticity vector is the enstrophy, linked to the dissipation rate, as discussed further in Section \ref{sec:prod_and_diss}.

\subsection{Effect of the wall oscillation on the near-wall vortical motions}
\label{sec:vortical}

The observation that the reduction in drag is accompanied by a reduction in the intensity of the turbulent velocity and vorticity fluctuations near the wall has stimulated the interest in the physical changes of the vortical motions induced by the wall oscillation. Therefore, while Section \ref{sec:rms} discusses the changes of the turbulence statistics brought about by the wall motion, the present section presents the results on the alteration of the vortical structures near the wall.

Some of the earliest papers on the effect of wall motion on the turbulent flows refer back to the study of \cite{sendstad-moin-1991} on the structural changes of a transient three-dimensional flow to gain insight on how the near-wall turbulent structures are altered (the simulation results for this analysis were taken from the DNS by \cite{moin-etal-1990}). \cite{sendstad-moin-1991} noticed that, when a spanwise mean flow develops, the quasi-streamwise vortices shift in the spanwise direction relatively to the near-wall streaky structures, weakening the streaks. They also observed that the vortices with the same sign as the longitudinal mean-flow vorticity became weaker. Their conclusion regarding the cause for the drag reduction is contrary to numerous subsequent studies, however. \cite{sendstad-moin-1991} claim that the streak attenuation, and thus the drag reduction, is caused by the spanwise shear pumping high-speed fluid down towards the low-speed streaks and upward from the high-speed streaks, both processes being typically associated with an increase of the wall-shear stress via an intensification of the Reynolds stresses.

\begin{figure}
  \centerline{\includegraphics[trim= 0 0 0 0, clip=true,width=0.6\textwidth]{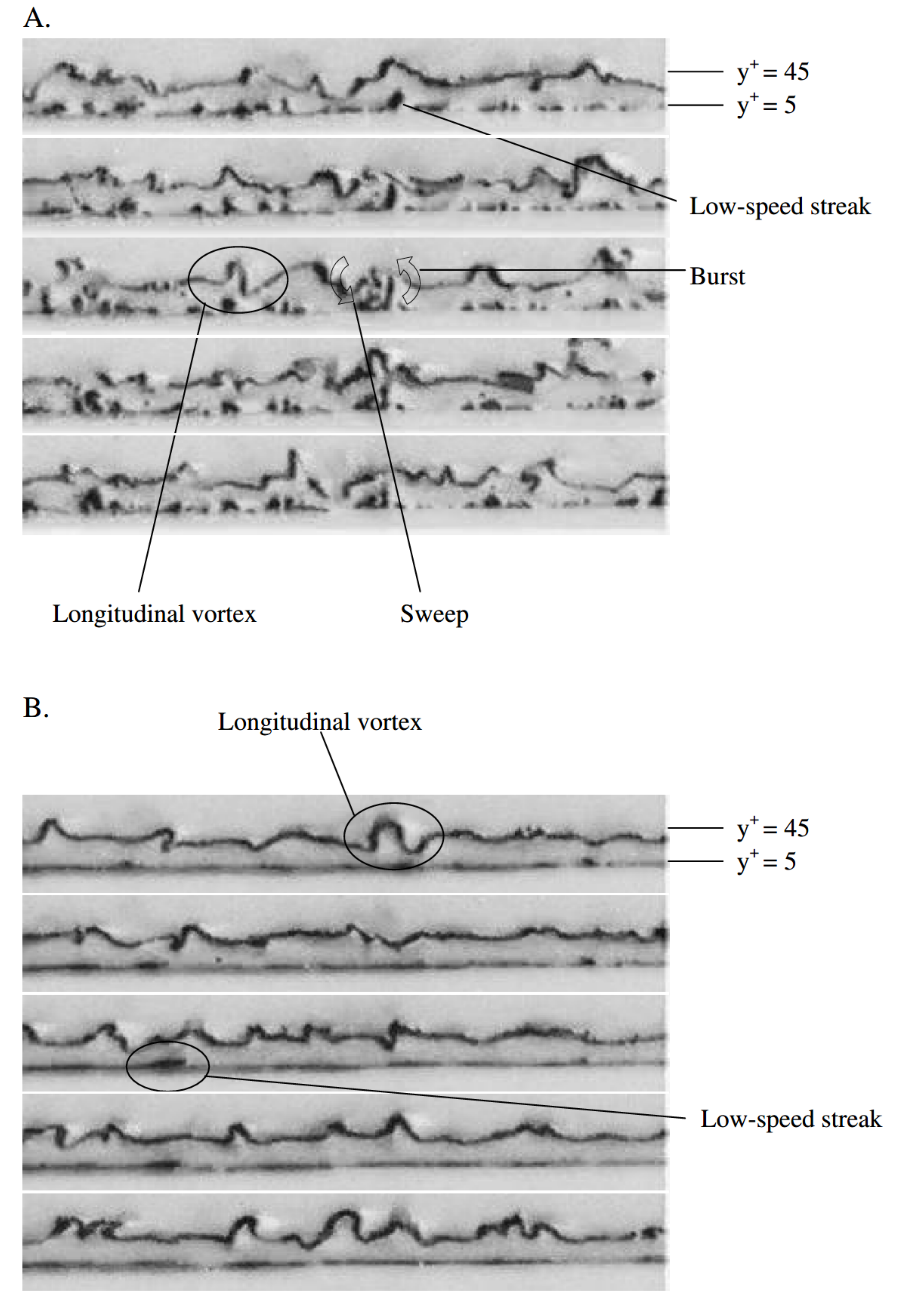}}
  \caption{Flow visualization of the near-wall turbulent structures by hydrogen bubbles generated by platinum wires at $y^+ = 5$ and 45 \citep{ricco-2004}. The views are from the end of the channel with the wall at the bottom of the pictures. The time between each frame (the images go from top to bottom) is $t^+$=5.4, while the width of the frames is $z^+$=900. The fixed-wall flow is shown in images A and the flow alterated by the oscillating plate is shown in images B. The circled regions indicate examples of the flow structures and their interactions, the former defined according to the criteria specified in \cite{ricco-2004}. The quasi-streamwise vortices are located above the streaks and in B the streaks are fewer and shifted in the spanwise direction compared to the vortices. Taken from \cite{ricco-2004}, with permission from Taylor \& Francis.}
\label{fig:R-vis}
\end{figure}

The analysis of the flow structures by \cite{akhavan-jung-mangiavacchi-1993}, derived from the simulations of \cite{jung-mangiavacchi-akhavan-1992}, first suggested that the suppression of turbulence was caused by a weakening of the low-speed streaks and by a decrease in the number and intensity of turbulent bursts in the oscillated flow compared to the unperturbed flow. A central observation was the continuous shift of the near-wall streamwise vortices, located above $y^+$=15, relative to their associated streaks, located below $y^+ = 15$, later confirmed by the water-channel flow visualizations of \cite{ricco-2000} and \cite{ricco-2004}, as illustrated in Fig. \ref{fig:R-vis}. 
\cite{akhavan-jung-mangiavacchi-1993} also speculated that similar energy-efficient methods for disrupting the near-wall turbulence could be developed by either passive methods, such as guiding vanes or surface grooves, or active methods, such as alternative spanwise-forcing actuators or wall jets. These interesting suggestions have been pursued by several researchers, as amply discussed in Section \ref{sec:extensions}. 

\cite{laadhari-skandaji-morel-1994} agreed with the physical paradigm proposed by \cite{sendstad-moin-1991}, in that the effect of the streamwise vorticity induced by the spanwise mean velocity gradient is to alter the quasi-streamwise vortices and the low-speed streaks, which leads to a reduction of the burst-sweep activity, an increase of near-wall viscous dissipation, and thus drag reduction. 

The explanation for the drag reduction proposed by \cite{akhavan-jung-mangiavacchi-1993} and \cite{laadhari-skandaji-morel-1994}, rooted in the disruption of the interactions between the vortices and the streaks, was elaborated by \cite{baron-quadrio-1996}, based on the argument that the Stokes layer must embed as many low-speed streaks as possible without influencing the quasi-streamwise vortices. This optimal forcing scenario allows the maximum relative spanwise displacement between the streaks and the vortices. \cite{baron-quadrio-1996} also estimated the optimal period of oscillation at $T^+ \approx 100$ with the aid of the penetration depth of the Stokes-layer, calculated from equation \eqref{eq:stokes}. 

Although \cite{akhavan-jung-mangiavacchi-1993} described the shift of the near-wall vortices by the oscillating flow driven by a spanwise pressure gradient, while \cite{baron-quadrio-1996} and subsequent investigators discussed the streaks as being moved by the induced shear caused by the oscillating wall, in both cases the consequence is the broken interaction due to the shift in the relative position of turbulence structures, which weakens the streaks. The relative displacement is visualized by the end-view flow visualizations of \cite{ricco-2004} in Fig. \ref{fig:R-vis}b, where the (much fewer) low-speed streaks are no longer correlated with the streamwise vortices when the wall oscillates.

The central idea that the lateral movement of the surface causes the low-speed streaks to move relatively to the streamwise vortices was further supported by the water-channel experimental study of \cite{trujillo-bogard-ball-1997}. Their measurements proved that the wall motion decreased the frequency and duration of bursts and sweeps near the wall. The stronger bursts and sweeps, i.e., with larger $\overline{uv}$ magnitude, were suppressed significantly more than the weaker ones. Along the same lines, \cite{choi-debisschop-clayton-1998} argued that the near-wall turbulence cycle is disturbed if the wall moves spanwise by a distance larger than the typical length of the spanwise correlation of the streamwise velocity fluctuations, i.e., about 50 wall units, thereby leading to a reduction in the turbulent energy production. In other words, the spatial coherence between the longitudinal vortices and low-speed streaks has to be disrupted to obtain drag reduction. This paradigm is in full agreement with the concept of the minimal spanwise displacement for drag reduction, advanced by \cite{quadrio-ricco-2011} and discussed in Section \ref{sec:channel-form}.

\begin{figure}
  \centerline{\includegraphics[trim= 0 0 0 0, clip=true,width=0.6\textwidth]{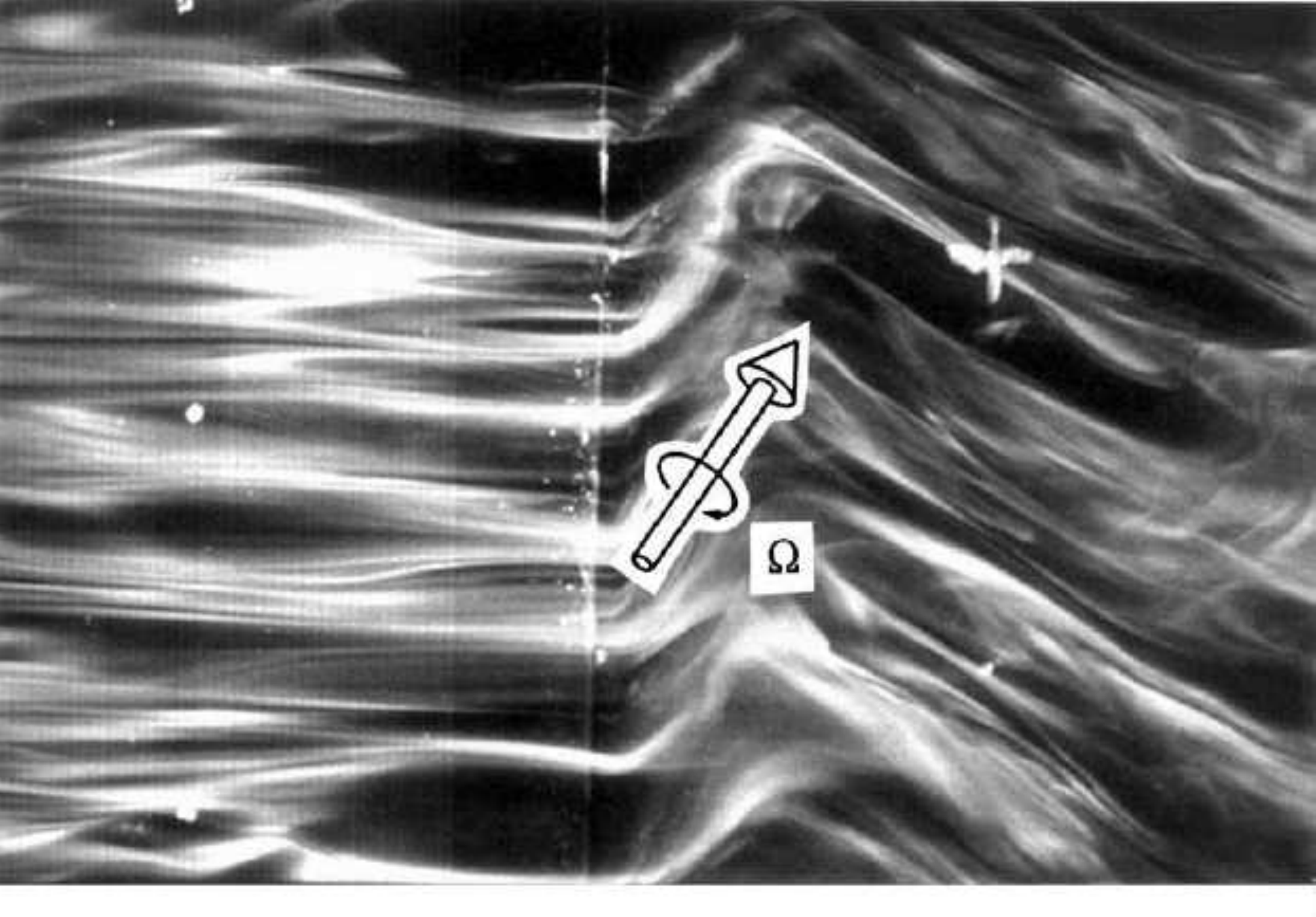}}
  \centerline{\includegraphics[trim= 0 0 0 0, clip=true,width=0.6\textwidth]{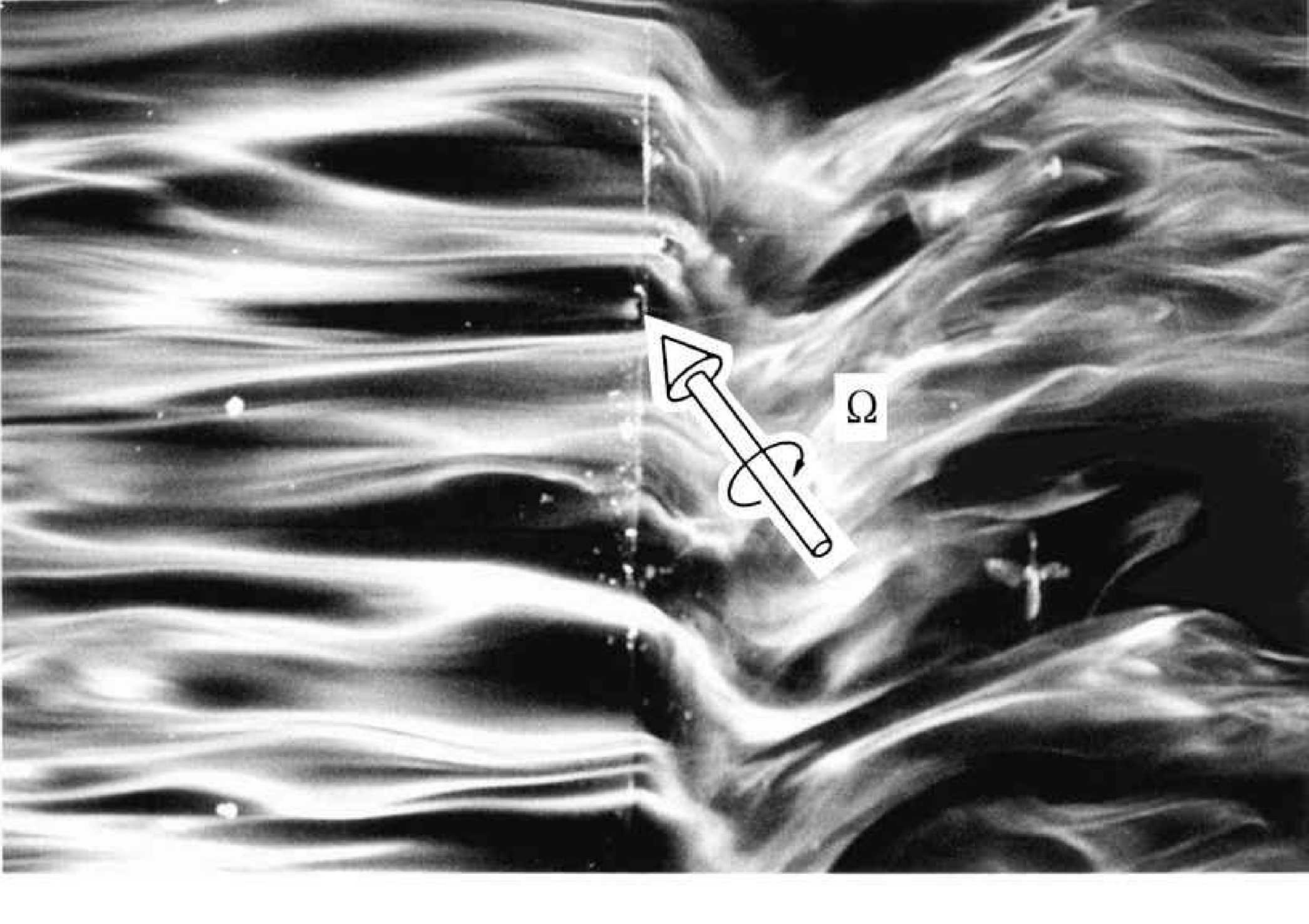}}
  \caption{Smoke flow visualization of the near-wall turbulent structures by \cite{choi-debisschop-clayton-1998}. The flow is from left to right with the leading edge of the oscillating plate visible near the center of the pictures. The plate is moving upwards in the upper panel and downwards in the lower panel. The arrow indicates the vorticity vector tilted in the spanwise direction. The vorticity vector is tilted to the upward direction in both cycles of the oscillation, hence creating a net spanwise vorticity in the flow, modifying the streamwise velocity profile. Taken from \cite{choi-debisschop-clayton-1998}, with permission from AIAA.}
\label{fig:choi-1998-flow-vis}
\end{figure}

\cite{choi-debisschop-clayton-1998}, \cite{choi-clayton-2001}, and \cite{choi-2002} argued that the effect of the spanwise motion is to reduce the mean spanwise vorticity by tilting the vorticity vector (or, as stated the authors, the ``vorticity sheet'') during both the positive and negative half-cycles of the actuation. This process induces a mean negative spanwise vorticity, related directly to the drag reduction. They also claim that the spanwise vorticity reduces the stretching of the longitudinal vortices in the viscous sublayer, thus decreasing their streamwise vorticity. As a result, the near-wall burst activity, which is associated with the downwash of high-momentum fluid near the wall, is weakened, causing the reduction in skin-friction drag. 
Fig. \ref{fig:choi-1998-flow-vis} shows the flow visualizations of the near-wall vortical structures by \cite{choi-debisschop-clayton-1998}. These structures inclined cyclically with respect to the streamwise direction as they were dragged laterally by the Stokes layer.  

The earliest explanations for the drag reduction offered by \cite{laadhari-skandaji-morel-1994}, based on the relative displacement of the quasi-streamwise vortical structures and low-speed streaks, was criticized by \cite{nikitin-2000} who maintained that the vortices and streaks were wrongly treated as separate entities by \cite{laadhari-skandaji-morel-1994}. \cite{nikitin-2000} instead argued that the vortices were directly affected by the wall oscillation, rather than being simply displaced, a view that contradicts the interpretation of \cite{quadrio-sibilla-2000}, as well as the flow visualizations by \cite{ricco-2004}, shown in Fig. \ref{fig:ricco-wu-2004-u}.

\cite{miyake-tsujimoto-takahashi-1997} presented an analysis that focused on the effects of the oscillatory spanwise wall motion  on the production of different turbulent-vorticity components in a channel flow. The drag fell monotonically as the log-law layer virtually disappeared. They examined the transport equation for the streamwise turbulent vorticity, which contains production terms associated with vortex ``stretching'' (in the streamwise direction), ``tilting''  and ``twisting'' (in the wall-normal direction). The actuation reduced all the production terms. They concluded that the extra terms associated with the spanwise strain are not directly responsible for the drag reduction, instead stating that the cause of drag reduction is the difference between the stretching and tilting production terms. Strong streamwise vortices, most closely associated with high-intensity turbulence and drag, were affected primarily by the vortex stretching, which tended to decline, relatively to the unforced case, especially when the Stokes strain varied rapidly. 

Leaning on arguments by \cite{orlandi-jimenez-1994} and \cite{dhanak-si-1999}, \cite{quadrio-sibilla-2000} proposed a phenomenological explanation based on the effect of the spanwise motion on the quasi-streamwise vortices and low-speed streaks in a circumferentially oscillating pipe flow. Focusing on the scenario where the vorticity of a streamwise vortex has the same sign as the vorticity in the Stokes layer, \cite{quadrio-sibilla-2000} argued that the spanwise motion drags high-speed fluid beneath low-speed fluid within and below the buffer layer, thus enhancing the wall-normal gradient of the streamwise fluctuations in the viscous sublayer. This mechanism should therefore increase the drag. As the half-cycle continues and the wall slows down but still moves in the same direction, \cite{quadrio-sibilla-2000} claimed that the high streamwise fluctuations close to the wall are eventually transported away from the wall by the ejection component of the streamwise vortex, generating a counter-gradient Reynolds stress, ``smoothing'' the streamwise  velocity gradient and diminishing the shear at the wall. Even though this sequence of processes is accepted, the question must however be posed as to why the exact opposite does not apply when the signs of the vorticity of the streamwise vortex and the Stokes layer are opposite, rather than the same. In other words, why should there be an asymmetry for vortices that rotate in different direction?

A set of arguments that appear congruent with those of \cite{quadrio-sibilla-2000} was advanced by \cite{choi-xu-sung-2002}. They considered the implications of DNS-derived conditionally averaged quasi-streamwise vortices around the buffer layer in a circumferentially oscillating pipe, and the interaction of the vortices with high- and low-speed regions on either side of the vortices. \cite{choi-xu-sung-2002} argued that the Stokes layer, when having the same vorticity sign as the vortices, drags high-speed fluid below the vortices towards the wall. When the wall reverses its direction, there is a tendency for low-speed fluid to be dragged towards the wall, but the effect is much less pronounced than in the first half of the cycle. They claim that this asymmetry explains the reduction in drag. As with the explanation of \cite{quadrio-sibilla-2000}, it can be argued, however, that for every streamwise vortex with one vorticity sign, there is another one with the opposite vorticity. Hence, it is unclear why there should be a net drag-reduction effect.

\cite{coxe-peet-adrian-2019} performed phase-wise analysis of vorticity statistics of their CPG pipe flow, and demonstrated that the vorticity skewness depends on the phase angle. This dependence, in turn, causes a skewness in the induced velocity fluctuations in the radial-azimuthal plane. These observations led to the conclusion that the oscillating wall affects the near-wall streamwise counter-rotating vortex pairs unequally. Instead of two equally-sized counter-rotating vortices pumping fluid upward as in the unactuated flow, the two unequal sized vortices in the controlled flow lifted fluid in the wall-normal direction at an angle, hence reducing the momentum transferred upward and inhibiting the ejection and sweep events.

In their minimal channel-flow DNS, \cite{jimenez-pinelli-1997} advected the streaks in the spanwise direction directly and in an oscillatory manner, without moving the wall or affecting the streamwise vortices. They obtained decaying streaks by disrupting their structural coherence with the rest of the flow.

\subsection{Effect of the Stokes strain on the time modulation of the wall turbulence}
\label{sec:mean}

The distinctive periodic modulation of the near-wall turbulence has prompted researchers to examine the response of the viscous and buffer layers to the unsteadiness of the actuation cycle. \cite{touber-leschziner-2012} focused on the low-speed streaks as they are representative of the transport of streamwise momentum across the near-wall layer because they are regions of depressed and elevated streamwise-velocity fluctuations, corresponding to wall-normal sweeps and ejections, respectively. 
In fully agreement with the flow visualizations of \cite{choi-debisschop-clayton-1998} in Fig. \ref{fig:choi-1998-flow-vis} and \cite{skote-2011} in Fig. \ref{fig:skote-spatial}, \cite{touber-leschziner-2012} observed, especially for $T^+=200$, that the low-speed streaks oriented at an angle dictated by the ratio between the streamwise and the spanwise strain, in visual agreement with the hydrogen-bubble sheet utilized in the water-channel experiments of \cite{ricco-2004}, discussed in Section \ref{sec:exp-planar}.
An intense weakening of the streak strength was detected during the phase of the forcing cycle in which the Stokes strain in the buffer layer varied at a high temporal rate. The low-streaks instead tended to re-establish themselves when the Stokes strain changed rather slowly. These observations led \cite{touber-leschziner-2012} to conclude that a major mechanism at play is the time-dependent re-orientation of the shear-strain vector that inhibits the streak-generation mechanism.

\begin{figure}
\begin{center}
\subfigure[]{\label{fig:strain-mapa}\includegraphics[trim= 0 0 0 0, clip=true,width=0.6\textwidth]{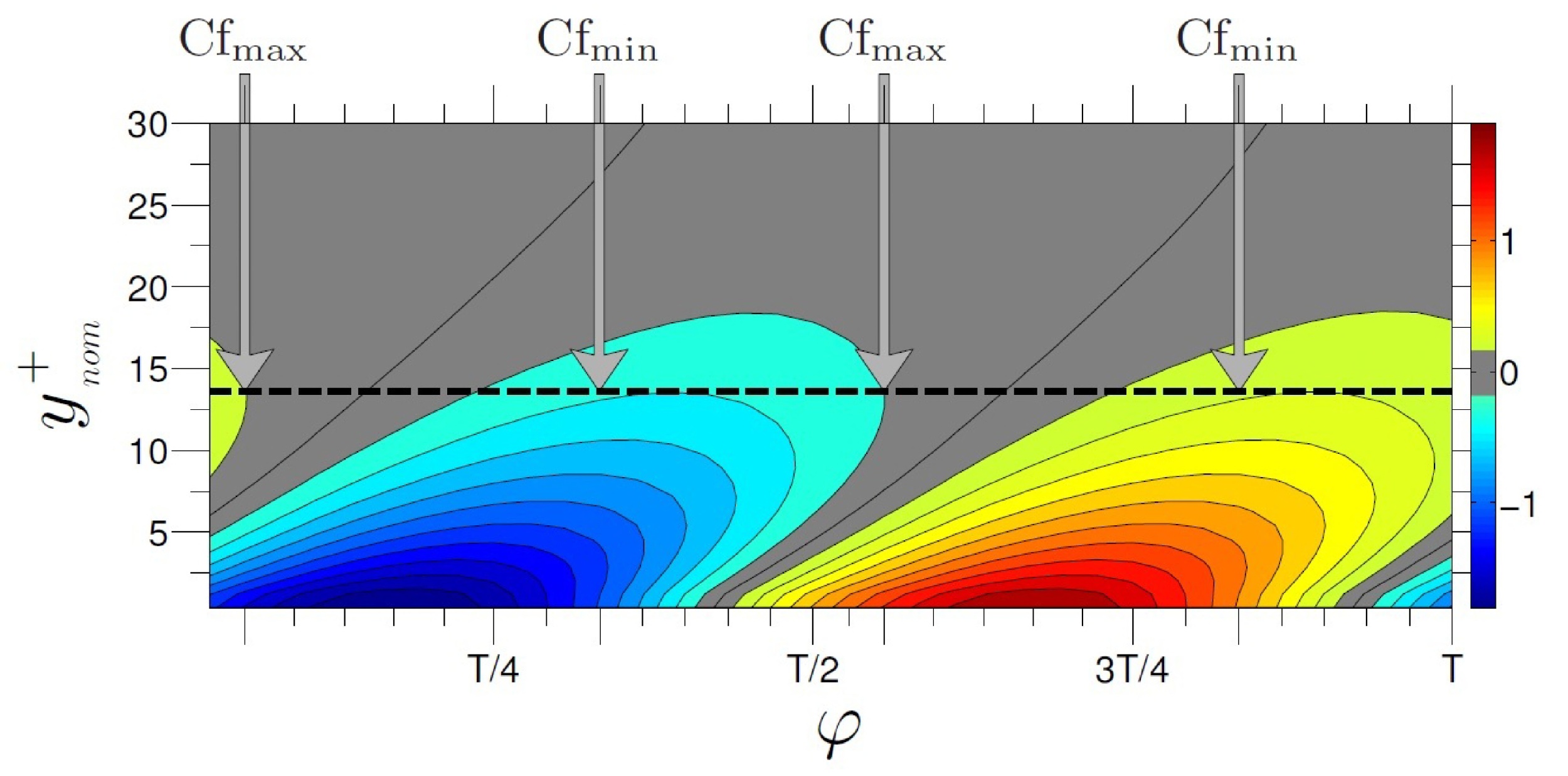}}
\subfigure[]{\label{fig:strain-mapb}\includegraphics[trim= 0 0 0 0, clip=true,width=0.6\textwidth]{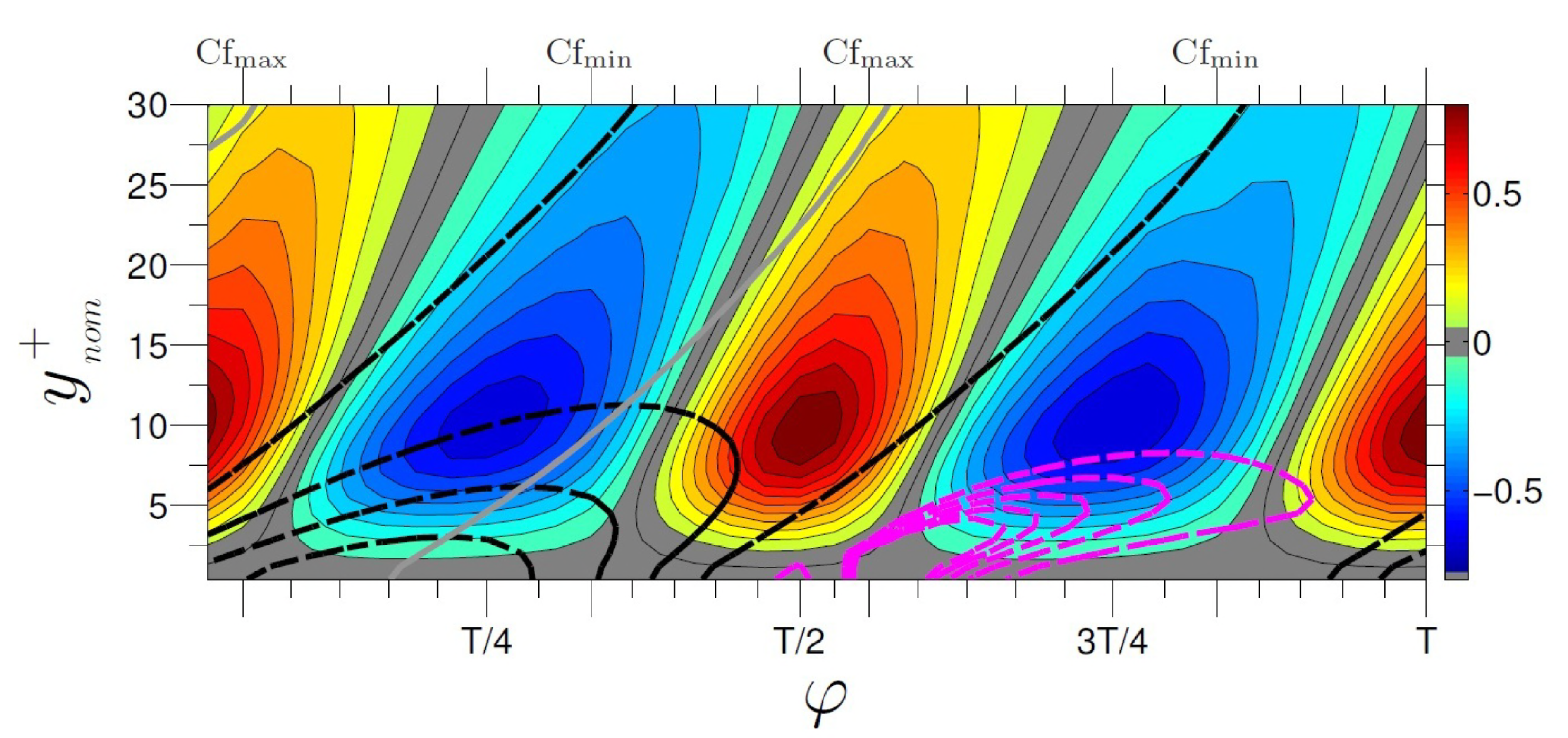}}
\caption{Contours of spanwise strain rate as function of phase during the actuation cycle (top) and corresponding contours of phase-averaged streamwise stress relative to time-averaged field (bottom) at $Re_\tau=1000$, $T^+=200$; black curved dashed lines indicate spanwise strain; gray curved lines are loci of maximum Stokes strain; magenta lines indicate skewness $\partial \theta/\partial y$. Reproduced from \cite{agostini-touber-leschziner-2014}, with permission from CUP.}
\label{fig:strain-map}
\end{center}
\end{figure}

\cite{agostini-touber-leschziner-2014} examined the response of the near-wall turbulence to the actuation cycle during the drag-decrease and drag-increase phases that occurred at the non-optimal period $T^+=200$, intentionally chosen because of the marked temporal modulation of the near-wall flow, observed in Fig. \ref{fig:DR}. 
Attention was focused on the phase-averaged properties to improve the understanding of the periodic reduction and increase in drag. This analysis was carried out using maps of the Stokes strain, the phase-averaged stress components, and their production rates in phase/wall-normal-distance planes. Fig. \ref{fig:strain-map}(a) shows contours of the spanwise strain rate and the phase locations at which the skin friction attains its maximum and minimum values. Fig. \ref{fig:strain-map}(b) shows contours of the phase-averaged streamwise stress relative to the time-averaged field, where the black dashed lines are iso-Stokes-strain values, the grey lines identify loci of maximum Stokes strain, and the magenta dashed lines represent the flow skewness -- namely $\partial \theta/\partial y$, where $\theta$ is the flow angle relative to the streamwise direction. 

Through these statistical results, \cite{agostini-touber-leschziner-2014} concluded that the observations of \cite{touber-leschziner-2012} on the relationship between the Stokes strain and the generation rate in the buffer layer were correct. \cite{agostini-touber-leschziner-2014} however remarked that another influential mechanism occurring much closer to the wall had to be considered, i.e., the substantial damping of turbulent stresses in the viscous and buffer layers caused by the high amplitude and temporal rate of change of the flow skewness, shown by the magenta lines in Fig. \ref{fig:strain-map}. They also explained that this modulation was the cause of a hysteresis (or asymmetry) in the phase-wise rate of reduction in the skin friction relative to the rate of increase, illustrated by the sequence $C_{f,max}~\mbox{-}~C_{f,min}~\mbox{-}~C_{f,max}$ in Fig. \ref{fig:strain-map}.
\cite{yuan-etal-2019} adopted the same phase/wall-normal-distance maps used by \cite{agostini-touber-leschziner-2014} to display the phase-dependence of the turbulent transport term in the turbulence-energy budget and some vorticity-related properties, but at $Re_\tau=200$, rather than at $Re_\tau=1000$ as in \cite{agostini-touber-leschziner-2014}.  

While the above discussion addresses the drag response to the Stokes strain via the relationship of the latter with the turbulent stresses, there remain the questions as to what drives the drag towards its low-valued state and why the actuation-induced drag fluctuations around the low-drag state are relatively weak, especially at the optimum actuation period $T^+=100$. In other words: why is it that, once the drag is depressed during some part of the actuation cycle, does it not rise back to the original level during the remaining portion of the cycle? A possible answer to this question lies in the fact that the regeneration time scale of the streaks is about $t^+=50$ \citep{ricco-2004,blesbois-etal-2013}. 
At the optimal period of oscillation $T^+=100$, the streaks do not have time to form again before the wall starts moving in the opposite direction every $t^+=50$. 
\cite{agostini-touber-leschziner-2014} suggested that a further potential contributor to the failure of the drag to recover was the skewness $\partial \theta/\partial y$, shown in Fig. \ref{fig:strain-map} (dashed magenta line), which causes a hysteresis in the drag-decline and drag-increase phases, with the former more prolonged than the latter.

\subsection{Changes in production, enstrophy, and dissipation}
\label{sec:prod_and_diss}

\begin{figure}
  \centerline{\includegraphics[trim= 0 0 0 0, clip=true,width=0.7\textwidth]{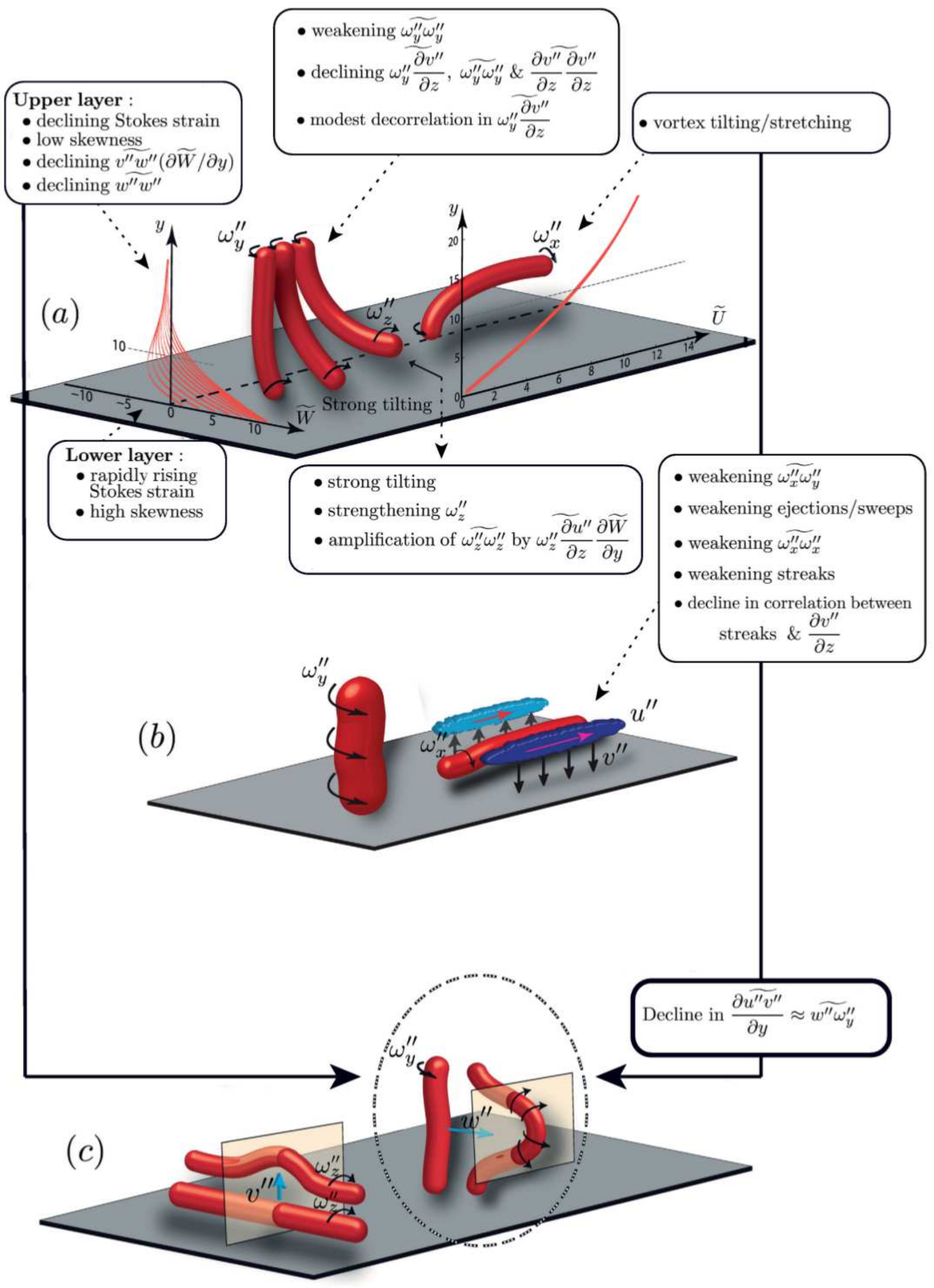}}
  \caption{Paradigm of interactions during the drag-reduction interval of the actuation cycle, proposed by \cite{agostini-touber-leschziner-2014}; (a) influence of Stokes motion (strain) on the tilting of wall-normal vortices, amplifying spanwise vorticity very close to the wall and weakening the wall-normal vorticity higher up (representative of streak strength); (b) interaction between weakening wall-normal vorticity (streak strength) and reduction in sweeps/ejections (streamwise vorticity); (c) result of declining wall-normal vorticity and spanwise fluctuations (sketch (a)), leading to decline in shear-stress gradient (for detailed interpretation, refer to \cite{agostini-etal-2015}). Taken from \cite{agostini-etal-2015}, with permission from Elsevier.}
\label{fig:sk1}
\end{figure}

\begin{figure}[t]
  \centerline{\includegraphics[trim= 0 0 0 0, clip=true,width=0.55\textwidth]{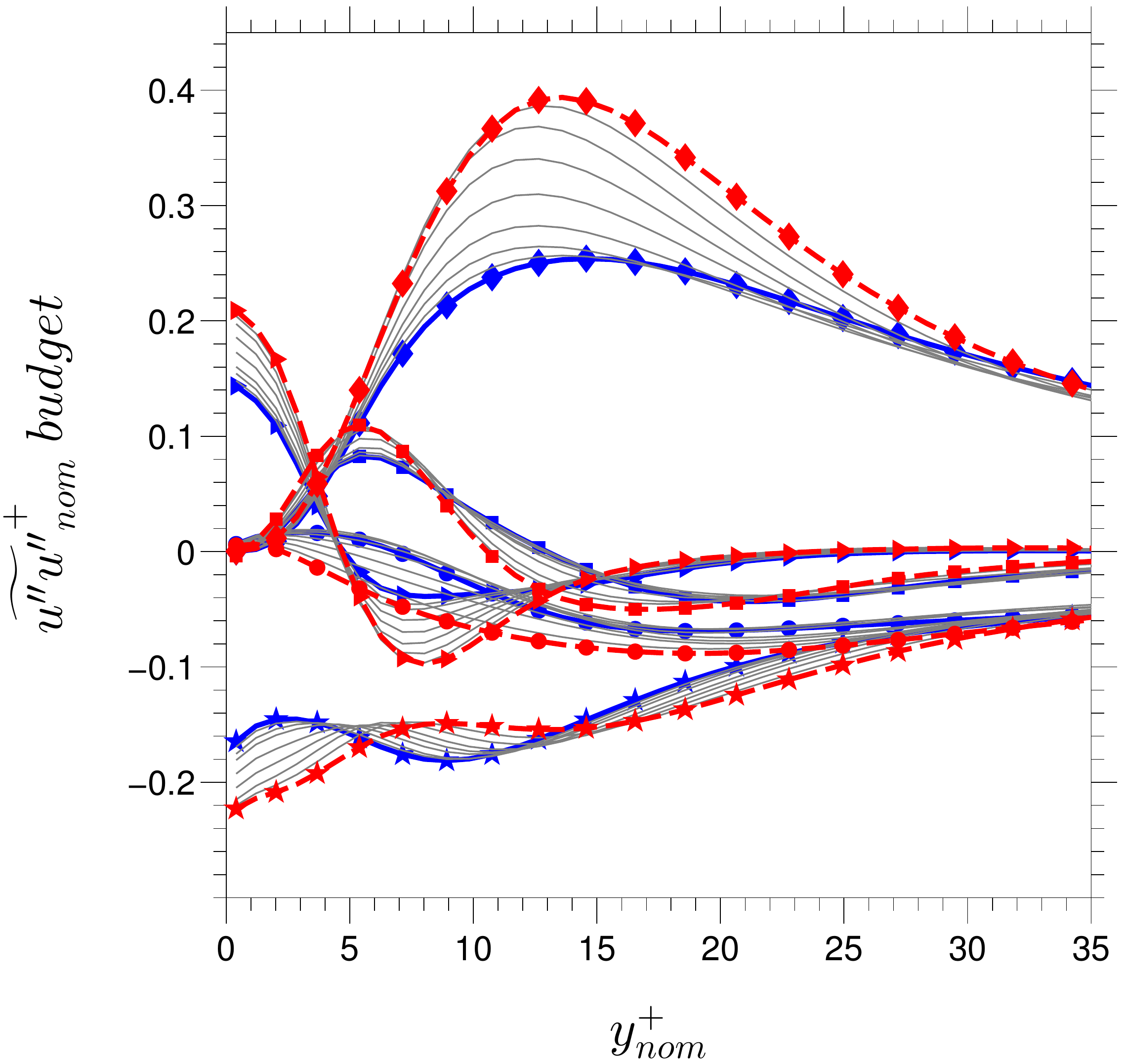}}
  \caption{Phase-wise variation of budget contributions for the streamwise stress at $Re_\tau=1000$,  $T^+=200$, during the drag-reduction phase, starting from red curves at $C_{f,max}$ down to blue curves at $C_{f,min}$; uppermost curves are for production; lowest curves are for dissipation. Reproduced from \cite{agostini-touber-leschziner-2014}, with permission from CUP.}
\label{fig:budget-uu-periodic}
\end{figure}

The study by \cite{miyake-tsujimoto-takahashi-1997}, reviewed earlier, made a valiant attempt to identify the reason for the reduction in the streamwise enstrophy fluctuation by examining the vorticity-production terms and related them to the actuation during the transient drag-reduction phase, following the start of the actuation. The study by \cite{agostini-touber-leschziner-2014} also considered the enstrophy components, but in different conditions and at higher Reynolds number, focusing on the non-optimal actuation case, $T^+=200$, for which the surface-averaged drag was oscillatory, as shown in Fig. \ref{fig:DR}. Although \cite{agostini-touber-leschziner-2014} observed the enstrophy (in line with dissipation) to experience only modest phase-wise variations, as evidenced in Fig. \ref{fig:budget-uu-periodic}, the individual components of the turbulent enstrophy tensor, $\overline{\omega_i \omega_j}$ showed substantial variations that were also phase shifted. Motivated by the link between the shear stress and the turbulent vorticity components, i.e.,

\begin{equation}
\label{eq:vort}
-\frac{\partial{\overline{uv}^+}}{\partial y}=\overline{v\omega_z}^+ -\overline{w\omega_y}^+,
\end{equation}
in a turbulent channel flow under CFR conditions with spanwise wall oscillations, \cite{agostini-etal-2015} examined the phase-averaged variations of the enstrophy components, their production rates and the variations of the right-hand-side correlations in equation \eqref{eq:vort}, with the objective of gaining yet another perspective on the drag-reduction mechanism.   
The wall-normal and spanwise enstrophy components, as well as their respective production rates, were phase-shifted by a quarter of an actuation period, and a reduction in the wall-normal component in the buffer layer, $\overline{\omega_y \omega_y}$, progressed hand-in-hand with an increase in the spanwise component $\overline{\omega_z \omega_z}$ close to the wall.   
This observation, alongside others, led \cite{agostini-etal-2015} to propose a scenario, summarised in Fig. \ref{fig:sk1}, in which an actuation-induced spamwise tilting of wall-normal vorticity provoked a strong increase in the skewness near the wall, leading to the reduction in the shear stress and wall-normal vorticity, the latter closely linked to the strength of the streaks and thus the drag (a second scenario pertaining to the increase in drag in the actuation phase following the decrease during the periodic drag variation can be found in \cite{agostini-touber-leschziner-2014}). The increase in the spanwise enstrophy component in the viscous sublayer above $y^+ \approx 5$ coincides with the minima of the wall-shear stress and the wall-normal enstrophy. This agreement appears to support, prima facie, \cite{ricco-etal-2012}'s proposition, already mentioned earlier, that the drag reduction is driven by increased dissipation. Indeed, in the layer $y^+=5-15$, Fig. \ref{fig:budget-uu-periodic} also shows the dissipation to rise with decreasing friction. The main effect is, however, a reduction in turbulence production and in turbulent dissipation. Moreover, below $y^+=5$, the variation of the spanwise enstrophy is in phase with $C_f$, as also observed in Fig. \ref{fig:budget-uu-periodic}.  

The study by \cite{yuan-etal-2019}, also conducted under CFR conditions, is also pertinent to the vorticity-based paradigm of \cite{agostini-etal-2015}. The study reported phase-space maps of the enstrophy components and the right-hand-side terms of equation \eqref{eq:vort}, also considered by \cite{agostini-etal-2015}. Both studies showed that the streamwise and normal enstrophy components were shifted by $T/4$, which was explained by \cite{agostini-etal-2015} using a reference to the production of the respective components. The shear-stress variations were shown to be driven primarily by the spanwise transport of the wall-normal vorticity $\overline{w\omega_y}^+$, a result consistent with the paradigm of \cite{agostini-etal-2015}. 

\cite{agostini-touber-leschziner-2014} further examined the relative role of production and dissipation in dictating the drag behaviour during the drag-decrease and drag-increase phases. This mechanism is exemplified by Fig. \ref{fig:budget-uu-periodic}, which shows profiles of the phase-averaged contributions to the budget of the streamwise stress, starting from the red lines for the $C_{f,max}$ state and moving progressively towards the blue lines corresponding to the $C_{f,min}$ state. The uppermost curves relate to production and the lowest curves to dissipation. These budgets show, consistently with other results, that the principal mechanism is the change of the production rate, while variations in the dissipation rate are of subordinate importance. The decline in production leads to a weakening in the streaks, reflecting the detrimental influence of the rapidly varying strain on a sustained turbulence re-generation process. \cite{agostini-touber-leschziner-2014} also showed that dissipation and enstrophy are closely co-related and that both tend to decrease in the lower part of the viscous sublayer in harmony with the decrease in production in the buffer layer. 

\cite{yuan-etal-2019} confirmed the intense reduction in turbulence production to be a distinctive feature of the drag-reduction behaviour, but also reported, seemingly in contrast with \cite{touber-leschziner-2012}, an increase in enstrophy very close to the wall, implying a corresponding increase in dissipation. Further analysis is therefore required on this point, i.e., to elucidate when and how an increase of enstrophy near the wall is linked to the drag reduction and to discern whether and how the local enstrophy differs from the local dissipation in such a highly sheared flow, recalling that the two quantities coincide when volume-averaged.

These studies all agree that, under CFR conditions and in fully-established drag-reduced conditions, the volume-averaged and time-averaged turbulent enstrophy, which coincides with the total turbulent dissipation, decreases with respect to the reference fixed-wall case.

\cite{ricco-etal-2012} used DNS to investigate a channel flow with spanwise oscillating walls under CPG conditions and focussed on the mean and turbulent kinetic energies and on the turbulent enstrophy dynamics. As the power $\mathcal{P}_x$ used along the streamwise direction increased because of the larger bulk velocity when the walls moved, both the mean streamwise dissipation and the total turbulent dissipation increased, while the mean spanwise-flow dissipation almost balanced the power $\mathcal{P}_z$ needed to move the walls because the Stokes layer agreed with the laminar solution \eqref{eq:stokes} very well. By an analysis of the transport equation of the turbulent enstrophy, they demonstrated that the increase of turbulent enstrophy was mainly caused by a production term of the turbulent enstrophy, $\overline{{\omega_z}{\omega_y}}^+ {\partial \overline{W}^+}/ {\partial y}^+$,  where ${\omega_z}$ and ${\omega_y}$ are the spanwise and wall-normal vorticity fluctuations. This enstrophy production term was shown to have a maximum at about $y^+=6$, and it was interpreted physically as the intensification of turbulent enstrophy due to the action of the Stokes layer on the turbulence structures located at the interface between the low-speed streaks and high-velocity regions. \cite{ricco-etal-2012} also showed that the time-averaged volume integral of this production term was linearly proportional to the drag reduction and also revealed a new physical interpretation of the drag-reduction scaling parameter used by \cite{quadrio-ricco-2004}, as further discussed in Section \ref{sec:models}.

\cite{ge-jin-2017} repeated the simulation under CPG conditions by \cite{ricco-etal-2012} at the lower Reynolds number of $Re_\tau=180$. Their oscillation parameters, $W_m^+=12$ and $T^+=100$, resulted in $\mathcal{R}=31\%$. They analysed the transient increment of dissipation from the start of the wall movement and corroborated the results by \cite{ricco-etal-2012}. With the help of the enstrophy transport equation, \cite{ge-jin-2017} connected the subsequent decline of the dissipation to the response of the vorticity components to the wall oscillation.

\begin{figure}
  \centerline{\includegraphics[trim= 0 0 0 0, clip=true,width=0.65\textwidth]{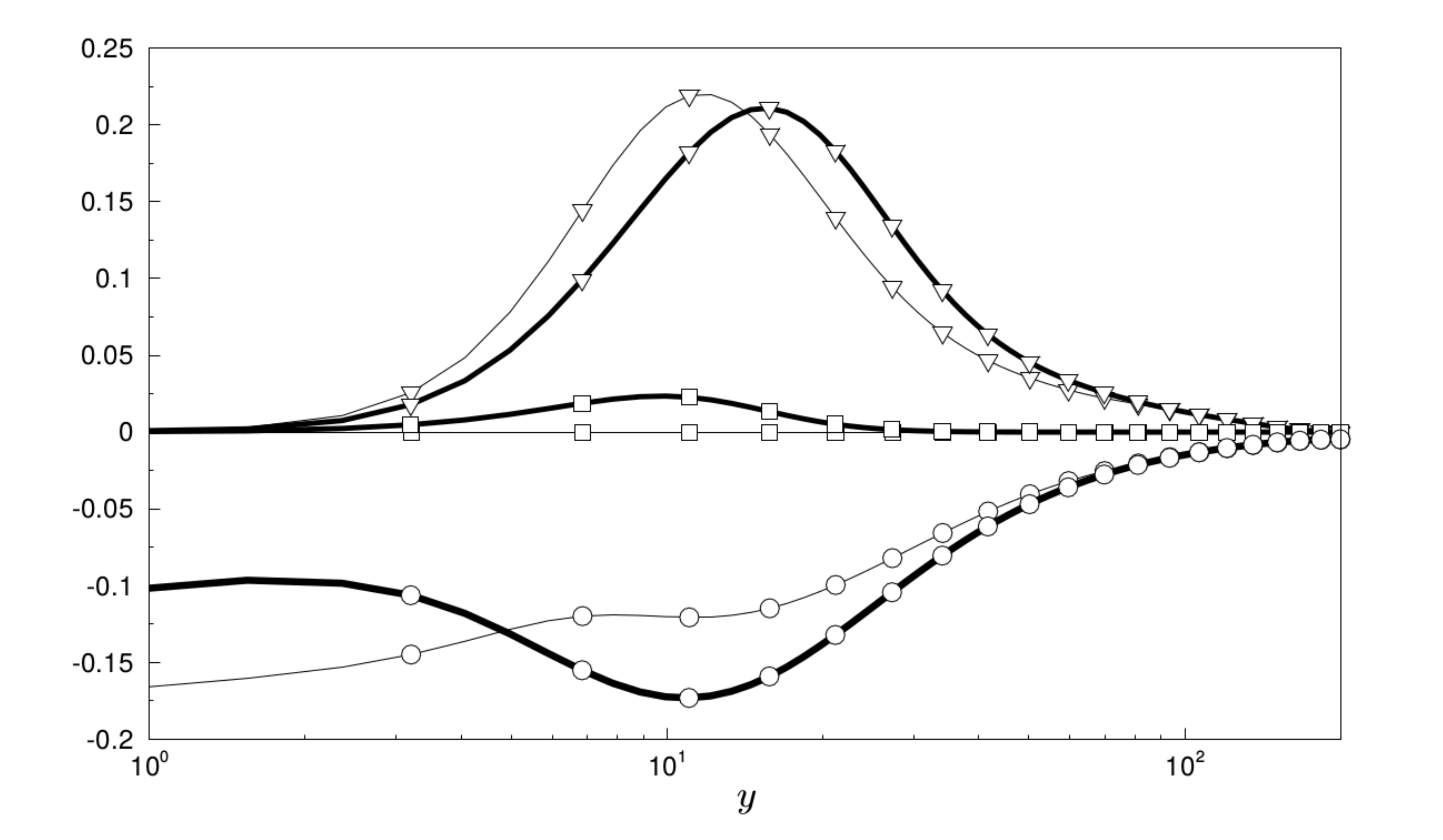}}
  \caption{Turbulent kinetic energy balance for CPG flow at $Re_\tau=200$. Triangles and squares represent the streamwise and spanwise production terms, respectively, while circles represent the total dissipation. Thick lines for the oscillating case and thin lines for the fixed-wall case. Taken from \cite{ricco-etal-2012}, with permission from CUP.}
\label{fig:ricco-energy}
\end{figure}

\begin{figure}
  \centerline{\includegraphics[trim= 0 0 0 0, clip=true,width=0.55\textwidth]{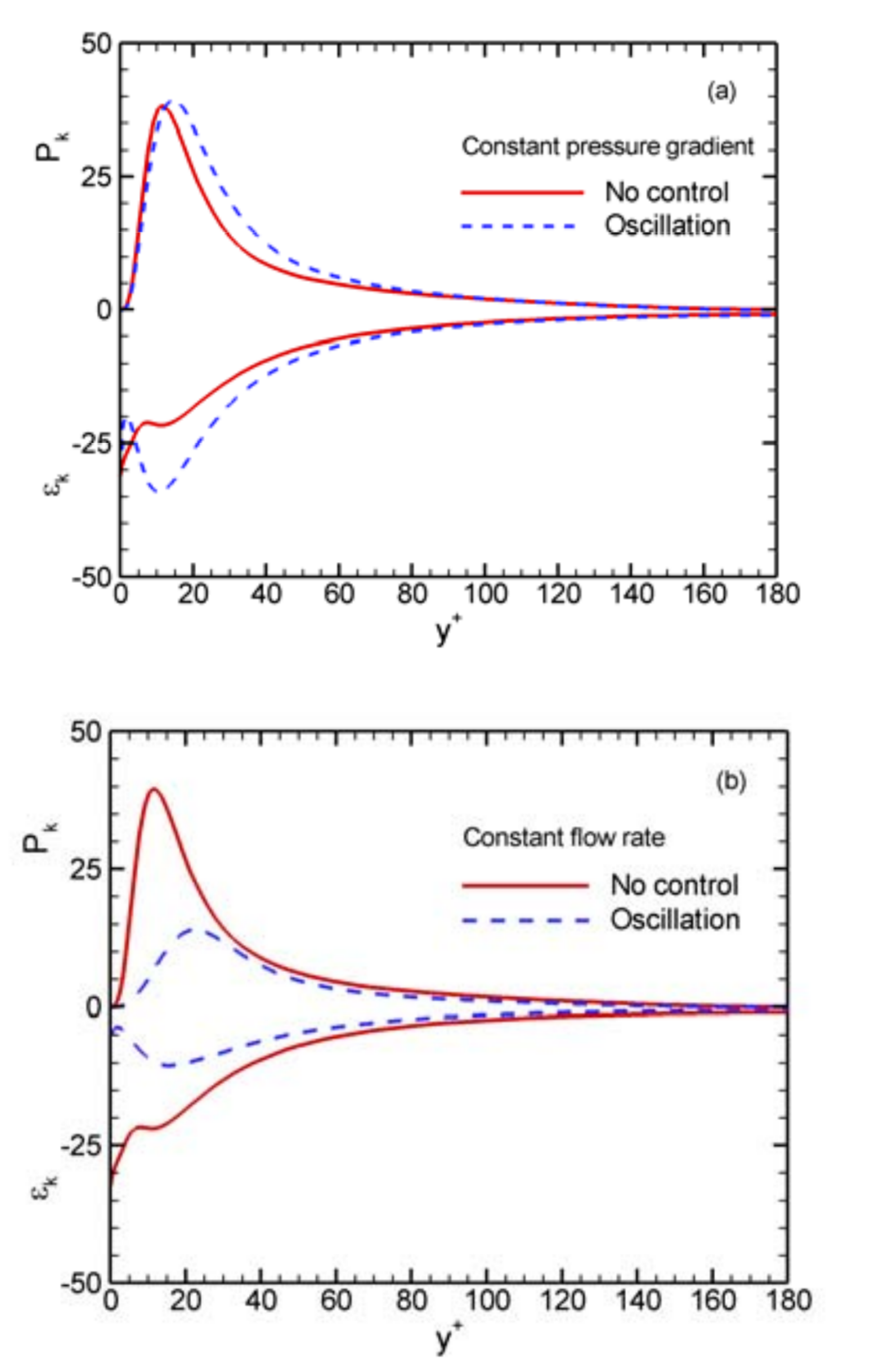}}
  \caption{Total production and dissipation in the CPG case (upper panel) and the CFR case (lower panel). Taken from \cite{zhou-ball-2008}, with permission from IJE Transactions.}
\label{fig:zhou-ball-energy}
\end{figure}

An apparent contradiction emerges when comparing the changes of the global turbulent enstrophy reported by \cite{touber-leschziner-2012}, \cite{agostini-touber-leschziner-2014}, and \cite{yuan-etal-2019} under CFR conditions, with those computed by \cite{zhou-ball-2008}, \cite{ricco-etal-2012}, \cite{yakeno-etal-2014}, and \cite{ge-jin-2017} under CPG conditions. 
In the former studies, the enstrophy decreases, while the latter present an increase of enstrophy. Fig. \ref{fig:ricco-energy} indeed shows that the intensities of the turbulent production and turbulent enstrophy, obtained by \cite{ricco-etal-2012}, increase when the walls are in motion. These results also match with the ones of \cite{zhou-ball-2008}, shown in Fig. \ref{fig:zhou-ball-energy}. In the top graph, the turbulent production and dissipation under CPG conditions are fully consistent with the profiles of \cite{ricco-etal-2012} in Fig. \ref{fig:ricco-energy} (note that logarithmic scale in the latter): the dissipation decreases in the viscous sublayer, but it is larger than in the fixed-wall case from $y^+=5$ up to $y^+=90$, which implies that the volume integral increases when the walls oscillate. The bottom graph of Fig. \ref{fig:zhou-ball-energy} instead reports that the turbulent production and dissipation decrease under CFR condition for the whole extent of the channel, results that agree with those of \cite{agostini-touber-leschziner-2014}.


The different physical constraints, the CFR and CPG conditions, lead to this apparent controversy. To explain this point, we use the numbers in Table \ref{tab:dissipation}, where $\phi_M$ is the total dissipation due to the streamwise mean flow and $\phi_T$ is the total dissipation due to the turbulent fluctuations. The ``+'' symbol denotes scaling with the wall-friction velocity of flows 1 and 2.
Flow 1 is the fixed-wall reference flow studied by \cite{ricco-etal-2012}, flow 2 is the optimal drag-reduction flow studied by \cite{ricco-etal-2012}, and flow 3 is the fixed-wall turbulent flow that has the same bulk velocity as flow 2.
For flow 2, the spanwise mean-flow velocity profile matches the laminar flow solution \ref{eq:stokes} very well and thus the dissipation due to the spanwise mean flow is balanced accurately by the work spent by moving the walls. It follows that the total dissipation is given by the sum of $\phi_M$ and $\phi_T$.

For flow 2, the bulk velocity increases by 22\% and therefore a larger bulk Reynolds number ensues, while the friction Reynolds number of flow 2 is the same as flow 1 because the streamwise mean pressure gradient is the same in the two flows. More power $\mathcal{P}_x$ must be used to drive flow 2 along the streamwise direction because the bulk velocity increases. The additional power must be dissipated by both $\phi_M$ and $\phi_T$, i.e., they both must increase in absolute terms with respect to the fixed-wall reference case 1.

A first observation is that $\phi_M$ for flow 2 accounts for a larger percentage of the total dissipation than in flow 1, indicating that flow 2 is closer to a laminar flow (for which $\phi_T=0$) than flow 1.
The second observation is that one must judge the energetic performance of flow 2 in terms of the power spent along the streamwise direction by comparing flow 2 with the fixed-wall flow 3, which has the same bulk velocity as flow 2 but a larger friction Reynolds number because the wall-shear stress is larger.
The output of flow 3 is the same of flow 2, i.e., the same increased mass flow rate is obtained, but less power is used to propel flow 2 than flow 3 along the streamwise direction because the wall-shear stress of flow 3 is larger. This result is the advantage brought about by the motion of the wall.
Furthermore, the turbulence dissipates in percentage more power than the mean flow in case 3 than in case 2, i.e., the oscillating-wall flow has a less turbulent dissipative nature than flow 3, which is consistent with the observation of a lower turbulent dissipation when the wall moves under CFR conditions.

\begin{table}
\begin{tabular}{|c|c|c|c|c|c|c|c|c|c|}
\hline
Flow      & Walls          & $Re_b$ & $Re_\tau$ & $U_b^+$ & $\mathcal{P}_x^+$ & $\phi_M^+ (\%)$ & $\phi_T^+ (\%)$ & $\phi_M^+$ & $\phi_T^+$ \\
\hline
1                 & Fixed          & 3176   & 200       & 15.9 & 31.7 & 59 & 41 & 18.8 & 12.9  \\
\hline
2                 & Oscillating    & 3875   & 200       & 19.4 & 38.7 & 62 & 38 & 24.0 & 14.7  \\
\hline
3                 & Fixed          & 3875   & 238       & 19.4 & 54.8 & 55 & 45 & 30.1 & 24.6  \\
\hline
\end{tabular}
\caption{Contributions to the total dissipation from $\phi_M$ and $\phi_T$ for the three flows described in the text. Wall-scaling is obtained using the friction velocity of flows 1 and 2.}
\label{tab:dissipation}
\end{table}

\begin{figure}
\centering
\includegraphics[width=0.8\textwidth]{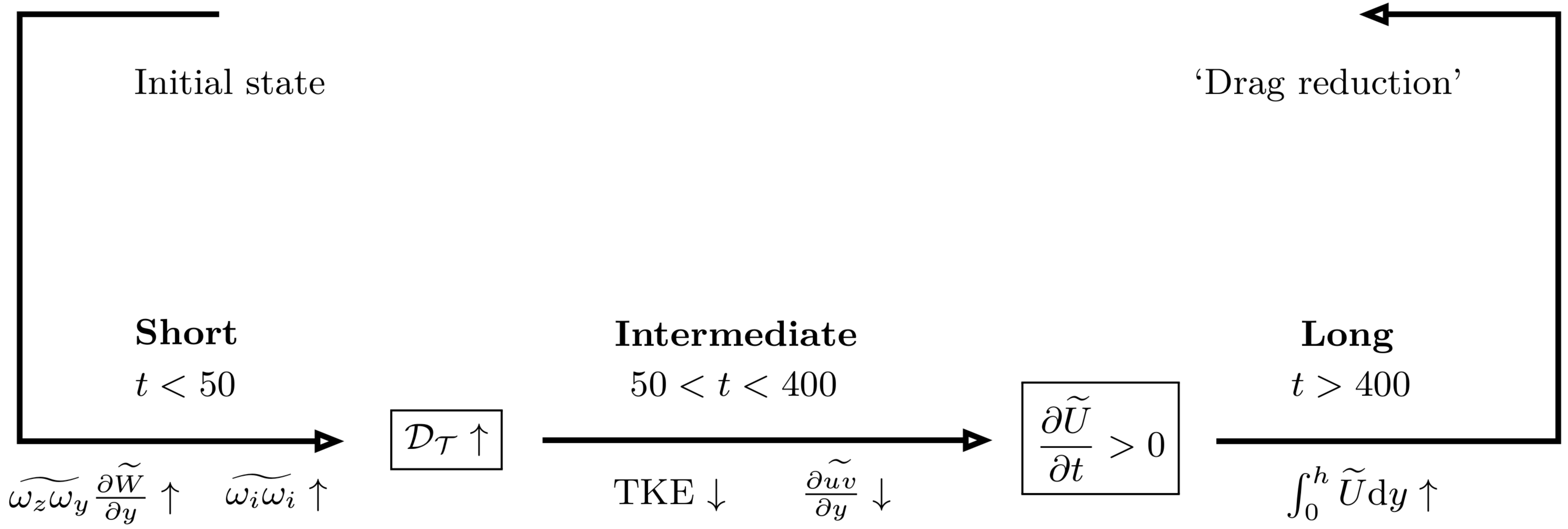}
\caption{Paradigm of the transient process leading to drag reduction under CPG conditions. Quantities are scaled in viscous units, the symbol $\widetilde{}$ denotes averaging over the homogeneous directions $x$ and $z$, $\mathcal{D}_{\mathcal{T}}$ denotes the total dissipation and TKE denotes the turbulent kinetic energy. Taken from \cite{ricco-etal-2012}, with permission from CUP.}
\label{fig:QRparadigm}
\end{figure}

\cite{ricco-etal-2012} studied the time-evolution of the terms in the enstrophy equation and showed that, due to the increased production term, the dissipation was initially enhanced, which in turn caused the turbulent kinetic energy and the Reynolds stresses to decline. The temporary enhancement of the volume-averaged enstrophy (which coincided with the total turbulent dissipation) was due to the enstrophy production term discussed above. The initial growth of the enstrophy was demonstrated by the production term growing significantly and at the same time of the peak of the turbulent enstrophy. The temporary enhancement of dissipation causes the turbulent activity to weaken. As a consequence, the vorticity and the production term decreased after this initial phase. By use of the volume-integrated momentum equation, \cite{ricco-etal-2012} proved that the streamwise mean flow accelerated because the turbulent kinetic energy decreased. This transient process continued because the dissipation was larger than the production. This paradigm was confirmed by the numerical calculations of \cite{ge-jin-2017} and is illustrated in Fig. \ref{fig:QRparadigm}. The initial drop of the wall-shear stress due to the transient increase of dissipation also occurs under CFR conditions, as proved by \cite{zhou-ball-2008} when they showed that the time evolutions of the space-averaged wall-shear stress under CFR and CPG coincided during the initial transients. It follows that, under CPG conditions, the bulk flow reacts to the imposed constraint after the initial interval, and specifically after the space-averaged wall-shear stress has reached its instantaneous minimum.

Further support that the initial transient interval is equivalent under CFR and CPG conditions was given by the DNS of the transient flow by \cite{agostini-leschziner-2014} under CFR conditions. They documented, like \cite{ricco-etal-2012}, the increase of turbulent enstrophy after the start of the oscillation phase. The temporary increase was however not given the physical significance assigned by \cite{ricco-etal-2012}, which therefore represents an interesting open point for future research.

\subsection{Other proposed mechanisms}
\label{sec:phys-others}

Another mechanism, in which the pressure-strain process is claimed to be the key drag-reduction driver, was proposed by \cite{xu-huang-2005}.  The process was represented by correlations between pressure and turbulent-strain fluctuations in the stress-transport equations. Their role was to drive turbulence towards a state of isotropy by transferring energy from the elevated streamwise stress to other, lower, normal stresses and, simultaneously, depressed the shear stress. The proposed mechanism was derived from observations of the budgets for the wall-normal stress, $\overline{v^2}$, and shear stress, $\overline{uv}$, during the transient drag reduction that occurred within the first actuation cycle at $Re_\tau=170$ and $T^+=90$. \cite{xu-huang-2005} argued, in particular, that the onset of actuation provoked a reduction in  $\overline{v^2}$ through a decrease in energy transfer from $\overline{u^2}$ by the pressure-strain term. This finding then led to a decline in the production of $\overline{uv}$ and thus a reduction in drag. 
This chain of events is similar to the description of the transient drag reduction in three-dimensional flow forced by a impulsively imposed spanwise pressure gradient by \cite{moin-etal-1990}. It could however be argued that the reduction in the pressure-strain term is a consequence of a reduction in the production of $\overline{u'^2}$ and of a commensurate decline in the streamwise stress itself, which then inevitably leads to a diminution of the energy-redistribution process. These scenarios illustrate the difficulty of resolving the ``chicken-and-egg'' conundrum, in which the majority of the contributors to the budgets of the stresses decline together as the actuation progresses. 

As already mentioned in Section \ref{sec:pipe-temp}, \cite{duggleby-ball-paul-2007} studied a turbulent pipe flow altered by circumferential wall oscillations and examined the flow statistical properties to gain insight into the drag-reduction mechanism. To explain the physical origin of drag reduction, \cite{duggleby-ball-paul-2007} resorted to the Karhunen-Lo\'{e}ve mode decomposition, which brought to the fore the energy modes that represent coherent regions of vorticity. For the first time, a dynamical-system model was used to study the physics of drag reduction, leading to the explanation that the dimensions of the chaotic attractor describing turbulence are reduced when the pipe oscillates. They focused on the eigenvalues and eigenvectors of the spatially integrated two-point correlation tensor to extract information on how the energy of different modes is changed by the wall forcing. The modes, representing coherent motions, were classified as ``propagating'' and ``non-propagating'', the former identifying travelling waves with a finite streamwise wavelength, while the latter had an infinite wavelength. The oscillation resulted in a significant reduction in the energy of the major propagating mode -- the ``wall mode'' -- which has a streamwise wavelength larger than the azimuthal wavelength. The flow structures were shifted away from the wall, hence attaining a higher convection velocity and having less time to interact with the streamwise vortices. This process was argued to shorten the bursting period, i.e., the duration of a stable state separated by unstable, intermittent, disruptions, thus leading to a reduction in turbulence intensity. The propagating modes were, therefore, less effective in interacting with the roll modes to produce Reynolds stresses, thus leading to drag reduction. 

\cite{yakeno-etal-2014} applied a creative conditional sampling technique which yielded a statistical representation of the quasi-streamwise vortices of which sweeps and ejections were constituents. This analysis allowed them to examine the properties of the vortices across the range of actuation periods, including the distribution of the Reynolds stresses linked with the sweeps and ejections, the enstrophy and its production. The low drag occurred together with an increased enstrophy in the viscous sublayer. This increase, or rather the increased dissipation, was thus deemed responsible for the reduction in drag. This interpretation chimes with the results of \cite{ricco-etal-2012}. 

\cite{yang-hwang-tsfp-2019} varied the oscillation parameters in simulations of wall-normal localised ``exact coherent states'' at various Reynolds numbers, motivated by the conjecture that the extracted coherent structures generate a large portion of the skin friction. They concluded that the study of the modulation of the turbulent structures given by the wall control at high Reynolds numbers can give insights into the drag-reduction mechanism.

\section{Effect of Reynolds number}
\label{sec:redep}

Although it is now well established that the drag-reduction margin declines as the Reynolds number increases, the challenges related to numerical simulations and experiments of controlled flows at Reynolds numbers higher than $Re_\tau \approx 500$ has made it difficult to arrive at clear-cut conclusions about the rate of decline. The decrease of the drag-reduction level is not only observed in flows where the spanwise forcing is confined in the very proximity of the wall, as in the wall-motion cases, but also when the spanwise forcing targets flow structures existing in the buffer layer and beyond, as in the body-force cases studied by \cite{canton-etal-2016-ftc}, \cite{canton-etal-2016}, and \cite{yao-etal-2017}.

Moreover, as it transpires in Section \ref{sec:re-issues}, it is likely that Reynolds-number effects will become increasingly profound at values well in excess of those studied so far by means of DNS and experiments. The purpose of this section is, therefore, to report the research path through and the state of the art of the effects of the Reynolds number on the drag behaviour of flows subject to spanwise forcing. In Section \ref{sec:re-issues}, we first present computational and experimental issues arising from increasing the Reynolds number, expanding the discussion in Section \ref{sec:domain} on computational requirements. In Sections \ref{sec:re-exponent} and \ref{sec:re-deltaB}, we summarize previous studies on two alternative view points on how the drag is affected by the Reynolds number, and in Section \ref{sec:re-physics} we present some aspects of the physical changes of the controlled flows as the Reynolds number grows.

\subsection{Issues related to studies at high Reynolds number}
\label{sec:re-issues}

As discussed in previous sections, there is ample and unambiguous evidence from DNS studies of channel and boundary-layer flows that an oscillatory spanwise wall motion yields gross drag-reduction levels approaching $\R=50\%$ and may yield net levels of power saved of the order of $\mathcal{P}_{net}=20\%$, depending upon the parameters of the actuation. The bulk of the drag-reduction data derived from DNS are for low Reynolds-number values, which, however, are not pertinent to realistic operational conditions. The reason is simple: the computational costs of DNS rise approximately in proportion to $Re_\tau^3$, given a constant domain size, because the length scale of the turbulent eddies that need to be resolved at the dissipative end of the eddy-size spectrum diminishes in proportion to $Re_\tau^{3/4}$. 

\begin{sloppypar}
A case in point, illustrating the predominance of low-Reynolds-number data, is the much-referenced ``Quadrio map'' \citep{quadrio-ricco-viotti-2009}, shown in Fig. \ref{fig:QRVmap} and discussed in Section \ref{sec:channel-form}. The map was derived by covering the frequency/wavenumber domain with hundreds of DNS computations for the streamwise-travelling waves given by equation \eqref{eq:waves}, all at $Re_\tau=200$ under CFR conditions. It is instructive to point out the limited real-world applicability of flows at this Reynolds number by considering a boundary layer developing on a flat plate: assuming air at standard conditions, $Re_\tau=200$ approximately corresponds to a boundary-layer thickness of about 1 cm and a free-stream velocity of about 4 m/s. Typical values for large civilian aircraft operating at cruise condition are instead a boundary-layer thickness of $30$ cm, a speed of $300$ m/s and a friction Reynolds number in the region of $Re_\tau =5$$\cdot$$ 10^4$. The question of how the drag-reduction effectiveness -- quantified by the gross level of drag reduction and the net power saved --  varies as the Reynolds number increases is therefore of major interest from a practical perspective. 
\end{sloppypar}

At the time of writing, the highest Reynolds-number values computed for a channel-flow simulation with spanwise wall motion are $Re_\tau=2100$ \citep{gatti-quadrio-2013} and $Re_\tau=2000$ \citep{yao-etal-2019}, while there have been several other studies reporting simulations in the range $Re_\tau=1000-1600$ \citep{agostini-touber-leschziner-2014,agostini-etal-2015,hurst-etal-2014,chung2016dns}. These simulations have however been performed for isolated or narrow ranges of the actuation parameters and over substantially different box sizes. For example, \cite{gatti-quadrio-2013}'s simulation was carried out over a very small streamwise-spanwise box of only $1.2h \times 0.6h$, while \cite{yao-etal-2019}'s box size was approximately $12h \times 6h$. Such computations, when carried out with box sizes as generously proportioned as that of \cite{yao-etal-2019}, require meshes of the order of 3 billion points to ensure sufficiently high fidelity. The costs are further increased by the need to simulate the flow for at least ten actuation periods and/or accommodate several travelling waves within the computational domain.

The only way to reduce the extremely high costs is to restrict the box size, as done by \cite{gatti-quadrio-2013}, but this adjustment is only possible for relatively short wavelengths, because the streamwise domain length must accommodate at least two waves. A substantial reduction in box size also leads to a loss of fidelity through the exclusion of the medium- and large-scale motions in the upper portion of the boundary layer, which become increasingly relevant to the turbulence dynamics as the Reynolds number increases, as already discussed in Section \ref{sec:domain}. 

As an aside, it is pertinent to note that the highest computed Reynolds numbers of unforced channel flow are $Re_\tau=4000$ \citep{bernardini2011inner}, $Re_{\tau}=4200$ \citep{lozano2014effect}, $Re_\tau=5200$ \citep{lee-moser-2015} and $Re_\tau=8000$ \citep{yamamoto2018numerical}, while for unforced boundary layers, the maximum value reported is $Re_\tau=2000$ \citep{sillero2013one,sillero2014two,pirozzoli2013probing}. To convey an inkling of the cost, it is remarked that the grid used for the highest value, $Re_\tau=8000$, contained 100 billion points, and grids of the order of $10-50$ billion points have been adopted even at half of this value in efforts to achieve sufficiently high accuracy to capture the dissipative small-scale motions at the Kolmogorov end of the eddy-size spectrum.

In experimental campaigns, the increase of the Reynolds number also poses severe problems when the wall friction and its change due to the wall control have to be obtained via the direct measurement of the mean-flow gradient at the wall. As discussed in Section \ref{sec:exp}, this computation is the only option in free-stream boundary layers, while measurements of the mean pressure gradient or of the bulk velocity are more convenient alternatives in pressure-driven confined flows. The central difficulty lies in measuring the mean flow in the viscous sublayer, i.e., where the velocity is linearly related to the wall-normal distance, which becomes thinner as the Reynolds number grows. For example, at $Re_\tau = 1000$, the thickness of the viscous sublayer is approximately 0.5\% of the boundary layer thickness, i.e., about 0.5 mm in a water channel with $U_\infty=35$ cm/s or in a wind tunnel with $U_\infty=3.5$ cm/s. Further complications arise in the context of spanwise forcing because the wall motion renders the streamwise mean-flow measurements in the viscous sublayer even more demanding. In boundary layers, obtaining accurate experimental data of the streamwise variation of the drag reduction as the Reynolds number increases is especially challenging and this result has not been achieved at any Reynolds number.

\subsection{Exponential decay of the drag-reduction margin}
\label{sec:re-exponent}

\begin{figure}
\begin{center}
\subfigure[]{\label{fig:yp3}\includegraphics[trim= 0 0 0   0, 
clip=true,width=0.53\textwidth]{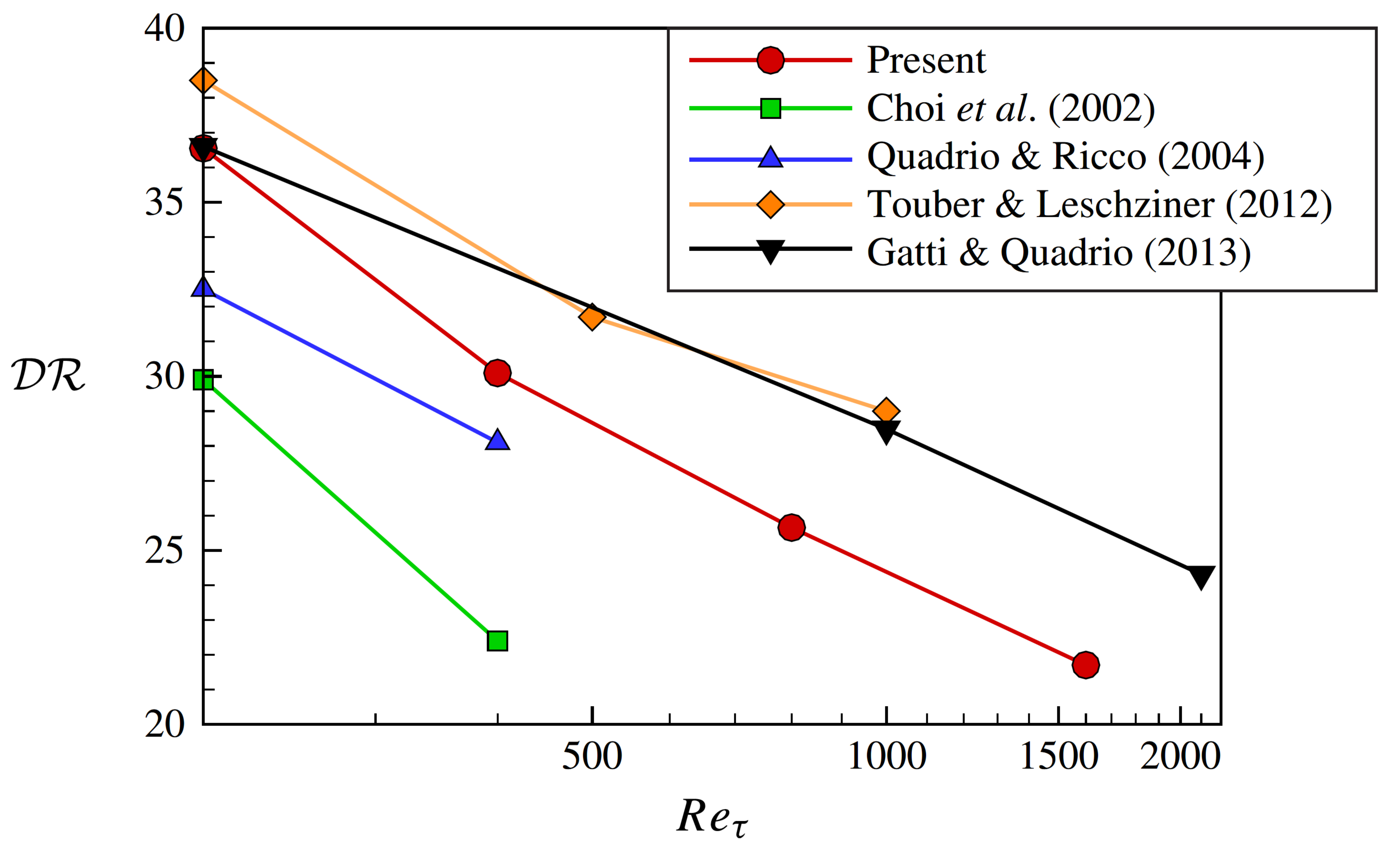}}
\subfigure[]{\label{fig:yp18}\includegraphics[trim=   0  0 0 0, 
clip=true,width=0.45\textwidth]{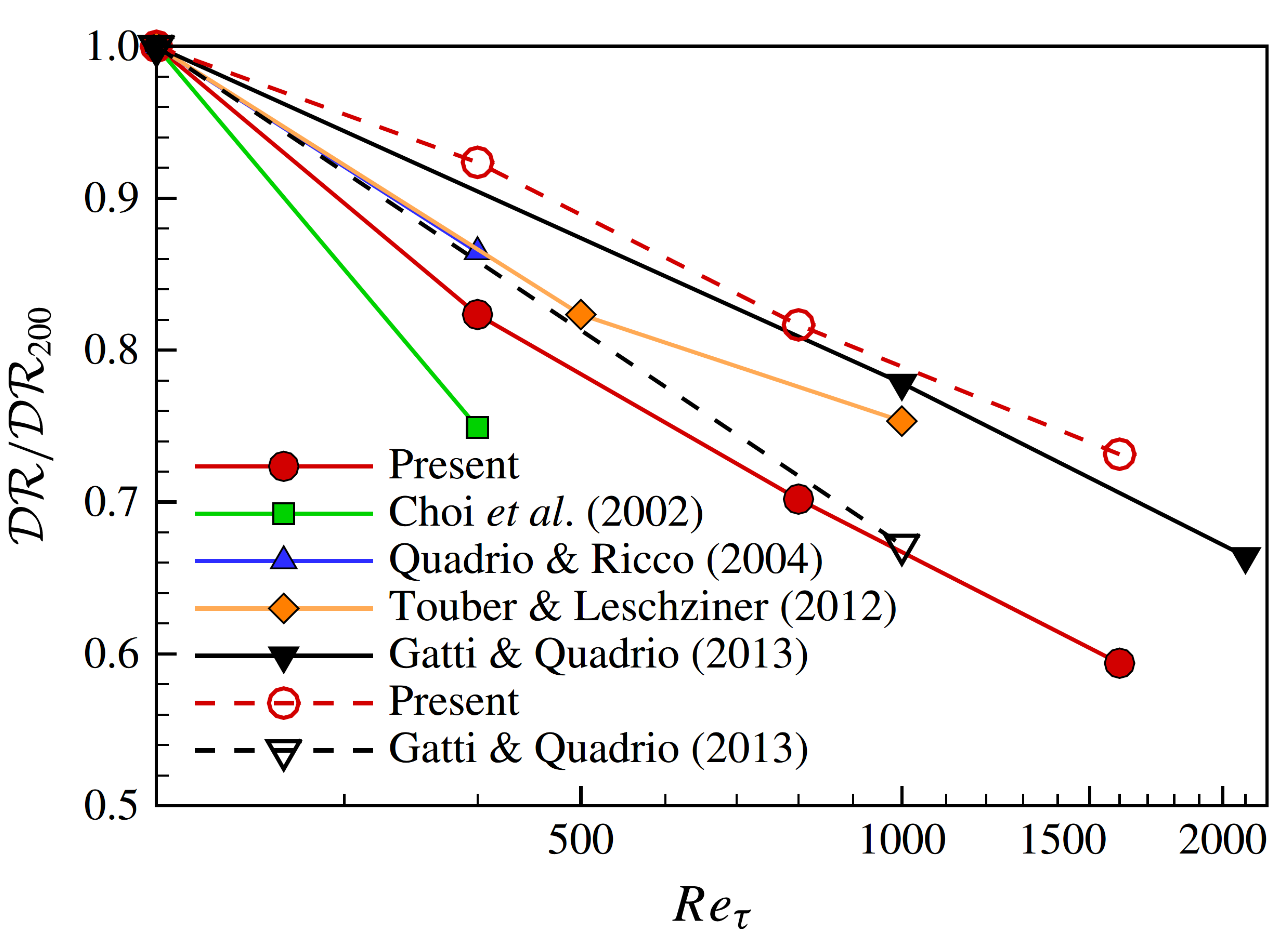}}
\caption{Compilation of five sets of gross drag-reduction levels as a function of the unactuated-friction Reynolds number derived from computations for channel flow with various implementations of spanwise wall motion; (a) percentage drag reduction; (b) drag reduction relative to the set-specific level at $Re_\tau=200$. Taken from \cite{hurst-etal-2014}, with permission from CUP.}
\label{Fig:1}
\end{center}
\end{figure}

\begin{figure}
\begin{center}
\includegraphics[trim= 0 0 0   0, 
clip=true,width=0.6\textwidth]{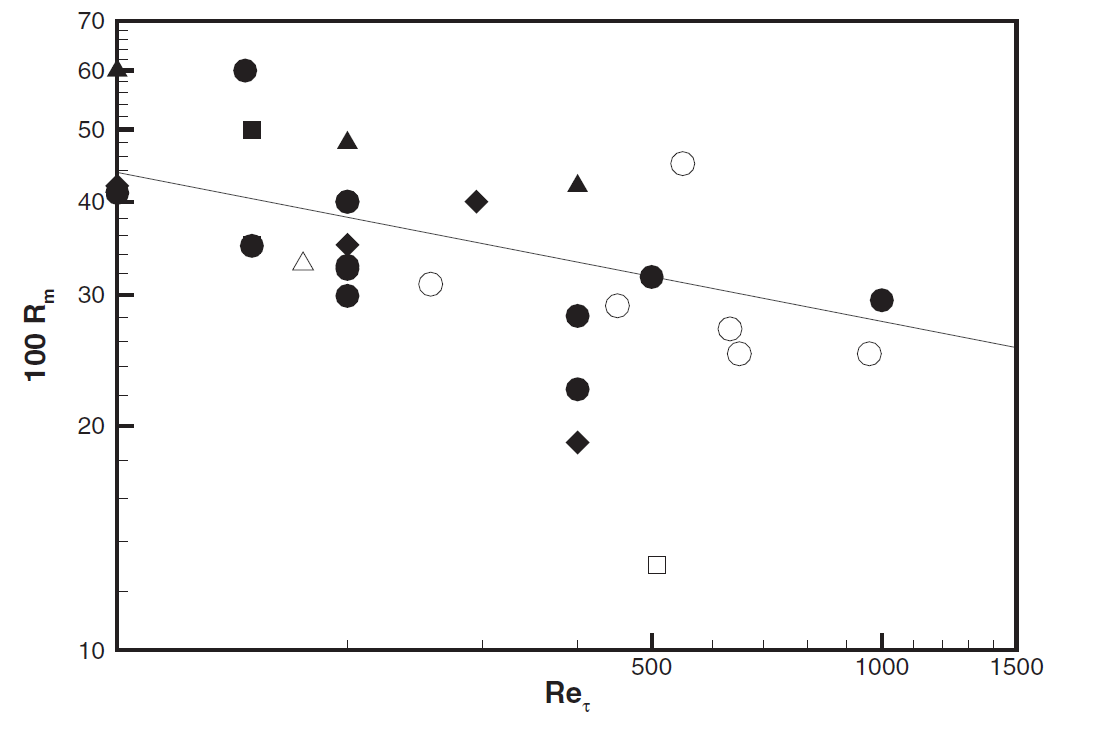}
\caption{Compilation of gross drag-reduction levels obtained in computational (empty symbols) and experimental (full symbols) studies in a variety of geometries and for various types of forcing (spanwise wall motion, streamwise- and spanwise travelling waves, Lorenz forcing).  The straight line represents $R_m \propto Re_\tau^{-0.2}$. Taken from \cite{gatti-quadrio-2013}, with permission from AIP Publishing.}
\label{Fig:extra}
\end{center}
\end{figure}

Until around 2016, the general view, based on DNS results in channel flow, was that the drag-reduction decline with the Reynolds number could be adequately expressed by the simple relationship $\mathcal{R} \propto Re_\tau^{-\alpha}$, with the exponent varying in the range $\alpha=0.1-0.4$. 
Fig. \ref{Fig:1} from \cite{hurst-etal-2014} and Fig. \ref{Fig:extra} from \cite{gatti-quadrio-2013} show the decrease of the drag-reduction margin as a function of the Reynolds number.


Several studies have attempted to find how the drag-reduction margin depends on the Reynolds number for wide ranges of the forcing parameters.
\cite{choi-graham-1998} first reported measurements of drag reduction via spanwise forcing at two different Reynolds numbers in an oscillating pipe, but no information could be extracted because the amplitude and the period varied simultaneously.
The algebraic dependence $\mathcal{R} \propto Re_\tau^{-\alpha}$ was first suggested by \cite{choi-xu-sung-2002} in a DNS study at $Re_\tau=$100, 200, and 400. By fitting the numerical data, $\alpha=0.4$ was obtained. 
The experimental data by \cite{wu-2000} and \cite{ricco-wu-2004-a} instead reported much more encouraging results in free-stream boundary layers with spanwise wall oscillations up to $Re_\theta=2400$ ($Re_\tau \approx 1000$). 
The drag-reduction margins were within the experimental uncertainty range as the Reynolds number increased for periods of oscillations lower than the optimum $T^+=100$.
The DNS study by \cite{ricco-quadrio-2008} of a channel flow with wall oscillations chimed with the findings of \cite{ricco-wu-2004-a} as the drag-reduction level was unvaried when the Reynolds number increased from $Re_\tau=$ 200 to 400 for periods of oscillations lower than the optimum and decreased with the Reynolds number for periods comparable or larger than the optimum. They estimated a maximum $\alpha=0.2$ at large periods. 
Very similar values of the exponent were computed via DNS by \cite{touber-leschziner-2012} for uniform spanwise motion and by \cite{quadrio-ricco-viotti-2009} for streamwise travelling waves, while modelling approaches based on Reynolds-Averaged-Navier-Stokes (RANS) equations led to lower values, i.e., $\alpha=0.15$ \citep{moarref-jovanovic-2012} and $\alpha=0.06$ \citep{belan-quadrio-2013}, or to a negligible Reynolds-number effect when linearized Navier-Stokes equations were employed \citep{duque-etal-2012}.

A problem with the simple representation $\mathcal{R} \propto Re_\tau^{-\alpha}$ is that the exponent $\alpha$ depends greatly on the actuation frequency and wavelength, and also on the Reynolds number itself, as implied by some of the data sets included in Fig. \ref{fig:yp18}. For purely oscillatory actuation at $Re_\tau=2000$, \cite{yao-etal-2019} found that $\alpha$ varied in the range $0.14-0.25$ when the forcing period increased from $T^+=63$ to $125$, with $\alpha \approx 0.18$ at the near-optimum period $T^+=100$. The extensive DNS study undertaken by \cite{gatti-quadrio-2016}, involving 4000 simulations with various box sizes covering the range $Re_\tau=200-1000$ and numerous combinations of the actuation parameters, suggested that, for purely oscillatory actuation, the decline was at a more modest rate, $\alpha \approx 0.15$, as illustrated by the black line in Fig. \ref{Fig:1}. This exponent was reported to decline to a minimum of around 0.1 when the actuation was in the optimum range of frequency and streamwise wavelength. Two different alternative propositions, where the drag-reduction margin asymptotically approaches a finite value (instead of zero) at infinite Reynolds number, were proposed by \cite{belan-quadrio-2013} for channel flow and \cite{skote-mishra-wu-2015} for boundary-layer flow. Both formulations were however based purely on curve fitting of the data and hence no theoretical foundation was provided in either case.

Suffice it to say that the most optimistic decline, if boldly extrapolated to flight Reynolds numbers, implies the potential of achieving net drag-reduction levels of the order of 10\% \citep{gatti-quadrio-2016}, assuming -- unrealistically, of course -- that mechanical losses in activating the wall motion are negligible. 

\subsection{Upward shift of the mean-flow profile}
\label{sec:re-deltaB}

As already discussed, \cite{gatti-quadrio-2016} first attempted to extract the exponent $\alpha$ from their extensive DNS database for channel flows altered by streamwise-travelling waves at $Re_\tau=200$ and $Re_\tau=1000$. It was evident that the map of $\alpha$ as a function of the wavenumber and frequency was ambiguous  and it did not constitute a credible correlation. \cite{gatti-quadrio-2016} therefore judged the algebraic fitting as unsuited to describe the Reynolds-number dependence. 

\begin{figure}
    \centering
    \includegraphics[trim= 0 0 0 0, clip=true,width=0.6\textwidth]{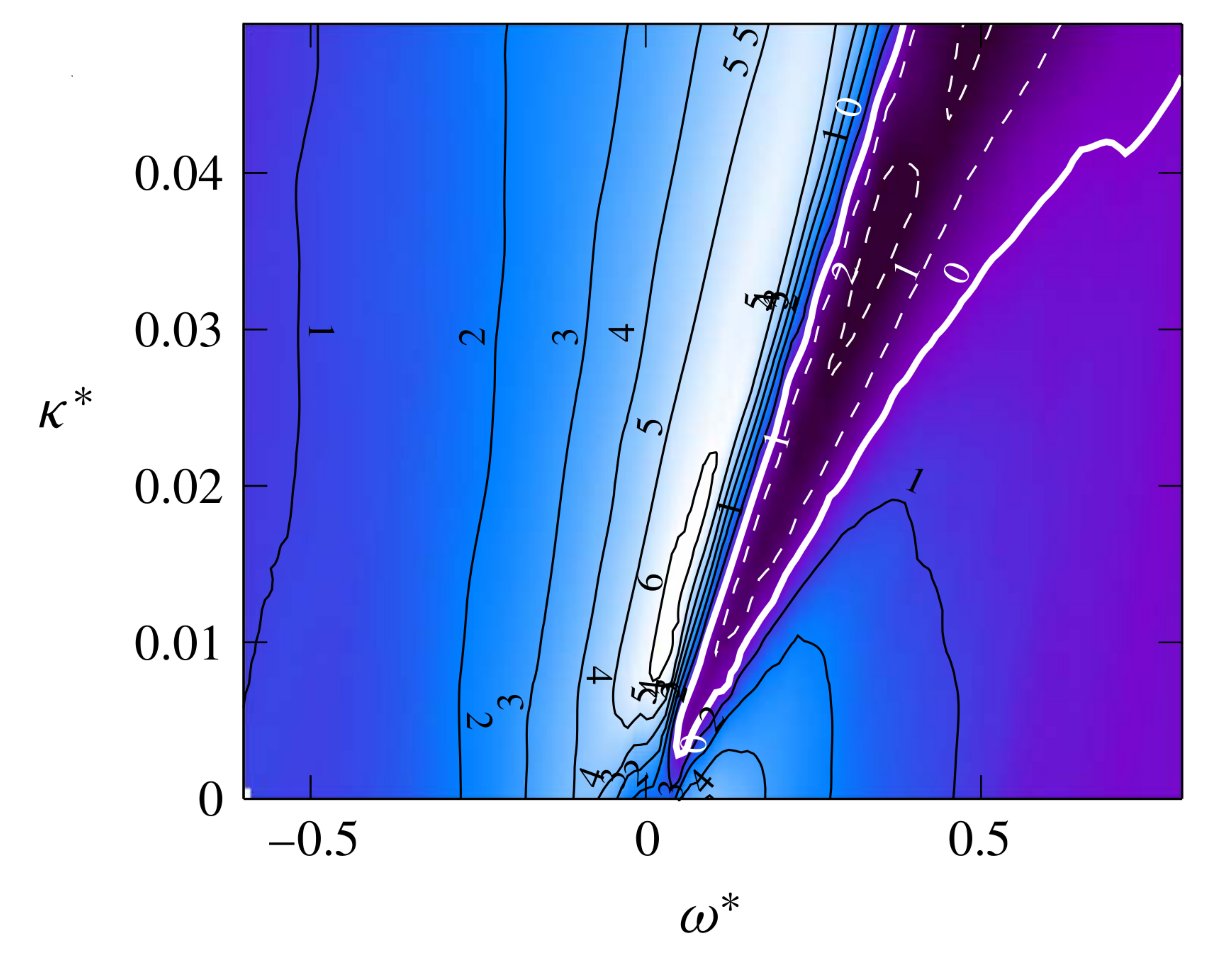}
    \caption{Map of $\Delta B^+$ for streamwise-travelling waves with $W_m^+=12$ at $Re_\tau=200$. Taken from \cite{gatti-quadrio-2016}, with permission from CUP.}
    \label{fig:deltab}
\end{figure}

They also reported that, as expected, the additive constant $B$ in equation \eqref{eq:log-law} increased with respect to the uncontrolled case for drag-reduction regimes, indicating an upward shift of the profile, while $B$ decreased when drag increase occurred, i.e., a downward shift of the profile occurred. 
These shifts were rigid: the constant $\kappa$ and the wake outer-flow constant $B_1$, i.e., the difference between the mean centreline velocity and the velocity found by the velocity-defect law up to the centreline \citep{pope-2000}, did not vary when the wall moved. It was therefore possible to relate the drag-reduction level to the shift $\Delta B$. 
This connection was first proposed by \cite{luchini-1996b} for flows over riblets, elaborating on the well-established idea that, while the logarithmic behaviour depends only on the balance between turbulent production and turbulent dissipation, the upward-shift constant $B$ is fixed by the wall-shear stress and is thus related to the near-wall control \citep{jimenez-2004}. 
Fig. \ref{fig:deltab} confirms this hypothesis: the map of $\Delta B^+$ as a function of the forcing parameters displays a striking similarity to the drag-reduction map shown in Fig.\ref{fig:QRVmap}. \cite{gatti-quadrio-2016} further showed that the relationship between $\Delta B^+$ and the drag-reduction margin was almost linear, as shown by their figure 12.

\begin{figure}
    \centering
    \includegraphics[trim= 0 0 0 0, clip=true,width=0.6\textwidth]{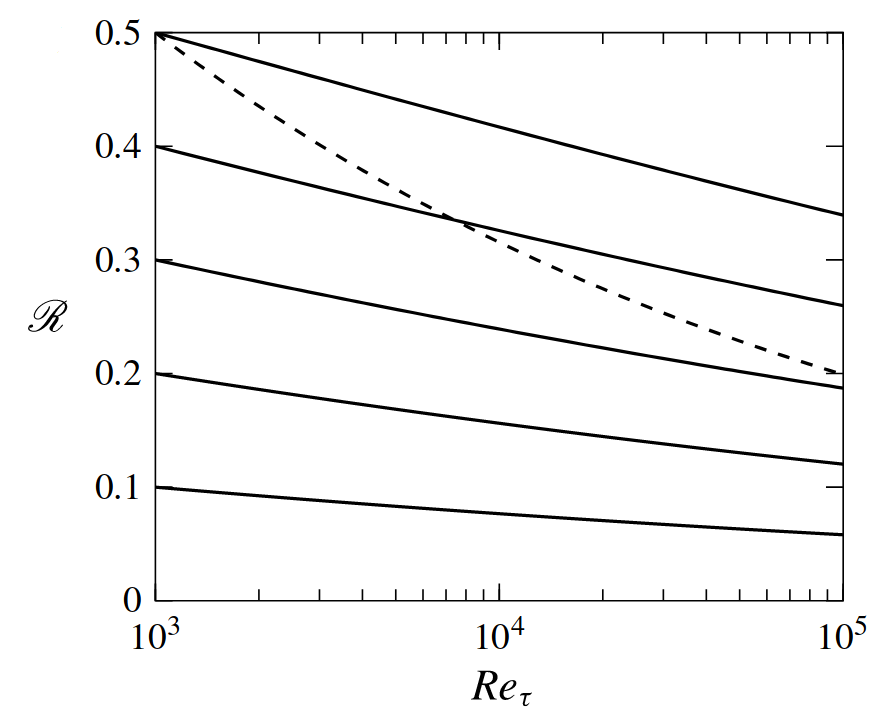}
    \caption{Predicted Reynolds-number dependence of the decline in drag reduction based on the assumption of constant upward shift in the log-law; solid curves are for different actuation parameters (e.g., spanwise wall-velocity amplitude); dashed line represents $\mathscr{R} \propto Re_\tau^{-0.2}$; solid lines follow roughly $\mathscr{R} \propto Re_\tau^{-0.08}$. Taken from \cite{gatti-quadrio-2016}, with permission from CUP.}
    \label{Fig:6}
\end{figure}

By using the log-law scaled in inner units and the defect logarithmic law scaled in outer units, and by taking the differences between these expressions for the fixed-wall case and the actuated case, \cite{gatti-quadrio-2016} obtained

\begin{equation}
\sqrt{\frac{2}{C_f}}-\sqrt{\frac{2}{C_{f,0}}}
=\frac{1}{\kappa} 
\ln \left({\frac{Re_\tau}{Re_{\tau,0}}}\right) 
+ \Delta B^+.
\label{eq:cfre}
\end{equation}
Under CFR conditions, equation \eqref{eq:cfre} was expressed as 
\begin{equation}
\Delta B^+ = 
\sqrt{\frac{2}{C_{f,0}}} \left[ \left(1-\mathscr{R}\right)^{-1/2} - 1 \right] 
- \frac{1}{2\kappa} \ln \left( 1 - \mathscr{R}\right),
\label{eq:gatti-quadrio-2016}
\end{equation}
and a similar expression was found under CPG conditions. Note that $\mathscr{R}=\mathcal{R}/100$ in equation \eqref{eq:gatti-quadrio-2016} and Fig. \ref{Fig:6}, where the drag-reduction margin $\mathcal{R}$ is defined in \eqref{eq:drag-reduction}. As $C_{f,0}$ is a known function of the bulk Reynolds number only, the drag-reduction margin $\mathcal{R}$ can be computed at any Reynolds number once the dependence of $\Delta B^+$ on the forcing parameters $\omega$ and $\kappa_x$ is known at a given Reynolds number, provided that the Reynolds number is large enough for the logarithmic behaviour to hold and the constants $\kappa$ and $B_1$ are not affected by the Reynolds number and the wall actuation.

Fig. \ref{Fig:6}, also taken from \cite{gatti-quadrio-2016}, shows the variation of the drag-reduction levels derived from expression \eqref{eq:gatti-quadrio-2016}, starting from $Re_\tau=1000$. The various levels arise from different actuation scenarios in terms of frequency, wavenumber, and wall-velocity amplitude. The majority of studies have adopted a wall-velocity amplitude $W_m^+=12$. For this wall velocity and at $Re_\tau = 1000$, \cite{gatti-quadrio-2016} reported a maximum gross drag-reduction margin of 38\% for the best travelling-wave scenario and a net power saved of 19\%. Extrapolated to $Re_\tau=10^5$, this level reduced to around 10\%. The solid lines in Fig. \ref{Fig:6} imply a rate of decline of $\alpha=0.08$, which is optimistic when compared to the more conservative $\alpha=0.15-0.2$ estimated by other researchers.

The results of \cite{gatti-quadrio-2016} were discussed by \cite{agostini-leschziner-2018}, who pointed out that a direct consequence of the implicit Reynolds-number dependence \eqref{eq:gatti-quadrio-2016} was that the outer structures, that become increasingly important as the Reynolds number grows, affect the controlled flow in the same way as they affect the uncontrolled flow, i.e., their influence is only present in the dependence of the fixed-wall $C_{f,0}$ on the bulk Reynolds number and not in the shift $\Delta B^+$.
\cite{yao-hussain-2019} carried out DNS of channel flows with purely temporal oscillations up to $Re_\tau=2000$. They first remarked that the disagreement between the DNS-computed drag-reduction margins and those obtained via \eqref{eq:gatti-quadrio-2016} at $Re_\tau=200$ was expected, because the log-law relation is not valid at such low Reynolds number. However, they reported a significant $\Delta \mathcal{R}=3\%$ overprediction of the DNS-computed $\mathcal{R}=22\%$ at $Re_\tau=2000$ for nearly optimal conditions. By including a correction to the assumption of constant $B_1$ they obtained much better agreement, therefore advocating models that include the dependence of $\kappa$ and $B_1$ on the flow-control parameters and the Reynolds number.

\subsection{Flow physics of controlled flows at high Reynolds number}
\label{sec:re-physics}

The question of the origin of the Reynolds-number dependence, shown in Figs. \ref{Fig:1} and \ref{Fig:extra}, is undisputed, at least in broad principle. The turbulent fluctuations above the viscous sublayer become more intense, so that their impact on the viscous sublayer, and thus on the wall-shear stress, becomes more significant. As a consequence, the drag change caused by the control also becomes less influential, because of the declining effectiveness with which the sublayer is modified by the wall motion. 

\begin{figure}
\begin{center}
\subfigure[]{\label{fig:2a}\includegraphics[trim= 2cm  0 2cm 0, clip=true,width=0.45\textwidth]{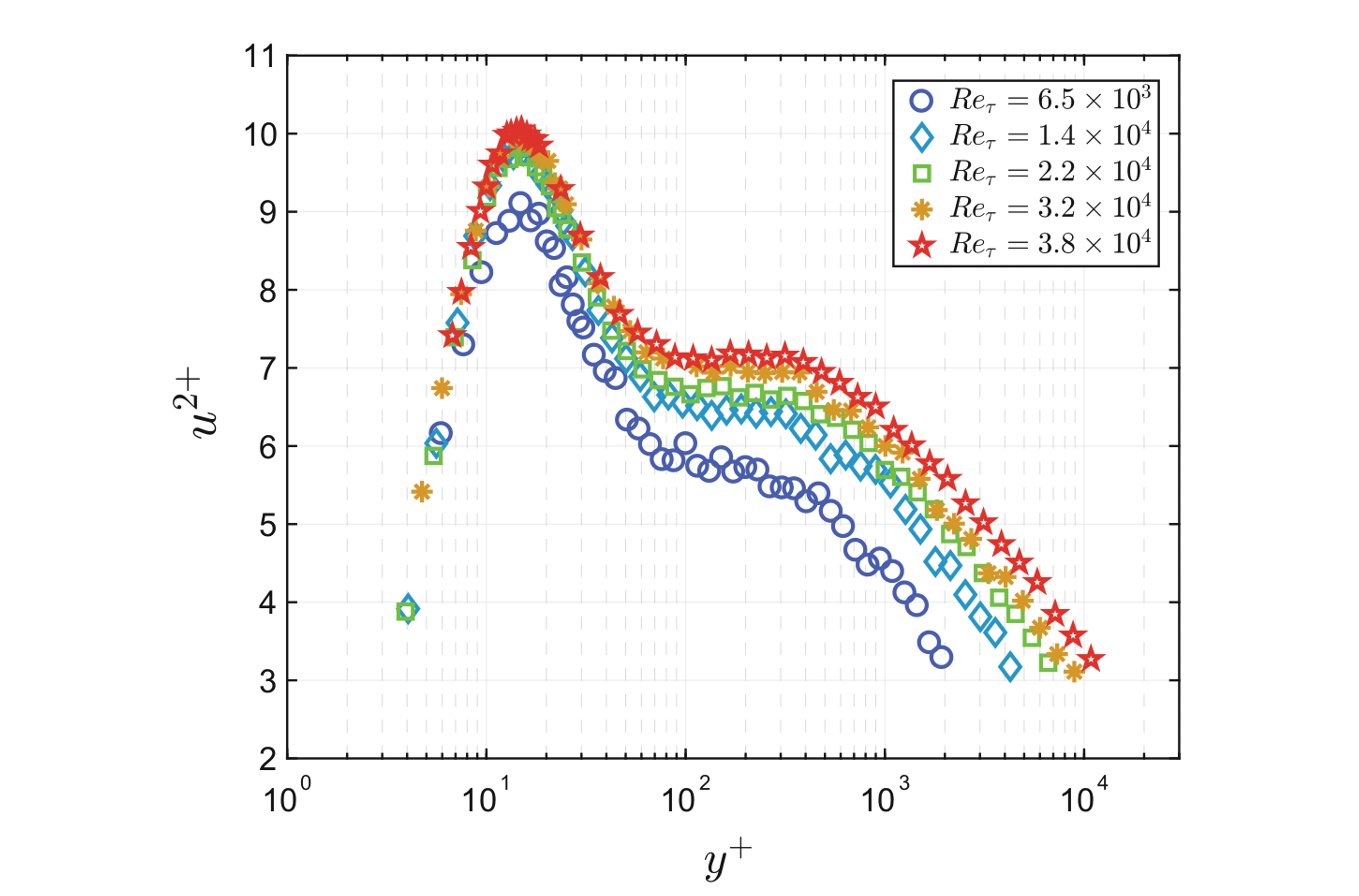}}
\subfigure[]{\label{fig:2b}\includegraphics[trim= 0 0 0 0, clip=true,width=0.5\textwidth]{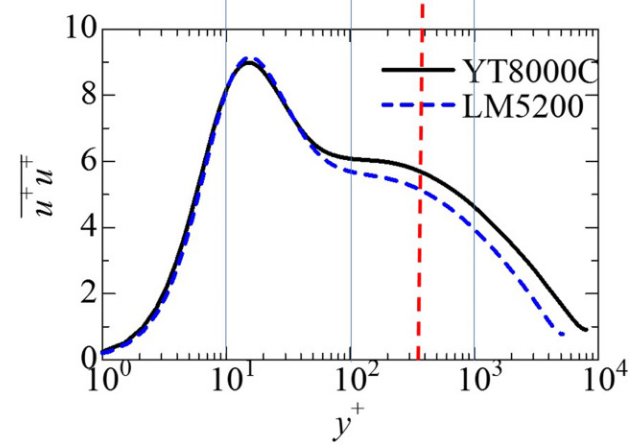}}
\caption{Profiles of streamwise turbulence energy for unactuated flows at difference friction Reynolds numbers; (a) experimental data for pipe flows; (b) DNS data for channel flow at $Re_\tau=5200$ and 8000. Plot (a) taken from \cite{fiorini2017turbulent}, with permission from Springer. Plot (b) taken from \cite{yamamoto2018numerical}, with permission from American Physical Society (APS).}
\label{Fig:2}
\end{center}
\end{figure}

Despite this generally accepted view of more intense turbulent intensity above the viscous layer, the detailed physical mechanisms that drive the dependence of the drag and of the drag-reduction effectiveness on the Reynolds number are a matter of ongoing debate and are the subject of much current research. As it transpires from the discussion below, a quantitatively incontestable statement on the effects of the rising turbulence energy in the outer layer, shown in Fig. \ref{Fig:2}, cannot be provided at the time of writing, except for the contribution of the energetic outer scales, relative to that of other scales, to the drag at one Reynolds number. One predominant line of thought is that the formation of streamwise-stretched energetic structures in the logarithmic layer plays a critical role. Experiments by Marusic and collaborators at the University of Melbourne \citep{marusicrole2001,hutchins-marusic-2007b,mathispredictive2011,Baars2018data}, and by Vallikivi, Hultmark, Rosenberg, and Smits \citep{smits2011high,hultmark2013logarithmic,rosenberg2013turbulence,vallikivi2015turbulent} in the Princeton University `Superpipe' have shown that the streamwise turbulence-energy component, scaled with the wall-shear stress, rises substantially with the Reynolds number towards a plateau or even, as sometimes claimed, towards a second maximum. This result is illustrated in Fig. \ref{fig:2a} for unactuated boundary layers. \cite{marusicpredictive2010} proposed the correlation $y^+=3.9 \sqrt{Re_\tau}$. This rise in energy reflects the presence of large, streak-like, energetic structures. DNS have confirmed the presence of the outer plateau, as shown by the profiles in Fig. \ref{fig:2b}, obtained from channel-flow simulations at $Re_\tau = 5200$ and 8000 \citep{lee-etal-2014,yamamoto2018numerical}. This figure confirms that the influence of the outer structures is rather weak at $Re_\tau = 500-1000$. 

It is important to point out that the rise in the outer streamwise energy, exemplified by Fig. \ref{Fig:2}, occurs in profiles that are scaled with the wall-shear stress, i.e., scaling that presumes a universal behaviour of the near-wall turbulence. While the near-wall maximum, around $y^+=15$ is not quite universal either, its value varying between 7.8 at $Re_\tau=1000$ and 8.8 at $Re_\tau=20000$ in channel flow, the dependence of the energy in the outer layer is much more pronounced and indicative of departures from universality in terms of inner scaling. It is also this intense rise in energy that appears to offer a promising route to explaining the Reynolds-number-dependent decline in the drag-reduction effectiveness.

Unfortunately, only a very small part of this literature deals with actuated flows, most studies analysing canonical channels flows and boundary layers. These studies are nevertheless pertinent to controlled flows because they have explained interactions that must also occur in actuated flows. An argument supporting this assertion is that the wall motion only (or mostly) affects the turbulence within the viscosity-affected sublayer, but not the outer flow -- except perhaps through global effects that arise from the reduction in the friction Reynolds number provoked by the actuation for a constant level of bulk Reynolds number. The only study of the influence of outer structures on the near-wall physics in a drag-reduced flow is that of \cite{agostini-leschziner-2018} for the case of streamwise homogeneous oscillatory spanwise motion at $Re_\tau=1000$ and $T^+=100$ and 200. 

\begin{figure}[t]
\begin{center}
\subfigure[]{\label{fig:foota}\includegraphics[trim= 0 0 0 0, clip=true,width=0.335\textwidth]{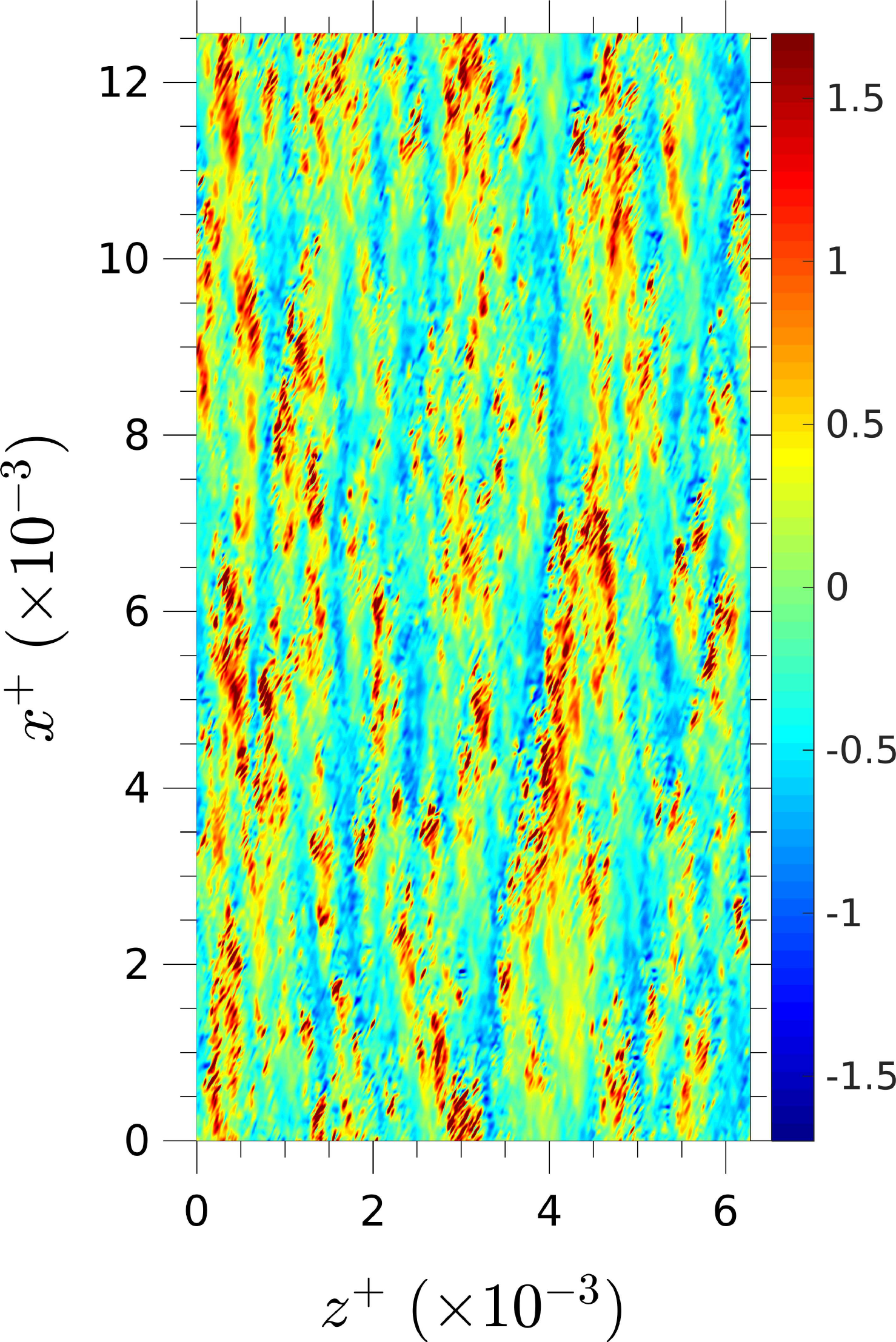}}
\subfigure[]{\label{fig:footb}\includegraphics[trim= 0 0 0 0, clip=true,width=0.32\textwidth]{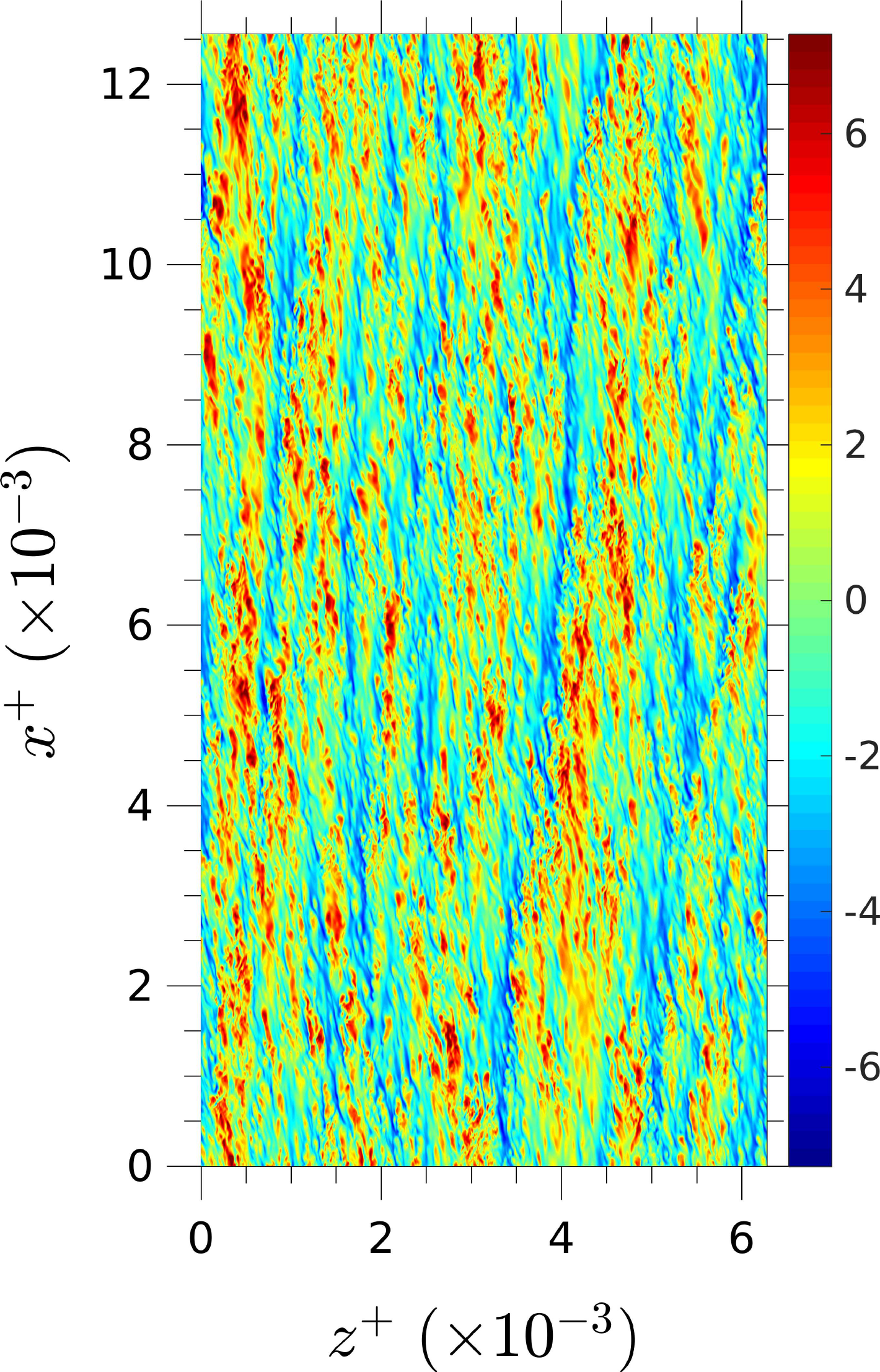}}
\subfigure[]{\label{fig:footc}\includegraphics[trim= 0 0 0 0, clip=true,width=0.32\textwidth]{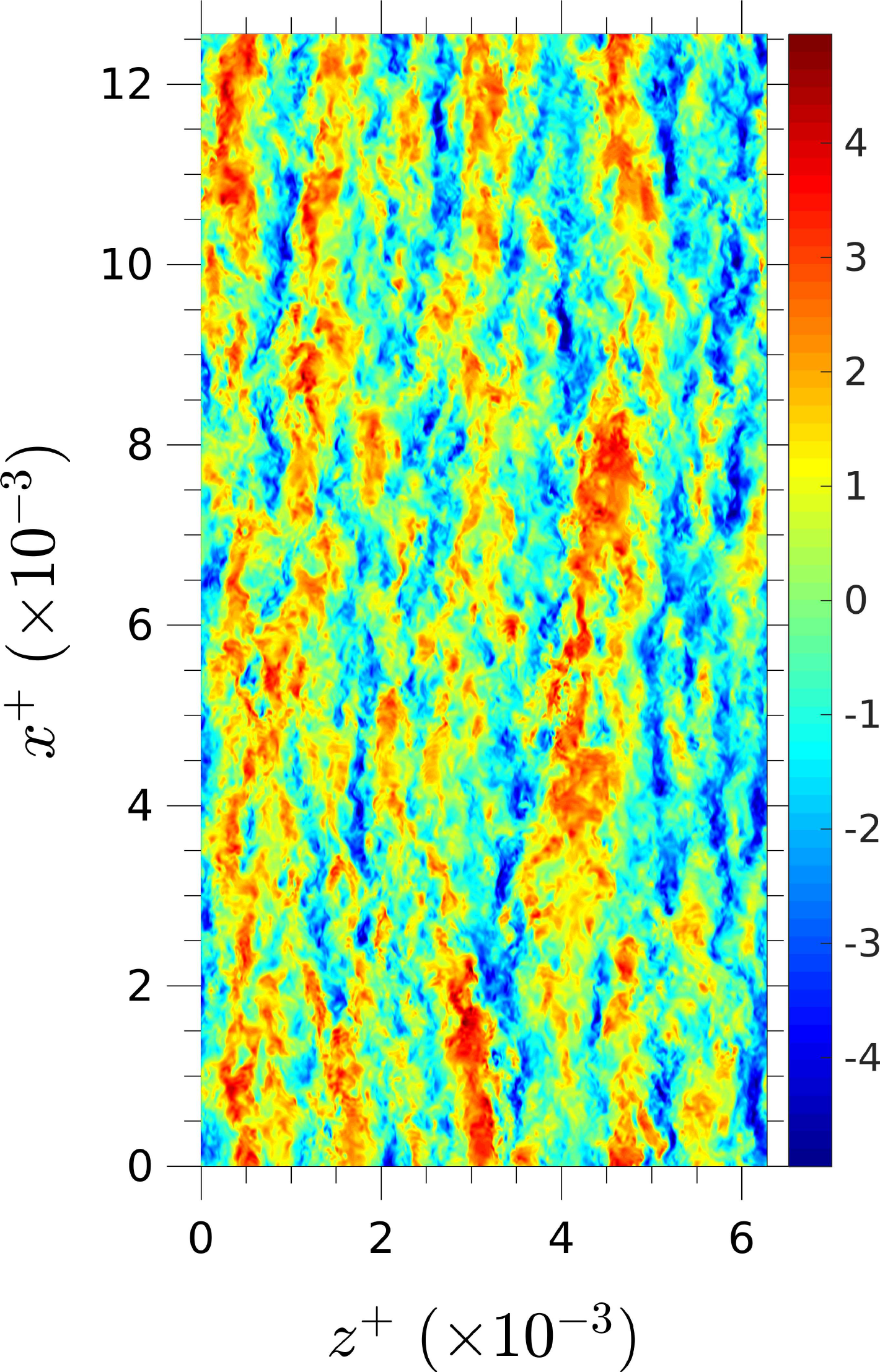}}
\caption{Illustration of footprinting by way of three snapshots of streamwise velocity fluctuations across three wall-parallel planes from a channel flow at $Re_\tau=1000$ with oscillatory wall motion ($T^+ = 100$ and $W_m^+ = 12$); (a) $y^+=3$; (b) $y^+=18$; (c) $y^+=200$. Taken from \cite{agostini-leschziner-2018}, with permission from Springer.}
\label{Fig:4}
\end{center}
\end{figure}

The question of why the outer structures should play a role in a control method, the effects of which are confined to the viscous sublayer, requires a detailed examination of the near-wall turbulence structure. 
Fig. \ref{Fig:4}, taken from  \cite{agostini-leschziner-2018}, shows three instantaneous fields of streamwise velocity fluctuations at $y^+=3$, 18 and 200, shown from left to right. There is a visually obvious correlation between the field at $y^+=200$ and the near-wall fields, which demonstrates that the outer structures leaves a marked ``footprint'' on the near-wall structures. Besides ``footprinting'', ``modulation'', introduced by \cite{mathispredictive2011}, and ``splatting'', proposed and demonstrated by \cite{agostini-leschziner-2014b}, have also been advanced as relevant mechanism by which the outer structures impact the near-wall flow.

In the context of the Reynolds-number dependence of drag reduction by wall motion, these processes are expected to be influential because they all operate on the dynamics of the streaks, which are significantly reduced in strength by the wall control.
Modulation entails an attenuation of the near-wall turbulence by the action of negative footprints. The streaks are, however, already substantially weakened by the actuation when efficiently implemented, so that the attenuation is marginal. On the other hand, the amplification by the positive footprints tends to be strong.  Hence, the asymmetric modulation in relation to splatting can be expected to be considerably more pronounced.

\begin{figure}[t]
\begin{center}
\subfigure[]{\label{fig:7a}\includegraphics[trim=   0  0 0 0, 
clip=true,width=0.45\textwidth]{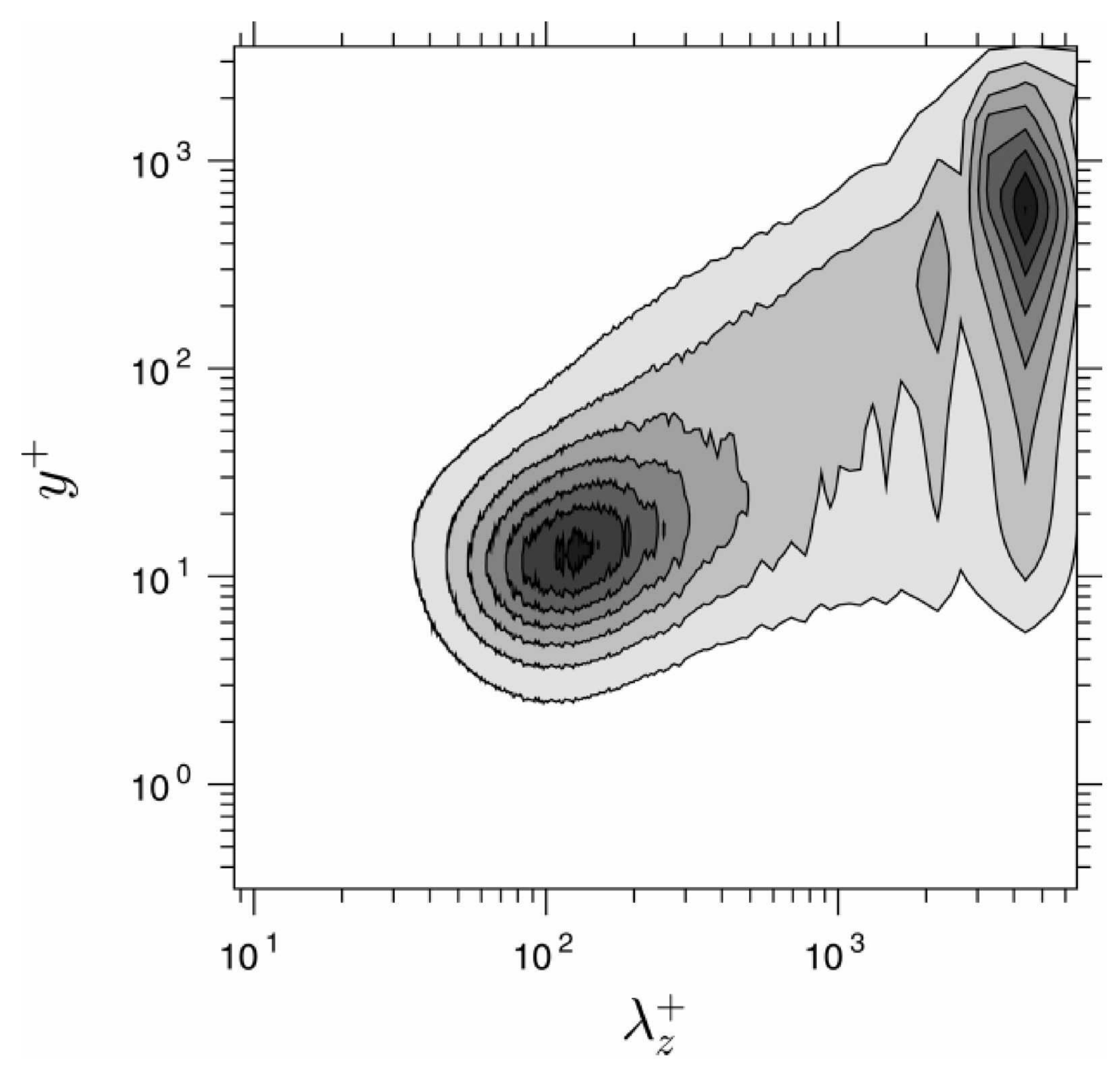}}
\subfigure[]{\label{fig:7b}\includegraphics[trim= 0 0 0   0, 
clip=true,width=0.50\textwidth]{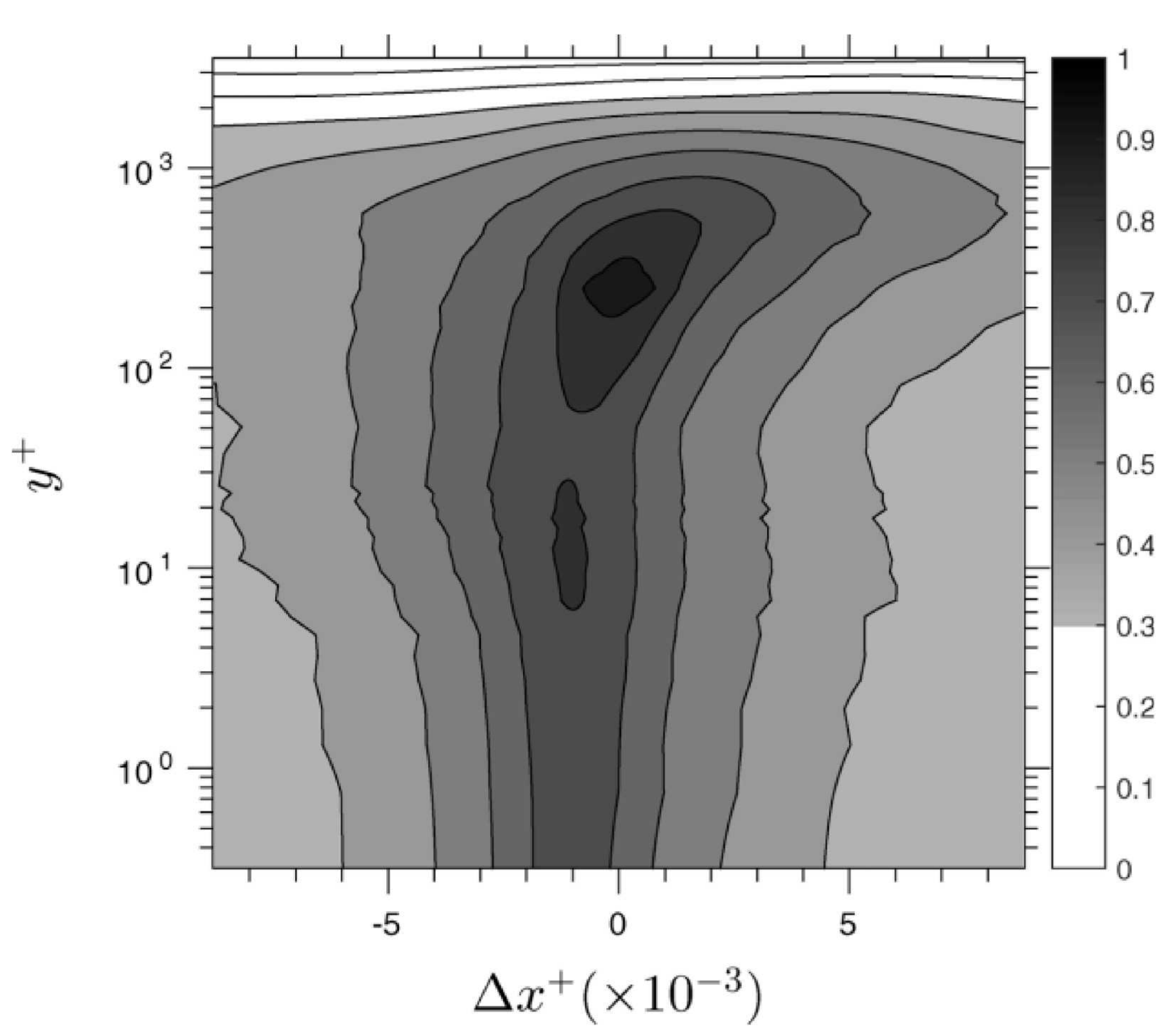}}
\caption{Manifestations of large-scale outer structures in channel flow at $Re_\tau=4200$ (DNS by \cite{lozano2014time}); (a) spectral map of streamwise energy in wall-normal-distance/spanwise-wavelength plane: peak of large-scale energy at $y^+ \approx 400$, $\lambda_z^+ \approx 4000$; (b) map of two-point correlation of large-scale fluctuations relative to location $y^+=250$. Taken from \cite{agostini2016validity}, with permission from AIP Publishing.}
\label{Fig:7}
\end{center}
\end{figure}

Any attempt to unravel the details of the interactions between the outer structures and the near-wall region -- and hence the effect on the drag -- requires access to detailed measurements or computational data extracted from DNS.  An illustration of this requirement has already been given by Fig. \ref{Fig:4}, derived from DNS. Detailed experimental data targeting the role of outer structures have only been obtained for unactuated boundary layers with the aid of hot-wire probes
\citep{marusicrole2001,hutchins-marusic-2007b,mathispredictive2011,mathisestimating2013,chung-etal-2015,Baars2018data}. The data so extracted are in the form of time-series of streamwise turbulent signals at different wall-normal distances. When these data are subjected to Fourier filtering, for example, the signal can be decomposed into large-scale and small-scale components.  This decomposition then allows the interaction between small-scale and large-scale turbulence to be studied.  In particular, the manner in which the small-scale fluctuations near the wall are augmented, amplified or attenuated by large-scale motions can be quantified. 

The choice of the small-scale/large-scale filter is not unambiguous, but not arbitrary either, as exemplified by Fig. \ref{fig:7a}, which shows the streamwise-energy density as a function of spanwise eddy size (wavelength) and wall-normal distance for a channel flow at $Re_\tau=4200$ \citep{agostini2016validity,lozano2014effect}. This spectrum provides clear evidence of the existence of two distinct regions of elevated energy -- one, close to the wall, at around $y^+=15$, associated with the near-wall streaks, and the other in the outer region, at around $y^+ \approx 400$, indicative of the energetic outer structures.  In terms of length scale, the inner peak, at around $\lambda_z^+ \approx 100$, is characteristic of the inter-streak distance in the buffer layer, while the outer peak is at $\lambda_z^+ \approx 4000$, identifying ``super-streaks'' that are roughly one channel height apart.  This spectrum thus suggests that a filter at around $\lambda_z^+ \approx 1000$ would be a reasonable choice.  More elaborate approaches to scale decomposition are those of \cite{zhang2016quasisteady}, based on a Pareto-front optimisation of the Fourier filter, of \cite{baars2016spectral}, based on a coherence function, in Fourier space, between the velocity signal at the outer location at which the large-scale structures are observed and the velocity signal at any other wall-normal location, and of \cite{agostini-leschziner-2014b}, in which a modal-decomposition formalism called ``Empirical Mode Decomposition'' has been used. This latter approach decomposes any spatial two-dimensional field, derived from full-volume DNS realizations, into a series of narrow-wavelength-spectrum fields or ``intrinsic modes''. These modes are then collected into two groups of small-scale and large-scale contributions, a process that require judgement much like the choice of a single Fourier filter to separate small-scale from large-scale components in a temporal signal. 
A fourth, though rather qualitative method, is one in which the same flow is computed on a range of decreasing box sizes so as to progressively exclude large-scale structures and observe the dependence of this exclusion on the drag.  This approach was adopted by \cite{hwang2016self} who used LES for an unactuated channel flow at $Re_\tau=1000$ to investigate the validity of Townsend's Attached-Eddy Hypothesis. 

Once the flow has been decomposed into small-scale and large-scale fields, it is possible to investigate how large-scale fluctuations in the log layer are correlated with large-scale fluctuations close to wall -- the latter being ``footprints'' of the former, and how the intensity of the small-scale fluctuations near the wall responds to the outer large-scale fluctuations -- referred to as ``modulation'' above.  An illustration of the former is given in Fig. \ref{fig:7b} \citep{agostini2016validity}, which gives a map, in the $x-y$ plane, of the two-point correlation of large-scale motions.  The essential message to be taken from this map is that these large-scale fluctuations at the wall are closely correlated with large-scale fluctuations at $y^+ \approx 500$.  

\cite{hurst-etal-2014}, for $Re_\tau=200 - 1600$, and \cite{yao-etal-2019}, for $Re_\tau=200 - 2000$, both separated the contributions to the drag reduction arising from the flow in the inner (near-wall) and outer regions, with the boundary between the two regions being at the height where the Reynolds shear stress attained its maximum value. The contributions to the drag reduction was determined using the identity derived by \cite{fukagata-iwamoto-kasagi-2002} upon a $y$-wise integration of the streamwise-momentum equation: 

\begin{equation}
\label{eq:fik}
C_f=\frac{6}{Re_{b}}+6\int_0^1\ \left(1-\frac{y}{h}\right)\frac{-\overline{uv}}{U_b^2}\mathrm{d}\left(\frac{y}{h}\right),
\end{equation}
in which the first term represents the laminar contribution and the second term represents the contribution due to the turbulent fluctuations. The separate contributions from the inner and outer region arise simply upon applying the latter term in equation \eqref{eq:fik} to the two layers separately. It is important to point out, however, that this approach does not distinguish between small scales and large scales, but merely considers the contributions of two wall-normal regions to the total drag, with all scales included in the two layers. 
Both investigations came to the conclusion that the drag reduction from the outer part remains constant as the Reynolds number increases, while the contribution of the inner part declines. The implication is, therefore, that the decrease of drag reduction with increasing Reynolds number is due to the decline of the effectiveness of the control to reduce the near-wall turbulence. 
\cite{yao-etal-2019} also performed a Fourier-based division of spectrum of the Reynolds shear stress into small and large scales, with a wavelength of $\lambda_z^+ \approx 1000$ separating the two ranges of scales, which is an arbitrary but broadly reasonable choice according to the previous discussion. This analysis showed that the large-scale turbulence near the wall remained largely unaffected by the control within the Reynolds-number range considered.
\cite{yao-etal-2019} also examined the response of the near-wall structure -- in particular, the prevalence of coherent vortical motions, as elucidated by the second eigenvalue, $\lambda_2$, of the tensor $S^2+\Omega^2$. They observed that the suppression of the drag-enhancing small-scale coherent structures by the control declined with increasing Reynolds number. This led them to conjecture that the observed shift of the optimum frequency towards higher values with increasing Reynolds number is due to the fact that forcing at higher frequency, at a given forcing amplitude, amounts to forcing the near-wall layer at a smaller spanwise length scale, which thus enhances the effectiveness of the forcing in depressing the small near-wall vortical scales primarily responsible for the drag.

\begin{figure}[ht]
\begin{center}
\subfigure[]{\label{fig:9a}\includegraphics[trim=   0 0 0 0, 
clip=true,width=0.45\textwidth]{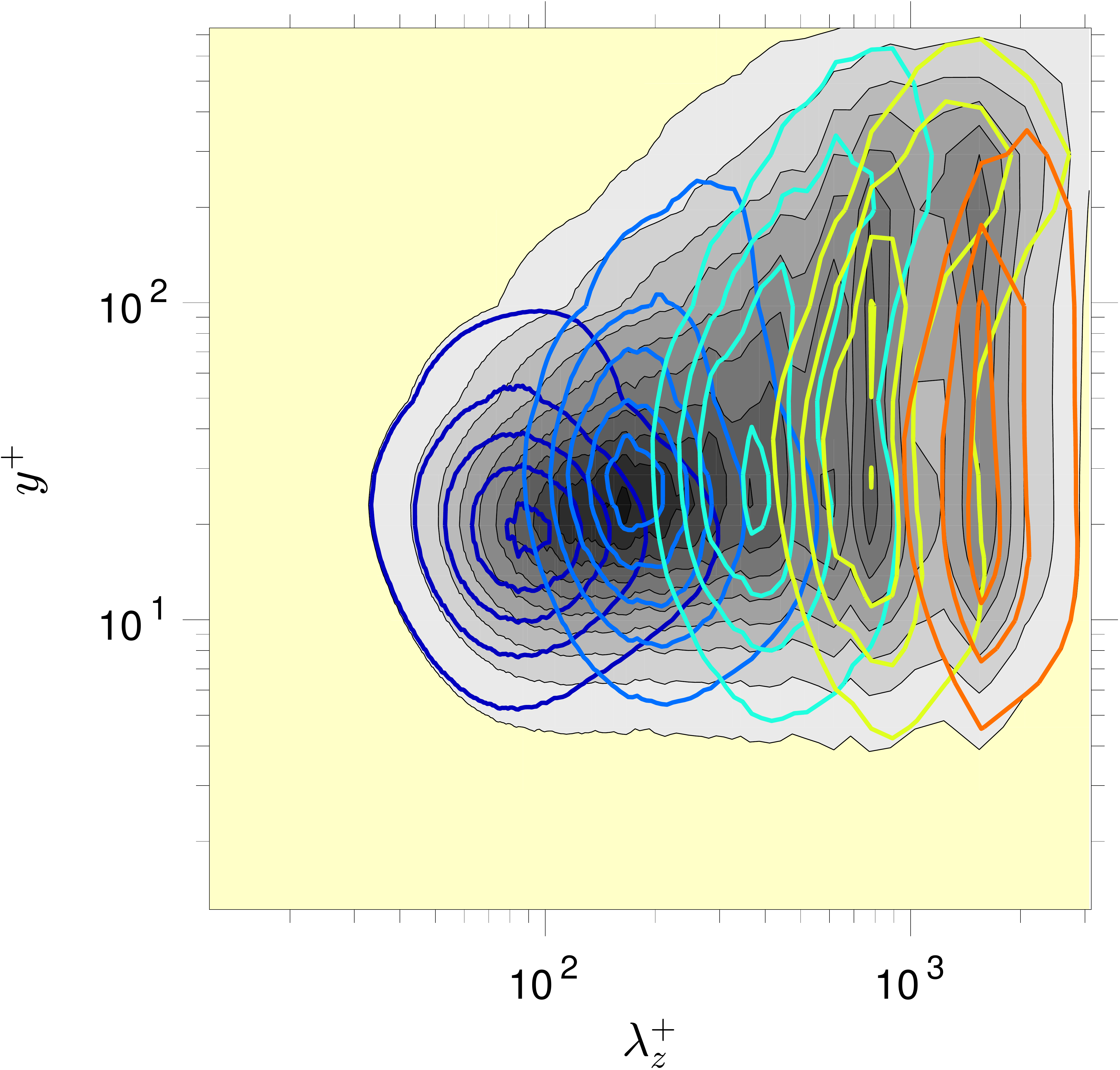}}
\subfigure[]{\label{fig:9b}\includegraphics[trim= 0 0 0   0, 
clip=true,width=0.50\textwidth]{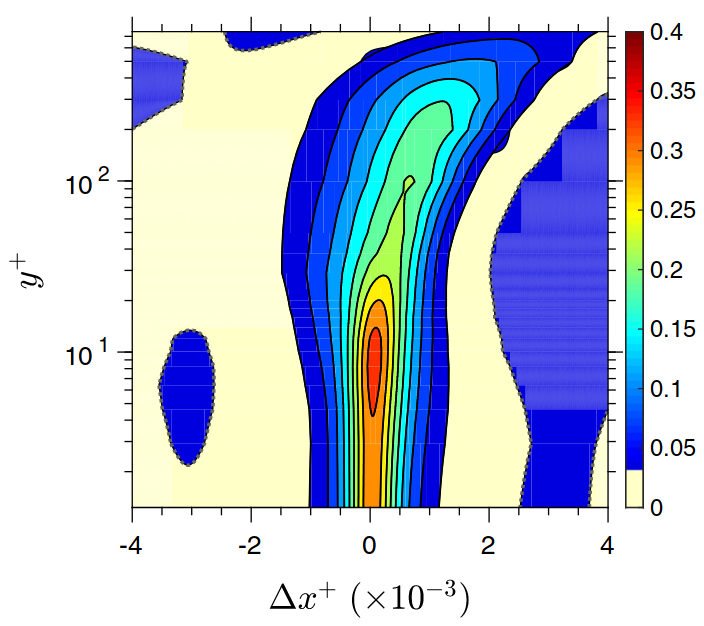}}
\caption{Decomposition of streamwise energy of an actuated flow at $T^+=100$ into spectral modes using the Empirical Mode Decomposition; (a) Energy density in wall-normal/spanwise-wavelength plane: grey contours identify the entire spectrum, coloured contours indicate sub-spectra for five modes (the 6'th mode is the residual, not shown); (b) map of two-point correlation of large-scale fluctuations (modes 4+5+6) demonstrating wall-normal coherence. Taken from \cite{agostini-leschziner-2018}, with permission from Springer.}
\label{Fig:9}
\end{center}
\end{figure}

\cite{agostini-leschziner-2018} examined DNS data for channel flow at the unactuated Reynolds-number value $Re_\tau=1000$ at two actuation periods, $T^+=100$ and 200. Although this case is, energetically, entirely unpromising -- as the actuation power is larger than the power saved by the gross drag reduction of $\R=30\%$ and $\R=20\%$ at the two actuation periods, respectively -- this actuation mode is simple in terms of the statistical processing that is needed to derive information on the direct and indirect contributions of footprinting and modulation of the outer structures on the drag. 
\cite{agostini-leschziner-2018} used the Empirical Mode Decomposition (EMD) to separate the large-scale from the small-scale fields.  Fig. \ref{Fig:9} shows, for $T^+=100$, the decomposition into 6 (5+residual) modes by way of the streamwise-energy spectra in the spanwise-wavelength/wall-normal plane.  Each mode covers a fairly narrow range of scales.  Modes 1 and 2 -- the left-most modes in Fig. \ref{Fig:9} -- are deemed to represent the small scales, while modes 4-6 are regarded as representing the large scales.  Mode 3 represents intermediate scales and is excluded from the attribution. Fig. \ref{Fig:9} highlights one limitation of the study -- namely, the absence of a distinct scale-separation region, which is present in Fig. \ref{Fig:7} for $Re_\tau=4200$. It is arguable that $Re_\tau \approx 1000$ is the lowest Reynolds number at which the interactions at issue can be studied. It needs to be noted that the actual Reynolds number is even lower here, at $Re_\tau \approx 850$, because the actuation causes a 30\% reduction in the wall-shear stress.  
A feature that is apparent in Fig. \ref{fig:9a} is that the large-scale-energy contours extend from $y^+ \approx 500$ down to $y^+ \approx 8$, qualitatively indicating the footprinting process. Fig. \ref{fig:9b} shows this result in a more quantitative manner by contours of the two-point-correlation map of the large scales fluctuations across the boundary layer.  This map also identifies the streamwise lag in the footprinting process, which is of order $0.5Re_\tau$.

Through the EMD decomposition, it is possible to quantify the direct contribution of the large scales (identified by the subscript ``LS'') on the skin friction. The first step is to determine the wall-normal profiles of the large-scale shear stress $\overline{u_{LS}v_{LS}}$, obtained by averaging the product of large-scale velocity fields for $u$ and $v$ across the statistically homogeneous $x-z$ planes and over however many temporal realizations have been stored. This distribution can then be inserted into the FIK identity (\ref{eq:fik}).

The result from the EMD analysis by \cite{agostini-leschziner-2018} of the channel flow undergoing wall oscillations with $T^+=100$ is that the large scales contribute 26\% of the total skin friction, a contribution that is further increased by mixed-scale shear-stress terms that correlate large-scale motions with small-scale and intermediate-scale fluctuations. This percentage is remarkable in view of the relatively low Reynolds number, and a recent study by \cite{agostini-leschziner-2019-b} of the unactuated channel flow at the same nominal Reynolds number has also yielded a very similar large-scale contribution, at 24\%.  
This value is also considerably larger than the estimate of 10\% given by \cite{hwang2016self} for unactuated channel flow and derived from a series of LES computations of domains of decreasing dimensions, designed to progressively exclude outer large-scale motions.
\cite{agostini-leschziner-2019-b} also showed that, if the contributions of the large scales is derived from an $y$-wise integrated form of the energy equation, involving turbulence energy production and dissipation, the implied direct contribution of the large scales for $Re_\tau$ is only 8\%. 
This same value is also estimated by \cite{Ruan_2019} from experiments in which closed-loop-controlled wall jets, actuated by recording upstream footprints, were used to target the outer structure and reduce their energy. Ultimately, it is likely that the momentum-based estimate of the large-scale contribution is more appropriate and credible. 

A quantification of the indirect contribution of the large scales, by the process of modulation, is more difficult to undertake. \cite{agostini-leschziner-2018} approached this challenge by constructing a PDF of the large-scale fluctuations of velocity, $u_{LS}$, or skin friction, $C_{f,LS}$, dividing this PDF into 5\% bins, by area, and then conditionally sampling the intensity of the small-scale motions of skin friction within the bins, so as to bring out the response of the small-scale fluctuations to the intensity of the large-scale footprints. 

\begin{figure}[ht]
    \centering
    \includegraphics[width=0.9\textwidth]{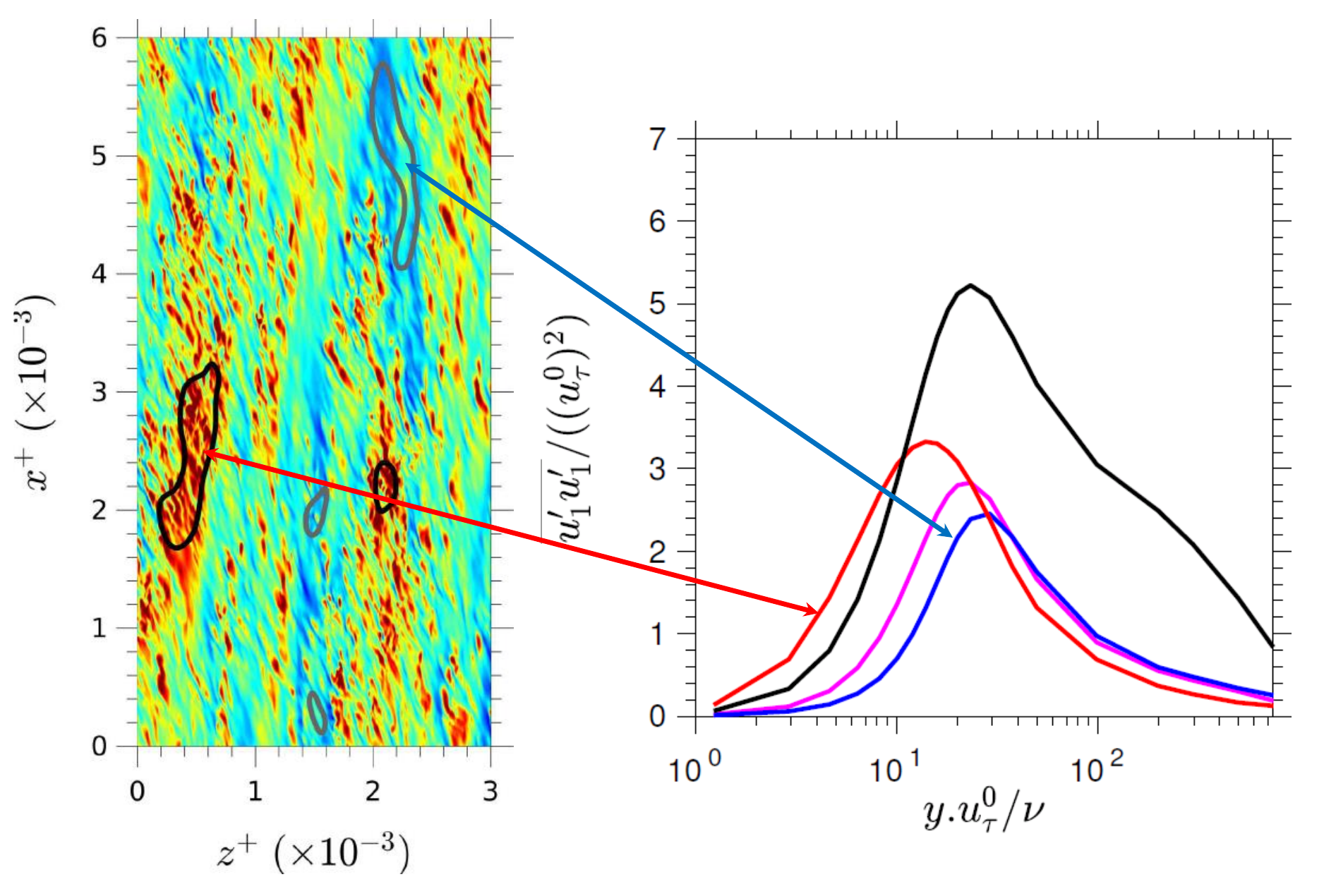}
    \caption{ Profiles of streamwise energy of small-scale fluctuations conditional on extreme 5\% positive (red curve), 5\% negative (blue curve) and 5\% weakest (magenta curve) large-scale fluctuations in actuated channel flow at $Re_\tau=1000$ and $T^+=100$; LHS plot is a snapshot illustrating the conditional sampling for large positive and large negative footprints. Taken from \cite{agostini-leschziner-2018}, with permission from Springer.}
    \label{Fig:10}
\end{figure}

\begin{figure}
    \centering
    \includegraphics[trim= 0 0 0 0, clip=true,width=0.85\textwidth]{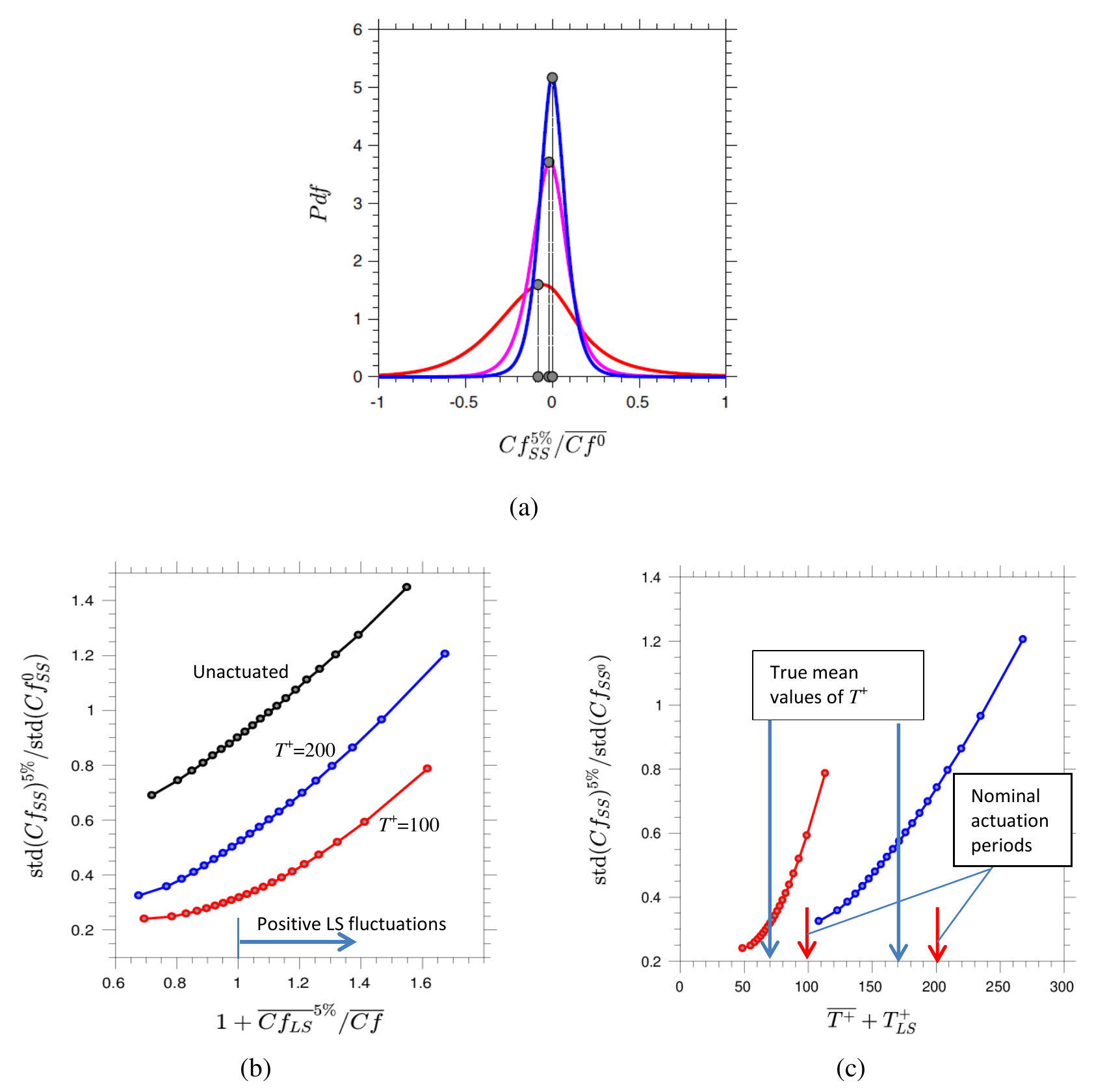}
\caption{Asymmetry in the modulation of small scales by large scales in actuated channel flow at $Re_\tau=1000$ and $T^+=100$; (a) PDFs of small-scale skin-friction fluctuations conditional on extreme 5\% positive (red curve), 5\% negative (blue curve) and 5\% weakest (magenta curve) large-scale skin-friction fluctuations; (b) standard deviation of PDFs of small-scale skin-friction fluctuations conditional on 5\% bins of of PDF for large-scale skin-friction fluctuations; (c) as (b), but plotted in terms of actuation period scaled with large-scale friction velocity to show the range of $T^+_{LS}$ for nominal actuation periods $T^+=100$ and 200. Taken from \cite{agostini-leschziner-2018}, with permission from Springer.}
\label{fig:11}
\end{figure}

The manner in which the small-scale streamwise energy responds to large-scale fluctuations is shown in Fig. \ref{Fig:10} for the actuated flow at $T^+=100$. The red and blue profiles characterize the response to the extreme 5\% positive and 5\% negative large-scale fluctuations, respectively, while the magenta line arises from the 5\% central PDF bin for which the large-scale fluctuations are negligible. The profiles suggest that positive fluctuations causes a substantial thinning of the viscous sublayer, while negative fluctuations cause a thickening. The sensitivity is asymmetric, with positive large-scale fluctuations having a substantially larger impact, thus implying a damaging influence on the drag-reduction process.     

The asymmetric response is further illustrated by Fig. \ref{fig:11}(a). It shows three PDFs for the small-scale skin-friction fluctuations, derived for the same three large-scale PDF bins for which the profiles in Fig. \ref{Fig:10} were obtained.  These distributions show that positive footprints cause a substantial amplification of small-scale fluctuations, while negative footprints only cause a mild damping of the skin-friction fluctuations. This asymmetry, peculiar to actuated cases, is most distinctly brought out in Fig. \ref{fig:11}(b) which shows three distributions of the standard deviation of the small-scale skin-friction fluctuations as a function of the intensity of the large-scale footprints. The black curve is for the unactuated flow, the blue for $T^+=200$ and the red for $T^+=100$.  A first observation is that the actuation results in a general decline in the intensity of the small-scale fluctuations. This decrease is expected, of course, because the actuation causes a reduction in drag and near-wall turbulence activity.  A feature that is not expected is the increasing asymmetry in the modulation intensity, wherein large-scale footprints cause a higher rise in the fluctuations in positive large-scale footprints level than the attenuation in the negative footprints. This asymmetry may well reflect the mechanism by which increasingly intense outer structures affect the drag-reduction effectiveness, as the strong amplification of the small-scale fluctuations reflect a corresponding destabilization of the near-wall streaks and thus a tendency to increase the drag. Unfortunately, these results, while pertinent and informative in a general sense, do not allow a quantitative statement to be made on the effects of this process on the drag.  

Fig. \ref{fig:11}(c) indicates a secondary mechanism by which large scales influence the drag-reduction effectiveness.  Here, the red and blue curves in Fig. \ref{fig:11}(a) are replotted as functions of the large-scale period $T^+_{LS}=T u_{\tau,LS}^2/\nu$. This plot therefore shows, for a given actuation period, the range of locally scaled periods that arise due to the strong variations of large-scale wall-shear stress provoked by footprinting. The scaled actuation period varies significantly around the nominal value.  For the optimum case $T^+=100$, this variability, $50<T^+_{LS}<115$, results in local near-wall conditions deviating from the optimal actuation conditions. The larger the deviations, the higher the likelihood of detrimental effects on the drag-reduction effectiveness will be.

\section{Modelling approaches}
\label{sec:models}

Modelling approaches for understanding and predicting the drag-reduction effect do not rely on direct and large-eddy simulations or experiments, but involve simplified reduced-order versions of the Navier-Stokes equations or empirical scaling parameters. Key simplifications have been, first, the use of linearized versions of the Navier-Stokes equations to describe the principal interactions involved, and, second, the resolution of cross-flow processes without recourse to the streamwise flow.

A two-dimensional model was implemented numerically by \cite{dhanak-si-1999} by solving simplified versions of the Navier-Stokes equations based on the assumption of a weak streamwise variation of the streamwise vortices. The flow was solved in the cross-flow plane to qualitatively reveal the typical features observed in the DNS, such as the optimal period for drag reduction at $T^+=100$ and the sustained attenuation of the intensity of the vortical structures and of the Reynolds stress.
The streamwise structures were modelled by a two-dimensional vortex undergoing axial stretching, distortion, and translation, subject to streamwise-homogeneous straining, which accounted for the flow induced by the near-wall coherent structures in the outer flow field. When the oscillatory spanwise cross-flow was imposed, the coherent structures deformed and interacted more energetically with the wall, which led to their rapid annihilation. \cite{dhanak-si-1999} also noted that the low-speed streaks were significantly distorted which resulted in a reduction of the rate of momentum convection normal to the wall. This change of momentum transfer, in turn, had a direct attenuating effect on the Reynolds stress, the production of kinetic energy, and the skin friction. 

\cite{nikitin-2000} also solved simplified flow equations in the cross-flow plane, but adopted the theoretical framework of \cite{nikitin-chernyshenko-1997} to study the linearized dynamics of the perturbations of the cross-plane velocity components. An optimal wall-oscillation frequency produced the maximum vortex attenuation, similar to the DNS results. This result confirmed that the linearized approach can reveal some key features of the dynamics observed in fully turbulent conditions, although no information about the change in wall-shear stress could be extracted because the streamwise flow was assumed unchanged by the wall motion.

\cite{bandyopadhyay-2006} utilized the experimental data from \cite{choi-debisschop-clayton-1998} to build a model based on the kinematic reorientation of the vorticity. By introducing an attenuation factor in the formulation of the Stokes layer, the spanwise fluid displacement was shown to follow the laminar solution, although the parameter varied through the boundary layer. \cite{bandyopadhyay-2006} also linked the velocity profiles from the oscillating-wall experiments to the maximum drag reduction obtained in polymer-added flows. 

The linearized equations were also used by \cite{duque-etal-2011,duque-etal-2012} to study the effect of the wall motion given in \eqref{eq:waves} on the transient growth of the perturbation energy of localized initial disturbances. They found that the contour map of the energy reduction and amplification showed a distinctive qualitative similarity to the drag-reduction map by \cite{quadrio-ricco-viotti-2009}, shown in Fig. \ref{fig:QRVmap}, thus suggesting that the linearized dynamics can be used to infer the origin of drag reduction. 
A further step in this direction was taken by \cite{blesbois-chernyshenko-2011} and \cite{blesbois-etal-2013}, who used a linearized optimal-growth approach to study the streak dynamics in near-wall turbulence subjected to uniform spanwise wall oscillations. The rationale was that the linearized system acts as a filter and amplifier of certain spectral signals, giving preference to some specific perturbations, amplifying them, and thus providing information on the natural tendency of the system to sustain them. These structures were argued to be the low-speed streaks in the buffer layer, the formation of which was especially sensitive to shear-induced amplification of the optimal perturbation.
The slow variation, the inclination angle, and the spanwise spacing of the streaks compared reasonably well with DNS results of \cite{touber-leschziner-2012}. Another outcome of the study was that the streak-amplification time scale was $t^+ \approx 50$.  The implication was that, once the streaks are weakened, they require this time to regenerate. If the time scale of the wall actuation is sufficiently short, the weakened streaks do not have time to strengthen before being weakened further.
This interpretation, first advanced by \cite{quadrio-ricco-2004} and discussed in Section \ref{sec:channel-temp}, allowed a prediction of the optimal period. At $T^+=100$, the streaks undergo weakening and strengthening twice and thus the time scales of the actuation and the streak regeneration are very similar. This analysis provided a plausible mechanism for the progressive depression of the turbulence activity and therefore of the drag towards a stable low-drag state by a sequence of actuation cycles that progressively weaken the streaks (refer also to page 44 of \cite{touber-leschziner-2012}).  

\cite{belan-quadrio-2013} utilized a perturbation method under the assumption of small amplitude of wall forcing \eqref{eq:waves} and of validity of the Boussinesq hypothesis, which models the Reynolds stresses as proportional to the wall-normal gradient of the streamwise velocity. They obtained a decay of drag reduction that was milder than the one extracted by DNS at that time and also proposed that the drag-reduction margin reaches a constant as the Reynolds number increases beyond a threshold value.

The study of \cite{moarref-jovanovic-2012} employed Reynolds-Averaged-Navier-Stokes (RANS) equations to study a turbulent channel flow with spanwise wall oscillations. The numerical procedure started from assuming that the eddy viscosity was not initially influenced by the wall motion and that it coincided with the eddy viscosity in the fixed-wall case. The mean velocity was computed with this eddy viscosity and it was used, in turn, to solve the stochastically-forced linearized Navier-Stokes equations for the case with modified moving-wall boundary conditions. The turbulent kinetic energy and dissipation were then obtained from the RANS solution using a $k-\epsilon$ model, and were utilized to modify the eddy viscosity, through which the altered mean velocity was finally found. The reduction of drag was found from the modified mean-velocity profile. The mean velocity was updated only once, avoiding expensive iterations. Both the drag reduction and the power spent for enforcing the wall motion were computed accurately for periods of oscillations smaller than, or comparable with, the optimum. 
The study of \cite{sharma-etal-2018} is the only one where a commercial code, Ansys Fluent, was employed to investigate drag reduction by spanwise wall oscillations. A RANS simulation using the $k-\epsilon$ model was carried out and only a limited section of the wall was set in motion. Drag-reduction margins up to $\R=20\%$ were computed, although the uncertainty was not reported.
The only study that employed the Reynolds-stress model (based on the Reynolds-stress transport equations) to investigate an actuated flow was carried out by \cite{duan-etal-2015}. They studied a turbulent channel flow at $Re_\tau=180$ altered by spanwise wall oscillations and showed that the turbulence intensity was attenuated by the wall motion through a change of the pressure-strain term, although the authors admit that the quantitative agreement with DNS results was not satisfactory.

\cite{choi-xu-sung-2002} first proposed an empirical parameter function of $T$ and $W_m$ to which the amount of drag reduction could be related. The function involved the product of the Stokes-layer thickness, given in \eqref{eq:delta}, and the acceleration of the Stokes layer obtained from the laminar solution, given in \eqref{eq:stokes}. It was found that this function correlated quadratically with the drag reduction. \cite{quadrio-ricco-2004} showed that the same function related linearly to the drag reduction as long as the Stokes-layer acceleration was computed in the lower portion of the buffer layer, i.e., at $y^+=6.3$, and the typical interaction time between the oscillating wall and the near-wall turbulence, i.e., half of the oscillation period, was shorter than the typical longitudinal lifetime of the turbulent structures. 
The function proposed by \cite{choi-xu-sung-2002} was further employed by \cite{ricco-quadrio-2008} to study the optimum period at fixed spatial displacement, to predict the net power saved, and to quantify the minimal wall displacement below which the wall friction is unperturbed. \cite{ricco-quadrio-2008} showed that this parameter offers a reliable means for predicting the drag-reduction margin when $W_m^+ \leq 40$ and $30 \leq T^+ \leq 150$.
\cite{ricco-etal-2012} observed that, at fixed $W_m$, the scaling parameter was linearly linked to the wall-normal gradient of the Stokes layer at $y^+=6.3$. They further noted that, at this same height, the most intense enstrophy-production term, being central to the oscillating turbulence dynamics and depending directly on the Stokes-layer shear, reached its maximum.

The empirical model by \cite{choi-xu-sung-2002} was not able to predict the drag-reduction margin for a general non-sinusoidal oscillation as shown by \cite{cimarelli-etal-2013}, who instead developed a predictive model based on a penetration length defined as the distance from the wall where the induced variance of the oscillating velocity drops below a threshold value. The drag reduction was shown to be correlated to the penetration length to the power of $3/2$ as long as $T^+ \leq 150$. \cite{yakeno-etal-2014} developed another empirical model, based on the Stokes strain at two vertical positions, that provided successful predictions of the drag-reduction margin for $T^+$ up to 250 and for $W_m^+$ up to 12.

\section{Wall-motion-inspired drag-control methods}
\label{sec:extensions}

In this section, the discussion focuses on control methods of wall-bounded shear flows that lie outside, but are closely related to, the main research subject being reviewed, i.e., the study of wall-shear-stress reduction of wall-bounded incompressible turbulent flows by spanwise wall motion. The studies included here thus present spanwise forcing on laminar-flow stability and transition to turbulence, axial oscillations on Taylor-Couette flows, and the turbulent skin-friction reduction by spanwise-oriented body forces and jets, rotating discs and rings, passive methods (oblique wavy walls, dimples, and spanwise-forcing riblets), and spanwise-travelling waves of wall-normal wall displacement. Publications on the role of compressibility and heat transfer on the properties of drag reduction by spanwise wall motion are also discussed.

\subsection{Perturbed laminar flows and transitional flows}

Inspired by the success of spanwise wall oscillations to reduce the turbulent friction drag, scientists have naturally pursued research aimed to modify pre-transitional and transitional wall-bounded flows by wall motion, although these flows have received considerably less attention than fully-developed turbulent flows. All the studies, except the one by \cite{wang-liu-2019}, report a beneficial effect of the wall motion, i.e., an attenuation of the growing perturbations or a delay of transition.

\begin{figure}
\centering
\includegraphics[width=0.7\columnwidth]{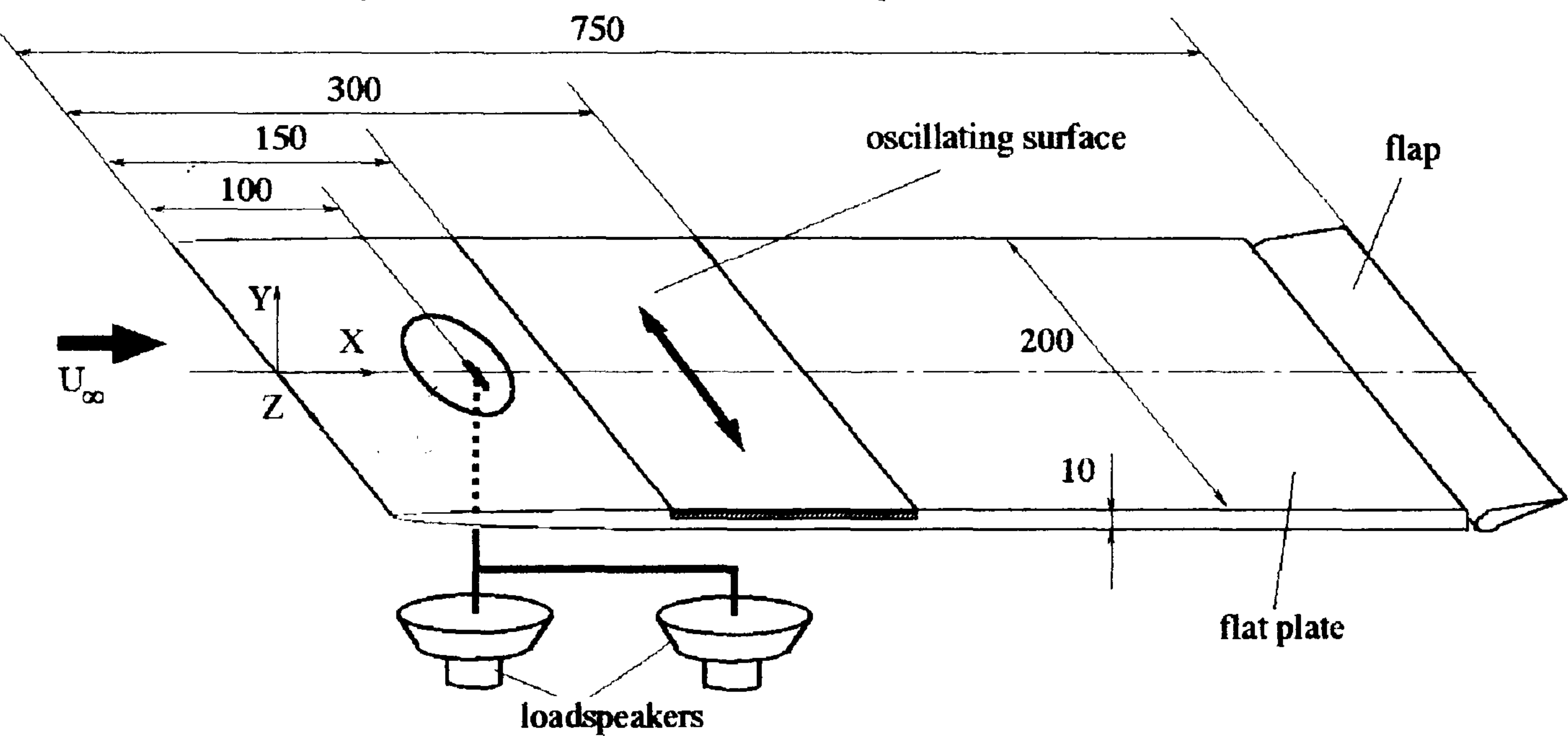}
\caption{Schematic of wind-tunnel apparatus used by \citet{katasonov-kozlov-1998,katasonov-kozlov-2000} to study the effect of spanwise wall oscillations on a transitional boundary layer. Taken from \citet{katasonov-kozlov-2000}, with permission from Springer.}
\label{fig:katasonov-kozlov}
\end{figure}

Most of the research studies have focused on the alteration of the initial perturbation growth, either under the small-amplitude assumption, thus examining the linearized dynamics, or during the subsequent nonlinear evolution. To the best of the writers' knowledge, the work of \citet{katasonov-kozlov-1998,katasonov-kozlov-2000} is the only experimental study on spanwise wall oscillations beneath a transitional boundary layer. The most important outcome of their study was that the wall motion attenuates the intensity of streaky ``puff'' structures and widens their spanwise spacing. Fig. \ref{fig:katasonov-kozlov} shows a schematic of their experimental apparatus.

\cite{berlin-etal-1998} controlled oblique transition with spanwise forcing in the form of both wall oscillations and volume force given in the second row of Table \ref{table:body-forcing}. The latter forcing was more efficient in delaying transition, although care is needed since unsuitable parameters can cause earlier transition. Random disturbances were shown to be controlled efficiently as long as the amplitude of the wall motion was large enough. \cite{galionis-hall-2005b} studied the effects of the wall motion on the growth of the G\"{o}rtler vortices and showed that the growth rate is significantly reduced as the vortices move away from the wall. Later studies have consistently reported that spanwise wall oscillations lead to a decrease of perturbation energy, the differences being the origins and types of the disturbances existing in various pre-transitional shear layers. Through a nonlinear variational method, \cite{rabin-etal-2014} showed that laminar Couette flow can be efficiently stabilized by wall oscillations, while \cite{jovanovic-2008}, and more recently \cite{zhao-chen-li-2019}, computed the reduced linear growth of perturbations in laminar channel Poiseuille flows as a consequence of the spanwise forcing. 

Open transitional boundary layers subjected to free-stream vortical disturbances have also been stabilized by steady and time-periodic spanwise wall oscillations \citep{ricco-2011,hack-zaki-2012,hicks-ricco-2015,hicks-ricco-2015a,hack-zaki-2015}. The simulations by \cite{hack-zaki-2014} with uniform spanwise oscillations, \cite{negi-mishra-skote-2015} with steady sinusoidal spanwise-velocity waves, \cite{negi-etal-2019} with both temporal and spatial oscillations demonstrated a delayed onset of transition and a lengthened transition region. \cite{hack-zaki-2014} proved that, differently from the response of wall-bounded turbulence, there exists an optimal amplitude of oscillation for transition delay and that square-wave spanwise oscillations are more efficient than sinusoidal oscillations. The existence of an optimal frequency and amplitude was confirmed by \cite{negi-etal-2019}. Suppression of controlled Tollmien-Schlichting waves and a consequent transition delay of 55\% were measured in a low-speed water channel where an otherwise Blasius boundary layer was altered by surfaces mimicking biomimetic fish scales arranged in typical periodic overlapping arrangement \citep{muthuramalingam-etal-2019,muthuramalingam-etal-2020}. Although the near-wall spanwise flow was not measured, it was conjectured that the wall scales generated a streamwise repeating lateral flow which, combined with the observed streamwise-elongated streaky structures, led to a more stable flow.

\cite{wang-liu-2019} simulated a boundary layer altered by spanwise wall oscillations, in which the transition to turbulence was induced by Tollmien-Schlichting waves. Differently from all the other studies, they reported that the instability was energized rather than inhibited, and the transition location moved upstream. It remains to be verified whether this contrasting result is due to different transition scenarios, i.e., classical stability eigenmodes in the case of \cite{wang-liu-2019} or longitudinal streaks generated by free-stream disturbances in most of the other simulations. 

\subsection{Taylor-Couette flows}

A few studies have also focused on the effects of axial oscillations of the inner cylinder and of axially-oscillating pressure gradients in Taylor-Couette flows. By adopting the narrow-gap approximation and by assuming small-amplitude perturbations, \cite{hu-kelly-1995} showed that, at finite Reynolds numbers, stabilization can be achieved by oscillations of the pressure gradient and the inner cylinder, with the notable difference that the optimal conditions are obtained in the zero-frequency limit in the former case, while an optimal finite frequency is found when the inner cylinder oscillates. 
The problem was also studied experimentally by \cite{weisberg-etal-1997}, who found that the stabilization results were only in qualitative agreement with those of \cite{hu-kelly-1995}. 
\cite{marques-lopez-1997} resolved this controversy by extending \cite{hu-kelly-1995}'s theory to include the effects of curvature and of the end walls, thus elucidating that the reason for the disagreement was due to the end walls present in the experiments (termed ``enclosed-system''), while \cite{hu-kelly-1995}'s narrow-gap assumption was valid. \cite{marques-lopez-1997} obtained excellent agreement with \cite{weisberg-etal-1997}'s experimental results and with \cite{hu-kelly-1995}'s data when the latter ``open-system'' results were corrected by introducing an axial pressure gradient to model the presence of the side walls. \cite{marques-lopez-1997} also reported that the optimal finite frequency found by \cite{hu-kelly-1995} disappeared when the side walls were modelled.

The study of \cite{marques-lopez-1997} was extended to a visco-elastic fluid by \cite{ramanan-kumar-graham-1999}, who proved that the axial oscillations could also destabilize the flow in some cases, for example when the forcing frequency was close to the inverse of the relaxation time of the visco-elastic fluid, thus highlighting a resonance mechanism. Further resonance mechanisms in the Newtonian case were computed by \cite{marques-lopez-2000} for the case of Naimark-Sacker bifurcations when the forcing frequency matched the natural frequency of the system and the forcing amplitude was sufficiently large. Further results for the nonlinear regime were obtained via DNS by \cite{avila-etal-2007}, who showed that the high-frequency perturbations measured in the experiments of \cite{sinha-etal-2006} were not measurement noise, but were instead caused by a sequence of symmetry-breaking bifurcations that stemmed from the primary onset of instability.

\subsection{Spanwise-oriented body forces}  
\label{sec:bf}

There have been several studies on the effects of spanwise-oriented body forces on wall-bounded turbulent flows, aiming to emulate the drag-reducing action of the oscillating wall, while avoiding the motion of the surface. 

\subsubsection{Numerical studies}

{
\footnotesize
\begin{sidewaystable}
\begin{tabular}{|c|c|c|c|c|}
\hline
\centering
\textbf{Body force/velocity} & \textbf{Geometry} & \textbf{Reynolds number} & {\bf max} $\boldsymbol{\mathcal{R}(\%)}$ & {\bf Authors}\\
\hline
\makecell{
$f_z=-\alpha w$
}
& Channel flow & up to $Re_\tau=150$ & $20\%$ & \citet{satake-kasagi-1996} \\
\hline
\makecell{
$V=-A\beta\cos(\beta z) [1+\cos(\pi(y/h-1))]$ \\ $W=-A \pi \sin(\beta z) \sin(\pi(y/h-1))$
}
& Channel flow & up to $Re_\tau=150$ & $50\%$ & \citet{schoppa-hussain-1998} \\
\hline
\makecell{
$f_z=-\frac{\partial w}{\partial y}\mid_{y=0}ye^{-\Delta y}$
}
& Channel flow & up to $Re_\tau=110$ & $35\%$ & \citet{lee-kim-2002} \\
\hline
$f_z=\mbox{A}e^{-\Delta y} \sin(\omega t)$ & Channel flow & up to $Re_\tau=400$ & $40\%$ & \makecell{\citet{berger-etal-2000} \\ \citet{du-karniadakis-2000} \\ \cite{huang-etal-2012} \\ \citet{altintacs-davidson-2017}} \\
\hline
$f_z=A e^{-\Delta y} \sin(\kappa_x x)$ & Channel flow & up to $Re_\tau=400$ & $58\%$ & \makecell{\citet{berger-etal-2000} \\ \citet{wu-etal-2015} \\ \citet{han-etal-2017} \\ \citet{jiang-etal-2019}} \\
\hline
$f_z=A e^{-\Delta y} \sin(\kappa_x x - \omega t)$ & Channel flow & up to $Re_\tau=200$ & $47\%$ & \makecell{\citet{huang-fan-dong-2010} \\ \citet{quadrio-xie-2015}} \\
\hline
$f_z=A e^{-\Delta y} \sin(\kappa_z z - \omega t)$ & Channel flow & $Re_\tau=180$ & $30\%$ & \makecell{\citet{du-karniadakis-2000} \\ \citet{du-symeonidis-karniadakis-2002} \\ \citet{zhao-wu-luo-2004} \\ \citet{huang-etal-2011}} \\
\hline
$f_z=A e^{-\Delta y} \sin(\kappa_x x + \kappa_z z - \omega t)$ & Channel flow & $Re_\tau=180$ & $45\%$ & \citet{huang-etal-2014} \\
\hline
Navier-Stokes and Maxwell equations & Boundary layer & $Re_\theta=300$ ($Re_\tau=164$) & $20\%$ & \citet{lee-sung-2005} \\
\hline
$f_z=A x \exp(-\Delta x^2) \exp(-\lambda y)$ & Channel flow & $Re_\tau=200$ & $15\%$ & \citet{yang-etal-2015} \\
\hline
\makecell{
$f_y=A\beta\cos(\beta z) [1+\cos(\pi y/h)]$ \\ $f_z=(A\pi/h) \sin(\beta z) \sin(\pi y/h)$
}
& Channel flow & up to $Re_\tau=550$ & $17\%$ & \citet{canton-etal-2016}  \\
\hline
$f_z=A x \exp(- (x-\mu)^2/2\sigma^2) f(y) \sin(\omega t)$
& Channel flow & $Re_\tau=180$ & $33\%$ & \citet{mahfoze-laizet-2017}  \\
\hline
$f_z=Ay \exp(-\Delta y^2) \sin(b z)$
& Channel flow & up to $Re_\tau=550$ & $19\%$ & \citet{yao-etal-2017,yao-etal-2018}  \\
\hline
Navier-Stokes and Maxwell equations
& Cylinder & up to $Re_\tau=272$ & $43\%$ & \citet{zhao-etal-2019}  \\
\hline
\end{tabular}
\caption{Numerical studies on the effects of spanwise-oriented body forces. 
The spanwise-oriented body force is denoted by $f_z$, $V$ and $W$ indicate the imposed wall-normal and spanwise velocities, $A$ is the amplitude of the body force, and $\Delta$ is a parameter defining the exponential decay. The Reynolds numbers are defined in the caption of table \ref{table:experimental-studies}. The friction Reynolds number in parenthesis is estimated according to the formula $Re_\tau=1.118Re_\theta^{0.875}$ \citep{ricco-quadrio-2008}.}
\label{table:body-forcing}
\end{sidewaystable}
}

Table \ref{table:body-forcing} summarizes the numerical investigations according to the type of body force added to the Navier-Stokes equations.
The general picture of these numerical studies is that the drag-reduction margins are comparable with, and even exceed, those produced by oscillatory wall motion, although the power required to generate these body forces is substantially larger than the power saved through the drag reduction. It is further noted that a systematic study across the whole parameter space is still lacking, and that all the simulations are at relatively low Reynolds numbers, preventing any assessment on the applicability of these techniques at high Reynolds numbers.

A spanwise body force was implemented for the first time by \cite{satake-kasagi-1996} in a thin near-wall layer in a channel flow at $Re_{\tau}=150$. The force, proportional to the spanwise velocity, led to an attenuation of all the velocity components and to a maximum drag-reduction margin of $20\%$. 
\cite{schoppa-hussain-1998} produced colliding spanwise wall jets in their DNS of a channel flow at $Re_{\tau}=200$ by imposing a large-scale control and obtained 50\% drag reduction.
\citet{berger-etal-2000} numerically tested the flow response to temporal and spatial, sinusoidally-varying body forces that decay exponentially along the wall-normal direction, obtaining drag-reduction levels as large as 40\%, while \cite{lee-kim-2002} enforced a spanwise near-wall body force proportional to the spanwise wall-shear stress, obtaining a drag-reduction margin of $35\%$ in a channel flow at $Re_{\tau}=110$.

In \citet{berger-etal-2000} and in most of the studies listed in Table \ref{table:body-forcing}, the body force has been added to the spanwise momentum equation as a given function of space and/or time without taking into account the coupling with the equations describing the force generation. Notable exceptions are the studies of \cite{lee-sung-2005} and \citet{zhao-etal-2019}, who solved the Navier-Stokes equations coupled with the Maxwell equations. The former publication is also the only analysis of developing boundary layers altered by a spanwise-oriented body force, while the latter is the only study of a turbulent axial flow along the external surface of a cylinder. All the other studies have instead focused on channel flows. The spanwise body force travelling at an angle with respect to the streamwise direction, used by \citet{huang-etal-2014}, is particularly interesting, but its wall-motion analogue has never been tested. 
\cite{chagelishvili-etal-2014} introduced a weak near-wall volume forcing (optimal seed velocity perturbations) that led to the generation of a spanwise mean flow. The turbulence level and the production of turbulent kinetic energy was reduced when the control was implemented in a plane Couette flow using DNS.

The idealized concept of opposition control was implemented for the first time by \cite{choi-moin-kim-1994} along the wall-normal and spanwise directions. When the spanwise velocity was detected above the wall, at $y^+ \approx 10$, and an opposite spanwise velocity was imposed at the wall, \cite{choi-moin-kim-1994} obtained $\mathcal{R}=30\%$ in a channel flow at $Re_{\tau}=98$. The same technique was used by \cite{jozsa-etal-2019}, who achieved $\mathcal{R}=24\%$ at $Re_{\tau}=180$ and $\mathcal{R}=19\%$ at $Re_{\tau}=1000$. 

\subsubsection{Experimental studies}

\citet{wilkinson-2003} was the first to generate a near-wall spanwise oscillating body force experimentally by producing plasma jets. Although this study was not completely successful as a drag-reduction method because the jets could not be confined near the wall and the produced frequency was lower than the optimal one, it has paved the way for further investigations.

\cite{breuer-park-henoch-2004} and \citet{park-etal-2004} carried out the first experimental campaign of drag reduction by a spanwise oscillating Lorentz force implemented through spanwise-aligned rows of magnets and surface-mounted electrodes. The maximum drag-reduction margin was 10\%, thus lower than in the DNS of \citet{lee-sung-2005}, albeit the measured spanwise velocity profiles showed the expected oscillatory trend. Further experimental tests by \citet{pang-choi-aessopos-2004,pang-choi-2004} and \citet{xu-choi-2008} on spanwise Lorentz-force actuation, both temporally oscillating and streamwise travelling, led to drag-reduction margins as high as $40-45\%$, but also drag increase. K.-S. Choi's group \citep{jukes-etal-2004,jukes-etal-2006,jukes-etal-2006-b,choi-etal-2011,whalley-choi-2011,wang-etal-2013,whalley-choi-2014,choi-etal-2014-plasma} reported several results on the methodology and realization of drag-reduction techniques based on spanwise plasma forcing. The measured drag-reduction levels were as high as 45\%, and the reported optimum periods were as low as $T^+=15$, thus remarkably different from the oscillating-wall scenario. 

\citet{li-etal-2014-plasma}, \citet{wong-etal-2015}, \citet{wong-etal-2017} utilized wall-mounted electrodes arranged at different angles with respect to the streamwise direction to generate plasma actuation on a wind-tunnel turbulent boundary layer. The plasma forcing produced spanwise forcing in the form of streamwise vortices and a maximum drag reduction of 50\% was found.

Researchers led by T.S. Corke at the University of Notre Dame obtained turbulent drag reduction in a wind tunnel tunnel by pulsed-DC plasma actuation along the spanwise direction \citep{corke-thomas-aiaa-2018,thomas-etal-2019,duong-etal-soton-2019,corke-thomas-patent-2020}. At their lowest momentum-thickness Reynolds number, $Re_\theta=4538$, drag-reduction margins as high as 75\% were measured by a direct force balance. The plasma-induced spanwise velocity was induced by extremely short pulses at time intervals of $T^+=71$ that generated peak instantaneous values as large as $W_{max}^+=12$ that decayed exponentially fast, and average values of about $W^+=1$. The spacing between the electrodes was about 1000 wall units and, thanks to the spiky current, the power spent for the actuation, calculated as the time average of the product between the voltage and the current, was very low, i.e., not larger than 1 W/m. The low power spent led to the interesting result that a positive net power saved was attained. 

Recent experiments by \citet{hehner-etal-2019} and \citet{hehner-etal-2020} successfully produced a clean Stokes-layer-type boundary layer by means of near-wall plasma forcing, although the main difference from the theoretical profile given in formula \eqref{eq:stokes} is that the wall is stationary and thus the maximum spanwise velocity occurs away from the wall. 

\subsection{Indirect spanwise-forcing techniques}
\label{sec:indirect}

Wall-bounded turbulence has also been modified by wall-normal surface motion with the intention to induce perturbations into the turbulence self-generating cycle that result in a damping similar to that provoked by spanwise forcing, thereby reducing the wall friction. We herein only discuss studies on flows modified by wall-normal deformations that induce a spanwise motion indirectly. In most experimental and numerical studies, the maximum achievable drag-reduction margin does not exceed 15\%. Exceptions are the study by \cite{fernex-etal-2019} and \cite{albers-etal-2020}, who reported a maximum margin of 31\%, and the wind-tunnel study by \cite{bai-etal-2014,bai-etal-2018}, who measured a 50\% reduction.

Using DNS, \citet{mito-kasagi-1998} attempted to first favourably alter the near-wall turbulence-producing cycle by moving the wall in the form of a spanwise standing wave, i.e., $y_w=A\sin(\beta z) \sin(\omega t)$, but they only recorded an increase in drag. An increase of turbulence intensity via the same actuation was computed by \citet{wang-etal-2007} using LES. \citet{segawa-etal-2002} implemented a series of wall-based actuators that oscillated in the wall-normal direction to create a spanwise traveling wave beneath a turbulent channel water flow. Although the drag reduction was not measured, the streaky structures appeared more homogenized with respect to the canonical case. \citet{itoh-etal-2006} were the first to experimentally demonstrate that a flexible wall moving vertically by less than $y^+=20$ and generating a spanwise travelling wave could reduce the friction drag by up to 7.5\% at $Re_\theta=1000$, although the optimal conditions were not found. 

\citet{tamano-itoh-2012} examined a turbulent boundary layer in a wind tunnel where the flexible-wall motion was generated by a crank-slide mechanism and reported drag-reduction margins up to 13\%. The maximum drag reduction was measured for $T^+=110$, a period comparable to that of the oscillating-wall technique. \citet{roggenkamp-etal-2015,meysonnat-etal-2016,roggenkamp-etal-2019} and \citet{li-etal-2019b} measured drag-reduction levels that were lower than those obtained by \citet{tamano-itoh-2012} because of constraints related to the deformability of the aluminum plate used for creating the wavy motion, which caused a low deformation amplitude. A drag-reduction level of 5\% was measured in a wind tunnel by \citet{musgrave-tarazaga-2017,musgrave-etal-2019}.

\citet{itoh-etal-2006}'s wind-tunnel results were reproduced numerically by the group at RWTH Aachen University \citep{klumpp-etal-2010,klumpp-etal-2011,meysonnat-etal-2012,meysonnat-etal-2013,meysonnat-etal-2016c,albers-etal-2019,fernex-etal-2019,albers-etal-2020}. They computed drag-reduction levels up to 31\% at low Reynolds numbers by moving the wall as $y_w=A \sin(\beta z - \omega t)$. \citet{koh-etal-2015,koh-etal-2015b} however reported a drastic drop in drag reduction at higher Reynolds numbers, i.e., only 1\% at $Re_\tau=2250$. The results of the RWTH group were reproduced via DNS by \citet{tomiyama-fukagata-2013}, who also computed a net energy saving up to 12\% by accounting for the power spent to generate the wavy motion.

The wind-tunnel study by \cite{bai-etal-2014,bai-etal-2018} is the only one that has reported drag-reduction margins up to 50\% by spanwise-traveling waves of wall-normal displacements, thus significantly exceeding the reductions obtained in all the other similar experimental and numerical studies. The wall-shear-stress reduction was calculated by direct measurements of the streamwise mean velocity in the viscous sublayer, conducted via hot-wire anemometry. The wall location was estimated by moving the hot-wire probe towards the wall, down to a location where the measurements were affected by the wall cooling. The actuation was only confined to a strip of limited streamwise extent, about 140 wall units, a distance along which spanwise-wall oscillations can instead only produce drag-reduction margins of a few percent because the flow requires a much longer distance to fully adapt to the wall actuation \citep{ricco-wu-2004-a}. Furthermore, in \cite{bai-etal-2014} and \cite{bai-etal-2018}'s experiments no smooth transition occurs between the short actuated surface and the non-actuated portion of the wall. This abrupt change may have caused local flow separation, thereby affecting the drag-reduction measurements.

\subsection{Spanwise-oriented jets}

A number of experimental studies have focused on forcing the wall turbulence along the spanwise direction by jets located at the bottom or the side walls, mimicking the actuation of the spanwise wall motion. Significant drag-reduction margins have been reported, but these measurements were only at locations close to the jet apertures and no study exists on distributed jet actuation that has achieved drag reduction over a large surface area.

\citet{iuso-etal-2002} and \citet{iuso-etal-2007} experimentally investigated the response of a channel flow in a wind tunnel to spatially discrete synthetic jets located at a fixed downstream position and distributed along the span. The jets were alternatively inclined at $\pm45^\circ$ with respect to the wall-normal direction. Drag-reduction margins as high as 30\% were measured, presumably as a result of a combination of the effects of the induced large-scale streamwise vortices, similar to the numerical study of \citet{schoppa-hussain-1998}, and the local flow separation close to the location of the jets. A variant of \citet{iuso-etal-2002}'s problem was studied in a wind tunnel by \cite{tay-etal-2007}, wherein the holes, instead of being alternately inclined, were instead all directed at the same angle, thereby resulting in a net spanwise flow. The inclined jets generated large-scale streamwise vortices that persisted downstream. The drag-reduction margin measured below the core of the streamwise vortices was 50\%, although it was unclear what portion of this percentage was caused by the spanwise forcing and what portion was due to the local separation. In a related study, \citet{cannata-iuso-2008-a,cannata-iuso-2008-b} positioned the synthetic-jet holes along the upper half of the side walls of the duct to enforce the spanwise blowing and obtained a maximum drag reduction of 22\%. 

\citet{segawa-etal-2005} tested synthetic jets in a water channel to emulate the spanwise forcing of an oscillating wall indirectly. The jets were distributed along the span and oriented in the wall-normal direction, so that their out-of-phase oscillation caused the wall turbulence to be altered along the spanwise direction, similarly to the spanwise-travelling waves of wall-normal wall displacement discussed in Section \ref{sec:indirect}. The drag-reduction levels were as high as 30\% and, differently from the wall-motion case, were detected even for large and low oscillating periods.

To the best of the writers' knowledge, there are no numerical studies on the flow modifications caused by spanwise oriented jets discharging from wall apertures.

\subsection{Rotating actuators}

\begin{figure}
\psfrag{D}{$D$}
\psfrag{W}{\hspace{0 mm} $W$}
\centering
\includegraphics[width=0.49\columnwidth]{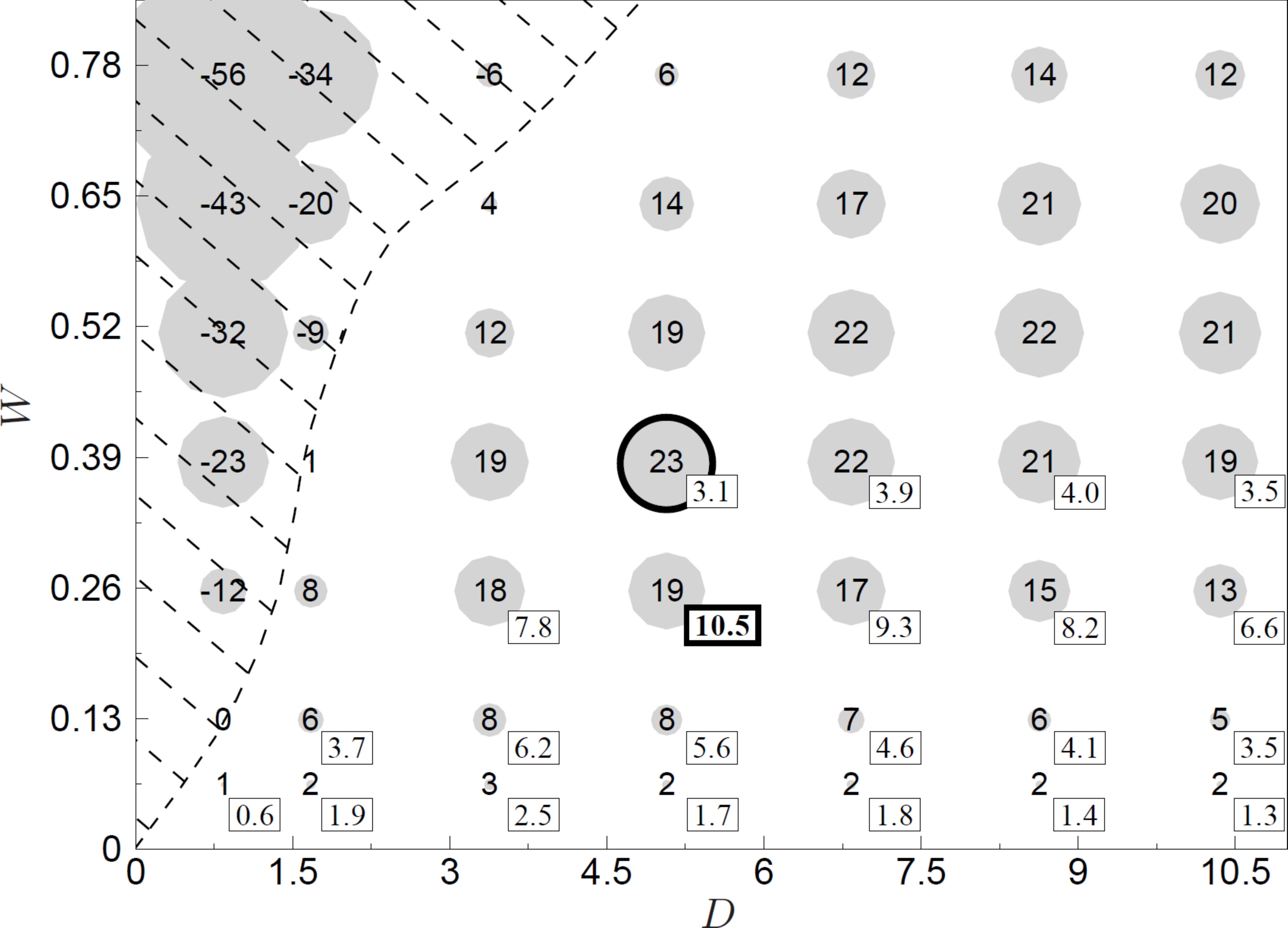}
\includegraphics[width=0.49\columnwidth]{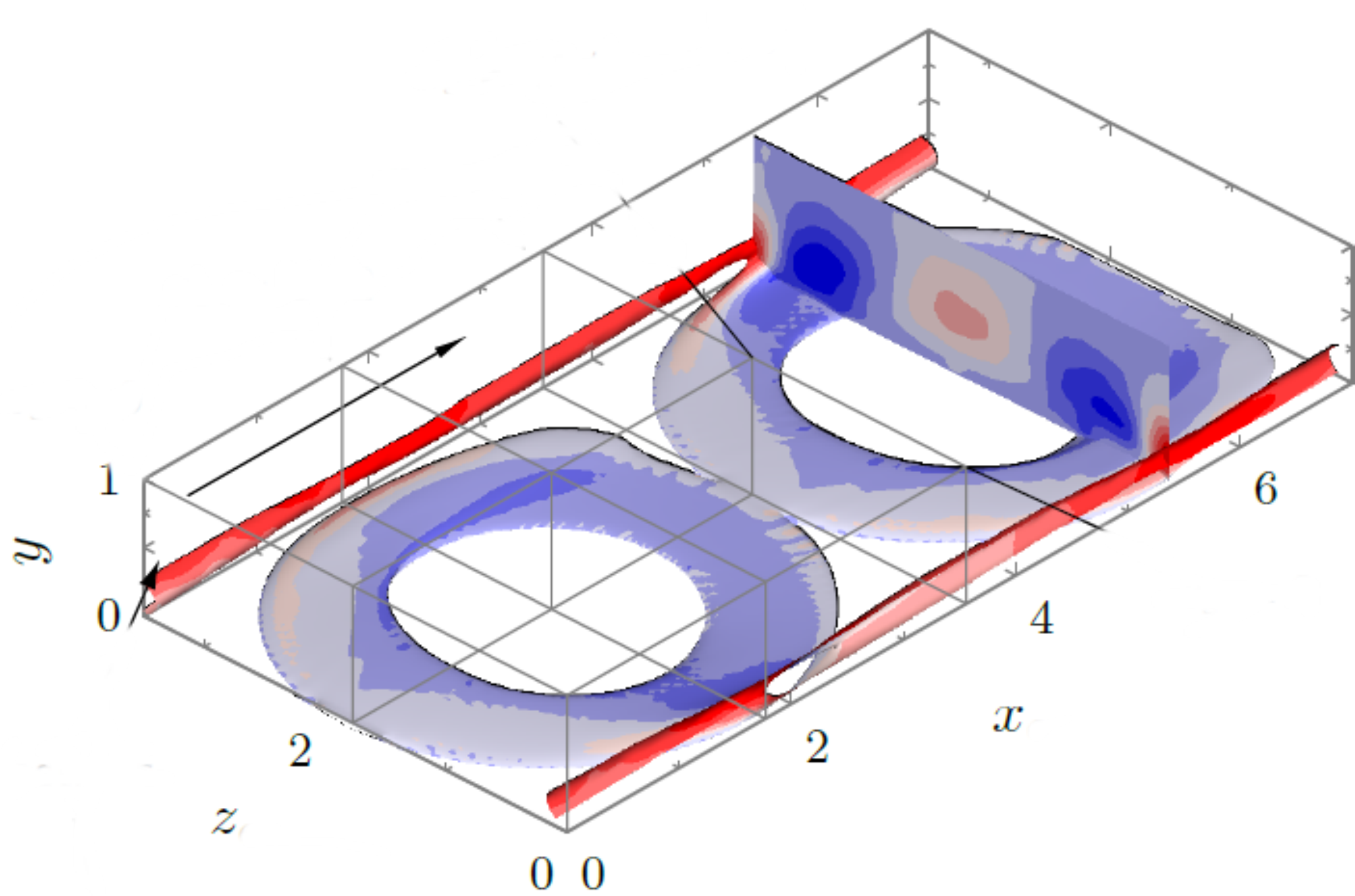}
\caption{Turbulent drag reduction through rotating discs. Left: map of the drag-reduction level given by rotating discs as a function of the disc diameter $D$ and the tip disc velocity $W$ \citep{ricco-hahn-2013}. The size of the circles is proportional to the absolute value of drag reduction. The shaded areas highlight the drag-increase cases and the boxed values report the positive net power saved, computed via \eqref{eq:power-net}; right: isosurfaces of fixed amplitude of the time-averaged flow produced by spinning rings \citep{olivucci-ricco-aghdam-2019}. The colours indicate the intensity of the wall-normal ring-flow velocity (blue is negative and red is positive). Taken with permission from CUP (left) and APS (right).}
\label{fig:rings}
\end{figure}

A question that is frequently raised in relation to drag-reduction methods involving spanwise-wall motion is how the actuation could be imposed in a practical scenario. Although several experimental studies have demonstrated that there are possible answers to this question, it is undoubtedly true that this type of actuation presents major engineering challenges, related to the small spatial and temporal scales and the curvature of the surfaces, among a number of factors. Against this background, several studies have been undertaken to examine the efficacy of using rotating discs embedded in the wall. While the fluid motions created by the discs are different from those induced by a spanwise-moving wall, the discs impose transverse motions within rotational near-wall shear layers, and these alter the baseline flow by interacting with the near-wall-turbulence dynamics. Numerical calculations have shown that drag-reduction margins as large as 25\% can be achieved, although no experimental evidence exists to confirm these results.

A flush-mounted rotating disc was first used by \cite{hill-klewicki-1996} in a wind tunnel to perturb a laminar boundary layer by producing high near-wall velocity gradients. The objective of their study was to test their measuring technique and not to control the boundary layer. More detailed measurements of this experimental campaign were published in  \cite{klewicki-hill-2003}, where the main focus was on the vortical structures detaching from the disc edges.

\citet{keefe-1997,keefe-1998} first proposed the idea of utilizing flush-mounted rotating disc actuators to achieve turbulent drag reduction. Keefe was inspired by the pioneering spanwise-oscillating work of \citet{jung-mangiavacchi-akhavan-1992} and conjectured that spinning discs could be a viable method for prescribing a near-wall three-dimensional field by injecting wall-normal vorticity, although he did not report experimental or numerical results.

\citet{ricco-hahn-2013} carried out DNS of a turbulent channel flow at $Re_\tau=180$ altered by steadily rotating discs located over the two opposite walls. As shown by the map in Fig. \ref{fig:rings} (left), the parametric study revealed that an optimal diameter and an optimal angular velocity for a maximum drag-reduction level of 23\% exist, although smaller diameters and larger tip velocities produced drag-increase levels up to 56\%. The power balance that accounted for the energy spent against the fluid frictional resistance showed that positive net energy savings as large as 10\% could be achieved. \citet{ricco-hahn-2013}'s simulations revealed that a thin spinning boundary layer, analogous to the von K\'arm\'an laminar shear layer over an infinite spinning plate, forms over the core of the discs, while persistent streamwise-elongated structures appear at the sides of the discs, generating Reynolds stresses that contribute negatively to the drag reduction \citep{wise-olivucci-ricco-2018}.

\citet{wise-ricco-2014} extended \citet{ricco-hahn-2013}'s research by exploring the influence of sinusoidal oscillation of the discs, thereby demonstrating that low-frequency actuation leads to comparable drag-reduction levels. The spatial arrangements of the discs was shown by \cite{wise-alvarenga-ricco-2014} to have a remarkable influence on the friction reduction. The removal of half of the discs led to an increase in the drag-reduction margin because regions of stationary wall between discs were subjected to a beneficial radial flow induced by the wallward motion of outer fluid toward the rotating disc. This result demonstrated that the spanwise shearing action of the near-wall disc flow is only partially responsible for the drag-reduction mechanism and therefore that the disc-flow physics is considerably more complex than that involved in flows altered by purely spanwise oscillations or streamwise-travelling waves of spanwise velocity, which do not entail any radial flow or wallward large-scale fluid motion. \cite{wise-alvarenga-ricco-2014} further reported that half-disc and annular-spinning actuators guarantee improved performance because the detrimental downstream radial flow and the disc-centered fluid motions were partially suppressed. The spectral analysis of the disc velocity highlighted that the large scales were the primary cause of drag reduction.

The spinning action of rings was further explored by \cite{olivucci-ricco-aghdam-2019} by strengthening the drag-reduction effect through wall-transpiration and wall-hydrophobicity imposed over the stationary wall that was not occupied by the rings. The unfavourable impact on the reduction of wall-shear stress of the elongated structures occurring between spanwise-adjacent rings, depicted in Fig. \ref{fig:rings} (right), was quantified by the localized increase of the Reynolds stresses in a three-dimensional version of the FIK identity \citep{fukagata-iwamoto-kasagi-2002}. The inner-ring regions featured quasi-uniform and intense drag reduction, despite the surface being stationary.

\begin{figure}
\centering
\includegraphics[width=0.49\columnwidth]{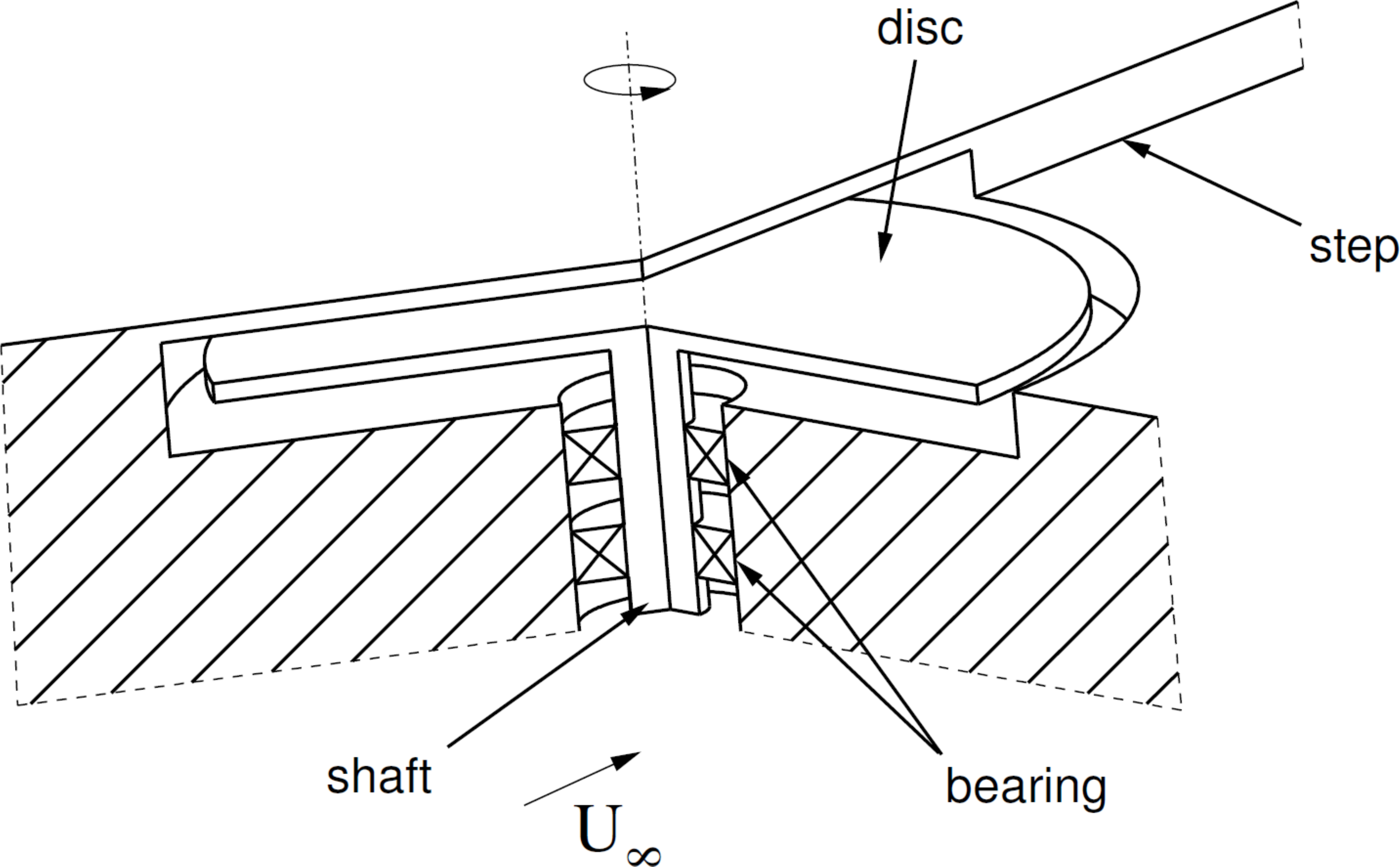}
\caption{Schematic of the passive rotating-disc apparatus used by \citet{koch-kozulovic-2013,koch-kozulovic-2014}. Taken from \cite{koch-kozulovic-2013} with permission from American Society of Mechanical Engineers.}
\label{fig:koch-kozulovic}
\end{figure}

Independently from Ricco's group, \citet{koch-kozulovic-2013,koch-kozulovic-2014} experimentally studied the drag-reduction properties of discs that were free to rotate under the action of a turbulent boundary layer. The maximum tip velocity reached a quarter of the free-stream velocity, although the discs were not flush mounted because part of the discs was covered by the flat surface, as shown in Fig. \ref{fig:koch-kozulovic}. 
\citet{koch-kozulovic-2013} did not measure the wall-shear stress over the disc surface, but were able to estimate it by assuming that the boundary-layer thickness did not vary when the disc moved. They used the difference in velocity between the free-stream flow and the disc surface in a correlation for the skin-friction coefficient to 
predict a maximum drag-reduction margin of 17\%.

\subsection{Passive techniques}

Passive drag-reduction techniques, i.e., those that do not require an external energy input, have been inspired by the more widely studied active methods based on spanwise forcing, or by naturally-occurring surfaces, such as bird feathers and shark skins. These methods involve modifications to the surface geometry, and they can be grouped as wavy walls, dimples, and spanwise-oriented riblets. 

\subsubsection{Oblique wavy walls} 
\label{sec:oblique-waves}

\begin{figure}
\centering
\includegraphics[width=0.59\columnwidth]{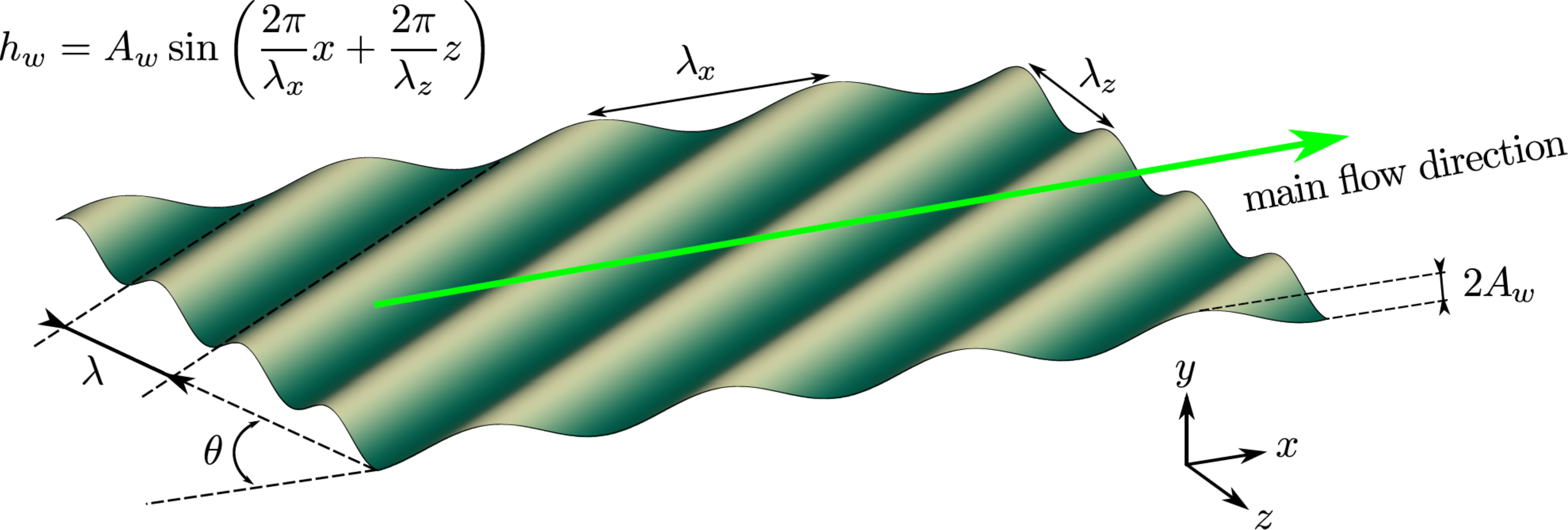}
\includegraphics[width=0.39\columnwidth]{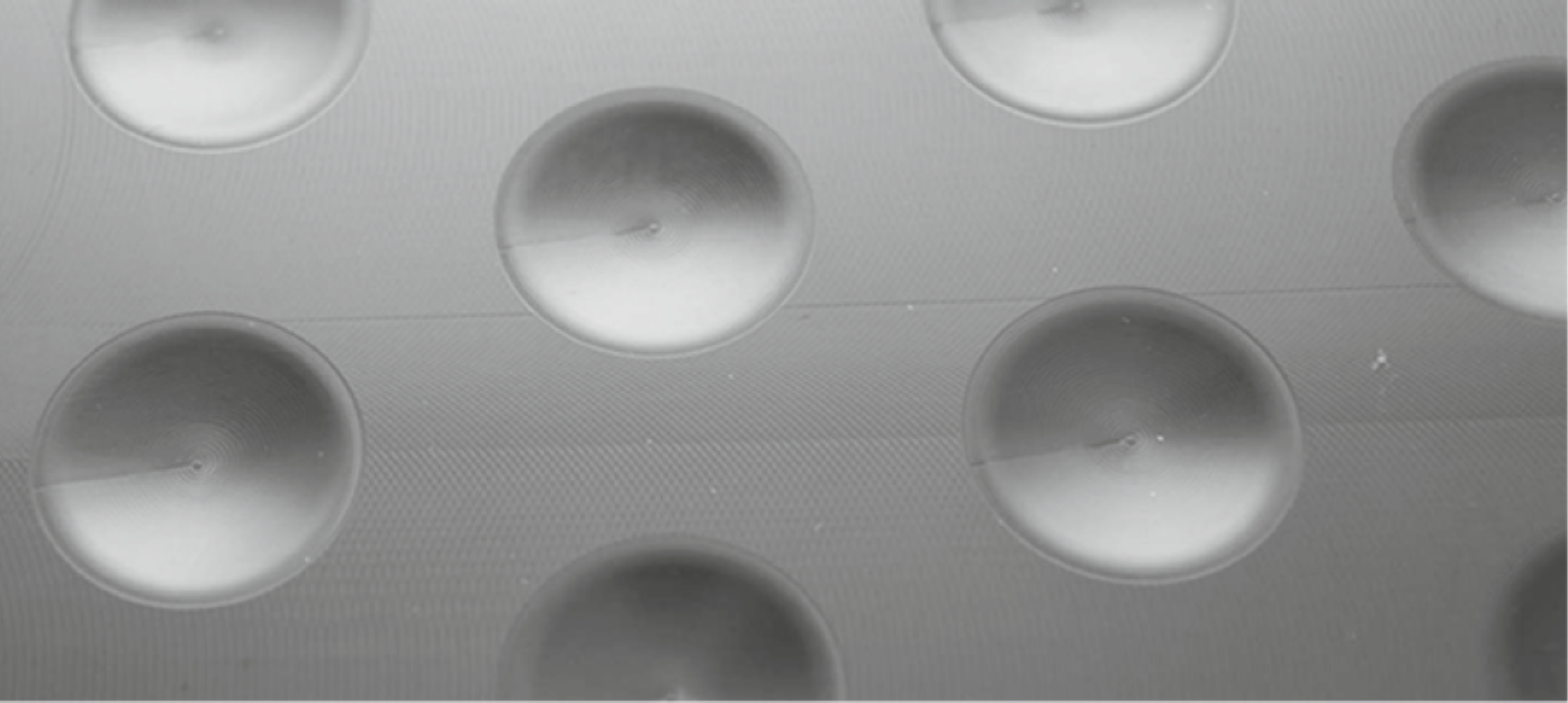}
\caption{Examples of passive drag-reduction methods. Left: oblique wavy wall used by \citet{ghebali-etal-2017,ghebali-etal-2017b,ghebali-etal-2018,ghebali-etal-2019} to emulate the effect of standing spanwise-forcing waves; right: dimpled surface employed by \citet{vannesselrooij-etal-2016} in wind-tunnel experiments. Taken with permission from AIP Publishing (left) and from Springer (right).}
\label{fig:ghebali-dimples}
\end{figure}

\citet{lekoudis-sengupta-1986} and \citet{chernyshenko-2013} proposed to reduce the turbulent skin-friction drag by stationary wavy walls with sinusoidal form along the streamwise and spanwise directions, shown in Fig. \ref{fig:ghebali-dimples} (left).
\citet{lekoudis-sengupta-1986}'s modelling approach led to skin-friction drag reduction, although this benefit was countered by an increase of pressure drag and no overall drag reduction was computed. \citet{chernyshenko-2013}'s idea was to mimic, via the wall undulations, the effective spanwise-forcing action of the steady Stokes layer studied by \citep{viotti-quadrio-luchini-2009}, thereby generating an alternating wall-normal shear of spanwise velocity without the undesirable energy expenditure due to the surface motion.
His laminar analysis predicted a drag-reduction margin of $2.4\%$, which took into account the decrease of wall friction and the increase of pressure drag. These surfaces were tested by \citet{denison-etal-2015} experimentally in a wind tunnel and numerically through DNS, but the changes of drag were within the uncertainty range and therefore no definite conclusions could be drawn on the effectiveness of this technique.
\citet{ghebali-etal-2017,ghebali-etal-2017b,ghebali-etal-2018,ghebali-etal-2019} first used DNS to study the drag-reducing properties of a turbulent channel flow at $Re_\tau=375$ bounded by the wavy walls proposed by \citet{chernyshenko-2013}. The objective of wall-friction reduction was achieved in principle, as the friction drag was reduced by $3-4\%$. The pressure-drag penalty due to the leeward flow undulations however almost neutralized the friction-reduction effect, leading to an overall maximum drag-reduction margin of 0.6\%. This work thus led, for the first time, to a reduction of the overall drag by these wavy surfaces.

This method is more attractive than classical riblets in terms of practical implementation and durability, because the scales of the sinusoidal wall modulations are two order of magnitude larger than those of riblets, i.e., about 1000 wall units (comparable with the wavelength of the steady Stokes layer in \cite{viotti-quadrio-luchini-2009} and the rotating-disc diameter in \cite{ricco-hahn-2013}). Another advantage is the smoothness of the surface form, in contrast to the sharp crests typical of efficient riblets.

\subsubsection{Dimples} 

As in the case of oblique wavy walls reviewed in Section \ref{sec:oblique-waves}, dimples are intended to provoke, passively, spanwise flow motions and thus reduce drag. These dimples are shallow, either circular, tear-drop, or diamond-shaped recesses over flat surfaces. Fig. \ref{fig:ghebali-dimples} (right) depicts the shallow dimples used by \citet{vannesselrooij-etal-2016} in a wind tunnel. The original aim of dimples was, however, to enhance the heat transfer while inducing a small drag penalty \citep{chyu-etal-1997,lin-etal-1999,turnow-etal-2012}. In this section, only the publications on flows over dimpled surfaces that aim at drag reduction are discussed. The general consensus is that drag-reduction margins of at least 5\% can be achieved by dimples with characteristic size comparable with the largest scale of the wall turbulence, although the optimal shape and spatial arrangement have not been found and the effect of Reynolds number is unknown.

The drag-reduction properties of dimples was first suggested by researchers at the Kurchatov Institute of Atomic Energy in Russia \citep{kiknadze-etal-1984}. They later also advanced hypotheses on the physical mechanism behind these surface modifications, such as the creation of streamwise ``tornado-like'' vortical structures, and explored practical applications, such as large dimples over the surface of a high speed train \citep{kiknadze-etal-2009,kiknadze-etal-2012} (refer also to \cite{blood-1995} for further possible applications on other transport vehicles). The drag-reduction margins reported by this group are the largest ever reported for dimples, namely 20\% by \citet{alekseev-etal-1998}, 17\% by \citet{kiknadze-etal-2006}, and 15\% by \cite{kiknadze-etal-2009}, although insufficient details on the methodology and the uncertainty of the overall measured drag-reduction levels were given in order to assess the validity and reproducibility of these experimental campaigns. \citet{lienhart-etal-2008} carried out experiments and numerical simulations over dimples and negated the conclusions by the previous drag-reduction findings, as they discovered that the induced pressure drag over the dimples exceeded the skin-friction drag reduction. 

Further experimental and computational studies on dimpled surfaces were performed by Veldhuis's group at Delft University of Technology. They first showed that overall drag-reduction margins up to 15\% were possible \citep{veldhuis-vervoort-2009}, although these large values were not confirmed by the numerical computations. Later investigations by the same group suggested that drag-reduction levels up to 4\% could be obtained \citep{veldhuis-vervoort-2009,vancampenhout-etal-2016,vannesselrooij-etal-2016,vancampenhout-etal-2018}, concluding that the absence of drag reduction reported by \citet{lienhart-etal-2008} could have been due to the low Reynolds number of the flow. The Delft University results were confirmed and extended by Tay's group at the National University of Singapore. They reported drag-reduction margins of 3\% for round axial-symmetrical dimples \citep{tay-2011,tay-etal-2015,tay-etal-2017}, 3.5\% for round dimples where the lowest point of the non-axial-symmetric indentation was moved further downstream \citep{tay-etal-2016,tay-lim-2017}, 6\% for teardrop-shaped dimples \citep{tay-etal-2018}, and 7.5\% for diamond-shaped and elliptical dimples \citep{tay-etal-2019}. The recent experimental results on round dimples by \cite{gowree-etal-2019} confirmed the drag-reduction results by \cite{tay-2011}, i.e., around 3\%.

It is now recognized that the overall drag modification generated by shallow staggered dimples is caused by the combination of friction drag reduction due to the localized deceleration and separation, the alternate spanwise velocity, and the increased pressure drag produced by the high pressure on the downstream part of the dimple and the low pressure in separated region \citep{lienhart-etal-2008}. The drag downstream of a dimple is also strongly affected by primary, often unsteady, counterrotating vortices formed in the centre of the dimple and by secondary vortices at the dimple sides \citep{won-zhang-ligrani-2005}. Several flow patterns, summarized by \citet{vannesselrooij-etal-2016}, can contribute to the increase of the friction drag and/or the pressure drag, thereby determining whether an overall drag reduction can be achieved. The staggered dimple arrangement has been proved to be decisive for producing the alternate spanwise near-wall pattern, called ``zig-zag'' by \citet{lin-etal-1999} and ``meandering'' by \citet{vancampenhout-etal-2016} and \citet{vannesselrooij-etal-2016}, a feature bearing similarities with the flow modifications generated by the wavy wall studied by \citet{ghebali-etal-2017} and the steady and traveling waves investigated by \citet{viotti-quadrio-luchini-2009} and \citet{quadrio-ricco-viotti-2009}. 

\citet{tay-etal-2015} also observed the spanwise-forcing effect of the dimples and measured a shift of the energy spectrum to lower frequency. They argued that the reduction in skin friction is caused by this transfer of energy from the small to the large scales, which disrupts the canonical energy cascade of the turbulent energy to dissipation. 

The importance of spanwise motions for the efficiency of the dimples was further confirmed by \citet{vannesselrooij-etal-2016}, who measured drag increase for aligned dimples and drag reduction for staggered dimples. They also reported a loss of drag reduction with increase coverage, a result that chimes with the detrimental disc-side interactions that increase the Reynolds stresses, discovered by \citet{wise-alvarenga-ricco-2014} and \citet{olivucci-ricco-aghdam-2019}. \citet{vancampenhout-etal-2018} predicted the drag reduction by dimples using the drag-reduction map by \citet{quadrio-ricco-2004} for oscillatory wall forcing, highlighting the importance of a Stokes-layer-like effect on the near-wall turbulence. They hence rejected the hypothesis that the drag reduction was created by the ``vortical bearing'' concept \citep{kiknadze-etal-2006} or by the upstream friction caused by flow reversal \citep{isaev-etal-2000}. 
The absence of drag reduction reported by \citet{lienhart-etal-2008} was thought to be due to a dimpled-area coverage ratio that was higher than that used by \citet{tay-etal-2015} and by the low Reynolds number in the study of \citet{vannesselrooij-etal-2016}. Interestingly, the latter studies both showed that the drag-reduction margin increased with the Reynolds number.

\subsubsection{Riblets with spanwise orientation}

Riblets are thin, streamwise-elongated surface protrusions and indentations that have been shown to reduce the turbulent skin-friction drag up to about $8-10\%$ \citep{walsh-1982,walsh-1983}. Various riblet cross sections have been tested, with the sharp-crested geometry being optimal. Crucially, the riblet dimensions are very small -- their height and separation distance being of the order of the thickness of the viscous sublayer, i.e., $y^+\approx 15$. Inspired by the success of the wall-oscillating technique, several studies have explored modifications of the canonical streamwise-aligned geometry to induce spanwise distortions to the near-wall flow, with the intent of enhancing the drag reduction in spite of the possibility of undesirable pressure-drag effects. With the notable exception of the experimental study of \citet{chen-etal-2014}, who reported a drag-reduction level of 21\% by bird-feather-like inclined riblets, researchers concur that riblets that vary sinusoidally along the streamwise direction may only offer a small percent improvement of drag reduction when compared with straight riblets.
\\

\noindent
{\em Inclined riblets}

A near-wall spanwise velocity is readily generated by orientating straight riblets with uniform tip height at a yaw angle with respect to the streamwise direction, as described by \citet{koeltzsch-dinkelacker-grundmann-2002} and \citet{chen-etal-2014}. 
The consensus is that the drag reduction achieved with streamwise-oriented riblets remains unchanged for yaw angles up to 10$^\circ-15^\circ$, it is close to zero for yaw angles in the range 20$^\circ-25^\circ$, and it reverses to drag increase at higher yaw angles. 
More elaborate variants have led to large drag reductions that have not yet found confirmation, such as the drag-reduction levels of 10\% by inclined bird-feather riblets \citep{feng-etal-2015}, 19\% by riblets inspired by the shortfin ``mako'' shark skin \citep{martin-bhushan-2016}, and 7\% by the inclined riblets studied by \cite{gu-etal-2017}. A potentially effective variant was proposed by \citet{zheng-yan-2010}, whereby the inclined riblets are covered by a compliant elastic material that morphs due to the wall-shear stress. A low-drag laminar flow thus experiences a smooth surface, while the surface deforms gradually when transition occurs to induce the desired wall-shear stress reduction.
\\

\noindent
{\em Converging-diverging riblets with uniform tip height}

The pipe-flow experiments by \citet{koeltzsch-dinkelacker-grundmann-2002} revealed unexpected large-scale flow modification by thin convergent-divergent ``herringbone'' riblets. Half of the inner pipe surface was covered by riblets inclined at a 45$^\circ$ yaw angle, while riblets over the other half were oriented at a $-45^\circ$ yaw angle. Over the divergent riblet patterns, the time-averaged streamwise velocity increased, while the streamwise velocity fluctuations decreased. The opposite effects were observed over the convergent riblet patterns, and no overall drag reduction was reported.

\begin{figure}
\centering
\includegraphics[width=0.54\columnwidth]{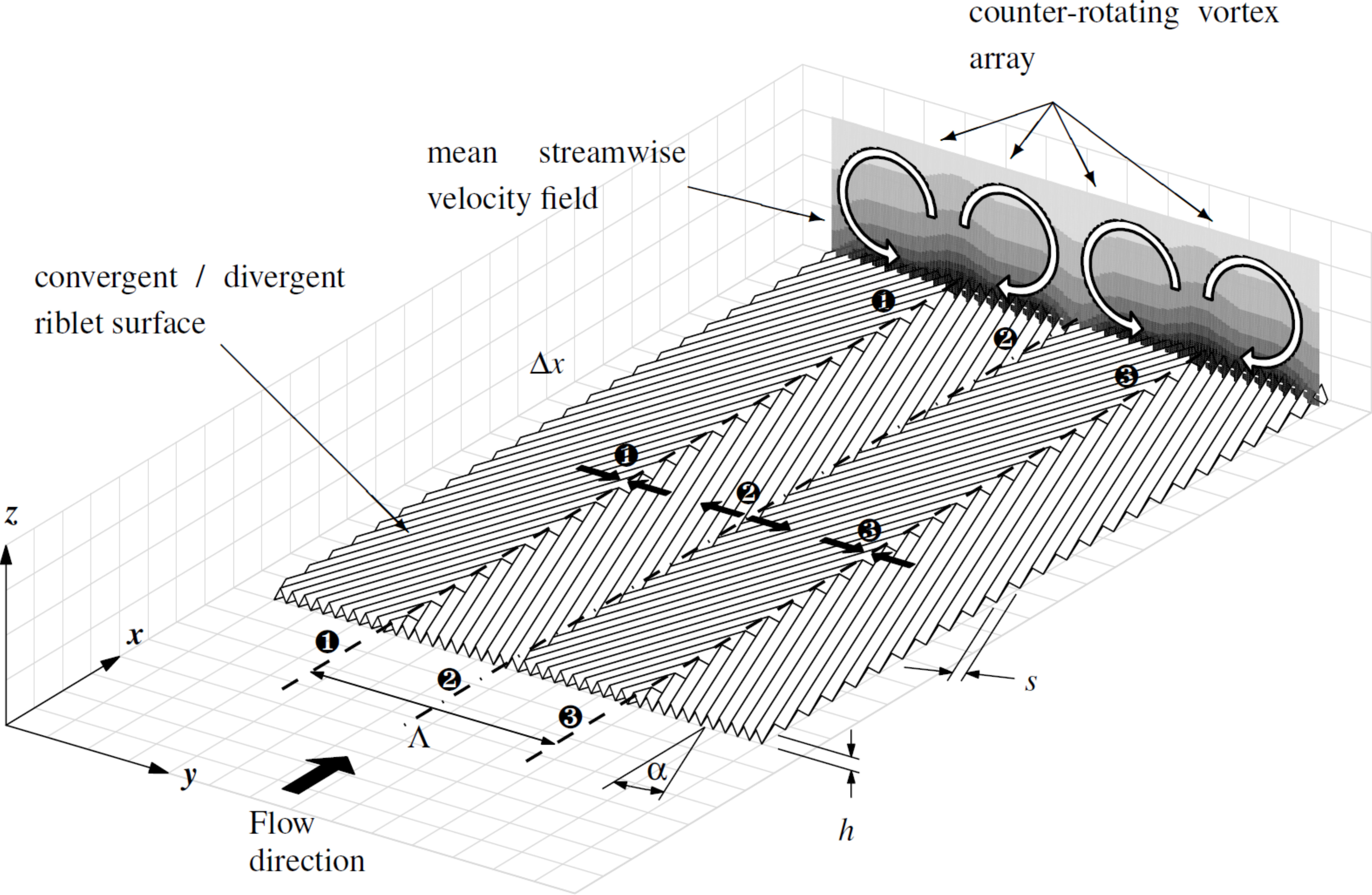}
\includegraphics[width=0.45\columnwidth]{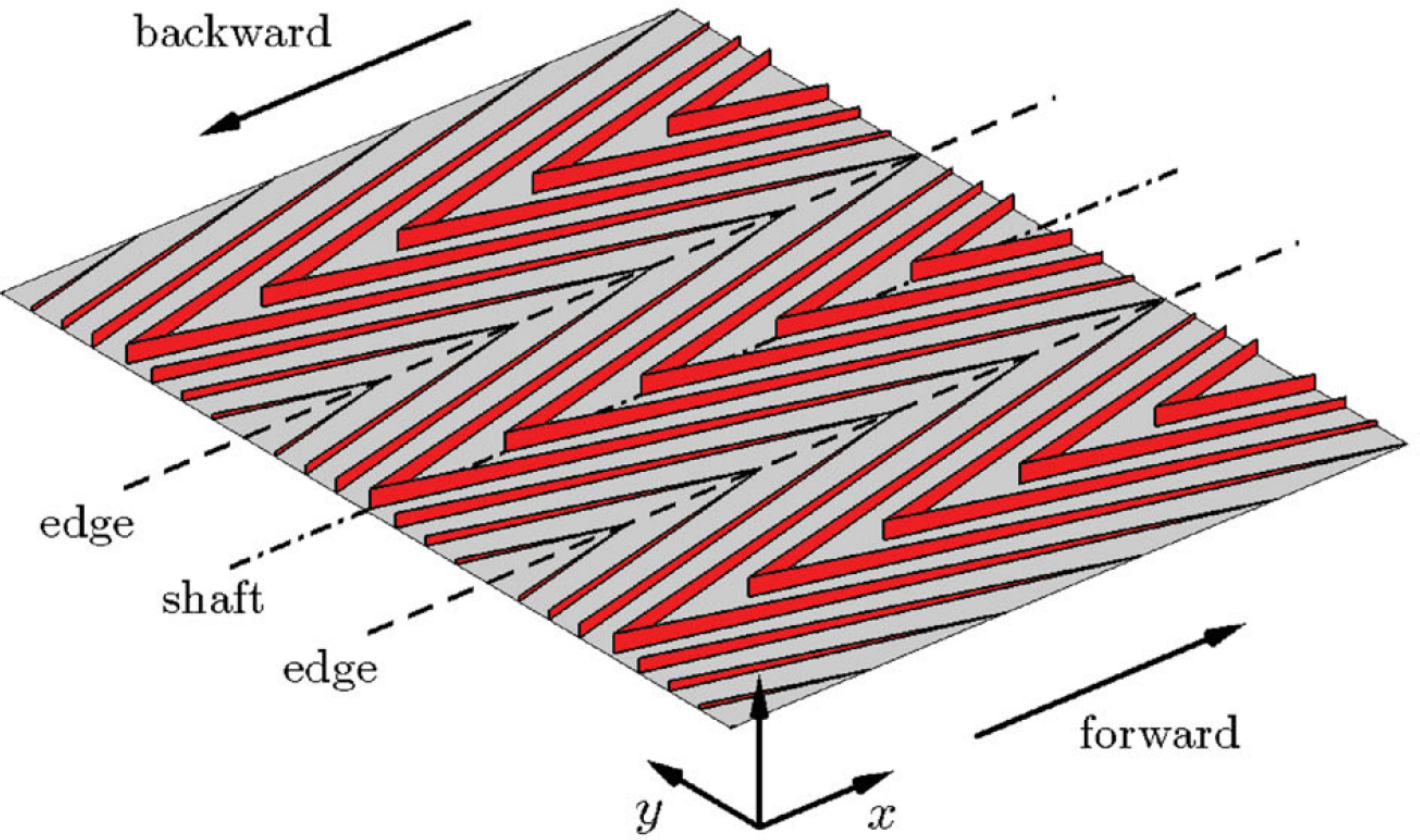}
\caption{Examples of herringbone riblets. Left: triangular converging-diverging riblets with uniform tip height, employed in the experiments of \citet{nugroho-etal-2010}; right: sharp-crested converging-diverging riblets with large-scale spanwise modulation, used by \citet{benschop-breugem-2017} to study the bird-feather-like surface tested experimentally by \citet{chen-etal-2014}. Taken with permission from the authors (left) and from Taylor \& Francis (right).}
\label{fig:riblets-conv-div}
\end{figure}

The large-scale alterations of the flow and the absence of drag reduction have been confirmed by experimental campaigns conducted by three groups. Nugroho, Kevin and co-workers \citep{nugroho-etal-2010,nugroho-etal-2012,nugroho-etal-2013,nugroho-etal-2014,kevin-etal-2014,kevin-etal-2017,bai-etal-2018-b,harun-etal-2019} proved that herringbone riblets at yaw angles in the 10$^\circ-30^\circ$ range, shown in Fig. \ref{fig:riblets-conv-div} (left), provoke very large streamwise vortices with a pronounced spanwise modulation, thereby strongly affecting the streamwise turbulence intensity. They found that the drag reduces over the diverging regions and increases over the converging ones, while the overall effect was a slight increase in drag. The vortices on herringbone riblets were also visualized by \citet{nadesan-etal-2014}, who confirmed the absence of drag reduction, and were investigated further by Zhong's group at the University of Manchester \citep{xu-etal-2018,xu-etal-2019}. 
\\

\noindent
{\em Converging-diverging riblets with varying tip height}

Tests in a water-pipe flow by \citet{chen-etal-2014} led to the remarkable result of a maximum drag-reduction margin of 21\% over bird-feather-like herringbone triangular riblets that featured large-scale undulations of the height along the spanwise direction. \citet{benschop-etal-2015} and \citet{benschop-breugem-2017} carried out DNS of similar sharp-crested converging-divergent riblets, shown in Fig. \ref{fig:riblets-conv-div} (right), and found only a maximum drag-reduction level of 2\% or even drag increase. They suggested that possible reasons for this disagreement were that the Reynolds number in \citet{chen-etal-2014}'s experiments was much larger, the spanwise size of the converging-diverging pattern was much larger in their calculations, and that the shape and yaw angles of the riblets were different in the experiments and the simulations. At present, there is no consensus on whether the experimental results of \citet{chen-etal-2014} are correct or reflect an error in the experimental procedure.
\\

\noindent
{\em Sinusoidal riblets}

\begin{figure}
\centering
\includegraphics[width=0.49\columnwidth]{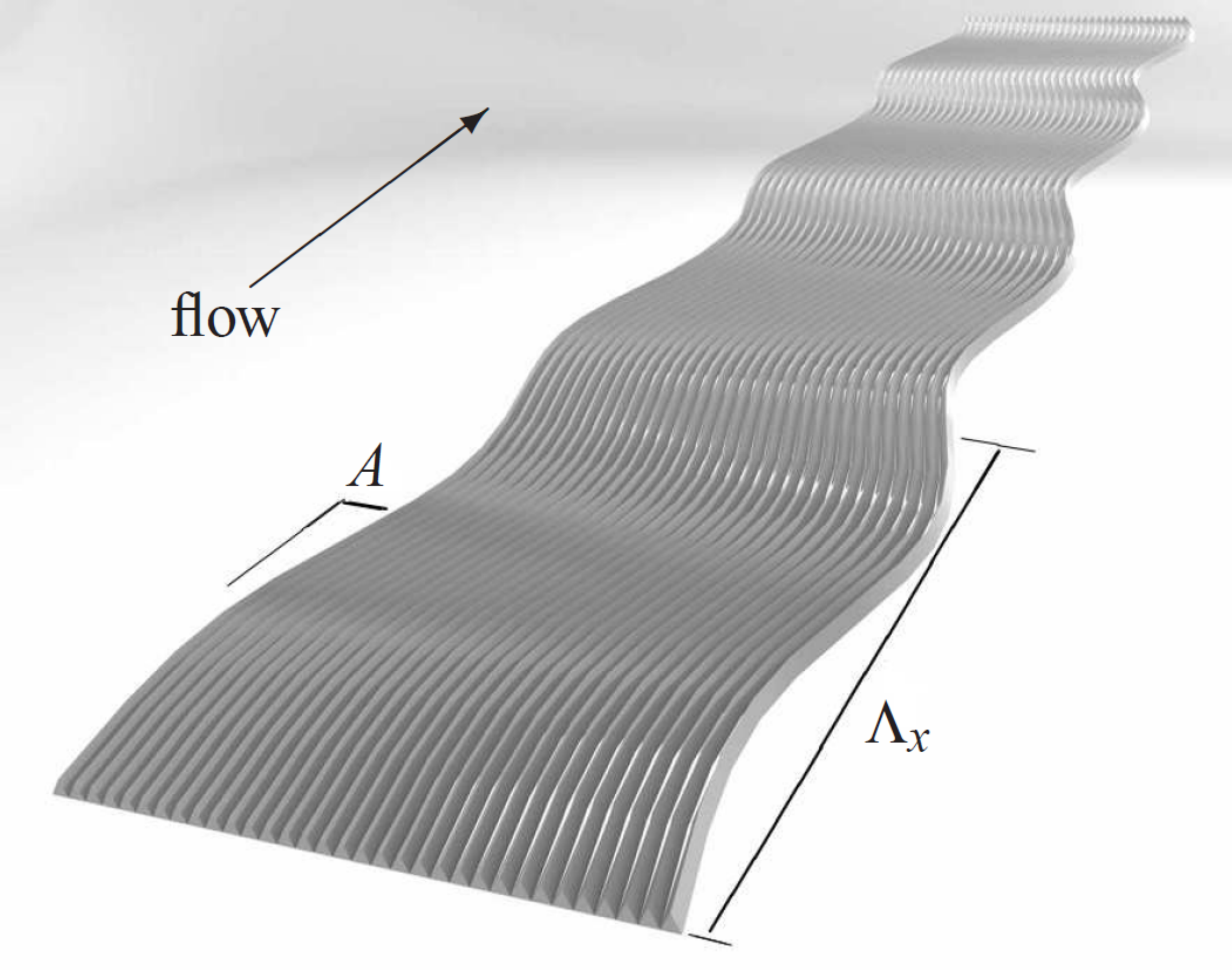}
\includegraphics[width=0.26\columnwidth]{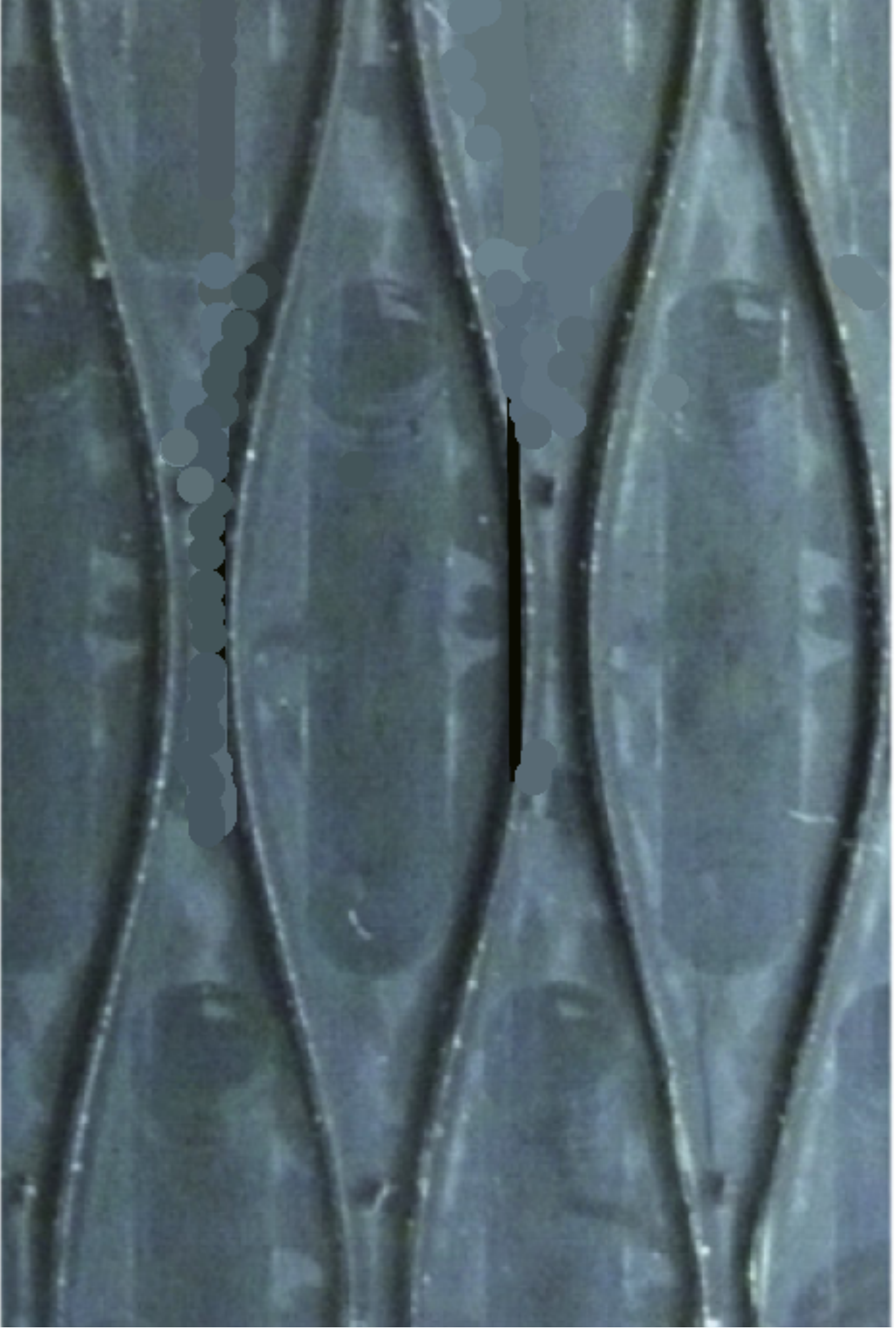}
\caption{Examples of sinusoidal riblets. Left: sinusoidal riblets with uniform spanwise spacing, studied by \citet{oh-etal-2012}; right: riblets with sinusoidally varying spanwise spacing, studied by \citet{mamori-etal-2019}. Taken with permission from the authors (left) and from Elsevier (right).}
\label{fig:riblets-sinusoidal}
\end{figure}

Riblets with a streamwise sinusoidal modulation and constant spanwise spacing, such as those shown in Fig. \ref{fig:riblets-sinusoidal} (left), were first studied numerically by \citet{peet-etal-2007,peet-etal-2008,peet-etal-2009}. They showed that the weaker wall-normal and spanwise turbulent fluctuations led to a maximum increase of the drag-reduction margin from 5.4\% in the straight case to 7.4\% in the sinusoidal case, i.e., an almost 40\% improvement, although the uncertainty of these computations was not reported.
Their result was contradicted by the DNS and experiments of \citet{kramer-etal-2010} and \citet{gruneberger-etal-2012} at low Reynolds numbers and by the measurements at a higher Reynolds number by \citet{oh-etal-2012}, who reported changes of drag reduction that fell within the experimental uncertainty range. Further results suggesting slight drag-reduction improvements were reported by \citet{okabayashi-2016,okabayashi-2017}.

Extensive experiments and numerical simulations by \citet{miki-etal-2011}, Sasamori, Mamori et al. \citep{sasamori-etal-2014,yamaguchi-etal-2014,iihama-etal-2014,sasamori-etal-2015,sasamori-etal-2017,mamori-etal-2019} showed that sinusoidal riblets with varying spanwise spacing, such as those depicted in Fig. \ref{fig:riblets-sinusoidal} (right), were more efficient than the riblets with constant spacing, yielding drag-reduction values up to $10-12\%$, thus larger than $8-9\%$ for straight riblets.
\\

\noindent
{\em Other spanwise-forcing surface protrusions}

The additional studies on riblet geometries that do not fit under previous headings are briefly mentioned here. \citet{lang-2010,lang-2012} proposed sharp spanwise staggered protrusions as possible vehicles for drag reduction, although these microcavities have not been tested either numerically or experimentally. \citet{okabayashi-etal-2018}'s ``Miura fold'' riblets have similarities with Lang's protrusions as the zig-zag modulation is along the two spatial dimensions and led to a maximum 3\% overall drag reduction. Both Lang's and Okabayashi's surfaces are thought to cause drag reduction by inducing alternate near-wall spanwise motions, although further tests are necessary to assess the drag-reduction performance of these surfaces. Converging-diverging protrusions were also proposed as a drag-reduction method in a Nature paper and in U.S. patents by L. Sirovich and co-workers \citep{sirovich-karlsson-1997,sirovich-etal-1998-a,sirovich-etal-1998-b}.
\cite{fuaad2019enhanced} used a streamwise sinusoidal microgroove structure that induced a stationary distribution of spanwise velocity that oscillated in the streamwise direction. Their DNS of a channel flow with superhydrophobic surfaces at $Re_\tau=180$ demonstrated improved drag reductions compared with the sinusoidal surface structure.

\subsection{Active dimples and ribbed surfaces}
\label{sec:active-geometry}

J. Morrison and co-workers at Imperial College London experimentally studied the streamwise vortical flow  induced by round dimples that oscillated in the wall-normal direction, although the drag-reduction potential of these actuators was not reported \citep{lambert-etal-2005,arthur-etal-2006,dearing-lambert-morrison-2007,dearing-etal-2010,morrison-2013}. \citet{sun-etal-2011} and \citet{zhang-etal-2015} instead computed drag-reduction margins by active dimples up to 15\% and 5\%, respectively, and confirmed that the spanwise-velocity shear can be effective in reducing the wall-shear stress.

Riblets have also been combined with active spanwise forcing, in the following scenarios.

\begin{itemize}

\item \cite{lenza-2001} conducted experiments in the water flume utilized by \citet{trujillo-bogard-ball-1997}, \citet{ricco-wu-2004-a} and \citet{ricco-2004}, showing that the drag-reduction margin generated by a spanwise-oscillating wall with streamwise-aligned riblets is a few percent larger than the value achieved with a smooth oscillating wall.

\item \cite{wassen-etal-2008,kramer-thiele-wassen-2009-germany} and \cite{gruneberger-etal-2013} investigated the drag-reducing properties of streamwise-aligned sharp-crested riblets that were mounted on a fixed wall at their base and oscillated in the spanwise direction. Contrary to the effect of wall oscillation, the measurements revealed that the oscillation period had no impact on the drag in the range $25<T^+<190$, where an improvement on the drag-reduction margin of 1.5\% with respect to the smooth-wall case was found through numerical calculations. The more accurate numerical calculations by \citet{kramer-thiele-wassen-2009-korea} reported a drag-reduction level of 11\% by decreasing the oscillating period to $T^+=35$.

\item \cite{vodopyanov-etal-2013} found that a spanwise-oscillating wall with fixed sharp-crested riblets led to a net power saving of 13.5\%, i.e., higher than a maximum 8\% in the case of a smooth oscillating wall. These results chimes with \cite{lenza-2001}'s observations, although the shape of the riblets was different. 
It is evident from the results of \cite{vodopyanov-etal-2013} and \cite{wassen-etal-2008} that the effects of the two methods, the riblets and the oscillating wall, are not additive, and that the interaction is detrimental to the drag-reduction performance. This finding is in line, qualitatively, with \citet{olivucci-ricco-aghdam-2019}'s results on the combination of spinning rings and wall-distributed actuations. \citet{olivucci-ricco-aghdam-2019}'s simple model on the combination of two drag-reduction techniques could be employed in the cases of \cite{vodopyanov-etal-2013} and \cite{wassen-etal-2008} to predict the reduction in drag.

\item Wind-tunnel experiments by \cite{li-etal-2015,li-etal-2017} and \cite{meysonnat-etal-2016b} showed that a ribbed surface with wall-normal deformations travelling in the spanwise direction generated drag-reduction margins up to 9.4\%, thus higher than the 4.7\% drag-reduction level obtained for the fixed ribbed surface.

\end{itemize}

\subsection{Compressibility and heat-transfer effects}

The first investigation that exploited temperature gradients to induce a drag-reduction effect was carried out by \citet{yoon-etal-2006} via DNS. The spanwise action was generated by the buoyancy forces induced in a turbulent channel flow by alternate heating and cooling wall strips. The maximum reported drag-reduction level was 35\%.

A compressible turbulent flow altered by spanwise wall oscillations was studied via LES by \citet{fang-etal-2009,fang-etal-2010} and \citet{fang-etal-2010-b}. Both the Reynolds number and the free-stream Mach number were low, i.e., $Re_\tau=150$ and $M_\infty=0.5$, respectively. The Reynolds analogy between the wall-shear stress and the heat transfer was confirmed as the velocity and the temperature fluctuations were both reduced by similar amounts. Although the Stokes layer introduced additional dissipation, the average wall heat flux was reduced by a careful choice of the oscillating parameters.

\citet{ni-etal-2016} extended Fang et al.'s work to the supersonic regime at $M_\infty=2.9$ and noted that the heat transfer could also increase significantly because of the additional aerodynamic dissipation of kinetic energy into heat due to the supersonic Stokes layer. They thus concluded that, in the supersonic case, the selection of the oscillation parameters requires more attention than in the incompressible case. 

\cite{yao-hussain-2019} used DNS to study the effect of Mach number on the drag reduction in turbulent channel flows with spanwise wall oscillations at a fixed bulk Reynolds number $Re_b=3000$ and Mach numbers $M_b=U_b/c_w=0.3, 0.8, 1.5$, where $c_w$ is the speed of sound at the wall temperature. The drag-reduction margin grew with $M_b$ and relaminarization occurred, which was never observed for uniform flat-plate wall oscillations in the incompressible regime. They also noted that, contrarily to the incompressible case, the drag-reduction level increases monotonically with the oscillation period for $0<T^+<300$ and no optimal period exists in this range.

We close this section by mentioning the unconventional study of \cite{dong-etal-2015}, who used the molecular dynamics Lennard-Jones method to simulate a Couette flow of liquid argon forced by spanwise wall oscillations. It was found that the wall oscillations decreased the density of argon near the wall, which apparently caused a lower fluid viscosity and therefore a lower friction force at the wall. 

\section{Future research directions}
\label{sec:future}

This final section presents ideas about possible future directions for research in the area of turbulent friction reduction, mainly, but not exclusively, by actuation along the spanwise direction. It is important to acknowledge that, despite more than thirty years of intense research in this area of flow control, the optimistic results of large drag-reduction margins, even exceeding 50\%, pertain to idealized scenarios. The future application of these methods to actual engineering systems is thus undoubtedly bound to be limited to lower levels of drag reduction, due to numerous factors, such as the impacts of the alteration of skin-friction drag on the other drag forms, mainly pressure drag and wave drag, of transonic flow, and of three-dimensionality and surface curvature \citep{albers-schroder-2021,albers-etal-2019,atzori-etal-2020}.

In Section \ref{sec:future-experiments}, a few examples of current issues that could be targeted by experimental campaigns are listed. In Section \ref{sec:future-physics}, we discuss research avenues to be pursued in order to improve our understanding of the modification of the flow physics and of the mechanism behind the reduction of wall-shear stress. In Section \ref{sec:future-technology}, we explore relevant issues that must be addressed in the light of technological applications and examine the two spanwise-actuation techniques that, in our view, are the most promising ones if a step change towards industrial implementation is to be achieved.

\subsection{Experimental studies}
\label{sec:future-experiments}

Future experiments should preferably focus on key results that have only been obtained by simulations, which would therefore benefit from experimental verification. Particular scenarios of interest include: i) the Reynolds-number dependence of the drag-reduction margin in parts of the actuation parameter space (especially in the free-stream boundary-layer case), ii) the drag increase first reported by \cite{quadrio-ricco-viotti-2009}, but never measured in a laboratory, iii) the measurements of the power budget for an improved understanding of the energy-saving properties of spanwise-forcing techniques, and iv) the effect of external mean streamwise pressure gradients on the drag-reduction performance with a view to long-term technological applications. A step change is necessary towards efforts to create and experimentally test novel drag-reduction methods that are inspired by the existing spanwise-forcing knowledge, but with a clearer vision towards technological applicability. 

\subsection{Physics of drag-reduced flows}
\label{sec:future-physics}

A central, extremely challenging, aim is to obtain a mathematical relation linking the wall-shear stress to the wall-oscillation parameters, possibly by manipulation of the Navier-Stokes equations, or some other transport equations derived from the Navier-Stokes equations, along the lines of the FIK identity. It would be equally valuable to relate the Reynolds stresses to the forcing parameters because the FIK identity in the wall-oscillating case has the same form as in the fixed-wall case. Such relation would offer unique insight into the physical mechanism leading to the drag reduction and would also serve the purpose of predicting the drag-reduction margin. 

A clear and unambiguous explanation of the drag-reduction mechanism from the perspective of turbulence structures and their interactions would be welcomed. A complete description of the energy transfer between the Reynolds-stress budget terms and the final outcome of causality behind the decreased turbulence kinetic energy would lay the pathway for a turbulence modelling approach targeted at investigations of flows in more complex geometries and higher Reynolds numbers with technologically realizable actuators as the ultimate goals.

More theoretical analysis into the physical modifications engendered by the streamwise-traveling waves given by equation
\eqref{eq:waves} is certainly needed, as less attention has been devoted to the travelling-waves case than to turbulent flows altered by streamwise-uniform wall motion. Two main unresolved issues are to explain why standing waves lead to higher drag reduction than purely temporal forcing and to unravel the intriguing physics behind the drag increase that occurs when the phase speed of the travelling waves is comparable with the convection velocity of the near-wall turbulent structures. A curious issue to assess is the difference between two flows that have the same drag, i.e., the canonical turbulent channel flow and the channel flow modified by the spanwise-wall-velocity waves travelling forward at the phase speeds that bound the drag-increase region in Fig. \ref{fig:QRVmap}.

The cardinal point of the effect of Reynolds number still needs to be fully understood. When considering applications of spanwise forcing for drag reduction in real flows, the trends of the drag-reduction level for channel flows as a function of the Reynolds number, such those depicted in Fig. \ref{Fig:6}, are encouraging. Before reliable experiments or accurate DNS computations at large Reynolds numbers become available, more analysis is needed to guarantee the reliability of the extrapolation of low-Reynolds-number results to high-Reynolds-number cases. The recent investigation of \cite{yao-etal-2019} has shed doubt on the validity of the underlying assumptions leading to the results of Fig. \ref{Fig:6}, showing a non-negligible discrepancy between the DNS value and the predicted value at $Re_\tau=2000$.
The impact of the outer large structures on the Reynolds-number effect must be accurately quantified as they become increasingly important as the Reynolds number grows, as discussed in Section \ref{sec:re-physics}. The modelling approach leading to the encouraging results in Fig. \ref{Fig:6} is based on the assumption that these outer structures influence the flow implicitly, through the fixed-wall relation between the Reynolds number and the skin-friction coefficient. 
Another crucial point is to quantify how the geometry of a free-stream boundary layer affects the drag reduction at high Reynolds number, compared with the channel-flow geometry. The required changes in the physical-oscillation period with the Reynolds number is accentuated by the shift of the optimal $T^+$ and $\lambda_x^+$ to smaller values for increasing Reynolds numbers, while the opposite effect occurs for spatially developing boundary layers, for which the requirements on the physical-oscillation parameters are alleviated by the shift. For the latter flow, it would be of importance for cost-benefit estimations to determine whether the downstream decline in the drag reduction is due to spatial transients or to an intrinsic effect of the boundary-layer flow.

\subsection{Vision towards technological applications}
\label{sec:future-technology}

Another central problem for future technological implementations is related to the scaling of the optimal parameters in inner viscous units as the Reynolds number grows, irrespectively on the influence of this parameter on the drag-reduction level. 
If this scaling applies, the characteristic spatial dimensions become smaller and the temporal scales shorter, which means that the physical maximum spanwise displacement and the physical oscillation period both decrease as the Reynolds number grows. This change occurs because, as a larger streamwise pressure gradient in a channel flow is imposed or the free-stream velocity of a boundary layer grows, the wall-friction velocity also increases.
Therefore, keeping $W_m^+$, $T^+$, or $\lambda_x^+$ constant around their optimal values, while neglecting any shift in these parameters as the Reynolds number changes, requires $W_m$ to be larger, and $T$ or $\lambda_x$ to be smaller. \cite{ricco-hahn-2013} amply discuss this issue for the rotating-disc technique, concluding that at high Reynolds numbers proper of commercial flight conditions, specifically for flows over aircraft wings, the optimal ring diameter would be of the order of 1mm and the typical period of rotation would be of the order of 0.1ms. These scales are, respectively, extremely small and fast, not only for the admittedly unrealistic application of rotating patches, but for any drag-reduction actuators that operates effectively with similar space and time scales. One viable option to obviate this problem would be to devise drag-reduction methods that instead scale in outer units in order to design actuators that are technologically realizable. Spanwise forcing combined with streamwise actuation, such as the spinning discs, could be a route to pursue further as \cite{wise-alvarenga-ricco-2014} showed that nearly optimal drag reduction conditions are guaranteed even though diameters as large as three times the optimal one are enforced, suggesting that these actuators operate on the largest scales of the turbulence.

Although the decline of the drag-reduction margin with the Reynolds number is an important topic, there are other fundamental questions to be answered before any practical implementations can be contemplated. Future research endeavours must also certainly be directed at flows in geometries that are more realistic -- and thus more complicated -- than the widely-studied canonical flows. The importance of this point has been accentuated by the few studies that have demonstrated the sensitivity of the drag reduction to the flow geometry, such as the duct-flow experimental campaign \citet{straub-etal-2017}.
The effects of wall curvature, variable streamwise and spanwise pressure gradients, compressibility in external flows, and the influence of spanwise forcing on form drag all represent fertile ground for exciting and useful research. Spanwise forcing on the shock/boundary-layer interaction has also never been investigated. For external flows, the drop of the drag-reduction margin in the streamwise direction is of concern, as are the sensitivity of the altered boundary layer to separation due to the reduced wall-shear stress and the impact of the wall-forcing on other aerodynamic performance indicators, such as lift. 

It is also imperative to think about how spanwise forcing can be implemented in flows of industrial interest. The experimental results by \cite{thomas-etal-2019} of a wind-tunnel turbulent boundary layer excited by spanwise plasma forcing will undoubtedly generate interest, both because of the remarkable maximum $\mathcal{R}=75\%$ and the small power input that led to a positive net energy balance. As the current was applied in almost instantaneous impulses, the key behind such a low energetic expenditure, reliable direct numerical simulations of this flow will be challenging because of the required temporal discretization. It must be however recalled that the non-sinusoidal wall motion studied via DNS by \cite{cimarelli-etal-2013}, where the forcing was only activated during short-time intervals and thus somewhat resembled \cite{thomas-etal-2019}'s spiky plasma actuation, leads to considerable lower drag-reduction margins than the more power-consuming sinusoidal motion. The type of actuation, volume forcing versus wall motion, could be the reason behind this apparent discrepancy. 
It must also be remarked that the use of plasma generates intense small-scale pressure and heat waves, which could raise serious safety issues when such technology is employed on an aircraft because of the proximity of the inflammable fuel.
Research on dimples and sinusoidal riblets should also be continued, primarily because these are passive techniques and therefore very attractive for technological purposes. The parameter spectrum is very wide and largely unexplored, while the effect of Reynolds number on these methods is unknown.

\section*{Acknowledgements}
\label{sec:ack}

Part of this research was carried out by PR and MAL thanks to the support of the H2020 EU-China project ``DRAGY: Drag Reduction in Turbulent Boundary Layer via Flow Control'' (grant agreement ID: 690623) on turbulent drag reduction.

\bibliographystyle{plainnat}
\bibliography{pr}

\end{document}